\documentclass[a4paper,12pt,leqno,english,titlepage]{book}
\usepackage{cite}

\usepackage[latin1]{inputenc}
\usepackage[T1]{fontenc}
\usepackage[german,danish,english]{babel}

\usepackage{amsmath}
\usepackage{amsfonts}
\usepackage{dsfont}
\usepackage{longtable}

\usepackage{ae,aecompl}
\usepackage{amssymb}
\usepackage{textcomp}
\usepackage{bm}
\usepackage{multicol}
\usepackage{algorithmic}
\usepackage{algorithm}
\usepackage{ifthen}
\usepackage{layout}

\usepackage{cite}
\usepackage{fancyhdr}
\usepackage{amssymb}
\usepackage{epic}
\usepackage{epsfig}

\usepackage{multirow}
\usepackage{hhline}
\usepackage{calc}
\usepackage{slashed}
\usepackage[bf,footnotesize]{caption}
\usepackage{babel}
\usepackage{ae,aecompl}
\usepackage{textcomp}
\usepackage{latexsym}
\usepackage{amsfonts}

\usepackage{amsmath,amsthm}
\allowdisplaybreaks[4]
\usepackage{graphicx}
\usepackage{array}
\usepackage{blkarray}
\usepackage{minitoc}
\usepackage{amsthm} 

\theoremstyle{definition}

\theoremstyle{plain}

\theoremstyle{remark}

\renewcommand{\chaptermark}[1]%
{\markboth{\chaptername\ \thechapter\ #1}{}}
\renewcommand{\sectionmark}[1]%
{\markright{\thesection\ #1}}

\newcommand{\tfkt}[3]{\vartheta\bigg[ {\begin{array}{*{20}c}#1  \\#2  \\\end{array}}\bigg]\left( #3 \right)}
\newcommand{\tfkto}[2]{\vartheta\bigg[ {\begin{array}{*{20}c}#1  \\#2  \\\end{array}}\bigg]}
\newcommand{\tfktvier}[3]{\vartheta^4\bigg[ {\begin{array}{*{20}c}#1  \\#2  \\\end{array}}\bigg]\left( #3 \right)}
\title{Aspects of Stability and Phenomenology in Type IIA Orientifolds with Intersecting D6-branes}
\author{Tassilo Ott\thanks{e-mail: ott@physik.hu-berlin.de}
\\ \\
Humboldt--Universit\"at zu Berlin, Institut f\"ur Physik\\
Newtonstra\ss e 15, D-12489 Berlin, Germany}
\date{}

\begin{document}
\frontmatter
\maketitle
\selectlanguage{english}
\chapter*{Abstract}
Intersecting branes have been the subject of an elaborate string
model building for several years. After a general introduction
into string theory, this work introduces in detail the toroidal
and $\mathbb{Z}_N$-orientifolds. The picture involving D9-branes
with B-fluxes is shortly reviewed, but the main discussion employs
the T-dual picture of intersecting D6-branes. The derivation of
the R-R and NS-NS tadpole cancellation conditions in the conformal
field theory is shown in great detail. Various aspects of the open
and closed chiral and non-chiral massless spectrum are discussed,
involving spacetime anomalies and the generalized Green-Schwarz
mechanism. An introduction into possible gauge breaking mechanisms
is given, too. Afterwards, both $\mathcal{N}$=1 supersymmetric and
non-supersymmetric approaches to low energy model building are
treated. Firstly, the problem of complex structure instabilities
in toroidal $\Omega R$-orientifolds is approached by a
$\mathbb{Z}_3$-orbifolded model. In particular, a stable
non-supersymmetric standard-like model with three fermion
generations is discussed. This model features the standard model
gauge groups at the same time as having a massless hypercharge,
but possessing an additional global $B$-$L$ symmetry. The
electroweak Higgs mechanism and the Yukawa couplings are not
realized in the usual way. It is shown that this model descends
naturally from a flipped $SU(5)$ GUT model, where the string scale
has to be at least of the order of the GUT scale. Secondly,
supersymmetric models on the $\mathbb{Z}_4$-orbifold are
discussed, involving exceptional 3-cycles and the explicit
construction of fractional D-branes. A three generation Pati-Salam
model is constructed as a particular example, where several brane
recombination mechanisms are used, yielding non-flat and
non-factorizable branes. This model even can be broken down to a
MSSM-like model with a massless hypercharge. Finally, the
possibility that unstable closed and open string moduli could have
played the role of the inflaton in the evolution of the universe
is being explored. In the closed string sector, the important
slow-rolling requirement can only be fulfilled for very specific
cases, where some moduli are frozen and a special choice of
coordinates is taken. In the open string sector, inflation does
not seem to be possible at all.

\pagestyle{fancy}
\fancyhead{}
\fancyhead[LE,RO]{\bfseries\thepage}
\fancyhead[LO]{\bfseries Contents}
\fancyhead[RE]{\bfseries Contents}
\tableofcontents
\cleardoublepage
\fancyhead{}
\fancyhead[LE,RO]{\bfseries\thepage}
\fancyhead[LO]{\bfseries List of Figures}
\fancyhead[RE]{\bfseries List of Figures}
\listoffigures
\cleardoublepage
\fancyhead{}
\fancyhead[LE,RO]{\bfseries\thepage}
\fancyhead[LO]{\bfseries List of Tables}
\fancyhead[RE]{\bfseries List of Tables}
\listoftables
\cleardoublepage

\lhead[\fancyplain{}{\bfseries\thepage}]%
{\fancyplain{}{\bfseries\rightmark}}
\rhead[\fancyplain{}{\bfseries\leftmark}]%
{\fancyplain{}{\bfseries\thepage}}
\chead{}
\lfoot{}
\cfoot{}
\rfoot{}
\renewcommand{\sectionmark}[1]{\markright{\thesection\ #1}}
\renewcommand{\chaptermark}[1]%
{\markboth{\chaptername\ \thechapter\ #1}{}}
\pagenumbering{arabic}
\mainmatter
\chapter{Introduction}\label{cha:Intro}
At the turn of the new century, the physical community suffers
from a similar crisis in spirit as it already did 100 years ago.
At that time, an older professor advised Planck against studying
physics, because the foundation of physics would be complete.
There would be not much to discover anymore as all observations
would have been explained already \cite{Planck:2001}.
Nevertheless, it was a great fortune that Planck still decided to
study physics. Indeed, some years after this unedifying statement,
the most exciting developments in physics so far, general
relativity and quantum mechanics, have taken place. Today, the
situation is quite similar: there are two phenomenological models
that seem to describe all empirical observations.

On the first hand, there is the standard model of particle
physics, which describes the microscopic structure of our world
very well, i.e. the observations that are done in particle
colliders up to the current limit of approximately 200 GeV. This
model is based on quantum field theory with gauge groups $SU(3)
\times SU(2) \times U(1)$. It has been discovered directly from
experiment and in its complex structure not just from fundamental
principles. There are 18 or more free parameters (depending on the
way of counting) \cite{Nachtmann:1990ta} in this model, just to
mention the masses of the fermions and bosons, the coupling
constants of the interactions and the coefficients of the
CKM-matrix. These parameters have to be measured, they cannot be
determined within the model. But there are even more open
questions: for instance, why are there exactly three families of
fermions? What is the reason for CP-violation?

On the other hand, there is the standard model of cosmology which
describes the macroscopic structure of the universe today
successfully, the galaxy formations and the global evolution of
the universe by the Hubble parameter. It is built on general
relativity combined with simple Hydrodynamics. But this model has
its problems, too. There are the Horizon and Flatness problems
and the small value of the Cosmological Constant, the last two
problems requiring an incredible finetuning of the parameters
within the model. These problems in the past have been addressed
by theories like inflation or quintessence which slightly alter
the phenomenological model but do not touch the underlying theory
of general relativity. But there are even more fundamental
shortcomings: one cannot determine the values for the Hubble
parameter from the model itself, it again is an input parameter.
Furthermore, if we interpolate the evolution of the universe back
in time, we reach a point at which the thermal energy of typical
particles is such that their de Broglie wavelength is equal or
smaller than their Schwarzschild radius. This energy is the Planck
mass $M_P=1.22 × 10^{19}$ GeV. It means nothing else than the
breakdown of general relativity at least at this scale, because
it relies on a smooth spacetime which would be destroyed by
quantum black holes.

This fact can be seen as a hint that neither quantum mechanics
(quantum field theory) nor general relativity on their own can
describe what has happened at the beginning of our universe. From
the philosophical point of view this maybe is the deepest question
physics might ever be able to answer. This inability within the
physical community motivated the idea that all physics might be
described by just one fundamental theory that unifies general
relativity and quantum field theory as its effective low energy
approximations.
\section{A unification of all fundamental forces}
Albert Einstein has started the program of searching for a unified
field theory more than 60 years ago \cite{Einstein:1938fk}.
Learning from Maxwells ideas that the electric and magnetic forces
are just two different appearances of one unified force, he
concluded that this might be also true for all other fundamental
forces. At his time, only the gravitational force was known in
addition to the electro-magnetic one. He extended the idea of
Kaluza \cite{Kaluza:1921tu} from 1921 and Klein
\cite{Klein:1926tv} from 1926 that within a 5-dimensional
classical field theory with one compact direction, gravity could
be understood as given by the 4-dimensional part of the metric
tensor $g_{\mu \nu}$ with $\mu,\nu=0..3$ and the compact subspace
contributing the massless photon as $g_{\mu 4}$. This theory later
was discarded, mainly because it predicted a new and unseen
massless particle, given by $g_{44}$. At the latest, when the weak
and strong forces were discovered, it has become apparent that this
imaginative idea has failed in its original formulation.

Unification can generally be understood in two different ways that
have to be distinguished carefully. Firstly, one could mean a
description of nature within the same theoretical framework. This
has indeed been achieved for the three forces excluding gravity by
the standard model of particle physics within the framework of
quantum field theory. The story is different considering
gravitation which is not quantizable, i.e. renormalizable, within
four-dimensional quantum field theory.

Secondly, by the term unification in a strong sense one could mean
that above a certain energy scale, the different forces dissolve
into just one fundamental one. For the electro-magnetic and weak
forces this was first achieved in the Salam-Weinberg model which
already is included within the standard model. It predicts an
electroweak phase transition which should have occurred at an
energy of approximately 300 GeV \cite{Peacock:1999} and has helped
to understand how our present matter has formed during the
cosmological evolution. But a direct evidence for the Higgs
particle, triggering this phase transition within the standard
model, still is missing. For the three forces without gravity,
unification in this sense is achieved in grand unification models.
Gauge coupling unification happens at a high energy, the so-called
GUT-scale of around $10^{16}$ GeV in typical models. The three
standard model gauge groups are getting replaced by one larger
simple group. The initial non-supersymmetric $SU(5)$ model has
been ruled out. This is due to the predicted proton decay that
does not happen, as Super Kamiokande has observed up to a limit
$\tau=4.4×10^{33}$ years \cite{Ganezer:1999fx,Ganezer:2001qk}. But
there are other models like $SO(10)$ with a larger gauge group
that still might give the right description of a
electroweak-strong unification, although there are many open
questions regarding the Higgs sector or the weak mixing angle that
cannot be answered correctly by these models so far.

Another important idea related to unification has entered particle
physics within the last thirty years: supersymmetry. It assumes a
fundamental symmetry between fermions and bosons, one can be
transferred into the other by an operator $Q$ that is called
supercharge. This idea has its origin in string theory but was
transferred by Wess and Zumino even to 4-dimensional field theory
\cite{Wess:1974tw, Wess:1974kz}. Supersymmetry predicts a
superpartner for every particle. But such superpartners of the
standard model particles have not yet been observed in accelerator
experiments. This means that at least below 200 GeV,
supersymmetry has to be broken, leading to a mass split between
the bosonic and fermionic partners that roughly is of the order of
the SUSY breaking scale. The additional light particles above this
scale in a specific and phenomenologically most favored
supersymmetric model, called the MSSM (Minimal Supersymmetric
Standard Model), imply (in analogy to GUT models) a unification
of the three couplings at a scale that might be around $10^{16}$
GeV, see for instance \cite{Ibanez:1984bd}. Therefore, one of the
main challenges of the LHC (Large Hadron Collider), which is being
built at CERN right now and is going to achieve center of mass
energies of up to 14 TeV, will be the search for supersymmetry. It
is even possible to build supersymmetric algebras with more than
one superpartner for every particle, this is generally called
extended supersymmetry. But it is phenologically disfavored
because it does not allow chiral gauge couplings like in the
standard model.

Due to its non-renormizability in four dimensions, the unification
with gravity in both senses still seems to be a much more
difficult problem. Indeed, there is just one prominent candidate
for a unifying theory: string theory. It unifies gravity with the
other forces within the same theoretical framework, not as one
unified force as in the second meaning of the word.

\section{String theory}
\begin{figure}
\centering
\includegraphics[width=9cm,height=4.5cm]{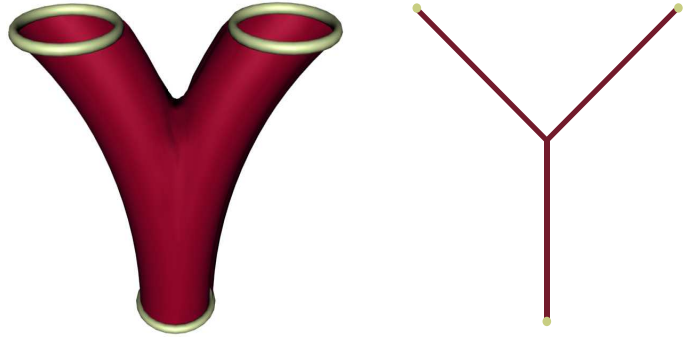}
\caption{Splitting Interaction of a string compared to the
corresponding field theory vertex, a coupling $g_{\mathrm{closed}}$
is assigned to this process.}\label{fig:splitting_int}
\end{figure}

String theory
\cite{Green:1987sp,Green:1987mn,Polchinski:1998rq,Polchinski:1998rr,Lust:1989tj}
manages to undergo the strong divergences of graviton scattering
amplitudes in field theory by replacing the concept of point
particles by strings. These strings are mathematical
one-dimensional curves that spread out a two-dimensional
worldsheet $\Sigma$ (which usually is parameterized by the two
variables $\sigma$ and $\tau$) when propagating in a higher
dimensional spacetime. The string has a characteristic length
scale of $\sqrt{\alpha'}$, where $\alpha'$ is the Regge slope
which is generally believed to be the only fundamental constant of
string theory. With this new concept, interaction does not take
place at a single point, but is smeared out into a region and is
already encoded in the topology of the worldsheet. In order to
include interaction, a first quantization is sufficient, which can
be performed quite easily. The difference between interaction
vertices in field theory and string theory is schematically shown
in figure \ref{fig:splitting_int} for a point particle (or a
closed string) that splits into two point particles (or closed
strings).
The characteristic energy scale of the string (the string scale)
is given by $1/\alpha'$. It constitutes the energy at which pure
stringy effects should be visible. Surely, this energy must be
very high, and for a long time, it was believed to be of the order
of the Planck scale \cite{Polchinski:1998rq}, but in more recent
scenarios sometimes even energies of 1 TeV are favored
\cite{Antoniadis:1999fj}. At energies much below the string scale
(corresponding to the limit $\alpha'\rightarrow 0$), the string
diagrams (like the left one of figure \ref{fig:splitting_int})
reduce to the usual field theoretic ones (the right one of the
same figure).
\subsection{The bosonic string}
On the worldsheet of the bosonic string, there exists a conformal
field theory. It is described by the bosonic fields
$X^{\mu}(\tau,\sigma)$ with $\mu=0,...,D-1$ and has the action of
a non-linear sigma model
\begin{equation}\label{eq:Polyakov}
\emph{S}_{\mathrm{Polyakov}}=\frac{1}{4 \pi \alpha'}\int_{\Sigma} d\tau d\sigma
\sqrt{-\gamma} \gamma^{a b}\partial_a X^\mu \partial_b X^\nu g_{\mu \nu}\ ,
\end{equation}
which is called the Polyakov action. Every one of the $D$ massless
scalar bosonic fields $X^{\mu}$ has the interpretation of an
embedding in a spacetime dimension, in analogy to the
point particle which moves in a curved space (with the usual
metric tensor $g_{\mu \nu})$. However $\gamma_{a b}$ is the
worldsheet metric (where a,b=0,1) that is introduced as an
additional field similarly to the tetrad of general relativity.
$\gamma_{a b}$ can be eliminated from the action using its
algebraic and therefore non-dynamical equation of motion. On the
classical level, the Polyakov action has the following three
symmetries:
\begin{enumerate}
    \item{$D$-dimensional Poincar\'{e} invariance}
    \item{2-dimensional Diffeomorphism invariance}
    \item{2-dimensional Weyl invariance}
\end{enumerate}
The Poincar\'{e} invariance is similar to that of usual special
relativity, in this case extended to all $D$ spacetime dimensions.
Diffeomorphism invariance is expected from the tetrad formalism of
general relativity. The Weyl invariance can be understood as a
local rescaling invariance of the worldsheet. It is crucial for
the fact that the 2-dimensional field theory on the worldsheet is
conformal.

Strings generally can be open or closed, corresponding to
different boundary conditions. In particular, for closed strings
the boundary conditions for the $X^\mu$ are periodic and so the
string forms a closed loop of length $l$:
\begin{equation}
X^\mu(\tau,l)=X^\mu(\tau,0)\ , \qquad \partial^\sigma
X^\mu(\tau,l)=\partial^\sigma X^\mu(\tau,0)\ , \qquad \gamma_{a
b}(\tau,l)=\gamma_{a b}(\tau,0)\ .
\end{equation}
By way of contrast, for open stings the endpoints of the string
are not being identified. Consequently, there are boundaries in
the conformal field theory:
\begin{equation}
\partial^\sigma X^\mu(\tau,0)=\partial^\sigma X^\mu(\tau,l)=0\ .
\end{equation}
These are Neumann boundary conditions and they are the only
possibility if we insist on $D$-dimensional Poincar\'{e}
invariance. Elsewhere an unwanted surface term would be
introduced in the variation of the action.

By using the simplest method of quantization for the theory, the
light cone quantization, one spatial degree of freedom and the
time are getting eliminated in the gauge fixed theory. This is
because the string is extended in one spatial direction. As a
consequence, not all spacial dimensions in spacetime can be
independent, just the transversal ones can oscillate.

In the process of quantization, another restriction arises by the
demand of a vanishing Weyl quantum anomaly: the total number of
dimensions must be $D=26$, the so-called critical dimension. If
one tries to define a meaningful quantum theory using strings,
this is the most severe break with usual quantum field
theory.\footnote{There is a close relation between a non-vanishing Weyl
anomaly on the worldsheet and a loss of Lorentz invariance in
spacetime which surely is unacceptable, see for instance
\cite{Polchinski:1998rq}.}. Hence one has to think about the
question, why the world that we observe is at least effectively
4-dimensional. We will soon return to this question.

The transversal oscillation modes on the string describe particles
in the usual sense. We want to describe them now in some more
detail:

One obtains a tachyon in both the closed and open string at the
lowest mass level, a particle with negative mass-squared. This
indicates in field theories that the vacuum, around which one
perturbs, is unstable. The same conclusion has been drawn for the
bosonic string: it is not a viable theory as it stands.

At the next mass level, one gets the zero-modes of $X^{\mu}$ which
correspond to massless bosonic fields in spacetime. In the closed
string, there are $(D-2)^2$ massless states forming a traceless
symmetric tensor, an antisymmetric tensor and a scalar. The
traceless symmetric tensor has been interpreted as the graviton by
Scherk and Schwarz in 1974 \cite{Scherk:1974mc, Scherk:1974zm}.
This lead to the big boom of string theory, because it was the
first quantum theory that seemed to naturally incorporate a
massless spin-2 particle. In the open string, one obtains a (D-2)
massless vector particle.

Furthermore, there is an infinite tower of massive states that are
organized in units of the string scale $\alpha'^{-1/2}$. Even the
lowest one of these modes are so heavy that they do not play any
important role if the string scale is really of the order of the
Planck scale. Therefore, in the rest of this work and generally within
phenomenological models, mainly the zero-modes are being
regarded as they highly dominate at energies much below the string
scale.

In analogy to quantum field theory, one would like to define a
path integral for string theory in order to describe interaction.
This then would allow to calculate scattering amplitudes for
certain incoming and outgoing string configurations. So far, we
have just spoken about one worldsheet with a certain metric
$\gamma_{a b}$ and a certain topology. To build up the path
integral, one would have to sum over all possible histories that
interpolate between the initial and final state. To do so, one
first has to classify all the different possible topologies of
2-dimensional Riemann surfaces. This can be done by determining
the genus $g$, corresponding to the number of handles and the
number of boundaries $b$, corresponding to holes within the
surface, and finally the number of crosscaps $c$, corresponding to
insertions of projective planes. From these three numbers, the
Euler number can be calculated by the simple equation
\begin{equation}
\chi=2-2g-b-c\ .
\end{equation}
In the simplest case of pure closed string theory, there are
neither boundaries nor crosscaps, so the perturbative expansion can
be directly understood just by the number of handles.
\begin{figure}
\centering
\includegraphics[width=13cm,height=2cm]{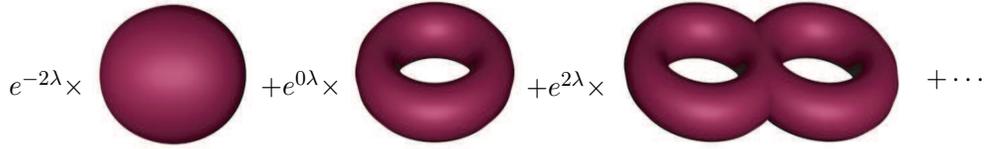}
\caption{Perturbative expansion for the closed string partition
function without insertion.}\label{fig:pert_exp}
\end{figure}
One assigns a coupling $g_{\mathrm{closed}}=e^{\lambda}$ to the
diagram that couples three closed strings (the left figure of
\ref{fig:splitting_int}) and then builds up the torus and
topologies with more handles by joining two or more of these.
Every diagram is then weighed with a factor $e^{-\lambda \chi}$ as
can be seen in figure \ref{fig:pert_exp}. This procedure also
works for topologies with boundaries or unoriented worldsheets.
The Polyakov path integral partition function schematically can be
defined in the following way:
\begin{equation}\label{eq:bos_partition_function}
\mathcal{Z}=\sum\limits_{\textrm{all compact topologies}}\int\frac{[dX\
d\gamma]}{V_{\textrm{diff}\times\textrm{Weyl}}}e^{-\mathcal{S(X,\gamma)}-\lambda\chi}\ ,
\end{equation}
$\mathcal{S(X,\gamma)}$ is the Polyakov action (\ref{eq:Polyakov})
and $V_{\textrm{diff}\times\textrm{Weyl}}$ stands for the volume
of the string worldsheet symmetry groups that carefully have to be
divided out. It is now possible to add asymptotic string states
(for instance for calculating the scattering amplitude between one
ingoing and two outgoing external closed string states, like in
figure \ref{fig:splitting_int}). This is done by adding so-called
vertex operators at a certain worldsheet position, one for each
external open or closed string state. One carefully has to fix the
gauge on every worldsheet topology, because the vertex operators
break some of the manifest symmetries. Just to mention, many
involved tools from conformal field theory, like operator product
expansion, are needed to perform these calculations.
\subsection{The superstring}
Besides the existence of the Tachyon, there is another problem:
the theory does not contain any spacetime fermions so far. This
can be cured if one adds fermionic degrees of freedom
$\psi^{\mu}(\tau,\sigma)$ and enlarges the symmetry algebra by
supersymmetry on the worldsheet at the same time. Instead of
(\ref{eq:Polyakov}), one starts with the Ramond-Neveu-Schwarz
action in superconformal gauge:
\begin{equation}\label{eq:RNSaction}
\emph{S}_{\mathrm{RNS}}=\frac{1}{4 \pi \alpha'}\int_{\Sigma} d\tau d\sigma
\left(\partial_\alpha X^\mu \partial^\alpha
X^\nu g_{\mu \nu}+i \bar{\psi}^\mu \rho^\alpha \partial_\alpha \psi_\mu\right)\ .
\end{equation}
The fields $\psi^\mu$ are Majorana spinors on the worldsheet, but
vectors in spacetime; $\rho^\alpha$ are the two-dimensional spin
matrices where the worldsheet spinor indices are suppressed. The
demand for a vanishing Weyl anomaly leads to a different
restriction on the total spacetime dimension as in the bosonic
string, and this is $D=10$. The $\psi^\mu$ of the open string can
have two different periodicities:
\begin{eqnarray}
\mathrm{Ramond (R)}&:\psi^\mu(\tau,\sigma+l)=+\psi^\mu(\tau,\sigma) \\
\mathrm{Neveu-Schwarz (NS)}&:\psi^\mu(\tau,\sigma+l)=-\psi^\mu(\tau,\sigma)\nonumber
\end{eqnarray}
The sign must be similar for all $\mu$. If one quantizes the
superstring in the same way as the bosonic string, one realizes
that the closed string always has two independent left and right
moving oscillation degrees of freedom, whereas the open string has
just one independent one. For the closed string, it is possible to
choose between Ramond and Neveu-Schwarz initial conditions
independently for the left and right moving spinors $\psi^\mu$ and
$\tilde \psi^\mu$. By doing so, one gets 4 different theories for
the closed string (NS-NS, R-R, NS-R and R-NS) and 2 different ones
for the open string (NS, R). The theories of NS-NS, R-R and NS
yield spacetime bosons, whereas NS-R, R-NS and R account for the
spacetime fermions. As explained further down, none of these
theories on their own are viable quantum theories, this is mainly
due to the demand of modular invariance of the one-loop amplitude
and possible non-vanishing tadpoles. Furthermore, some sectors
contain a tachyon as the ground state. Gliozzi, Scherk and Olive
have shown that it is possible to construct modular invariant,
tachyon-free theories from all these sectors. This is done by a
certain projection, which today is called GSO-projection
\cite{Gliozzi:1977qd}. It projects onto states of definite
world-sheet fermion number. There are five different string
theories known that can be constructed in this way. These are
summarized with several properties in table
\ref{tab:differentSUSYStrings}.
\begin{table}[t!]
\centering
\begin{tabular}{|p{40pt}||p{100pt}|p{38pt}|p{25pt}|p{28pt}|p{112pt}|}
  \hline
  Type & Strings & Gauge group & Chir. & SUSY\newline (10D) &Massless bosonic\newline spectrum\\
  \hline\hline
  IIA & closed\newline oriented & $U(1)$
  & non-\newline chiral & $\mathcal{N}$=2 & \textbf{NS-NS}: $g_{\mu\nu}$, $\Phi$,
  $B_{\mu\nu}$\newline \textbf{R-R}:$A_{\mu\nu\rho}$, $A_\mu$ in $U(1)$\\ \hline
  IIB & closed\newline oriented & none
  & chiral & $\mathcal{N}$=2 & \textbf{NS-NS}: $g_{\mu\nu}$, $\Phi$,
  $B_{\mu\nu}$\newline \textbf{R-R}:$A$, $A_{\mu\nu}$, $A_{\mu\nu\rho\kappa}$\\ \hline
  I & open\&closed \newline unoriented & $SO(32)$& chiral
  & $\mathcal{N}$=1 & $g_{\mu\nu}$, $\Phi$, $A_{\mu\nu}$,\newline
  $A_\mu$ in Ad[$SO(32)$]\\ \hline
  heterotic\newline $SO(32)$& closed\newline oriented & $SO(32)$
  & chiral & $\mathcal{N}$=1 & $g_{\mu\nu}$, $\Phi$,
  $B_{\mu\nu}$\newline $A_\mu$ in Ad[$SO(32)$]\\ \hline
  heterotic\newline $E_8\times E_8$& closed\newline oriented & $E_8\times E_8$
  & chiral & $\mathcal{N}$=1 & $g_{\mu\nu}$, $\Phi$,
  $B_{\mu\nu}$\newline $A_\mu$ in Ad[$E_8\times E_8$]\\ \hline
\end{tabular}
\caption{The five known consistent string theories in $D=10$.}
\label{tab:differentSUSYStrings}
\end{table}
It is possible to build a consistent theory either from just
closed strings (type II or heterotic) or from closed plus open
strings (type I). In contrast to this, it is not possible to
build an interacting theory just from open strings.\footnote{A
heuristic argument for this fact is that the joining interaction
of two open strings locally cannot be distinguished from the
joining of the two sides of just one open string, but this
produces a closed string.} To get a phenomenologically interesting
theory, one furthermore has to include non-abelian gauge groups
into the theory. This is not possible for the type II closed
string theories. However, for the open string one can attach
non-dynamical degrees of freedom to both ends of the string, the
so-called Chan-Paton-factors. The gauge groups are $U(n)$ in the
case of a oriented theory and $SO(n)$ or $Sp(n)$ in the unoriented
case, but only the case of $SO(32)$ is anomaly free, as a detailed
analysis shows. Therefore, one is also forced to include
unoriented worldsheets and the resulting theory is called type I.
Another possibility to include non-abelian gauge groups is the
heterotic string, where a different constraint algebra acts on the
left and right movers, spacetime supersymmetry acts only on the
right-movers; From the beginning of the 90s, a lot of research
effort has been put in these theories, but they seem to have a
serious problem: the gravitational and Yang-Mills-couplings are
directly related for the heterotic string and this produces a
4-dimensional Planck mass which is about a factor twenty too high.
In this work, the heterotic string will not be treated. The
superstring theories do not just have worldsheet supersymmetry,
but also extended spacetime supersymmetry. In 10 dimension, the
number of supercharges for the different theories varies in
between 32 for the type II theories and 16 for the other theories,
meaning $\mathcal{N}$=2 or $\mathcal{N}$=1 respectively in 10
dimensions.
\section{Compactification and spacetime supersymmetry}
It has to be explained within string theory why there are just
four so far observable dimensions. One hint has been given already
by Kaluza-Klein theories which assume a fifth compact and indeed
very small dimension. The total dimension for the supersymmetric
string theories of the last section  has been determined to be
$D=10$, meaning that if one expands the Kaluza-Klein idea to this
case, the compact subspace should have a dimension $D=6$. One
furthermore assumes that the spacetime has a product structure of
the following type:
\begin{equation}
g_{\mu\nu}=\left(%
\begin{array}{cc}
  g_{ij}^{(4)} & 0 \\
  0 & g_{a b}^{(6)} \\
\end{array}%
\right)\ ,
\end{equation}
where $g_{ij}^{(4)}$ is the 4-dimensional metric, ensuring
4-dimensional Poincar\'{e} invariance, and $g_{a b}^{(6)}$ the
internal metric of the compact subspace. $g_{a b}^{(6)}$ is
unknown. One could only hope to conclude some properties from
indirect considerations. In general, any kind of compactification
conserves a certain amount of spacetime supersymmetries and breaks
the others. Phenomenologically, extended supersymmetry in four
dimensions is disfavored, as mentioned already. Therefore, one
should end up with a theory having $\mathcal{N}$=1 in four
dimensions, meaning four conserved supercharges or even with
completely broken supersymmetry $\mathcal{N}$=0.

A first and simple try for such a compact space is given by the
six-dimensional torus. It can simply be parameterized by six radii
$R_a$ that are allowed to depend just on the $x_i$ of the
non-compact subspace. The metric is given explicitly by $g_{a
b}^{(6)}=\delta_{ab}{R_a}^2(x_i)$. The radii that indeed label
different string vacua are a first example of moduli that we will
encounter very often in the course of this work. Similarly to
Kaluza-Klein compactification, they can be understood as
additional spacetime fields. The case of toroidal compactification
already shows several important features of more general
compactifications: strings can move around the toroidal
dimensions, leaving a quantized center-of-mass momentum. The
induced spectrum is called the Kaluza-Klein-spectrum and an effect
that can be seen in field theory already. But strings can even
wind around the compact dimensions several times, they are then
described as topological solitons. The major problem of toroidal
compactification is that it conserves all supersymmetries and so
does not lead to $\mathcal{N}$=1 in the Minkowski-spacetime.

One simple resolution of this problem has been given in
\cite{Dixon:1985jw,Dixon:1986jc} by orbifolding the space.
Orbifolding is a classical geometrical method that divides out a
certain subspace $S$ of the original space $X$ and then makes the
transition to the quotient space $X/S$. For instance, $S$ can be a
discrete subgroup like $Z_n$. A complete classification has
been given in \cite{Dixon:1986jc} for toroidal orbifolds
$T^6/Z_n$. The procedure of orbifolding induces singularities on
the original space, which are unwanted. Still one can understand
orbifolds as limits of certain smooth manifolds that are called
Calabi-Yau manifolds, where the singularities have been resolved
by blowing up the fixed points.

Another last but very important phenomenon that generally occurs
in toroidal type compactifications shall be mentioned: T-duality.
This is a duality that leaves the coupling constants (and
therefore the physics) invariant, but exchanges the radius of the
compactified dimension with its inverse, or more precisely with
\begin{equation}\label{eq:T-duality}
R\rightarrow R'=\frac{\alpha'}{R}.
\end{equation}
Common sense, claiming that a large or small compactified
dimension should be related to very different physics, fails
within string theory. T-duality also exchanges Kaluza-Klein and
winding states. This fact will become very important in the following
chapters. T-duality also has an extension to Calabi-Yau manifolds
that are of major interest in string theory, this is called mirror
symmetry.

A general Calabi-Yau manifold \cite{Yau:1977ms} can be obtained by
demanding that its compact subspace has to be a manifold of
$SU(3)$-holonomy because this leaves a covariantly constant spinor
unbroken. This condition in mathematical language can also be
expressed as the requirement to have a Ricci-flat and K{\"a}hler
manifold. The surviving supercharges are the ones that are
invariant under the holonomy group, for $SU(3)$ (not a subgroup)
this leads to a minimal $\mathcal{N}$=1 supersymmetry for the
heterotic and type I string, but to $\mathcal{N}$=2 for the type
II theories in four dimensions.
\section{D-branes}
\begin{figure}
\centering
\includegraphics[width=6cm,height=5cm]{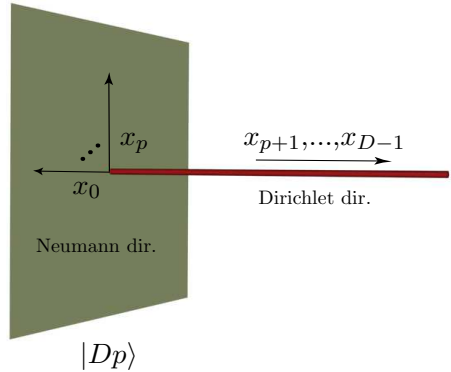}
\caption{A $Dp$-brane.}\label{fig:Dbrane}
\end{figure}

$Dp$-branes fulfill Dirichlet boundary conditions in $(D-p-1)$ directions:
\begin{equation}\label{eq:DbranesDir}
X^\mu(\tau,0)=x^\mu_1\ , \qquad X^\mu(\tau,l)=x^\mu_2 \qquad \text{for}\ \mu=(p+1),...,(D-1)
\end{equation}
and Neumann boundary conditions in the remaining $(p+1)$ directions:
\begin{equation}\label{eq:DbranesNeu}
\partial^\sigma X^\mu(\tau,0)=\partial^\sigma X^\mu(\tau,l)=0 \qquad \text{for}\ \mu=0,...,p\ .
\end{equation}
Here, $x^\mu_1$ and $x^\mu_2$ are fixed coordinates and $D$ is the
total dimension of spacetime. For the superstring it is $D=10$. Both
string endpoints are fixed transversally on a hyperplane with a
$(p+1)$ dimensional world-volume, a $Dp$-brane, but still can
move freely in the Neumann directions longitudinal to this
world-volume. This is schematically shown in figure
\ref{fig:Dbrane}.
It is also possible as a direct extension of (\ref{eq:DbranesDir})
and (\ref{eq:DbranesNeu}), that different boundary conditions
apply to both sides of the open string. This describes a string
ending on two different D-branes.

One first observation is the fact that the boundary conditions of
D-branes explicitly break Poincar\'{e} invariance along the
Dirichlet directions. But this is not a problem if the world
volume of the D-brane contains the observable 4-dimensional
Minkowski spacetime. This usually is assumed. D-branes first have
been discovered by Polchinski \cite{Dai:1989ua, Polchinski:1995mt} in
1995 by methods of T-duality, but nevertheless they can be understood
as non-perturbative objects. The reason is mainly that they can
carry certain conserved charges, the Ramond-Ramond (R-R) charges.
In other words, this means that they are sources for $(p+1)$-form
R-R gauge fields.

D-branes also can be found in supergravity theories (which will be
described in the following section), where they are solitonic
BPS states of the theory. They also have a mass, or tension,
determining their gravitational coupling.

To summarize, it can be said that they are dynamical objets that
can move, intersect or even decay into different configurations.
All of these properties will be described and extensively used in
the following chapters.

\section{Low energy supergravity}

The formulation of string theory by the RNS action
(\ref{eq:RNSaction}) is intrinsically two-dimensional, it is the
formulation on the worldsheet. But the worldsheet spreads out into
the 10-dimensional spacetime. Therefore, it should be possible to
find a complete formulation of the theory in spacetime, too.
Sadly, the full spacetime field theory action, describing all
massless and massive modes of string theory correctly, is unknown.
Still, there is a very important link: one of the earlier efforts
to generalize gravity from field theory was to simply generalize
the Einstein-Hilbert action in the obvious way to $D$ dimensions,
\begin{equation}\label{eq:einstein_hilbert_ddim}
    S_\text{gravity}=\frac{1}{2\kappa^2}\int d^D x\sqrt{g} R \ ,
\end{equation}
where $\kappa=(8 \pi)^{1/2}/M_P$ is the gravitational coupling and
$g_{\mu\nu}$ and $R$ are the D-dimensional metric and curvature
scalar. Then one can expand the metric tensor around Minkowski
space
\begin{equation}
g_{\mu\nu}=\eta_{\mu\nu}+\kappa h_{\mu\nu} \ ,
\end{equation}
and understand $h_{\mu\nu}$ as the $D$-dimensional graviton field.
As mentioned earlier, such a theory is non-renormalizable and thus
no meaningful quantum theory. On the other hand, for tree level
vertices, it should give a correct description. From the
perspective of string theory, this means that one takes the limit
of an infinitely high string scale, or equivalently
$\alpha'\rightarrow 0$, giving the massless states. Indeed in this
limit, the tree level amplitudes (like the three-graviton
scattering amplitude) which one calculates from the string
worldsheet (\ref{eq:Polyakov}) yields the effective action
(\ref{eq:einstein_hilbert_ddim}), see for instance
\cite{Polchinski:1998rq}. But this result is not valid anymore if
one takes into account the massive string modes. This is well
understandable as string theory is the correct quantum theory,
(\ref{eq:einstein_hilbert_ddim}) is it not.

Nevertheless, it is possible to construct a meaningful spacetime
action order by order in $\alpha'$. The most direct approach for
this construction is given by the matching of field and string
theory amplitudes. This method can be simplified by using the
symmetries to constrain the possible terms within the spacetime
action.

Another technique to determine the spacetime action is given by
looking at the Polyakov (\ref{eq:Polyakov}) or RNS action
(\ref{eq:RNSaction}) in a curved background spacetime by replacing
$\eta_{\mu\nu}$ by a general $g_{\mu\nu}$ in these equations and
also generalizing the other possible background fields, like the
antisymmetric B-field $B_{\mu\nu}$ or the dilaton $\Phi$ in a
similar way. For the supersymmetric theories, one has to proceed
in this fashion for all fields listed in table
\ref{tab:differentSUSYStrings}. If one now insists on
Weyl-invariance at a certain string loop order, one obtains
$\beta$-functions for every field which have to vanish. For
instance, the $\beta$-function for the metric is given by
\begin{equation}
\beta_{\mu\nu}^g=\alpha' R_{\mu\nu}+2\alpha'\nabla_\mu \nabla_\nu
\Phi-\frac{\alpha'}{4} H_{\mu\lambda\omega} H_\nu^{\
\lambda\omega}+O({\alpha'}^2)\ .
\end{equation}
The terms in this $\beta$-function reproduce exactly the possible
ones for a certain order in $\alpha'$ in the effective spacetime
action of interest.

To end this section, the type IIA lowest order effective action
is being listed, as it will be very useful in the following chapters:
\begin{align}\label{eq:typeIIAsupergravity}
    &\mathcal{S}_\text{IIA}=\mathcal{S}_\text{NS}+\mathcal{S}_\text{R}+\mathcal{S}_\text{CS}\ ,\\
    &\mathcal{S}_\text{NS}=\frac{1}{2{\kappa_{10}}^2}\int d^{10}x
\sqrt{g}e^{-2\Phi}\left(R+4\,\partial_\mu\Phi\,\partial^\mu\Phi-\frac{1}{2}|H_3|^2\right)\ ,\nonumber\\
    &\mathcal{S}_\text{R}=-\frac{1}{4{\kappa_{10}}^2}\int d^{10}x \sqrt{g}\left(|F_2|^2+|\tilde{F}_4|^2\right)\ ,\nonumber \\
    &\mathcal{S}_\text{CS}=-\frac{1}{4{\kappa_{10}}^2}\int B_2\wedge F_4\wedge F_4 \ ,\nonumber
\end{align}
This action by itself is called type IIA supergravity, and is the
most useful one if one wants to understand the low energy limit of
type IIA string theory. Note, that this action corresponds to a
different choice of coordinate system as compared to the
Einstein-Hilbert action (\ref{eq:einstein_hilbert_ddim}),
differing by the exponential dilaton factor. This is called the
string frame, but can be easily transformed into the so-called
Einstein frame. This is explained in much detail in chapter
\ref{cha:Inflation} and appendix \ref{cha:kinetic_terms}.

\section{How to understand low energy physics from string theory}
So far, we have discussed some of the major features of string
theory. On the other hand, we have not discussed yet in detail the
connection between these features and tools with low energy
physics, which is that kind of physics, one might observe in
future particle colliders (like LHC). As one can observe from table
\ref{tab:differentSUSYStrings}, there are several concurrent
perturbative string theories. From fundamental principles, it is
not possible to figure out which of these theories is the right
one to describe our world. In this context, it should also be
mentioned that Witten in 1995 realized that all these five
different string theories might stem from a 11-dimensional theory
called M-theory \cite{Witten:1995ex} which has 11-dimensional
supergravity as its low energy approximation. The different string
theories then are approximations in different corners of the
moduli space of M-theory. This result tells us that all string
theories are connected by dualities, unfortunately, it does not
help for the concrete construction of a phenomenological model.

Even worse, every one of these five theories has a very large
moduli space. These moduli parameters distinguish between
physically different background spaces in which the string
propagates. At this point, we approach the biggest problem of
perturbative string theory: it does not determine the background
space itself, this merely is an input parameter. This situation slightly
recalls the problem of the undetermined parameters (like the
masses) of the standard model and is somehow unsatisfactory. There
might be several ways out: some attempts are done to rebuild the
foundations of string theory in order to obtain a unique theory
\cite{Friedan:2002aa}, but so far, the result seems rather obscure.

This problem can be rephrased in such a way that the
non-perturbative formulation of string theory is unknown. String
field theory plays an important role in this context: perturbative
string theory is a first quantized theory. In contrast to this,
string field theory is a second quantized approach, incorporating
off-shell potentials for the contained fields. Here, the problem
arises that the field modes do not decouple and therefore, an
analytic solution often cannot be found.

The perspective which is taken in this work will be more
pragmatic. We will assume that perturbative string theory already
leads to a correct understanding of physics in this universe if
one makes some reasonable assumptions about the background space
and tries to verify these assumptions in a bottom-up approach,
which could be called a model building approach.

In the 1980's, the most efforts in this direction were made in
weakly coupled heterotic string theory \cite{Gross:1985dd}. At
that time, also the idea was formulated that one should have a
background allowing for $\mathcal{N}$=1 supersymmetry, see for
instance \cite{Candelas:1985hv}. This has been achieved for both
Calabi-Yau \cite{Candelas:1985en, Strominger:1985it} and orbifold
spaces \cite{Dixon:1985jw, Dixon:1986jc}.

Beside heterotic theory, the type I string also includes a gauge
group $SO(32)$. Some progress was made in \cite{Marcus:1982fr}
where it was proven that the gauge group of open strings must be a
classical group. Later also orientifold planes were discovered by
Sagnotti et al. (see for instance \cite{Sagnotti:1987tw,
Bianchi:1990yu}). Polchinski in 1995 reinterpreted some of these
results by D-branes \cite{Polchinski:1995mt} and
this started off an enormous amount of new model building
approaches. It was realized that the most important consistency
condition for meaningful quantum models in type II string theory
is given by the R-R tadpole cancellation equations \cite{Gimon:1996rq}.

\section{Intersecting brane worlds and phenomenological features}
In recent times, mainly two distinct paths have been treated, the
construction of spacetime supersymmetric and non-supersymmetric
models. These two possibilities originate from two different
philosophies of how to solve the hierarchy problem. One feature
common to both approaches is the existence of small gauge groups.

Non-extended $\mathcal{N}$=1 supersymmetry solves the hierarchy
problem by definition with its equality of bosons and fermions. As
this symmetry is not observed at present energies, it has to be
broken. Soft-breaking terms achieve this in an elegant way by
introducing logarithmic divergencies into the theory without
destroying the merits for solving the hierarchy problem
\cite{Girardello:1982wz}. The string scale can be very high in
supersymmetric scenarios, either close to the Planck scale or in
an intermediate regime \cite{Burgess:1998px}. In
\cite{Blumenhagen:1999md, Blumenhagen:1999ev, Blumenhagen:1999db}
some examples of $\mathbb{Z}_n$-orientifolds in six and four
dimensions were treated, but they did not give rise to chiral
fermions, the tadpoles were always cancelled locally. So far, the
only semi-realistic supersymmetric orientifolds with chiral
fermions have been constructed in a background
$T^6/(\mathbb{Z}_2\times\mathbb{Z}_2)$ in \cite{Cvetic:2001tj,
Cvetic:2001nr, Cvetic:2002qa, Cvetic:2002wh, Cvetic:2003xs,
Pradisi:2002vu}, for a background
$T^6/(\mathbb{Z}_2\times\mathbb{Z}_4)$ \cite{Honecker:2003vq} and
for $T^6/\mathbb{Z}_4$ in \cite{Blumenhagen:2002gw}. The result
that a realistic gauge coupling unification is possible for this
class of models has been obtained in \cite{Blumenhagen:2003jy}.
Nevertheless, the issue of supersymmetry breaking is not treated
satisfactory in this context, although it certainly needs an
understanding within string theory. At least some progress has
been made in \cite{Cvetic:2003yd}.

On the other hand, it is also possible to construct
non-supersymmetric intersecting D-brane models right from the
start. These models then are along the lines of an extra large
dimension scenario \cite{Arkani-Hamed:1998rs, Antoniadis:1998ig}.
In these papers, it was shown that the hierarchy problem also can
be solved by the assumption of additional dimensions (as compared
to the four of Minkowski space) if these are at the millimeter
scale. Such dimensions are not in contradiction with experiment.
Then spacetime supersymmetry is not required anymore. Some D-brane
model building examples being motivated by this idea have been
constructed using various types of branes \cite{Cremades:2002dh,
Bailin:2002fd, Bailin:2002gg}, or general Calabi-Yau spaces
\cite{Uranga:2002pg}.

In orientifold models, supersymmetry can be broken by either
turning on magnetic background fluxes in the picture of D9-branes,
or equivalently, by putting the D6-branes at angles in the T-dual
picture \cite{Blumenhagen:2000wh}, following the older ideas of
\cite{Bachas:1995ik}. For such models, it was even possible to
construct three-generation models \cite{Blumenhagen:2000ea} and
later, models with standard model-like gauge groups have been
obtained \cite{Ibanez:2001nd, Cremades:2002qm, Kokorelis:2002zz,
Kokorelis:2002wa}.

These first models were unstable due to some complex structure
moduli, this problem has been solved in
\cite{Blumenhagen:2001te} for the $\mathbb{Z}_3$-orbifold
background, although the dilaton instability still remains in
this class of models. Furthermore, issues like gauge breaking,
Yukawa-couplings and gauge couplings have been treated with some
success. In \cite{Blumenhagen:2002ua} the stability problems have
been reinterpreted in the context of cosmology.

Some other, recently discussed models just preserve supersymmetry
at some local D6-brane intersections, but not globally
\cite{Cremades:2002cs, Cremades:2002te}. Unfortunately, there
still remains a modest hierarchy problem in this type of models
\cite{Cremades:2002te}.

\section{Outline}
This work combines several results in the context of
$\mathbb{Z}_N$ orientifold models of type IIA with intersecting
D-branes under the two main subjects stability and phenomenology.
The organization is as follows.

Chapter \ref{cha:IntersectingBraneWorlds} gives a very detailed
introduction to orientifold type II models containing D-branes.
Both, the approach using D9-branes with B-fluxes and the approach
containing D6-branes at general angles are described. Much
attention is paid to the most important consistency requirement in
string theory, the R-R tadpole. The NS-NS tadpole is also discussed in
detail, as it delivers the scalar potential for the
dilaton and the complex structure moduli, being important to
understand several issues of stability. Then, the enhancement
towards orbifold models is discussed. The massless open and closed
string spectra, being important for low energy physics, are treated
besides anomaly cancellation and the possible gauge breaking
mechanisms of these models.

In chapter \ref{cha:Z3}, the specific $\mathbb{Z}_3$
orientifold, being especially suitable for the construction of a
non-supersymmetric standard-like model, is discussed in great
detail. The main attention is paid to the issues of model
building, but in the end, a detailed phenomenological model,
having the standard model gauge groups and the right chiral
fermions, is presented. Some phenomenological aspects are
discussed as well. This chapter is based on
\cite{Blumenhagen:2001te}.

Chapter \ref{cha:Z4} deals with the construction of a spacetime
$\mathcal{N}$=1 supersymmetric orientifold, being stable because
of supersymmetry. The $\mathbb{Z}_4$ orbifold model is discussed
in this context, where the main interest is paid to the
construction of fractional D-branes, being a new ingredient to
this type of models. These fractional branes allow for the
construction of a very interesting 3-generation Pati-Salam model
which even can be broken down to a MSSM-like model. Several
phenomenological aspects are treated.
This chapter is based on \cite{Blumenhagen:2002gw}.

In chapter \ref{cha:Inflation}, the problem of unstable closed
string moduli, is discussed from a very different point of view.
It is entered into the question if it is possible that these
unstable moduli in the beginning of our universe could have been
responsible for inflation. Inflation itself is a very successful
attempt for explaining the horizon and flatness problems, but the
key ingredient, a scalar field triggering off the very short
inflationary period, still has no fundamental explanation. Both,
the phenomenological aspects and the possible realization from
string theory, are discussed in much detail based on
\cite{Blumenhagen:2002ua}.

Finally, chapter \ref{cha:Conclusions} presents the conclusions
and gives a short outlook.

\chapter{Intersecting D-branes on type II orientifolds}\label{cha:IntersectingBraneWorlds}
This chapter provides a detailed introduction into intersection
D-branes on type II orientifolds, including both toroidal and
$\mathbb{Z}_N$-orbifolded models. The main concern is the
conformal field theory calculation. On the other hand, model
building aspects like the issue of gauge breaking mechanisms, are
treated as well, as they are especially important for
the concrete realizations in the following chapters.
\section{Intersecting D-branes on toroidal orientifolds}\label{cha:ORorientifold}
The starting point for our considerations is a general type I
model that has an amount of 16 supersymmetries. According to table
\ref{tab:differentSUSYStrings}, this string theory involves
non-oriented Riemann surfaces and is a theory of open plus closed
strings. Following Polchinskis picture, the endpoints of open
strings in general can be located on D-branes of a certain
dimensionality. This also leads to a new understanding of the type
I string with gauge group $SO(32)$: it is just the case of $N$=32
parallel D9-branes filling out the whole spacetime.

The closed string sector of type I string theory can be
represented by type IIB string theory having 32 supersymmetries
($\mathcal{N}$=2 in 10 dimensions), if the worldsheet parity
$\Omega$ is being gauged. This reduces the supersymmetry by half
of its amount, so afterwards one has $\mathcal{N}$=1 in 10
dimensions. Thus, one obtains the unoriented closed string
surfaces of type I. This procedure is not possible for the type
IIA string theory which does not have this particular worldsheet
symmetry, or in other words, the same chiralities for left and
right movers. Therefore, we will first consider
\begin{equation}\label{eq:toroidal_typeIIB}
{{\rm Type\ IIB}\ {\rm on}\ T^{6} \over {\Omega}}\ ,
\end{equation}
being a first example of a so-called orientifold.

It is possible to describe the projection of the theory formally
by the introduction of a so-called orientifold $O9$-plane. In the
language of topology this object is a cross-cap, because it
reverses the orientation, 
but in contrast to D-branes, the O-plane is non-dynamical.
But similarly to a D-brane, the orientifold
plane is localized and does not affect the physics far away from the
O-plane which still is described by the oriented string
theory.\footnote{Here, this argument does not apply because the
O-plane is spacetime-filling, but O-planes in general can have a lower
dimensionality as well, as we will see.}

In the following, we will consider a factorization of spacetime
$\mathcal{X}$ into a six-dimensional compact $\mathcal{M}$ and the
usual four-dimensional Minkowski space $\mathbb{R}^{1,3}$, so
\begin{equation}\label{eq:spactime}
\mathcal{X}=\mathbb{R}^{1,3}\times \mathcal{M}\ .
\end{equation}
The simplest example for such a compact space is given by a
6-torus $T^6$ which we well consider in most of this work. To
allow for simple crystallographic actions, we will assume that it
can be factorized in three 2-tori\footnote{This is of course a
special choice of complex structure.}, so
\begin{equation}
\mathcal{M}=T^6=T^2 \times T^2 \times T^2\ .
\end{equation}
It will useful to describe every 2-torus on a complex
plane, so one introduces complex coordinates
\begin{equation}\label{eq:complex_coordinates}
    Z_I=X_I+i Y_I ,
\end{equation}
where every torus has the two radii $R^{(I)}_x$ and $R^{(I)}_Y$
along its fundamental cycles.

In the open string sector of type I theory, there are also 16
unbroken supersymmetries. For a smaller dimensionality of the
D-branes, they locally break half of the original supersymmetry.

As a reminder, one can use the following formula for the relation
between dimensionality, number of unbroken supercharges and
supersymmetry:
\begin{equation}\label{eq:Nsusy}
    \#\ \text{supercharges}=\mathcal{N}\cdot\#\ \text{real comp. of min. representation in}\ D\ \text{dimensions}
\end{equation}
The minimal representations for several spacetime dimensions are indicated
in table \ref{tab:minimal_representation} together with their specific properties.
\begin{table}
\centering
\sloppy
\renewcommand{\arraystretch}{1.2}
\begin{tabular}{|c||c||c|c|c|}
  \hline
 $D$& Minimal rep.& Majorana & Weyl & Majorana-Weyl\\
  \hline\hline
  2&1&yes&self-conjugate&yes\\
  3&2&yes&$-$&$-$\\
  4&4&yes&complex&$-$\\
  5&8&$-$&$-$&$-$\\
  6&8&$-$&self-conjugate&$-$\\
  7&16&$-$&$-$&$-$\\
  8&16&yes&complex&$-$\\
  9&16&yes&$-$&$-$\\
  10&16&yes&self-conjugate&yes\\
  11&32&yes&$-$&$-$\\
\hline
\end{tabular}
\caption{Number of real components in the minimal representations
of $SO(D-1,1)$ spinors and the possible representations.}
\label{tab:minimal_representation}
\end{table}

Compactifying on the 6-torus, all supersymmetries are being
conserved. Therefore, this leads to $\mathcal{N}$=8
supersymmetries in four dimensions if one starts with type II
theory, or to $\mathcal{N}$=4 supersymmetry if one takes into
account the $\Omega$ projection. For phenomenological model
building this certainly is unacceptable.

\subsection{D9-branes with fluxes}
One possibility to solve the problem of too much supersymmetry is
to alter the theory by introducing various constant magnetic
$U(1)$ $F$-fluxes on the D9-branes. Turning on $F^{I}_{XY}=F^{I}$
on a certain 2-torus changes the Neumann conditions of the
D9-brane into mixed Neumann-Dirichlet boundary conditions as
\begin{align}\label{eq:boundary_conditions_flux}
    \partial_\sigma X_I+F^{I}\partial_\tau Y_I=0 \ ,\\
    \partial_\sigma Y_I-F^{I}\partial_\tau X_I=0 \ .\nonumber
\end{align}
It is possible that different D9-branes also have different
magnetic fluxes on at least one 2-torus. Using this property, one
can break supersymmetry further down. But the resulting 2-tori are
noncommutative rendering calculations difficult do do
\cite{Blumenhagen:2000wh, Bianchi:1992eu, Kakushadze:1998bw}.

Beside the introduction of $F$-fluxes, it is possible to switch on
an additional constant NS-NS 2-form flux $b$ \cite{Bianchi:1992eu,
Kakushadze:1998bw, Angelantonj:1999jh, Angelantonj:2000rw}. It has
to be discrete \cite{Blumenhagen:2000ea} with a value of either 0
or $1/2 \mod 1$ due to the restrictions that arise from the
orientifold $\Omega$-projection. This 2-form flux later will be
very important for phenomenological model building as it allows
for odd numbers of fermion generations in the effective theory
\cite{Blumenhagen:2000ea}.

\subsection{Intersecting D6-branes}
Because of non-commutativity in the flux-picture, a T-dual
description has been proven to be very useful
\cite{Blumenhagen:1999md, Blumenhagen:1999ev}. One applies a
T-duality (\ref{eq:T-duality}) to all three $Y_I$ directions of
the D9-branes and obtains D6 branes. These D6-branes fill out the
whole 4-dimensional Minkowski space and wrap a special Lagrangian
1-cycle on each torus, as a whole a special Lagrangian 3-cycle.

The T-duality maps the worldsheet parity $\Omega$
into $\Omega R$, where $R$ acts as a complex conjugation on the
coordinates of all 2-tori
\begin{equation}\label{eq:complexconj}
    R:\qquad Z_I\rightarrow \bar Z_I\ .
\end{equation}

 Note, that after having performed three
T-duality transformations subsequently, we are now considering a
type IIA string theory, as every single T-duality maps type
IIA into type IIB and vice versa. Accordingly, the initial orientifold
model (\ref{eq:toroidal_typeIIB}) has been mapped to
\begin{equation}\label{eq:our_orientifold}
{{\rm Type\ IIA}\ {\rm on}\ T^{6} \over {\Omega R}} \ .
\end{equation}
The most important difference to the case before T-duality has
been performed is that the internal coordinates now are completely
commutative and in contrary to the existence of fluxes, the picture
of intersecting branes is purely geometric.

\begin{figure}
\centering
\includegraphics[width=11cm,height=8cm]{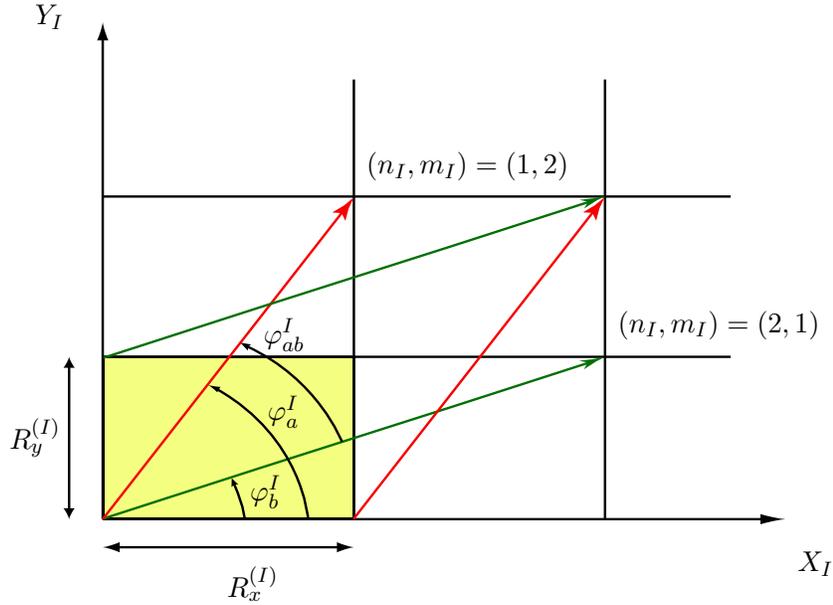}
\caption{Two exemplary branes $a$ and $b$ intersecting at angels
on one $A$-torus with a topological intersection
number $I_{a b}=3$.}\label{fig:branesatangles}
\end{figure}

Concretely, the T-duality transforms the F-flux into a certain
non-vanishing angle by which a stack of D6-branes now is arranged
relatively to the $X_I$-axis. This angle is given by
\begin{equation}\label{eq:angle_in_terms_of_F}
    \tan \varphi^{I}=F^{I} .
\end{equation}
If we have two different stacks of D-branes $D6_a$ and $D6_b$, then they
are intersecting at a relative angle
\begin{equation}\label{eq:relangle}
    \varphi^I_{ab} = {\rm \arctan} (F^I_a) - {\rm \arctan} (F^I_b) \ .
\end{equation}
This is schematically shown for one 2-torus in figure
\ref{fig:branesatangles}, where the fundamental region of the
torus has been hatched.

The action of the T-duality on the discrete $b$-flux is that the
torus is either transformed into a rectangular one for $b_I=0$, or
a tilted one for $b_I=1/2$. Commonly, the first possibility is
called A-torus, the second possibility $B$-torus, both are shown
in figure \ref{fig:torusAB}. As this choice can be taken
differently for every 2-torus, in general this leads to eight
distinct models.

The model is gauged under $\Omega R$, so under the worldsheet
parity symmetry together with a spacetime and hence geometrical
symmetry. To also hold within the presence of D-branes, this
symmetry requires the introduction of an $\Omega R$-mirror brane
for every D-brane.

Furthermore, we make the assumption that the branes do not densely
cover any one of the 2-tori. As a consequence, a set of two
integers $(n_I, m_I)$ is sufficient to describe the position of
any brane on the $I$-th 2-torus. $n_I$ and $m_I$ count the numbers
by which the 1-cycle is wrapping the two fundamental cycles
$\sqrt{1/2} R^{(I)}_x \mathbf{e}_1$, $m_I$ and $\sqrt{1/2}
R^{(I)}_y\mathbf{e}_2$ of the torus, respectively. For uniqueness,
one always has to choose the shortest length of any such brane
representation, or more concretely, $n_I$ and $m_I$ always have to
be coprime. Note, that by this definition, every brane has an
orientation on the torus. The intersection angle
(\ref{eq:relangle}) between two branes $D6_a$ and $D6_b$ then is given
by
\begin{equation}\label{eq:intangle in nmA}
    \varphi^{I,\text{A}}_{ba} = \varphi^I_{b}-\varphi^I_{a}={\rm arctan} \left[\frac{\left(n_I^a m_I^b-m_I^a
n_I^b\right) R^{(I)}_x R^{(I)}_y}{n_I^a n_I^b {R^{(I)}_x}^2+m_I^a m_I^b {R^{(I)}_y}^2}\right]
\end{equation}
for the $A$-torus, or by
\begin{equation}\label{eq:intangle in nmB}
\varphi^{I,\text{B}}_{ba} ={\rm arctan} \left[{\frac {{
R_x^{(I)}}\sqrt {4{{ R_y^{(I)}}}^{2}-{{ R_x^{(I)}}}^{2}} \left( {
n_I^a}{ m_I^b}-{ m_I^a}{ n_I^b} \right) }{{{ R_x^{(I)}}}^{2}{
n_I^a} { m_I^b}+{{ R_x^{(I)}}}^{2}{ m_I^a}{ n_I^b}+2{ n_I^a}{{
R_x^{(I)}}}^{2 }{ n_I^b}+2{ m_I^a}{{ R_y^{(I)}}}^{2}{ m_I^b}}}
\right]
\end{equation}
for the B-torus. Both can be parameterized in one equation by
\begin{equation}\label{eq:intangle in nm_general}
\varphi^{I}_{ba} ={\rm arctan} \left[{\frac {\frac{1}{2}{ R_x^{(I)}}\sqrt {4{{ R_y^{(I)}}}^{2}-2{
b_I}{{ R_x^{(I)}}}^{2}} \left( { n_I^a}{ m_I^b}-{ m_I^a}{ n_I^b}
\right) }{ { n_I^a}{ n_I^b}{{ R_x^{(I)}}}^{2}+{ m_I^a}{ m_I^b}{{
R_y^{(I)}} }^{2}+{ b_I}{{ R_x^{(I)}}}^{2} \left( { n_I^a}{
m_I^b}+{ m_I^a} { n_I^b} \right) }}\right]\ ,
\end{equation}
for either $b_I=0$ or $b_I=1/2$ on a certain torus.

Another important observation is that one can define a topological
intersection number $I_{a b}$ between two branes $a$ and $b$ by
\begin{equation}\label{eq:intersection_number}
I_{a b}=\prod\limits_{I=1}^{3}\left( n_I^a m^b_I-m^a_I n^b_I\right) \ .
\end{equation}
This number is topologically invariant and also has a very
intuitive meaning. It gives the number of orientated intersections
in between two branes, after all possible identified torus shifts
of both branes along the two fundamental cycles of the torus have
been regarded up to torus symmetries. A simple example is shown in
figure \ref{fig:branesatangles}, where the four differently
orientated intersection numbers totally add up to three.
Interestingly, this intersection number also can be derived just
from the consistency requirement of the boundary state formalism
with the CFT-loop channel calculations, as it is shown in appendix
\ref{cha:appendix_OR_annij}.

\subsection{Complex structure and K{\"a}hler moduli}
The torus moduli $R^{(I)}_x$, $R^{(I)}_y$ and the angle $\theta$
between them can be mapped to different ones, the complex
structure moduli $U^I$ and the K{\"a}hler structure moduli $T^I$.
Loosely speaking, the imaginary part of $T^I$ is related to the
volume of the torus and $U^I$ is related to the particular choice
of the second lattice vector of the torus. For the case of
D9-branes with $F$-fluxes, they can be defined in the following
way:
\begin{align}\label{eq:Torus_U}
    &U^I=U_1^I+i U_2^I=\frac{{R^{(I)}_y}}{{R^{(I)}_x}}\frac{\mathbf{e}_2}{\mathbf{e}_1}
    =\frac{R^{(I)}_y}{R^{(I)}_x}\ e^{i \theta}\ ,\\ \label{eq:Torus_T}
    &T^I=T^I_1+i T^I_2=b^I+i R^{(I)}_x R^{(I)}_y
\sin\left(\theta\right)\ ,
\end{align}
Note that in this equation, the discrete b-flux enters as well.
The real part of $U^I$ can be chosen to be zero, corresponding to
a rectangular torus with $\theta=\pi/2$. This actually is not a
restriction on the model because $U_1^I$ is a continuous modulus
of the theory.

Switching over to the T-dual description with
$R'^{(I)}_y=1/R^{(I)}_y$, $T^I$ and $U^I$ are getting mapped into
\begin{align}\label{eq:Tduality_T_U}
&T'^I=-\frac{1}{U^I}=-\frac{R^{(I)}_x}{R'^{(I)}_y}e^{-i \theta}     \ , \\
&U'^I=-\frac{1}{T^I}=-\frac{b+i{R^{(I)}_x} {R'^{(I)}_y}\sin\left(\theta\right)}{b^2+{R^{(I)}_x}^2
{R'^{(I)}_y}^2\sin^2\left(\theta\right)}\ . \nonumber
\end{align}
The torus now is tilted for the case $b=1/2 \mod 1$, but there is
no $B$-flux anymore. The significance of the tilt is that the
projection of the second torus basis vector with a length
${R'^{(I)}_y}\sqrt{2}$ onto the $X_I$-axis is exactly $1/2$ of the
length ${R'^{(I)}_x}\sqrt{2}$. Consequently, the angle $\theta$
between the torus vectors is fixed.

\begin{figure}
\centering
\includegraphics[width=12cm,height=6cm]{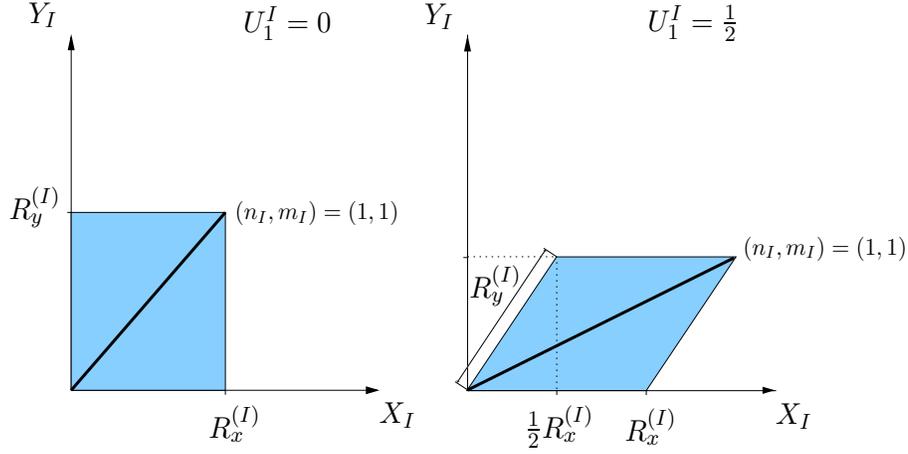}
\caption{The two inequivalent $A$- and $B$-tori, corresponding to
$b=0$ and $b=1/2$ in the flux picture.}\label{fig:torusAB}
\end{figure}

From now on, we will change the conventions on the branes at
angles side. These are denoted in appendix
\ref{cha:appendix_latcontrib_KB}, together the two sets of basis
vectors $\mathbf{e}_i^{\text{A}/\text{B}}$ for the two
inequivalent torus possibilities, being depicted in figure
\ref{fig:torusAB}. In common practice, these 2-tori are called A-
and B-torus \cite{Blumenhagen:1999ev}, here they are distinguished
by the two values $b_I=0$ and $b_I=1/2$ from the flux picture.
This notation will be kept, although there is no flux on the
branes at angles side anymore.

Then, the complex structure and K{\"a}hler moduli take the
following form
\begin{equation}\label{eq:Torus_U_T_AB}
    U^I=b_I+i\sqrt{\frac{{R^{(I)}_y}^2}{{R^{(I)}_x}^2}-{b_I}^2}\ ,
    \qquad \qquad T^I=T_1^I+i R^{(I)}_x R^{(I)}_y\sqrt{1-{b_I}^2\frac{{R^{(I)}_x}^2}{{R^{(I)}_y}^2}}\ ,
\end{equation}
where $b_I=0$ or $b_I=1/2$.

\subsection{One-loop consistency}
\begin{figure}
\centering
\includegraphics[width=12cm,height=7.5cm]{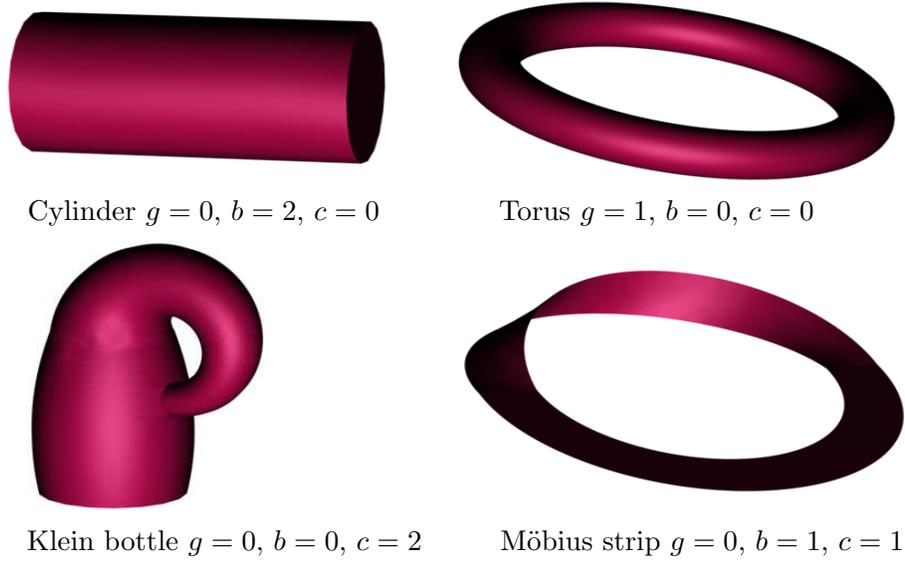}
\caption{All four Riemann surfaces with $\chi=0$.}\label{fig:oneloop_diag}
\end{figure}

The partition function for the bosonic string
(\ref{eq:bos_partition_function}) just contains the torus as a
world-sheet at the one loop level, corresponding to an Euler
number $\chi=0$. For our model, there are three additional
worldsheets shown in figure \ref{fig:oneloop_diag} that all have this
Euler number and so contribute at the same level.
Two of them are unoriented, the Klein bottle and the
M{\"o}bius strip, and two have boundaries, the cylinder (being conformally
equivalent to an annulus) and again the M{\"o}bius strip. The two
worldsheets with boundaries naturally are assigned to open strings
ending on these boundaries, whereas the torus and the Klein bottle
naturally are assigned to closed strings. The one-loop vacuum
amplitude $\mathcal{Z_{\text{one-loop}}}$
 is the sum of all 4 contributions coming from the different
$\chi=0$ worldsheets:
\begin{equation}\label{eq:part_fct_oneloop}
\mathcal{Z_{\text{one-loop}}}=\mathcal{T}+\mathcal{K}+\mathcal{A}+\mathcal{M}\ .
\end{equation}
Instead of the path integral representation of equation
(\ref{eq:bos_partition_function}), it is also possible to work
with the usual Hamiltonian formalism, where every
worldsheet integral can be written as a trace, this is for
instance for the cylinder amplitude up to normalization:
\begin{equation}\label{eq:Annulus_amplitude}
\mathcal{A}\thicksim\int\limits_{0}^{\infty}\frac{dt}{t}
\text{Tr}_{\text{open}}\left(\frac{1+(-1)^F}{2}e^{-2\pi t
\mathcal{H}_{\text{open}}}\right)\ .
\end{equation}
In this equation, $\mathcal{H}_{\text{open}}$ is the Hamilton
operator for the open string and the projector within the trace is
the usual GSO-projection, as discussed in the introductory
chapter. The trace is taken over both Neveu-Schwarz and Ramond
sectors and also includes the momentum integration
$V_{10}/(2\pi)^{10}\int d^{10}p$, where $V_{10}$ is the
regularized volume of a 10-torus. It is taken to be very large in
order to obtain the theory in the flat 10-dimensional spacetime. $t$
is the modular parameter of the cylinder.

\begin{figure}
\centering
\includegraphics[width=9cm,height=5cm]{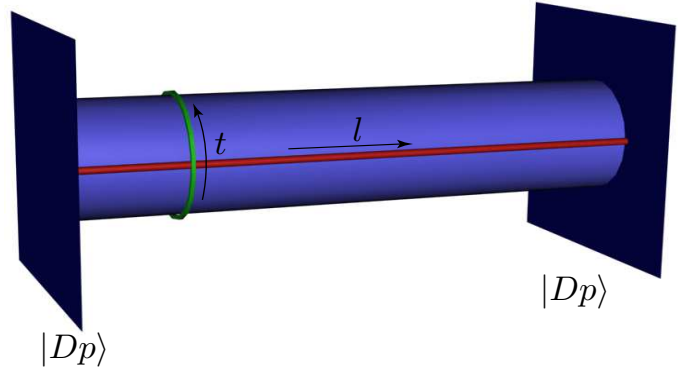}
\caption{The open-closed string duality for the cylinder.}\label{fig:openclosed}
\end{figure}

Taking a different point of view, the cylinder as a one-loop diagram
for open strings can also be understood as a tree level
propagation of a closed string. This is called open-closed string
duality and schematically shown in figure \ref{fig:openclosed}.

This duality can be understood via a modular transformation of the
worldsheet's modulus parameter $t$, schematically $t\rightarrow
l\sim1/t$. Usually the first point of view
(\ref{eq:part_fct_oneloop}) is called loop channel, the second one
tree-channel. The transformed amplitudes will be denoted by a
tilde and can be written with boundary states:
\begin{equation}\label{eq:Kleinbottle_amplitudetilde}
\widetilde{\mathcal{A}}\thicksim\int\limits_{0}^{\infty}dl\
\langle Dp| e^{-2\pi l
\mathcal{H}_{\text{closed}}}|Dp\rangle\ .
\end{equation}
The boundary states are coherent states in a generalized closed string
Hilbert space, fulfilling the transformed boundary conditions
which in the first place are being imposed on the open strings. In
this picture, two specific boundary state objects have to be
defined, $Dp$-branes and Orientifold $Op$-planes. Indeed, the loop
channel amplitudes together with the boundary conditions are
sufficient to completely specify the boundary states
\cite{Ishibashi:1989kg, Sen:1998ii}. We do not need their explicit
form at this point.

For the torus and the Klein bottle, the modular transformation
always transforms closed strings into closed strings, so the
modular transformation then is not strictly an open-closed string
duality for these worldsheets, but of course still possible to apply.
In order to obtain the correct string lengths for the different
amplitudes after the modular transformation, one has to take
different normalizing factors into the definition of the tree
channel modulus parameter $l$, this is summarized in table
\ref{tab:modulus_l}.
\begin{table}
\centering
\sloppy
\renewcommand{\arraystretch}{1.2}
\begin{tabular}{|c||c|}
  \hline
   Topology & Modular transformation\\
  \hline
  \hline
  Cylinder(Annulus) & $l=1/(2t)$ \\
  \hline
  Klein bottle & $l=1/(4t)$ \\
  \hline
  M{\"o}bius strip & $l=1/(8t)$\\
  \hline
\end{tabular}
\caption{The modular transformation parameters for the different topologies.}
\label{tab:modulus_l}
\end{table}

The torus amplitude $\mathcal{T}$ is modular invariant in type II
string theory, and by this reason finite. This statement remains
true for both our orientifold models, (\ref{eq:toroidal_typeIIB})
and (\ref{eq:our_orientifold}), because the torus amplitude stays
unaltered. If the theory is supersymmetric in spacetime, then
$\mathcal{T}$ by itself vanishes. This for instance is the case
for type II string theory.

The three remaining worldsheets do not have the property of
modular invariance, so for them it is not guaranteed that they do
not contain any divergencies which generally can spoil the whole
theory at the quantum level \cite{Green:1984sg, Green:1985ed}.
These divergencies are called tadpoles in analogy to the field
theory picture, where a single particle is generated from the
vacuum by quantum effects. In string theory, a non-vanishing
tadpole signals that the equations of motion of some massless
fields in the effective theory are not satisfied. Regarding the
different sectors of the superstring theory, both the R-R and the
NS-NS sector of the closed superstring theory in the tree channel
contribute to the overall tree channel tadpole. The two
contributions, coming from these two sectors are usually called
R-R and NS-NS tadpoles by themselves. One carefully has to
distinguish the notion of R-R and NS-NS sectors for loop and tree
channel, because the modular transformation maps one into the
other, depending also on the spin structure. These maps are
summarized for the different amplitudes without phase factors in
table \ref{tab:transformation_sectors}.
\begin{table}
\centering
\sloppy
\renewcommand{\arraystretch}{1.2}
\begin{tabular}{|c||c|c|}
  \hline
  Amplitude & Loop channel & Tree channel \\
  \hline
  \hline
    & (NS-NS,$+$) & (NS-NS,$+$) \\ \cline{2-3}
  Klein bottle & (NS-NS,$-$) & (R-R,$+$) \\ \cline{2-3}
   & (R-R,$+$) & (NS-NS,$-$) \\ \hline \hline
   & (NS,$+$) & (NS,$+$) \\ \cline{2-3}
  Cylinder & (NS,$-$) & (R,$+$) \\ \cline{2-3}
   & (R,$+$) & (NS,$-$) \\ \hline \hline
   & (NS,$+$) & (NS-NS,$-$) \\ \cline{2-3}
  M{\"o}bius strip& (NS,$-$) & (NS-NS,$+$) \\ \cline{2-3}
   & (R,$+$) & (R-R,$+$) \\ \hline
\end{tabular}
\caption{Correspondence between different sectors in loop and tree channel.}
\label{tab:transformation_sectors}
\end{table}

Although the two tadpoles do appear on the same grounds in the
partition function, their interpretation is quite different:
$Dp$-branes as well as an orientifold $Op$-planes are
p-dimensional hyperplanes of spacetime and therefore couple to R-R
$(p+1)$-forms $A_{p+1}$, as was first pointed out in
\cite{Polchinski:1995mt}. The orientifold plane by itself acts as a
background charge (what we will see soon in the Klein bottle R-R
contribution) which is a source term in the equations of motion for
the field $A_{p+1}$:
\begin{equation}\label{eq:eqmot_Ap1}
    dH_{p+2}=\ast J_{7-p} \ , \qquad d\ast{H_{p+2}}=\ast J_{p+1} \ .
\end{equation}
Here, $J_{p+1}$ and $J_{7-p}$ are the electric and magnetic
sources, respectively, and $H_{p+2}$ is the field strength of
$A_{p+1}$. If the field equations shall be consistent, then the
integral over the dual sources
\begin{equation}
\int\limits_{\Sigma(\chi)} \ast J_{9-\chi}
\end{equation}
for all surfaces $\Sigma(\chi)$ without boundaries has to vanish.
This is nothing but the analogue to the simple Gauss law of
electrodynamics. Using this picture, the field lines that are
originated from one charge must either go to infinity or lead to
another opposite charge. On a compact space, they cannot go to
infinity and so must end on an opposite charge. If there is no
such charge, the theory is inconsistent. This means for us that
the orientifold $Op$-plane R-R charge has to be cancelled. There
is just one possibility to do so, namely the introduction of open
sting sectors and therefore $Dp$-branes that do exactly cancel the
charge.\footnote{The argument first has been introduced for
D9-branes in type IIB that are spacetime-filling. Here the
restriction is even more severe: the 10-form potential does not
have a field strength in a 10-dimensional spacetime. This fact
implies that the R-R charges have to be neutralized locally, or in
other words, the orientifold planes and D-branes have to lie on top
of each other.}

This indeed is possible in many cases and imposes severe
restrictions on model building within orientifolds. Furthermore,
non-vanishing R-R tadpoles are related to non-vanishing gauge
anomalies in the effective field theory of the massless modes.
These are certainly unacceptable.

On the other hand, the NS-NS tadpole seems to be not as bad as the
R-R tadpole. The NS-NS sector contains the supergravity fields, in
particular the dilaton and the graviton, and the dilaton-graviton
interaction $\sqrt{-g}e^{-\Phi}$ is responsible for the tadpole
that is often even called dilaton tadpole. The term appears as an
overall factor in the effective action and having also a kinetic
term which is absent for the R-R tadpole. The theory consequently
is unstable (but not inconsistent).
There are two different possibilities to treat this problem: In
the first place, one can employ the Fischler-Susskind mechanism
that already has been invented in the context of the bosonic
string \cite{Fischler:1986ci, Fischler:1986tb}. The quantum
corrections coming from the NS-NS tadpole induce a source term
that gets incorporated into the equations of motion in this
mechanism, leading to a space-dependent background value for the
dilaton.

Secondly, there is the less ambitious approach: one might try to
solve the string equations of motions including the dilaton
tadpole in the effective field theory next to leading order. It
has been demonstrated in \cite{Dudas:2000ff,
Blumenhagen:2000dc} that this generally leads to warped geometries
and non-trivial profiles of the dilaton and other scalar
fields. In the non-supersymmetric type I string theory discussed
in these papers, the phenomenon of a spontaneous compactification
has occurred due to the NS-NS tadpoles. This perhaps can be
understood as a dynamical justification for a compactified
spacetime. Sadly, the non-linear sigma model on the worldsheet
then cannot be solved exactly in such a highly curved background
and furthermore, the procedure does not lead to a vanishing
tadpole at the next order of the perturbation theory. It merely is
a hope that the non-supersymmetric string theory self-adjusts its
background perturbatively order by order until eventually the true
quantum vacuum with a vanishing tadpole to all orders is reached
\cite{Dienes:2001se}.

By way of contrast, if the string theory is supersymmetric in
spacetime, then the sum of the two tadpoles vanishes for each
world sheet topology separately because the corresponding trace is
zero by supersymmetry and the NS-NS and R-R tadpoles are linked.
On the other hand, this is not sufficient to guarantee the absence of
divergencies, because it is just valid as long as no vertex operators
have been inserted near one end of each worldsheet surface.
Therefore, one demands that the two tadpoles are vanishing separately
(or strictly speaking the one independent one has to be zero).

We will make these general remarks now more precise for the case
of the $\Omega R$-orientifold containing D6-branes at angles. The R-R
tadpoles first have been calculated for the $A$-torus in
\cite{Blumenhagen:2000vk} and for the $B$-torus in the subsequent
paper \cite{Blumenhagen:2000ea}, the NS-NS tadpoles first have been
treated in \cite{Blumenhagen:2001te}.

\begin{figure}
\centering
\includegraphics[width=12cm,height=3cm]{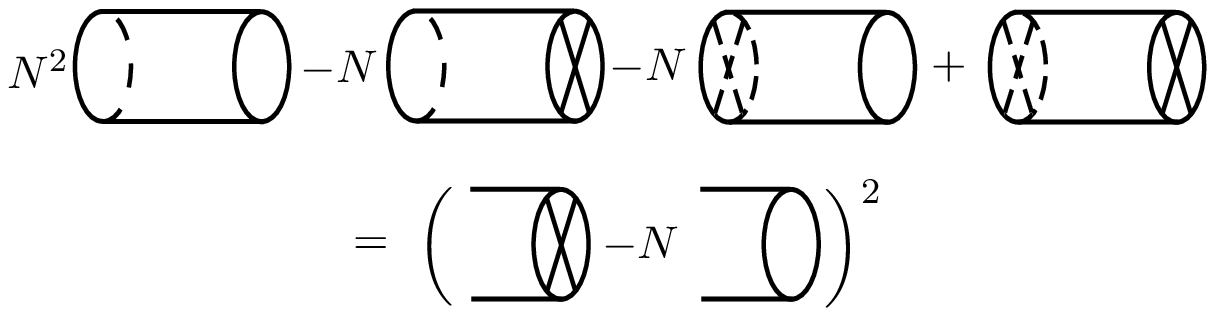}
\caption{Factorization of R-R and NS-NS tadpoles in the loop channel.}\label{fig:perfect_square}
\end{figure}

The orientifold plane is located at the fixed locus of the
geometric action of $\Omega R$, so on the $X$-axis in figure
\ref{fig:torusAB}.
In the tree channel, the total amplitude $\widetilde{\mathcal{Z}}$
for one certain stack of $N$ D6-branes is given by the
following equation:
\begin{equation}\label{eq:treechannel_partfct}
\widetilde{\mathcal{Z}}=\widetilde{\mathcal{K}}+\widetilde{\mathcal{A}}+\widetilde{\mathcal{M}}
=\int\limits_{0}^{\infty}dl\
\Big(\langle D6|N +\langle O6|\Big) e^{-2 \pi l
\mathcal{H}_{\text{closed}}}\Big(|O6\rangle+N|D6\rangle\Big) \ ,
\end{equation}
where $|D6\rangle$ and $|O6\rangle$ are the correctly normalized
boundary states of the $N$ D6-branes and the one orientifold
O6-plane. This sum of boundary states implies in the loop channel
that the contributions of the different amplitudes factorize into
a perfect square, what is schematically shown in figure \ref{fig:perfect_square}, where a
cross symbolically stands for a topological crosscap.
This factorization is very useful for actual computations, because
it implies that it is sufficient to calculate for instance the Cylinder and
Klein bottle amplitudes, and then use them to normalize the boundary
states, after a transformation in the tree channel. By doing so, the
M{\"o}bius strip amplitude is fixed unambiguously, without the
need of an explicit calculation.
\subsection{R-R tadpoles}\label{cha:RtadpolesOR}
\subsubsection{Klein bottle}
In order to find the correct normalization for the orientifold
plane $|O6\rangle$, we will first calculate the R-R part of the
Klein bottle tree channel amplitude which is given by
\begin{equation}\label{eq:Kleinbottle_amplitude}
\mathcal{K}^\text{(NS-NS,$-$)}=4c \int\limits_{0}^{\infty}\frac{dt}{t^3}
\text{Tr}_{\text{(NS-NS,$-$)}}\left(\frac{\Omega R}{2}\ \frac{1+(-1)^F}{2}e^{-2\pi t
\mathcal{H}_{\text{closed}}}\right) ,
\end{equation}
in the loop channel, being reminded that (R-R,$+$) in the tree channel corresponds to
(NS-NS,$-$) in the loop channel. The constant $c$ is given by
$c=V_4/(8\pi^2\alpha')^2$, where $V_4$ is the regularized volume of
the 4-dimensional Minkowski spacetime. In order to evaluate the
trace, one has to determine the Hamilton operator
$\mathcal{H}_{\text{closed}}$, which in the loop channel NS-NS
sector using (\ref{eq:Hamilton_closed}) is just given by
\begin{multline}\label{eq:Hamilton_closed}
    {\mathcal{H}}^{\text{NS-NS}}_{\text{closed}} = ({p^\mu})^2
+\sum_{\mu}{ \left( \sum_{n=1}^\infty{\left
      ( \alpha_{-n}^\mu \alpha_{n}^\mu + \tilde{\alpha}_{-n}^\mu
      \tilde{\alpha}_{n}^\mu \right)} \right. } \\
    \left. + \sum_{r\in\mathbb{Z}+1/2,\ r>0}{\left( r \psi^\mu_{-r}
        \psi^\mu_{r}+r
  \tilde{\psi}^\mu_{-r} \tilde{\psi}^\mu_{r} \right)} \right)  +
  E_0^{\text{NS-NS}}+\mathcal{H}_{\text{lattice, cl.}}\ .
\end{multline}
In order to obtain the zero point energy, we just have to
correctly count the number of complex fermions and bosons in the
sector and then use equation (\ref{eq:zeropoint_energies}), from which we get:
\begin{equation}\label{eq:Hclosed}
E_0^{\text{NS-NS}}=E_0^{\text{NS}, L}+E_0^{\text{NS}, R}=2\cdot4
\left(-\frac{1}{12}-\frac{1}{24}\right)=-1\ .
\end{equation}
The lattice contribution can be found in appendix \ref{cha:appendix_latcontrib_KB}.
The trace over the oscillators and the zero-point energy can be
treated separately from the lattice contribution, it just gives
the standard NS-NS sector $\vartheta$-functions, so altogether
\begin{equation}\label{eq:Kleinbottle_amplitude_OR_loop}
\mathcal{K}^{\text{(NS-NS,$-$)}}=c
\int\limits_{0}^{\infty}\frac{dt}{t^3}\frac{-\tfkto{0}{1/2}^4}{\eta^{12}}
\prod_{I=1}^{3}\left[\sum_{r_I,s_I} e^{-\pi t \left(\frac{\alpha'{s_I}^2}{{R_x^{(I)}}^2}\left(1+6b_I\right)+\frac{r_I^2}{\alpha'}
\left[\left(1+6b_I\right){R_y^{(I)}}^2-2b_I{R_x^{(I)}}^2\right]\right)}\right] ,
\end{equation}
where $b_I$ can be chosen separately for every torus to be $0$ or
$1/2$, meaning an $A$- or $B$-torus, respectively. The argument of
the $\vartheta$ and $\eta$-functions in this equation is $q=e^{-4\pi t}$. The
amplitude can be transformed to the tree channel using $t=1/(4l)$, where
equations (\ref{eq:modulartrans_theta}),
(\ref{eq:modulartrans_eta}) and
(\ref{eq:simple_poissonresummation}) have to be utilized. The result is given by:
\begin{multline}\label{eq:Kleinbottle_amplitude_OR_tree}
\widetilde{\mathcal{K}}^{\text{(R-R,$+$)}}=64c
\int\limits_{0}^{\infty}{dl}\frac{\tfkto{1/2}{0}^4}{\eta^{12}}\\
\prod_{I=1}^{3}\left[\frac{R_x^{(I)}}{\sqrt{1+6b_I}\sqrt{\left(1+6b_I\right){R_y^{(I)}}^2-2b_I{R_x^{(I)}}^2}}
\sum_{r_I,s_I} e^{-4\pi l\left(\frac{{s_I}^2{R_x^{(I)}}^2}{\alpha'\left(1+6b_I\right)}+\frac{r_I^2\alpha'}
{\left(1+6b_I\right){R_y^{(I)}}^2-2b_I{R_x^{(I)}}^2}\right)}\right]\ .
\end{multline}
Here, the $\vartheta$ and $\eta$-functions have the argument
$q=e^{-4\pi l}$. The equation directly allows to determine the
contribution of the Klein bottle to the tadpole, which will be
denoted by $T$. It is just given by the zeroth order
term in the $q$-expansion of the integrand in
(\ref{eq:Kleinbottle_amplitude_OR_tree}). Here, one has to use the
explicit series or product expansions of the $\vartheta$ and
$\eta$-functions (\ref{eq:defthetafktsum}) or
(\ref{eq:defthetafktprod}) and (\ref{eq:defeta}). The result is
given by
\begin{equation} \label{eq:tadpole_OR_KB}
T^{\text{R-R}}_{\widetilde{\mathcal{K}}}=1024 c\prod_{I=1}^{3}\left[\frac{R_x^{(I)}}{\sqrt{1+6b_I}
\sqrt{\left(1+6b_I\right){R_y^{(I)}}^2-2b_I{R_x^{(I)}}^2}}\right]\ .
\end{equation}
and also allows to fix the normalization of the corresponding orientifold plane
\begin{equation}\label{eq:crosscapstate_OR_normdef}
    |O6\rangle=\mathcal{N}_\text{O6}\Big(|O6_\text{NS}\rangle+|O6_\text{R}\rangle\Big)\ ,
\end{equation}
which is simply
\begin{equation}\label{eq:Norm_O6_OR}
\mathcal{N}_\text{O6}=\frac{1}{2}\sqrt{{{T}^{\text{R-R}}_{\widetilde{\mathcal{K}}}}/16}\ .
\end{equation}
\subsubsection{Cylinder}\label{cha:cylinder_OR}
Now we have to calculate the cylinder amplitude in the Ramond tree
channel, where just (R,$+$) contributes, corresponding to the
(NS,$-$) sector in the loop channel. For one stack of branes, the
Cylinder amplitude contains 4 different contributions:
\begin{equation}\label{eq:cylinder_amp_OR_allcontrib}
    \mathcal{A}=\mathcal{A}_{ii}+\mathcal{A}_{i'i'}+\mathcal{A}_{ii'}+\mathcal{A}_{i'i} \ .
\end{equation}
The first term stands for the sector of open strings that stretch
from a brane onto itself, the second one for the sector of strings
that stretch from the $\Omega R$-mirror brane onto itself and the
3rd and 4th term for strings that stretch from the brane to its
mirror brane and vice versa. The first two terms are easy to
obtain, because there is no angle in between the two involved branes.
The first one for the (NS,$-$) sector is given by
\begin{equation}\label{eq:Annulus_aii_general}
    \mathcal{A}_{ii}^\text{(NS,$-$)}=c\int\limits_{0}^{\infty}\frac{dt}{t^3}
\text{Tr}_{\text{D6i-D6i}}^\text{(NS,$-$)}\left(\frac{1}{2}\frac{1+(-1)^F}{2}e^{-2\pi t
\mathcal{H}_{\text{open}}}\right)\ .
\end{equation}
The different normalization factor in front of the integral in
comparison to (\ref{eq:Kleinbottle_amplitude}) comes from the
already performed momentum integration in the compact directions
that is different for open and closed stings. The open string
Hamiltonian is given by equation (\ref{eq:Hamilton_open}). Taking
the trace over the oscillator modes and the zero mode, again leads
to the standard NS-sector $\vartheta$- and $\eta$-functions,
whereas the Kaluza-Klein and winding contributions can be
determined using equation (\ref{eq:ABHamiltonian_lat_annulus}).
This yields altogether for one stack of $N$ D-branes:
\begin{equation}\label{eq:Annulus_amplitude_OR_loop}
\mathcal{A}_{ii}^\text{(NS,$-$)}=\frac{c}{4} N^2
\int\limits_{0}^{\infty}\frac{dt}{t^3}\frac{-\tfkto{0}{1/2}^4}{\eta^{12}}
\prod_{I=1}^{3}\left[\sum_{r_I,s_I} e^{-2\pi t
\frac{r_I^2\left(1-b_I\right)+s_I^2\left[\left(1+2b_I\right){R_x^{(I)}}^2{R_y^{(I)}}^2
-b_I{R_x^{(I)}}^4\right]}
{\left(n_I^2+2b_In_Im_I\right){R_x^{(I)}}^2+m_I^2{R_y^{(I)}}^2}}\right].
\end{equation}
The argument of the $\vartheta$- and $\eta$-functions here is
given by $q=e^{-2\pi t}$. The transformation to the tree channel
by using $t=1/(2l)$ leads to the amplitude
\begin{multline}\label{eq:Annulus_amplitude_OR_tree}
\widetilde{\mathcal{A}}_{ii}^\text{(R,$+$)}=\frac{c}{16} N^2
\int\limits_{0}^{\infty}{dl}\frac{-\tfkto{1/2}{0}^4}{\eta^{12}}\\
\prod_{I=1}^{3}\left[\frac{{\left(n_I^2+2b_In_Im_I\right)\frac{R_x^{(I)}}{R_y^{(I)}}+m_I^2\frac{R_y^{(I)}}{R_x^{(I)}}}}{
\sqrt{1-b_I}\sqrt{\left(1+2b_I\right)-b_I\frac{{R_x^{(I)}}^2}{{R_y^{(I)}}^2}}}\sum_{r_I,s_I}e^{-\pi l \frac{\left(n_I^2+2b_In_Im_I\right){R_x^{(I)}}^2+m_I^2{R_y^{(I)}}^2}
{r_I^2\left(1-b_I\right)+s_I^2\left[\left(1+2b_I\right){R_x^{(I)}}^2{R_y^{(I)}}^2-b_I{R_x^{(I)}}^4\right]} }\right],
\end{multline}
with an argument $q=e^{-4\pi l}$ of the $\vartheta$ and
$\eta$-functions. The expansion in $q$ again leads to the tadpole:
\begin{equation}\label{eq:tadpole_OR_Aii}
T^{\text{R}}_{\widetilde{\mathcal{A}}_{ii}}=-c N^2\prod_{I=1}^{3}\left[\frac{{\left(n_I^2+2b_In_Im_I\right)
\frac{R_x^{(I)}}{R_y^{(I)}}+m_I^2\frac{R_y^{(I)}}{R_x^{(I)}}}}{
\sqrt{1-b_I}\sqrt{\left(1+2b_I\right)-b_I\frac{{R_x^{(I)}}^2}{{R_y^{(I)}}^2}}}\right]\ .
\end{equation}
This is sufficient in order to determine the normalization of a general D6-brane as
\begin{equation}\label{eq:D6_OR_normdef}
    |D6\rangle=\mathcal{N}_\text{D6}\Big(|D6_\text{i, NS}\rangle+|D6_\text{i, R}\rangle\Big)\ ,
\end{equation}
which is given by
\begin{equation}\label{eq:Norm_D6_OR}
\mathcal{N}_\text{D6}=\frac{1}{2}\sqrt{{T^{\text{R}}_{\widetilde{\mathcal{A}}_{ii}}}/(16N^2)}\ .
\end{equation}
In general, there are two different possibilities how to further
proceed. Either, one can determine the other sectors of the
cylinder amplitude, which in general might get quite tedious, or
one can calculate the M{\"o}bius amplitude, which is fixed by the
two normalization factors, and then, by using the property of the
perfect square, directly obtain the tadpole equations. This
procedure indeed is sufficient, if all tadpoles receive
contributions from the orientifold planes\footnote{This usuallly
is the case for orbifold spaces that will be treated in the rest
of the work.}, in the present case of the $\Omega R$-orientifold,
some tadpoles are getting missed, and these are the ones that just
come from the cylinder amplitude.

The mirror brane in terms of $n'$ and $m'$ is related to the
original brane with wrapping numbers $n$ and $m$ by the map
\begin{align}\label{eq:ORmap_nm_ORorientifold}
    &n_I'=n_I+2 b_I m_I\ ,&\\
    &m_I'=-m_I\ .& \nonumber
\end{align}
This simply means that to obtain the amplitude
$\mathcal{A}_{i'i'}$, one just has to replace the $n_I$ and $m_I$
in the Kaluza-Klein and winding sum of
(\ref{eq:Annulus_amplitude_OR_tree}) by $n_I'$ and $m_I'$, because
the $\vartheta$-functions of the oscillator part, according to
equation (\ref{eq:Hamilton_open}) just depend on the relative
angle between the brane which is zero, as it was for
$\mathcal{A}_{ii}$. On the other hand, the Kaluza-Klein and
winding terms also remain unchanged after the map
(\ref{eq:ORmap_nm_ORorientifold}) has been applied, therefore
$\mathcal{A}_{i'i'}=\mathcal{A}_{ii}$. The next amplitude which has
to be calculated is $\mathcal{A}_{ii'}$. This in general is much
more difficult, because the two stacks of D-branes intersect at a
non-vanishing angle. The general amplitude $\mathcal{A}_{a b}$ for
any such angle is calculated in appendix
\ref{cha:appendix_OR_annij}. For our present purpose, we have
to insert the general winding numbers for the $a$-brane into the
tree channel tadpole contribution (\ref{eq:tadpole_OR_Aij}) and
for the $b$-brane the corresponding $\Omega R$-mirror wrapping
numbers (\ref{eq:ORmap_nm_ORorientifold}). Adding up all
contributions (\ref{eq:cylinder_amp_OR_allcontrib}), the overall
cylinder tadpole is given by
\begin{multline}\label{eq:tadpole_OR_Ages}
T^{\text{R}}_{\widetilde{\mathcal{A}}}=-16cN^2\left[\prod_{I=1}^{3}\left({\frac {\left( {{ n_I}}^{2}+2{
b_I}{ m_I}{ n_I} \right) {{ R_x^{(I)}}}^{2}+{{ m_I}}^{2}{{
R_y^{(I)}}}^ {2}}{\sqrt {4{{ R_y^{(I)}}}^{2}-2{ b_I}{{
R_x^{(I)}}}^{2}}{ R_x^{(I)}}}
}\right)\right.\\
\left.+\prod_{I=1}^{3}\left({\frac {\left( {{ n_I}}^{2}+2{
b_I}{ m_I}{ n_I}+2{{ b_I}}^{2}{{ m_I}}^{2} \right) {{
R_x^{(I)}}}^{2}-{{ m_I}}^{2}{{ R_y^{(I)}}}^{2}}{\sqrt {4{{
R_y^{(I)}}}^{2}-2{  b_I}{{ R_x^{(I)}}}^{2}}{
R_x^{(I)}}}}\right)\right]\ .
\end{multline}
With the two contributions
$T^{\text{R-R}}_{\widetilde{\mathcal{K}}}$ and
$T^{\text{R}}_{\widetilde{\mathcal{A}}}$ at hand, we are
able to write down the complete tadpole cancellation equation, which in our case is given
by
$T^{\text{R-R}}_{\widetilde{\mathcal{K}}}=T^{\text{R}}_{\widetilde{\mathcal{A}}}$,
explicitly:
\begin{multline}\label{eq:gestadpol_OR}
-N^2\bigg[{ {{ m_2
}}^{2}{{ m_1}}^{2}\left({ n_3} +{ b_3}{ m_3}\right) ^{2}\frac {{ R_x^{(3)}}{{ R_y^{(1)}}}^{2}{{
R_y^{(2)}}}^{2}} {{ R_x^{(1)}}{ R_x^{(2)}}}}\\
-{{
}{{ m_1}}^{2} {{ m_3}}^{2} \left( { n_2}+{ b_2} { m_2} \right) ^{2}\frac {R_x^{(2)}{{ R_y^{(1)}}}^{2}{{
R_y^{(3)}}}^{2}}{{
R_x^{(1)}}{ R_x^{(3)}}}}
-{{{ m_2}}^{2}{{ m_3}}^{2}
 \left( { n_1} +{ b_1}{ m_1}\right) ^{2}\frac {{
 R_x^{(1)}}{{ R_y^{(2)}}}^{2}{{
R_y^{(3)}}}^{2}}{{ R_x^{(2)}}{ R_x^{(3)}}
}}\\
+{{{m_1}}^{2} \big( {{ b_3}}^{2}{{ m_3}}^{2} \big( 2{{ b_2}}^{2 }{{
m_2}}^{2}+2{ b_2}{ m_2}{ n_2}+{{ n_2}}^{2}
 \big) +{{ b_2}}^{2}{{ m_2}}^{2}\big(2{ b_3}{
m_3}{ n_3}+{{ n_3}}^{2}\big)\big)\frac {{ R_x^{(2)}}{ R_x^{(3)}}{{ R_y^{(1)}}}^{2}}{{
R_x^{(1)}}}}\\
+{m_2}^2\big({b_1}^2{m_1}^2\big(2{b_3}^2{m_3}^2+2{b_3}m_3 n_3+{n_3}^2\big)
+{b_3}^2{m_3}^2\big(2b_1 m_1 n_1+{n_1}^2\big)\big)\frac {{ R_x^{(1)}}{ R_x^{(3)}}{{ R_y^{(2)}}}^{2}}{{ R_x^{(2)}}}\\
 +{{{ m_3}}^{2} \big( {{ b_1}}^{2}{{ m_1}}^{2}
 \big( 2{{ b_2}}^{2}{{ m_2}}^{2}+2{ b_2}{ m_2}{ n_2
}+{{ n_2}}^{2} \big) +{{
b_2}}^{2}{{ m_2}}^{2}\big(2{ b_1}{ m_1}{ n_1 }+{{ n_1}}^{2}\big)
\big) \frac {{ R_x^{(1)}}{ R_x^{(2)}}{{ R_y^{(3)}}}^{2}}{{
R_x^{(3)}}}}\bigg]\\
-\bigg[\Big({b_1}^2{m_1}^2\big(2{b_2}^2{m_2}^2+2b_2 m_2
n_2+{n_2}^2\big)\big(2{b_3}^2{m_3}^2+2b_3 m_3 n_3+{n_3}^2\big)+2n_1 m_1 b_1 \\
\cdot\big({n_3}^2\big(n_2+b_2 m_2\big)^2+2n_3 m_3
b_3\big(n_2+b_2
m_2\big)^2+{b_3}^2{m_3}^2\big({n_2}^2+2{b_2}^2{m_2}^2+2b_2 m_2 n_2
\big)\big)\\
+{n_1}^2\big({n_3}^2\big(n_2+b_2 m_2\big)^2+2m_3 b_3 n_3
\big(n_2+b_2 m_2\big)^2+{b_3}^2{m_3}^2\big(2{b_2}^2{m_2}^2 +2 b_2
m_2 n_2+{n_2}^2\big)\big)\Big)\\
-16\bigg]{R_x^{(1)}}{ R_x^{(2)}} { R_x^{(3)}}=0\ .
\end{multline}
In this equation, the products already have been evaluated and the
resulting terms with different volume factors have been separated.
Furthermore the substitution (\ref{eq:substitution_OR_ann}) has
been applied. It is only possible to solve this tadpole
cancellation equation in general, if all factors in front of the
different volume factors vanish separately. This gives 7 different
equations, but which are not all independent. Actually, the
equations coming from the 4th, the 5th and the 6th term already
are fulfilled if the first 3 equations are satisfied. By using
these first 3 equations on equation 7, this equation can be
drastically reduced and the final set of 4 tadpole equations is
just given by
\begin{align}\label{eq:tadpoles_OR}
    &\sum_{i=1}^{k}{N_i} \prod_{I=1}^{3}\left( { b_I^i}{ m_I^i}+{ n_I^i}
 \right)=16\ ,&\\
&\sum_{i=1}^{k}{N_i}{{ m_2^i}}{{ m_3^i}}\left( { n_1^i} +{ b_1^i}{ m_1^i}\right)=0\ ,&\nonumber\\
&\sum_{i=1}^{k}{N_i}{{ m_1^i}}{{ m_3^i}}\left( { n_2^i} +{ b_2^i}{ m_2^i}\right)=0\ ,&\nonumber\\
&\sum_{i=1}^{k}{N_i}{{ m_1^i}}{{ m_2^i}}\left( { n_3^i} +{ b_3^i}{ m_3^i}\right)=0\ .&\nonumber
\end{align}
This result here already has been generalized to the case of $k$
different stacks of D-branes, each consisting of $N_i$ parallel
branes, and it is equivalent to the one in
\cite{Blumenhagen:2000wh} for the A-torus and to the one in
\cite{Blumenhagen:2000ea} for the B-torus. To see this, one has to
transform the equations into the other chosen basis for the
B-torus via $n'_I\rightarrow n_I+b_I m_I$, $(m'_I+b_I
n'_I)\rightarrow m_I$ and also take into account the different
definition for the second radius $R_y^{(I)\prime}\rightarrow
\sqrt{{R_y^{(I)}}^2-{b_I}^2 {R_x^{(I)}}^2}$, where the unprimed
quantities are the ones of this work.

These tadpole equations also
have a direct interpretation in the T-dual type I theory: the
first equation in (\ref{eq:tadpoles_OR}) demands the
cancellation of the D9-brane and O9-plane charges against each
other, the other three demanding a vanishing of the three possible
types of D5-brane charges.

Connected to this, the R-R tadpole equations can even be
understood by means of topology. They can be basis-independently
written as
\begin{equation}\label{eq:RRtadpole_top}
\sum_{a=1}^k {N}_a\ \left(\pi_a+\pi'_a\right)+Q_q\pi_{\rm Oq}=0\ ,
\end{equation}
where  $\pi_a$ denotes the homological cycle of the wrapped
D$6_a$-branes and $\pi'_a$ that of its $\Omega R$-mirrors.
Furthermore, $\pi_{\rm Oq}$ denotes the cycle, the orientifold
planes are wrapping on all three 2-tori. $Q_q$ is the charge of the
orientifold plane that is fixed to be $Q_6=-4$ for four non-compact
dimensions.

From this observation, it was possible to generalize intersecting
brane model building to Calabi-Yau manifolds
\cite{Blumenhagen:2002wn, Blumenhagen:2002vp}that do not have a
simple description by conformal field theory, but still have known
homology.

\subsubsection{M{\"o}bius strip}\label{cha:moebius_OR}
In this chapter, we also write down the M{\"o}bius amplitude in
the tree channel, which is far simpler to obtain than the cylinder
amplitude. In particular, it will be needed for the
$\mathbb{Z}_N$-orbifolds. The M{\"o}bius amplitude can be
calculated directly from the overlap of a $|D6\rangle$ and a
$|O6\rangle$ boundary state to be
\begin{multline}\label{eq:Moebius_amplitude_OR_tree}
\widetilde{\mathcal{M}}_{[i]}^\text{(R,$+$)}=\pm N
\int\limits_{0}^{\infty}{dl}\ 2\cdot2\cdot2\ \mathcal{N}_\text{D6}\ \mathcal{N}_\text{O6}\ \\
 \cdot \gamma\
\frac{\tfkto{1/2}{0}\tfkto{1/2}
{-\kappa_1}\tfkto{1/2}{-\kappa_2}\tfkto{1/2}{-\kappa_3}}{\tfkto{1/2}{1/2-\kappa_1}
\tfkto{1/2}{1/2-\kappa_2}\tfkto{1/2}{1/2-\kappa_3}\eta^{3}}\ ,
\end{multline}
where the argument of the modular functions is again given by
$q=e^{-4\pi l}$. This amplitude needs some explanation: the three
factors of 2 come from, firstly, the two possible spin structures,
secondly, the two $\Omega R$-mirrors and finally, the
interchangeability of the bra- and ket-vector. The bracket $[i]$
indicates that the M\"obius amplitude is already taken over both
branes contained in the equivalence class of the brane under
consideration, the brane and its $\Omega R$-mirror.

Moreover, the product of the $\vartheta$-functions is formally
equivalent to the one of the cylinder amplitude which has been
derived in \ref{cha:appendix_OR_annij}, but the meaning of the
moding is different, the angle $\varphi_I=\pi \kappa_I$ means the
angle that the considered orientifold plane spans with the
specific D-brane. Finally, the constant $\gamma$ has been
introduced in order to cancel the contribution of the bosonic zero-modes by
hand. After the expansion in $q$ and the use of the two
simplifications (\ref{eq:expansion_theta_mitwinkel}), it turns out
that $\gamma=2^3\prod_I \sin (\pi \kappa_I)$ and the contribution
from the modular functions in terms of the wrapping numbers
together with $\gamma$ generally in lowest order is given by
\begin{multline}
16\prod_{I=1}^{3}\cos (\pi \kappa_I)=\\
16\prod_{I=1}^{3} \frac{{R_x^{(I)}}^2\left(n_I^\text{D}
n_I^\text{O}+b_I\left(n_I^\text{D} m_I^\text{O}+m_I^\text{D}
n_I^\text{O}\right)\right)+{R_y^{(I)}}^2m_I^\text{D} m_I^\text{O}}
{\sqrt{\left({n_I^\text{D}}^2+2b_I n_I^\text{D} m_I^\text{D}\right){R_x^{(I)}}^2+{m_I^\text{D}}^2{R_y^{(I)}}^2}
\sqrt{\left({n_I^\text{O}}^2+2b_I n_I^\text{O} m_I^\text{O}\right){R_x^{(I)}}^2+{m_I^\text{O}}^2{R_y^{(I)}}^2}},
\end{multline}
where the superscript $\text{D}$ stands for the D-brane and
$\text{O}$ for the orientifold plane. This procedure assumes that
the orientifold plane can be characterized by the 1-cycles it is
wrapping on the torus, similarly to the D6-branes. In the present
case, the wrapping numbers of the O6-plane are simply given by
$n_1^\text{O}=n_2^\text{O}=n_3^\text{O}=1$ and
$m_1^\text{O}=m_2^\text{O}=m_3^\text{O}=0$.

The resulting {M\"o}bius tadpole together with the Klein bottle
tapole lead exacly to the same tadpole equation as the first one
in (\ref{eq:tadpoles_OR}), but does not reproduce the other three
ones, as explained already.
\subsection{NS-NS tadpoles}\label{cha:NStadpoles_OR}
In the following, we are going to discuss the NS-NS tadpoles.
These will be deduced in much less detail, because the methods are
very similar. To keep the equations of manageable size, the case
$b_I=0$ will be chosen during the computation, but the final
result will be given for the general case.
\subsubsection{Klein bottle}
Starting again with the Klein bottle amplitude, we should first
take a look at table \ref{tab:transformation_sectors}. One
observes that the two different spin structures of the tree
channel NS-NS sector both contributing to the tadpole of interest,
correspond to the two loop channels (NS-NS,$+$) and (R-R,$+$).
Therefore, the only change as compared to
(\ref{eq:Kleinbottle_amplitude_OR_loop}) is that the theta
function $-\vartheta[0, 1/2]^4$ have to be replaced by the sum
$\vartheta[0, 0]^4-\vartheta[1/2, 0]^4$, so
\begin{multline}\label{eq:Kleinbottle_amplitude_OR_loopNS}
\mathcal{K}^{\text{(NS-NS,$+$)}}+\mathcal{K}^{\text{(R-R,$+$)}}=c
\int\limits_{0}^{\infty}\frac{dt}{t^3}\frac{\tfkto{0}{0}^4-\tfkto{1/2}{0}^4}{\eta^{12}}
\prod_{I=1}^{3}\left[\sum_{r_I,s_I} e^{-\pi t \left(\frac{\alpha'{s_I}^2}{{R_x^{(I)}}^2}
+\frac{r_I^2{R_y^{(I)}}^2}{\alpha'}
\right)}\right].
\end{multline}
The straightforward computation leads to the tree channel tadpole:
\begin{equation} \label{eq:tadpole_OR_KB_NS}
T^{\text{NS-NS}}_{\widetilde{\mathcal{K}}}=-1024{\frac {c{ R_x^{(1)}}{ R_x^{(2)}}{ R_x^{(3)}}}{{ R_y^{(1)}}{
R_y^{(2)}} { R_y^{(3)}}}}\ .
\end{equation}
\subsubsection{Cylinder}
Like for the R-R tadpole, the complete cylinder tree channel
NS-tadpole for one stack of branes is a sum of the four
contributions (\ref{eq:cylinder_amp_OR_allcontrib}). The two
contributions, where a string goes from one brane onto itself,
$\widetilde{\mathcal{A}}_{ii}$ and
$\widetilde{\mathcal{A}}_{i'i'}$, can be calculated like in
section \ref{cha:cylinder_OR}, if we again substitute the theta
function coming from the fermions  by the ones (NS,$+$) and
(R,$+$), the Kaluza-Klein and winding sum remains unchanged. After
the transformation into the tree channel and the expansion in $q$,
the cylinder tadpole from $\widetilde{\mathcal{A}}_{ii}$ is given by
\begin{equation}\label{eq:tadpole_OR_Aii_NS}
T^{\text{NS}}_{\widetilde{\mathcal{A}}_{ii}}=
T^{\text{NS}}_{\widetilde{\mathcal{A}}_{i'i'}}=
-c{N}^{2}{\prod_{I=1}^{3}\frac { \left( {{ R_x^{(I)}}}^{2}{{ n_I}}^{2}+{{
R_y^{(I)}}}^{2}{{ m_I }}^{2} \right)}{{R_x^{(I)}}{R_y^{(I)}}}}\ .
\end{equation}
The general contribution with non-vanishing angle
$\widetilde{A}_{a b}$ is more difficult to obtain, it is being
calculated in appendix \ref{cha:appendix_OR_annij_NS}. In our
case, the two contributions $\widetilde{\mathcal{A}}_{ii'}$ and
$\widetilde{\mathcal{A}}_{i'i}$ then directly can be written down from
the general expression (\ref{eq:tadpole_OR_AijNS}), if we proceed
precisely as for the R-tadpole and in particular use the map
(\ref{eq:ORmap_nm_ORorientifold}) for the $\Omega R$-mirror brane.
The final result for the cylinder NS-tadpole is given by:
\begin{multline}\label{eq:gestadpol_OR_NS}
T^{\text{NS}}_{\widetilde{\mathcal{A}}}=
-{\frac{c{N}^{2}}{\prod_{I=1}^{3}{R_x^{(I)}}{R_y^{(I)}}}}\left[
{L_1}^2 {L_2}^2 {L_3}^2+\frac{{L_2}^2{L_3}^2}{{L_1}^2}\left({n_1}^2{R_x^{(1)}}^2
-{m_1}^2{R_y^{(1)}}^2\right)\right. \\
\left. +\frac{{L_1}^2{L_3}^2}{{L_2}^2}\left({n_2}^2{R_x^{(2)}}^2
-{m_2}^2{R_y^{(2)}}^2\right)
+\frac{{L_1}^2{L_2}^2}{{L_3}^2}\left({n_3}^2{R_x^{(3)}}^2
-{m_3}^2{R_y^{(3)}}^2\right)\right]\ ,
\end{multline}
where
\begin{equation}\label{eq:def_w}
    L_I=\sqrt{{n_I}^2{R_x^{(I)}}^2+{m_I}^2{R_y^{(I)}}^2}\ .\nonumber
\end{equation}
In this equation, $L_I$ is the length of the D-brane in
consideration. Interestingly, the first term in the tadpole is
different from the three others. We can understand this easily, if
we are being reminded, of what are the massless fields in the
NS-NS-sector in our model. In general, there is the
four-dimensional dilaton and the 21 $\Omega{\cal R}$ invariant
components of the internal metric and the internal NS-NS two form
flux, but in our factorized ansatz with three 2-tori, only 9
moduli are evident. These are the six radions $R_x^{(I)}$ and
$R_y^{(I)}$, which are related to the size of the internal
dimensions and the two-form flux $b^I_{12}$ on each $T^2$.
Consequently, we can already guess at this point that the first
term in (\ref{eq:gestadpol_OR_NS}) is related to the dilaton,
whereas the three other terms come from the radions.

From the cylinder tadpole together with the Klein bottle tadpole,
we are now able to write down the full tadpole equations, using
the property that the full tadpole is a sum of perfect squares.
Naively, this seems to be problematic, because the volume factors
in the Klein bottle tadpole seem to be different from the 4
contributions of the cylinder tadpole. On the other hand, thinking
in terms of geometry, the location of the branes in the cylinder
tadpole (\ref{eq:gestadpol_OR_NS}) comes up in terms of the
winding numbers $n_I$ and $m_I$. The orientifold plane is located
on the X-axis, so it has the winding numbers $n_I=1$ and $m_I=0$.
If we insert this into the $L_I$ and the terms within the small
brackets in the cylinder tadpole, we see that all terms indeed
lead to the same volume factor. This also means, that we do have a
second problem: we do not know, to which term the Klein bottle
tadpole has to be assigned, such that the perfect square can be
written down. The only way to answer this question is by
calculating the M{\"o}bius amplitude, but the detailed calculation
is being omitted at this point. The result is that the Klein
bottle amplitude contributes equally to all four tadpoles. With
this information, the tadpoles finally can be written down,
starting with the dilaton tadpole and again allowing for both
cases $b_I=0$ and $b_I=1/2$ and generalizing for $k$ stacks:
\begin{equation}\label{eq:dilaton_tadpole_OR}
     \langle\phi\rangle_D=
 {1\over \sqrt{{\rm Vol}(T^6)}}\left( \sum_{a=1}^k  N_a\,
    {\rm Vol}({\rm D}6_a)
           -16\, {\rm Vol}({\rm O}6) \right)
\end{equation}
with
\begin{equation}\label{eq:volumea_OR}
{\rm Vol}({\rm D}6_a)=\prod_{I=1}^3 L_I({\rm D}6_a)=
        \prod_{I=1}^3 \sqrt{\left(\left(n_I^a\right)^2+2 b_I n_I^a m_I^a \right)
        {R_x^{(I)}}^2+ \left(m_I^a {R_y^{(I)}}\right)^2}
\end{equation}
and
\begin{equation}\label{eq:volumeb_OR}
{\rm Vol}({\rm O}6)=\prod_{I=1}^3 L_I({\rm O}6)=
        \prod_{I=1}^3 R_x^{(I)}\ .
\end{equation}
The interpretation of this tadpole is simple, it just is a
bookkeeping calculation of all volumes of both the D6-branes and
the orientifold planes, the latter ones entering with a negative
sign, as in the case of the R-tadpole. The dilaton tadpole is
nothing else but the four-dimensional total tension in appropriate
units. Interestingly, it is possible to express this tadpole
entirely in terms of the imaginary part of the complex structure
moduli $U_2^I$ of the three 2-tori:
\begin{equation}\label{eq:dilaton_tp_OR_U}
     \langle \phi \rangle_D= \left( \sum_{a=1}^k  N_a\, \prod_{I=1}^3
   \sqrt{\left(n_I^a+b_I m_I^a\right)^2\frac{1}{U_2^I}+\left(m_I^a\right)^2{U_2^I}}
          -16\, \prod_{I=1}^3 \sqrt{\frac{1}{U_2^I}} \right) .
\end{equation}
We can understand this in realizing that the boundary and
cross-cap states only couple to the left-right symmetric states of
the closed string Hilbert space. The complex structure moduli
are indeed left-right symmetric, whereas the K{\"a}hler
moduli appear in the left-right asymmetric sector, i.e.
D-branes and orientifold O6-planes only couple to the complex
structure moduli. This is reversed in the T-dual type I picture, where the
tadpole only depends on the K{\"a}hler moduli.

Now, we will discuss the remaining three radion tadpoles from
(\ref{eq:gestadpol_OR_NS}), which in the language of complex
structure and K{\"a}hler moduli correspond to tadpoles of the
imaginary part of the tree complex structures:
\begin{equation}\label{eq:radion_tp_OR}
    \langle U^I_2 \rangle_D=
 {1\over \sqrt{{\rm Vol}(T^6)}}\left( \sum_{a=1}^k  N_a\,
           v_I({\rm D}6_a) \frac{L_J({\rm D}6_a)\, L_K({\rm D}6_a)}{L_I({\rm D}6_a)}
           -16\, {\rm Vol}({\rm O}6) \right)\ ,
\end{equation}
with $I\neq J\neq K\neq I$ and
\begin{equation}\label{eq:def_v_I}
v_I({\rm D}6_a)=\left(\left(n_I^a\right)^2+2 b_I n_I^a m_I^a +2 {\left(b_I m_I^a\right)}^2\right)
        {R_x^{(I)}}^2- \left(m_I^a {R_y^{(I)}}\right)^2 \ .
\end{equation}
Also these tadpoles can be expressed entirely in terms of the
imaginary part of the complex structure moduli, $U_2^I$.
Concerning type II models which have been considered in similar
constructions \cite{Aldazabal:2000dg} , too,  one needs to regard
extra tadpoles for the real parts $U_1^I$, which cancel in type I.
Looking more closely at the 4 NS-tadpoles, we realize that it is
possible to express all of them as being derivatives from just one
scalar potential in the string frame:
\begin{equation}\label{eq:scalar_pot_OR_NS}
    V(\phi,U_2^I)=  e^{-\phi}\,\left( \sum_{a=1}^k  N_a\, \prod_{I=1}^3
   \sqrt{\left(n_I^a+b_I m_I^a\right)^2\frac{1}{U_2^I}+\left(m_I^a\right)^2{U_2^I}}
          -16\, \prod_{I=1}^3 \sqrt{\frac{1}{U_2^I}} \right)\ ,
\end{equation}
meaning
\begin{equation}\label{eq:ableintung_pot_OR_NS}
    \langle \phi \rangle_D \sim {\partial V \over \partial \phi}, \quad\quad
\langle U^I_2 \rangle_D \sim {\partial V \over \partial U^I_2}\ .
\end{equation}
Comparing with the type II potential, the only change would be in
erasing the term coming from the orientifold planes, and three
more tadpoles would appear due to $\langle U_1^I
\rangle_D\sim\partial V/\partial U_1^I$. As a side remark,
although this potential is leading order in string perturbation
theory, it already contains all higher orders in the complex
structure moduli. So it is exact in these moduli, though we have
only explicitly computed their one-point function, this fact needs
a careful interpretation.

The term of the scalar potential coming from the D-branes also
could have been anticipated from different considerations, as the
source for the dilaton is just given by the tension of the branes,
or to first order, by their volumes. The potential then arises by
the Dirac-Born-Infeld action for a D9$_a$-brane with a constant
$U(1)$ and two-form flux
\begin{equation}\label{eq:Born_Infeld}
{\cal S}_{\rm DBI} = -
\text{T}_p \int_{D9_a}{d^{10}x\ e^{-\phi} \sqrt{ {\rm det} \left ( G + (
F_a+B ) \right) } }\ ,
\end{equation}
where $\text{T}_p$ stands for the Dp-brane tension
\begin{equation}\label{eq:tension}
    \text{T}_p = {\sqrt{\pi} \over 16\kappa_0} \left( 4\pi^2
\alpha' \right)^{(11-p)/2} .
\end{equation}
If we assume that all background fields are block-diagonal in
terms of the two-dimensional tori, then they take the constant
values \cite{Blumenhagen:2000fp}
\begin{align}\label{eq:constvalues_background_fields}
&G^{ij} = \delta^{ij}, \quad \left( F_a^I
\right)^{ij} = {m_a^I \over n_a^I R_x^I
\sqrt{{R_y^{(I)}}^2-{b_I}^2 {R_x^{(I)}}^2}} \epsilon^{ij},&\\
&\left(B^I \right)^{ij} = { b^I \over R_x^I \sqrt{{R_y^{(I)}}^2-{b_I}^2 {R_x^{(I)}}^2}}
\epsilon^{ij}\ .& \nonumber
\end{align}
As a first step, one has to integrate out the internal six dimensions and take
care of the fact that the brane wraps each torus $n_I^a$ times.
Next, one only has to apply T-duality and then arrives at the
same D-brane term of the derived potential.

We now come to the conclusions. As mentioned in the introduction
already, a non-vanishing tadpole in perturbative field theory can
be understood in such a way that the tree level value of the
corresponding field has not been chosen at the minimum of the
potential. Thus, even if higher loop corrections are formally
computable, their meaning is very questionable, as fluctuations
might be large, no matter how small the coupling constant actually
is. As an resulting effect, the theory is driven away to a stable
minimum, where now a meaningful perturbation theory is possible.
Trying to apply this idea to our model, the only point where all
four tadpoles vanish is at $U^I_2$=$\infty$. This is due to the R-R
tadpole cancellation condition and the triangle inequality, and
shows that a partial breaking of supersymmetry in ${\cal N}$=4
vacua by introducing relative angles between the D6-branes (or by
the presence of magnetic fluxes on D9-branes) seems to be
impossible, although such possibilities have been worked out in
${\cal N}$=2 type II and heterotic vacua \cite{Antoniadis:1996vb,
Ferrara:1996xi, Taylor:1999ii, Curio:2000sc}.

The potential displays a runaway behavior, which is quite
typical for non-supersymmetric string models. The complex
structure is dynamically pushed to its degenerate limit; all
branes lie along the $X_I$ axes and the $Y_I$ directions shrink to
zero,but still keeping the volume fixed. In other words, the positive
tension of the branes pulls the tori towards the $X_I$-axes. The
slope of the runaway behavior should be set by the tension
proportional to the string scale, so we expect the system to decay
quickly. The possibility that the potential shows a slow-roll
behavior, as it is required by inflationary cosmological models
will be discussed in chapter \ref{cha:Inflation}.

Furthermore, there is a second, related problem: if the $Y_I$
shrink to zero, and the angle between certain branes decreases, at
some point open string tachyons, if not present already in the
construction, dynamically appear and start to propagate at the
open string tree level, this then indicates the decay of the brane
configuration one has started with. Several aspects of
this problem are discussed in some more detail in
\cite{Rabadan:2001mt}.

Our observation has strong consequences for all toroidal
intersecting brane world models. These usually require a tuning of
the parameters at tree-level and implicitly assume the global
stability of the background geometry. But this geometry is
determined by the closed string moduli, and if these display a runaway behavior,
there is a contradiction. The instabilities translate back via
T-duality into a dynamical decompactification towards the
ten-dimensional supersymmetric vacuum.

To end this discussion, every non-supersymmetric toroidal model
seems to be driven back to the degenerate supersymmetric vacuum.
This also matches the observations of \cite{Harvey:2001wm,
Vafa:2001ra, Dabholkar:2001wn, Adams:2001sv} within the context of
closed string tachyon condensation, where these problems are
discussed using elaborate tools from quiver diagrams, RG-flows and
the c-theorem.


\section{Intersecting D6-branes on $\mathbb{Z}_N$-orientifolds}\label{cha:Zn_orientifolds}

The models that we want to consider on the one hand shall be
simple and completely solvable. On the other hand, they shall have
the ingredients to break down supersymmetry in such a way that
either $\mathcal{N}$=1 or even completely broken $\mathcal{N}$=0
supersymmetry in the effective 4-dimensional theory is possible.

To better understand this from string theory, one carefully has to
distinguish in between the closed and open string sectors. In this
section, we are interested in the closed string sector which does
not notice the presence of D-branes. Therefore, the amount of
conserved supersymmetry in this sector depends first on the
specific model and a possible gauging (like $\Omega R$). Secondly,
it has to be taken into account, how much of the remaining
supercharges are conserved by the specific spacetime background.

Following \cite{Dixon:1986jc}, the classification of cyclic
orbifold groups preserving $\mathcal{N}$=1 in 4 dimensions for the
heterotic string is given in table
\ref{tab:Orbifold_classsification}. It is important to notice that
this classification corresponds to a preservation of
$\mathcal{N}$=2 in 4 dimensions in the case of type II models. For
$\Omega R$-orientifolds, the number of supercharges of the closed
string theory is reduced by half, so yielding $\mathcal{N}$=1 for
the orbifolded models of the given table.
Therefore, models of the type
\begin{equation}\label{ModeltypeIIA}
{{\rm Type\ IIA}\ {\rm on}\ T^{6} \over \{G+\Omega R G\} }
\end{equation}
will be discussed in the following chapters. The action for a
discrete $G=\mathbb{Z}_N$ group can be explicitly given by
\begin{equation}\label{eq:actOrbifoldIIB}
     Z^L_I\to e^{2\pi i {v_I}}\,  Z^L_I, \quad\quad
            Z^R_I\to e^{2\pi i {v_I}}\,  Z^R_I\ .
\end{equation}
These models are mapped under T-duality to asymmetric type I
orbifolds where the K\"ahler moduli are frozen.
\begin{table}
\centering
\sloppy
\renewcommand{\arraystretch}{1.2}
\begin{tabular}{|c||p{120pt}|}
  \hline
  Orbifold group & Action $(v_1, v_2, v_3)$\\
  \hline
  \hline
  $\mathbb{Z}_3$ & $(1/3,\ \ 1/3,\ \ -2/3)$ \\
  $\mathbb{Z}_4$ & $(1/4,\ \ 1/4,\ \ -1/2)$ \\
  $\mathbb{Z}_6$ & $(1/6,\ \ 1/6,\ \ -1/3)$ \\
  $\mathbb{Z}_6'$ & $(1/6,\ \ 1/3,\ \ -1/2)$ \\
  $\mathbb{Z}_7$ & $(1/7,\ \ 2/7,\ \ -3/7)$ \\
  $\mathbb{Z}_8$ & $(1/8,\ \ 3/8,\ \ -1/2)$ \\
  $\mathbb{Z}_8'$ & $(1/8,\ \ 1/4,\ \ -3/8)$ \\
  $\mathbb{Z}_{12}$ & $(1/12,\ 1/3, -5/12)$ \\
  $\mathbb{Z}_{12}'$ & $(1/12,\ 5/12,-1/2)$ \\
  \hline
\end{tabular}
\caption{$\mathbb{Z}_N$-orbifold groups that preserve $\mathcal{N}$=2 in 4 dimensions for type II theory.}
\label{tab:Orbifold_classsification}
\end{table}
Looking at the classification table
\ref{tab:Orbifold_classsification} in some more detail, it can be
observed that the sum of the angles on the three 2-tori for all
these orbifolds is zero. In the phenomenological model building of
chapters \ref{cha:Z3} and \ref{cha:Z4}, we will mainly concentrate
on the first two entries of the table, the $\mathbb{Z}_3$ and
$\mathbb{Z}_4$ groups, although an extension of the calculations
to higher groups could resolve several emerging problems, but
technically is even more difficult.

Orbifold constructions have not been first introduced in
intersecting brane world scenarios with arbitrary angles. The
orbifolding technique has already been used in some earlier
papers, where D-branes have been introduced on top of the
orientifold planes, see for instance \cite{Blumenhagen:1999md,
Blumenhagen:1999ev, Blumenhagen:1999db, Pradisi:1999ii,
Pradisi:2000dk, Forste:2000hx}. By doing so, the tadpoles have
been cancelled locally. These models of course are generally
supersymmetric.
\subsection{R-R and NS-NS tadpoles}\label{cha:tadpolesR_NS_ZN}
In order to determine the tadpoles of the model, one has to insert
the additional projector
\begin{equation}\label{eq:projectorZn}
    \mathbf{P}=\frac{1+\Theta+\ldots+\Theta^{N-1}}{N}
\end{equation}
into the trace of the four one-loop string amplitudes
(\ref{eq:part_fct_oneloop}). This means for instance for the
Klein-bottle amplitude \cite{Blumenhagen:1999md,
Blumenhagen:1999db, Blumenhagen:1999ev}:
\begin{equation}\label{eq:kb_amplitudeZ3}
\mathcal{K}=4c \int\limits_{0}^{\infty}\frac{dt}{t^3}
\text{Tr}_\text{U+T}\left(\frac{\Omega R}{2}\
\frac{1+\Theta+\ldots+\Theta^{N-1}}{N}\ \frac{1+(-1)^F}{2}e^{-2\pi
t \mathcal{H}_{\text{closed}}}\right) ,
\end{equation}
where in general, both untwisted (U) and twisted (T) sectors have
to be evaluated. In the tree channel, this can be understood in a
simple manner: Generally, under both the $\mathbb{Z}_N$ and
$\Omega R$ symmetry, the branes are organized in orbits of length
$2N$. Such an orbit constitutes an equivalence class $[a]$ of
D6$_a$-branes and in the following will be denoted by
$[(n_I^a,m_I^a)]$.

In order to determine the tadpoles in the untwisted sector, it is
possible to proceed very close to the $\Omega R$ orientifold, with
the important difference that we do not have to calculate the full
cylinder amplitude (\ref{eq:cylinder_amp_OR_allcontrib}), because
all tadpoles receive a contribution from the orientifold plane and
we can use the property of the perfect square and just need the
normalization of the $\widetilde{A}_{ii}$ tree level amplitude. It
would actually be very tedious to calculate the whole cylinder
amplitude, as the orbit here has a length $2N$, not just a length
two as for the $\Omega R$-orientifold.

Taking up the tree channel picture of orbits, it is very simple to
generalize the calculation procedure of the two sections
(\ref{cha:RtadpolesOR}) and (\ref{cha:NStadpoles_OR}). Some
important points have to be taken care of:
\begin{enumerate}
    \item In general, there will not be only one orientifold plane
anymore: in the discussed $\Omega R$-models, orientifold planes
geometrically have been defined as being the fixed loci
$\text{Fix}(R)|_{T^6}$ of the anti-holomorphic involution $R$ on
the 6-torus. If the background space now is orbifolded, the whole
orientifold group is generated by both $\Omega R$ and $\Theta$.
Then the fixed locus on this quotient space also can be understood
as being the orbit of the fixed locus on the torus in addition to
the orbit of $\text{Fix}(\Theta R)|_{T^6}$, or more explicitly,
\begin{equation}\label{eq:fixed_locus_orbifold}
\text{Fix}(R)|_\text{orbifold}=\bigcup_{i=0}^{N-1}\Theta^i
\left(\text{Fix}(R)|_{T^6}\right)\cup
\Theta^i\left(\text{Fix}(\Theta R)|_{T^6}\right)\ .
\end{equation}
This is due to the relation
\begin{equation}\label{eq:Thetaeinhalb}
\Theta^{1/2}R\Theta^{-1/2}=\Theta R\ .
\end{equation}
There is an important subtlety that should be mentioned: only if
$N\in 2\mathbb{Z}+1$, then $\Theta^{1/2}\in \mathbb{Z}$, and
consequently, the two orbits in (\ref{eq:fixed_locus_orbifold})
are identical. Hence the $\mathbb{Z}_3$ and the  $\mathbb{Z}_4$
models of the following chapters will include quite different
features.

Every one of the N terms $[\Theta^i
\left(\text{Fix}(R)|_{T^6}\right)\cup
\Theta^i\left(\text{Fix}(\Theta R)|_{T^6}\right)]$ can be treated
as a separated orientifold plane and leads to a certain volume
factor in the Klein bottle amplitude that in the end have to be
summed up. Then, one can assign a normalization to every
one of these $N$ orientifold planes, corresponding to a certain
boundary state $|O6_i\rangle$. The M\"obius amplitude in the tree
channel can be obtained by the sum of $N$ distinct amplitudes of
the form (\ref{eq:Moebius_amplitude_OR_tree}), where the
normalization factor for the orientifold plane
$\mathcal{N}_{\text{O6}_i}$ has to be substituted.

\item The Kaluza-Klein and winding contributions have to be
calculated separately for every one of these $N$ orientifold
planes, the reason lying in the fact that they have to respect
different symmetries $\Omega R \Theta^i$ for $i=0,\ldots (N-1)$.
\end{enumerate}
First, one has to calculate the tadpoles coming from the untwisted
sector in the loop channel. If there are additionally
contributions from the twisted sectors of the loop channel, a
twist operator has to be inserted. This leads to theta functions
with different characteristics $a$ in the loop channel, where $a$
simply is equivalent to the twist. The Kaluza-Klein and winding
contributions might alter as well. Some other subtleties will be
discussed for the specific example of the $\mathbb{Z}_3$- and
$\mathbb{Z}_4$-orientifolds in the chapters \ref{cha:RRz3},
\ref{cha:nsnsz3} and \ref{cha:z4modbuild}. The R-R tadpole often
can be obtained more easily using the description in terms of
homology, as can be seen for example for the
$\mathbb{Z}_4$-orientifold in chapter \ref{cha:Z4}.

\section{Massless closed and open string spectra}\label{cha:massless_chiral_nonchiral_general}
In order to search for interesting phenomenological models, it is
unavoidable to determine the low energy effective spectrum in four
dimensions. The massive string excitations are organized in units
of the string scale $\alpha'^{-1/2}$, so if we stick to the usual
picture of a very high string scale, then the massive modes should
be negligible at energies that today's colliders might possibly
achieve. From the string theoretical point of view, there are two
different sectors that differ fundamentally, the open and closed
string sectors. The closed string sector always includes the whole
dimension of 10-dimensional spacetime, the so-called bulk, that in
the picture of this work is factorized into a 4-dimensional flat
Minkowski and a 6-dimensional compact space. It includes the
supergravity multiplet and the dilaton, so it is fair to call it
the gravitational sector. In contrast to this, the open string
sector is determined by the specific D-brane content. In the
modern understanding of string theory, they carry the gauge fields
on their worldvolume (which in our case is 7-dimensional and
always covers the whole Minkowski space) and just at the
intersections on the compact space, there are chiral fermions
\cite{Berkooz:1996aa}, so the matter content of the standard model
and its possible extensions. All these different sectors will be
discussed now in more detail for the toroidal and orbifolded
$\Omega R$-orientifolds.
\subsection{Closed string spectrum}
The closed string sector does not notice the presence of D-branes.
Therefore, if one for now ignores the backreaction, the massless
spectrum only depends on the chosen spacetime background and
its moduli.

In order to find the massless states, one first has to compute the
overall ground state energy in the sector of interest (that might
be twisted as well) by the general formulae
(\ref{eq:zeropoint_energies}) for all bosons, NS- and R-sector
fermions (with a general moding $\kappa$ corresponding to the
twist).

In the untwisted sector, the left and right moving massless
states that are $\Theta$-invariant have to be symmetrized and
antisymmetrized under $\Omega R$ in both NS-NS and R-R sectors
separately\footnote{This statement is valid for the purely toroidal $\Omega
R$-orientifold as well, if one sets $\Theta=\text{Id}$.}. Then the NS-NS
sector always contributes the dilaton, the graviton and a certain
number of neutral chiral multiplets.

In the twisted sector, the transformation properties of the fixed
points, where the fields are localized, plays an important role:
the fixed points that are separately invariant under $\Theta$ and
$R$ are being treated as in the untwisted sector, they have to be
symmetrized and antisymmetrized, too. The fixed points that are
just invariant under a combination of $\Theta$ and $R$ require
less symmetrization and the ones that are not invariant under
$\Theta$ or $R$ no symmetrization at all. By this procedure, the
total number of chiral plus vector multiplets turns out to be the
sum of the two Hodge numbers $h_{1,1}+h_{2,1}$ , although the
distribution between chiral and vector multiplets depends on the
specific model, or in other words the choice of 2-tori. This is
summed up for the models of interest in table \ref{tab:closed_string_spectra_zn}.
\begin{table}
\centering
\sloppy
\renewcommand{\arraystretch}{1.5}
\begin{tabular}{|c|c|c|c|c|}
\hline
Orbifold group& Model & Untwisted & $\Theta+\Theta^{-1}$ twisted& $\Theta^2 \left(+\Theta^{-2}\right)$ twisted\\
\hline
\hline
$\mathbb{Z}_3$ & {\bf AAA} & 9{\mbox C}+$g_{\mu\nu}$+$\Phi$& 14{\mbox C}+13{\mbox V} & absent\\
$\mathbb{Z}_3$ & {\bf AAB} & 9{\mbox C}+$g_{\mu\nu}$+$\Phi$ & 15{\mbox C}+12{\mbox V} & absent\\
$\mathbb{Z}_3$ & {\bf ABB} & 9{\mbox C}+$g_{\mu\nu}$+$\Phi$ & 18{\mbox C}+9{\mbox V} & absent\\
$\mathbb{Z}_3$ & {\bf BBB} & 9{\mbox C}+$g_{\mu\nu}$+$\Phi$ & 27{\mbox C} & absent\\
$\mathbb{Z}_4$ & {\bf ABA} & 6{\mbox C}+$g_{\mu\nu}$+$\Phi$ & 16{\mbox C} & 15{\mbox C}+1{\mbox V}\\
$\mathbb{Z}_4$ & {\bf ABB} & 6{\mbox C}+$g_{\mu\nu}$+$\Phi$& 12{\mbox C}+4{\mbox V} & 15{\mbox C}+1{\mbox V}\\
\hline
\end{tabular}
\caption{The d=4 closed string spectra for some $\mathbb{Z}_N$-orientifolds.}
\label{tab:closed_string_spectra_zn}
\end{table}

All of this is understandable in the context of general Calabi-Yau
threefolds as well. There, the $\Omega R$-projection reduces the
supersymmetry from $\mathcal{N}$=2 down to $\mathcal{N}$=1 and the
bulk $\mathcal{N}$=2 superfields at the same time are truncated to
$\mathcal{N}$=1 superfields. Before this truncation takes place
there are $h_{1,1}$ abelian vector multiplets and $h_{2,1}$
hypermultiplets.

The $h_{1,1}$ vector multiplets consist of one
scalar field coming from the dimensional reduction of the metric,
the K\"ahler modulus $T$, and another scalar coming from the
reduction of the NS-NS 2-form and a 4-dimensional vector from the
reduction of the R-R 3-form along the 2-cycle on the specific
torus. If the (1,1)-form is invariant under $\Omega R$, then a
$\mathcal{N}$=1 chiral multiplet survives the projection. If
instead it is anti-invariant, a $\mathcal{N}$=1 vector multiplet
survives the projection.

The $h_{2,1}$ hypermultiplets consists of four scalar fields,
where two are coming from the metric, the complex structure
moduli $U$, and two arise from the dimensional reduction of the R-R
3-form along the two 3-cycles of $H^{2,1}(\mathcal{M})$ and
$H^{1,2}(\mathcal{M})$. The $\Omega R$-projection now divides out
one of two complex structure components and a linear combination
of the R-R scalars survives this projection, such that the
quaternionic complex structure moduli space is reduced to a
complex moduli space of dimension $h_{2,1}$. It has to be like
that because the $\mathbb{Z}_N$-orientifolds can be seen as singular
limits of the corresponding Calabi-Yau space.

\subsection{Open string chiral spectrum}\label{cha:chiral_spectrum_general}
The D6-branes carry the gauge fields via Chan-Paton factors at
both ends of the open string. One has to distinguish in between
branes that are not located along the fixed locus of the $\Omega
R$-projection, so along the O6-planes, and the ones that are. One
stack of the former ones supports a factor of $U(N_a)$ to the
total gauge group, a stack of the latter ones either an $SO(N_a)$
or $Sp(N_a)$ gauge factor \cite{Gimon:1996rq}. Here, we will just be
interested in the first more generic case, but still have to
mention that in the T-dual case of D9 branes, the $U(N_a)$ gauge
group corresponds to a stack of D9 branes with non-vanishing flux,
the $SO(N_a)$ to a stack of D9 branes with vanishing flux, and the
$Sp(N_a)$ factor to stack of D5-branes that are allowed in the
model as well \cite{Blumenhagen:2000wh}.

For $k$ stacks of branes not on top of the O-planes, the total
gauge group is given by
\begin{equation}\label{eq:tot_open_gauge_groups}
    \prod_{a=1}^{k}U(N_a)\ ,
\end{equation}
where these gauge groups are equipped with chiral matter in
bifundamental, symmetric and antisymmetric representations that
are located at the intersections on the compact space that break
supersymmetry. It first has been clarified in
\cite{Blumenhagen:1999ev} that the topological intersection number
(\ref{eq:intersection_number}) on the torus or the orbifold space
corresponds to the multiplicities of the certain representation,
or in other words the number of fermion families. Thus this
significant phenomenological model building property gets a
completely geometric interpretation within the discussed
orientifold models that arguably stays independent of continuous
deformations of the moduli space.

Generally, there might be intersections between the two different
types of branes, the branes and their $\Omega R$ mirror, each
coming from the same stack or a different one. Furthermore, there
are intersections with the orientifold plane. All these sectors
give rise to different representations of the gauge group and it
is important to mention that only the net intersection numbers on the
torus or orbifold play a role for this chiral spectrum.

The sectors between two distinct stacks of branes generally lead
to bifundamental representations. One has to distinguish between
the intersection of a certain stack with another stack's $\Omega
R$ mirror, giving rise to chiral fermions in the bifundamental
representation $(\overline{N}_a,N_b)$, and the sector between two
distinct stacks where not both are $\Omega R$ mirrors within their
stack. These sectors lead to $(N_a,N_b)$ representations. Formally
negative intersection numbers, corresponding to flipped
orientations on the torus or orbifold, simply enforce the
conjugated representations.

However, there are also the intersections between branes within
the same equivalence class. Open strings stretching between
two $\Omega R$ mirrors, or for the case of the orbifold, between
a brane and its $\Omega R \Theta^k$ mirror, lead to chiral fields
in the antisymmetric and symmetric representation. Naively, there
also could be intersections in between branes and their $\Theta$
mirrors that are no $\Omega R$ mirrors on the orbifold. These
sectors would lead to matter in adjoint representations of the
gauge group. However, in four flat dimensions they are absent, as
the topological self-intersection numbers always vanish because of
their antisymmetry for two 3-cycles.
Still these sectors are part of the non-chiral spectrum.

This chiral spectrum can be expressed in terms of homological
cycles too, and does not actually require a detailed CFT
computation. It is shown for intersection numbers between these
homological 3-cycles $\pi_a$ in table \ref{tab:massless_chiral_spectrum},
\begin{table}
\centering
\sloppy
\renewcommand{\arraystretch}{1.5}
\begin{tabular}{|c||c|}
  \hline
  Representation & Multiplicity\\
  \hline
  \hline
  $[{\bf A_a}]_{L}$ & ${1\over 2}\left(\pi'_a\circ \pi_a+\pi_{{\rm O}6} \circ \pi_a\right)$ \\
  \hline
  $[{\bf S_a}]_{L}$ & ${1\over 2}\left(\pi'_a\circ \pi_a-\pi_{{\rm O}6} \circ \pi_a\right)$ \\
  \hline
  $[{\bf (\overline{N}_a,N_b)}]_{L}$ & $\pi_a\circ \pi_{b}$ \\
  \hline
  $[{\bf (N_a, N_b)}]_{L}$ & $\pi'_a\circ \pi_{b}$ \\
  \hline
\end{tabular}
\caption{The massless chiral open string spectrum in 4 dimensions.}
\label{tab:massless_chiral_spectrum}
\end{table}
where the prime denotes
$\Omega R$ mirror cycles,
\begin{equation}\label{qe:ORmirror_cycles}
    \pi_a'\equiv \Omega R \pi_a\ ,
\end{equation}
and $\pi_{{\rm O}6}$ the homological cycle of the O6-plane. Note,
that for the case of the orbifold, one has to define the
homological cycle on the orbifold space by
\begin{equation}\label{eq:homological_Cycle_orbifoldspace}
\pi_a \equiv  \sum_{i=0}^{N-1} \Theta^i \pi^t_a \ .
\end{equation}
In this definition, the superscript $t$ denotes the toroidal
ambient space. All intersection numbers have to be computed on the
orbifold space, and the intersection between two 3-cycles then is
given by \cite{Blumenhagen:2002wn}:
\begin{equation}\label{eq:intersection_orbifold_space}
    \pi_a\circ\pi_b={1\over N} \left(\sum_{i=0}^{N-1} \Theta^i \pi^t_a
  \right) \circ \left(\sum_{j=0}^{N-1} \Theta^j \pi^t_b \right)\ .
\end{equation}
In this language, it is immediately clear that there are no
adjoint chiral representations in four dimensions, because $\pi_a
\circ \pi_a=0$.

The spectrum of table \ref{tab:massless_chiral_spectrum} holds for
more abstract general Calabi-Yau threefolds as well and in
\cite{Blumenhagen:2002wn, Blumenhagen:2002vp} even a standard
model on the Quintic has been constructed.

\subsection{Open string non-chiral spectrum}\label{cha:nonchiral_openstring_specrum}
The non-chiral open string spectrum can be obtained from the
conformal field theory calculation at the orbifold limit of a more
general Calabi-Yau space. In the language of boundary states, all the
sectors between the stacks of branes, where the net topological
intersection number is vanishing, give rise to non-chiral fields.
It is possible but quite tedious to calculate this explicitly if
one determines the open string partition function. To do so, the
cylinder amplitude (\ref{eq:cylinder_amp_z3_allcontrib}) has to be
calculated (plus the M\"obius amplitude) and the massless states
directly can be read off.

First, for a certain stack of branes, there are the fields in the
adjoint representation of the gauge group. They arise from the
sectors in between a brane and its $\mathbb{Z}_n$ image, and are
localized at this intersection point. In four compact dimensions
these fields can be chiral, because the orbifold space
intersection number $\pi_a\circ\pi_a$ can be non-vanishing. In six
compact dimensions, the self-intersection $\pi_a\circ\pi_a$ due to
its antisymmetry between two 3-cycles is always identically zero,
therefore, the adjoint fermions are inevitably non-chiral.

Generally, there can be all fields from table
\ref{tab:massless_chiral_spectrum} in the non-chiral spectrum if
the net intersection number for a given representation is
vanishing. For instance, for the bifundamental representations of
the type $[{\bf (\overline{N}_a,N_b)}]_{L}$ in between two
different stacks of branes, this definitely is the case if the
branes are parallel on all tori.

The $U(1)$-factors have to be treated carefully in this respect.
It is possible that they gain a mass through the
Green-Schwarz-mechanism, as we will see in chapter
\ref{cha:greenschwarz}, meaning that they drop out off the
massless spectrum.

\section{Anomalies}
String theory claims to be a consistent theory of quantum gravity.
Anomalies in simple terms indicate the breakdown of classical
symmetries at the quantum level. One carefully has to distinguish
in between anomalies in global and local symmetries. Anomalies in
local symmetries lead to inconsistencies, and this is often linked
to the breakdown of the renormalizability and unitarity of the
theory. On the other hand, anomalies in global symmetries are
unproblematic and just mean that the symmetry is no longer exact,
this for instance is the case for the parity violation of the
standard model.

Local gauge anomalies in four dimensions can be understood by one
specific diagram, the famous Adler-Bell-Jackiw anomaly triangle,
\begin{figure}
\centering
\includegraphics{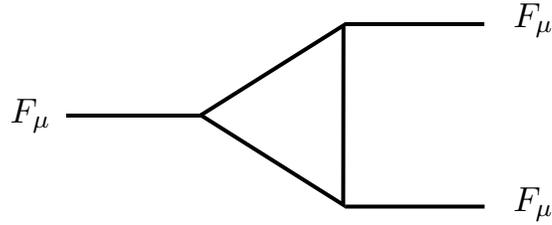}
\caption{Non-abelian triangle anomaly diagram.}\label{fig:nonabelian_anomaly}
\end{figure}
where a chiral fermion runs into a loop in between three external
gauge bosons, see figure \ref{fig:nonabelian_anomaly}. No higher perturbative
corrections occur. This statement has been proven by Adler and Bardeen
and also has an extension to higher dimensional theories: in a
$D$-dimensional theory, one just has to consider diagrams
involving $D/2+1$ external gauge bosons. For 10-dimensional string
theory, this means that one has to consider hexagon diagrams in
spacetime. For intersecting brane models with six compact
dimensions which are treated in this work, it is sufficient to use the
effective massless spectrum in four dimensions and treat triangle
diagrams, as this effective theory has to be consistent as well.

To understand spacetime anomalies from the viewpoint of string
theory, one has to distinguish in between the closed and open
string sectors of the theory.

The closed string sector basically is the sector of gravitational
interactions. Before the emergence of string theory, there were
consistent quantum field gauge theories, as for instance the most
simple one, Quantumelectrodynamics, but it was not possible to
construct a quantum field theory for gravity. So it was one of the
most appreciated successes of string theory in its early days to
show the absence of local gravitational anomalies
\cite{Alvarez-Gaume:1984ig}, this indeed is the case for a total
spacetime dimension $D=10$.

The open sector involves the specific content of D-branes and
therefore, freedom from gauge anomalies has to be explicitly
checked for any brane configuration. For the type I theory (or in
other words, space-time filling D9-branes), anomaly freedom has
been shown by Green and Schwarz in another ground-breaking work
\cite{Green:1984sg, Green:1985ed}. For intersecting brane models,
the freedom from gauge anomalies can be divided in two parts.
First of all, there are the potential cubic anomalies, the
non-abelian anomalies. Next, there are mixed anomalies involving
the abelian symmetries. Finally, there are also mixed anomalies
between the gravitational and the open string sectors.

As we already could have guessed, the freedom from gauge anomalies
in the effective space-time theory is tightly connected to the
consistence of the worlds-sheet formulation, or in other words to
the R-R tadpole cancellation. If this tadpole is cancelled, the
non-abelian anomalies  automatically vanish, too. This result will
be proven in the next section. In order to also see the
cancellation of the mixed gauge anomalies and the mixed
gravitational ones, one has to use the Green-Schwarz mechanism, this
will be treated in section \ref{cha:greenschwarz}.

\subsection{Non-abelian anomalies}\label{cha:nonab_anomalies}
The non-abelian gauge anomalies can be calculated in a simple
manner. One has to convert all occurring trace contributions of
the type that can be seen in figure \ref{fig:nonabelian_anomaly}
into traces over field strengths in the fundamental representation
of the gauge group, $\text{Tr}_\text{N} F^3$.
For the possible representations, this is shown in table
\ref{tab:nonabelian_reps_converted_to_fundamental}.
\begin{table}
\centering
\sloppy
\renewcommand{\arraystretch}{1.5}
\begin{tabular}{|c|c|}
  \hline
  $\text{Tr}_\text{S} F^3$ &$\left(N+4\right)\text{Tr}_\text{N} F^3$\\
  \hline
  $\text{Tr}_\text{A} F^3$ &$\left(N-4\right)\text{Tr}_\text{N} F^3$\\
  \hline
  $\text{Tr}_{\bar{\text{N}}} F^3$ &$-\text{Tr}_\text{N} F^3$\\
  \hline
\end{tabular}
\caption{Conversion of cubic traces in different representations (left)
into traces of the fundamental representation (right) of the gauge
group $U(N)$, $N>1$.}
\label{tab:nonabelian_reps_converted_to_fundamental}
\end{table}
With this information at hand, we can simply take the sum over all
contributions of the chiral spectrum from table
\ref{tab:massless_chiral_spectrum}, this explicitly reads for $k$ stacks
\begin{align}
    A_\text{non-abelian}=&\sum_{a=1}^k\bigg[\left(\frac{N_a}{2}+2\right)\left(\pi'_a\circ \pi_a-\pi_{{\rm O}6} \circ \pi_a\right)\\
    &+\left(\frac{N_a}{2}-2\right)\left(\pi'_a\circ \pi_a+\pi_{{\rm O}6} \circ \pi_a\right)
    \bigg]\nonumber\\
    =&\sum_{a=1}^k \Big[N_a \left(\pi_a+\pi'_a\right)-4\pi_{{\rm O}6}\Big]\circ\pi_a-\sum_{a=1}^k N_a\pi_a\circ\pi_a\nonumber\\
    =&-\sum_{a=1}^k N_a\pi_a\circ\pi_a=0\ ,\nonumber
\end{align}
where we have used the R-R tadpole cancellation condition
(\ref{eq:RRtadpole_top}) and the fact that $\pi_a\circ\pi_a=0$.
Indeed, the non-abelian gauge anomaly ${SU(N_a)}^3$ generally
vanishes.
\subsection{Generalized Green-Schwarz mechanism}\label{cha:greenschwarz}
In this section, we will discuss the mixed anomalies of the type
$U(1)_a-SU(N_b)^2$. These anomalies can be calculated again from
the chiral spectrum in table \ref{tab:massless_chiral_spectrum}
for the effective 4-dimensional theory, where we
have to use different conversions for the traces, given in table
\ref{tab:nonabelian_reps_converted_to_fundquad}.
\begin{table}
\centering
\sloppy
\renewcommand{\arraystretch}{1.5}
\begin{tabular}{|c|c|}
  \hline
  $\text{Tr}_\text{S} F^2$ &$\left(N+2\right)\text{Tr}_\text{N} F^2$\\
  \hline
  $\text{Tr}_\text{A} F^2$ &$\left(N-2\right)\text{Tr}_\text{N} F^2$\\
  \hline
  $\text{Tr}_\text{adj} F^2$ &$2\,N\,\text{Tr}_\text{N} F^2$\\
  \hline
  $\text{Tr}_{\bar{\text{N}}} F^2$ &$\text{Tr}_\text{N} F^2$\\
  \hline
\end{tabular}
\caption{Conversion of quadratic traces in different representations (left)
into traces of the fundamental representation (right) of the gauge
group $U(N)$, $N>1$.}
\label{tab:nonabelian_reps_converted_to_fundquad}
\end{table}
The result for $a \neq b$, where just the bifundamentals
contribute, is given by
\begin{equation}\label{eq:abelian_anomaly_aunglb}
    A_\text{mixed}^\text{$1^\text{st}$part}=\frac{N_a}{2}\left(\pi'_a-\pi_a\right)\circ \pi_b\ \qquad\text{for}\
{a\neq b}\ .
\end{equation}
On the other hand for $a=b$, the anomaly is given by
\begin{align}\label{eq:abelian_anomaly_aglb}
    A_\text{mixed}^\text{$1^\text{st}$part}=&-2\,\pi_{{\rm O}6} \circ \pi_a
    +N_a\,\left(\pi'_a\circ \pi_a\right)\\
    &-\frac{1}{2}\sum_{b\neq a}N_b \pi_a\circ\left(\pi_b+\pi'_b\right)\
\qquad\text{for}\ {a=b}\ .\nonumber
\end{align}
This equation can be reduced to the form
(\ref{eq:abelian_anomaly_aunglb}) by the use of the tadpole
equation (\ref{eq:RRtadpole_top}). This anomaly at first sight
seems to be existing. It was a very important result by Green and
Schwarz, who were discussing the type I string, that by a careful
study after all a term indeed can be identified cancelling this
formal anomaly \cite{Green:1984sg, Green:1985ed}. The general idea
is that besides the metric and the gauge fields, in string theory
there are also Chern-Simons interactions, being invariant under
gauge transformations of the vector potential because of their
construction from the field strength.

Here, we implement a generalized Green-Schwarz mechanism
analogously to \cite{Aldazabal:2000dg} to chancel the anomaly
(\ref{eq:abelian_anomaly_aunglb}). The Chern-Simons terms to be
considered are of the form
\begin{equation}\label{eq:chern_simons}
    \int_{D6_a}C_3\wedge \text{Tr}\left(F_a \wedge F_a\right)\qquad \text{and} \qquad
    \int_{D6_a}C_5\wedge \text{Tr}\left(F_a\right)\ .
\end{equation}
In this equation, $C_3$ is the antisymmetric 3-form and $C_5$ the
10-dimensional Hodge dual 5-form and $F_a$ denotes the gauge field
strength on the $D6_a$-brane. To further proceed, we have to
define a homological basis $e_I$, where $I=0,\ldots,h_{2,1}$ and its dual basis $e_I^\ast$, such that
\begin{equation}\label{eq:basis_dualbasis}
e_I\,e_J^\ast=\delta_{IJ}\ .
\end{equation}
Any 3-cycle (also fractional 3-cycles) and its mirror $\pi'_a$ now can be expanded in
the basis $(e_I, e_I^\ast)$,
\begin{align}
    &\pi_a=v_a^I\,e_I+v_a^{(I+h_{2,1}+1)}\, e_I^\ast\ ,\\
    &\pi'_a=v_a^{I \prime}\, e_I+v_a^{(I+h_{2,1}+1)\prime} \, e_I^\ast\ .\nonumber
\end{align}
One furthermore has to define the 4-dimensional axions $\Phi_I$
and 2-forms $B_I$, such that
\begin{align}
    &\Phi_I=\int_{e_I} C_3\ ,& &\Phi_{I+h_{2,1}+1}=\int_{e_I^\ast} C_3\ ,& \\
    &B_I=\int_{e_I^\ast} C_5\ ,& &B_{I+h_{2,1}+1}=\int_{e_I} C_5\ .& \nonumber
\end{align}
Then, the Chern-Simons couplings (\ref{eq:chern_simons}) can be
rewritten as
\begin{align}\label{eq:chern_simons_explicit}
    &\int_{D6_a}C_3\wedge \text{Tr}\left(F_a \wedge F_a\right)=\sum_{J=0}^{2h_{2,1}}
    \left(v_a^J+v_a^{J \prime}\right)\int_{\mathbb{R}^{1,3}}\Phi_I \wedge \text{Tr}\left(F_a \wedge F_a\right)\ ,\\
    &\int_{D6_a}C_5\wedge \text{Tr}\left(F_a\right)=N_a
    \sum_{I=0}^{h_{2,1}}
    \left(v_a^{(I+h_{2,1}+1)}-v_a^{(I+h_{2,1}+1)\prime}\right)\int_{\mathbb{R}^{1,3}}
    B_I\wedge F_a\nonumber\\
    &\qquad\qquad\qquad\qquad+N_a
    \sum_{J=0}^{h_{2,1}}\left(v_a^{J}-v_a^{J\prime}\right)\int_{\mathbb{R}^{1,3}}
    B_{J+h_{2,1}+1}\wedge F_a\ , \nonumber
\end{align}
where it has been used that the gauge field on the $\Omega
R$-mirror brane is equivalent to $F_a'=-F_a$. The prefactor $N_a$
arises from the normalization of the $U(1)$-generator.

\begin{figure}
\centering
\includegraphics{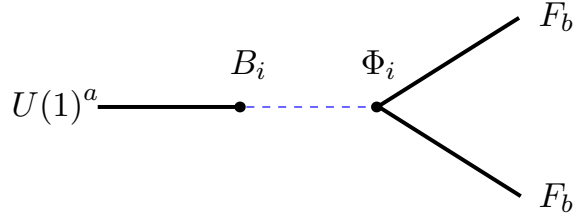}
\caption{The generalized Green-Schwarz mechanism.}\label{fig:green_schwarz}
\end{figure}

From these two couplings, one gets a contribution to the discussed
mixed anomaly between the $U(1)$ gauge factor from the first stack
of branes and its coupling to $B$ and the two $SU(N)$ gauge
factors and their coupling to $\Phi$. This is schematically shown
in figure \ref{fig:green_schwarz}.
This simply means that the $i$-th twisted field first couples to
the $U(1)^a$, then propagates and finally couples to $F_b^2$.
Therefore, the total anomaly contribution is given by
\begin{align}\label{eq:abelian_anomaly_aunglb2}
    A_\text{mixed}^\text{$2^\text{nd}$part}&\sim N_a \sum_{I=0}^{h_{2,1}}\Bigg[
    \left(v_a^{I}+v_a^{I\prime}\right)\left(v_b^{(I+h_{2,1}+1)}-v_b^{(I+h_{2,1}+1)\prime}\right)\\
    &\qquad\qquad\qquad\qquad\qquad\qquad+\left(v_a^{(I+h_{2,1}+1)}+v_a^{(I+h_{2,1}+1)\prime}\right)\left(v_a^{I}-v_a^{I\prime}\right)\Bigg]\nonumber\\
    &=2\,N_a\left(\pi_a-\pi'_a\right)\circ \pi_b\
    \qquad\text{for}\ {a\neq b}\ . \nonumber
\end{align}
This term has the right form to cancel the anomaly
(\ref{eq:abelian_anomaly_aunglb}).

The kernel of the matrix  $M_{aI}$, which can be defined by
\begin{equation}\label{eq:mass_matrix}
     M_{aI}= N_a (v_a^I - v_a^{I\prime} )\ ,
\end{equation}
yields the non-anomalous $U(1)$ gauge groups of the specific model,
which remain massless after the application of the Green-Schwarz mechanism.
\section{Gauge breaking mechanisms}
In this section several possible gauge breaking mechanisms will be
discussed in the context of intersecting D-brane configurations.
This is a very vivid topic, so we can just give a glimpse of the
different possibilities. Some more information can be found in
\cite{Blumenhagen:2000ea, Blumenhagen:2001te, Blumenhagen:2002gw,
Cremades:2002cs, Cremades:2002te, Cremades:2002qm, Ibanez:2001dj,
Aldazabal:2000sa, Aldazabal:2000cn}.
\subsection{Adjoint higgsing}\label{cha:adjointhiggsing}
The simplest possibility for gauge symmetry breaking is to
separate a stack of $N_a$ D-branes into two stacks, containing
$N_b$ and $N_c$ D-branes, where $b+c=a$. This breaks the gauge
groups from $U(N_a)$ down to $U(N_b)\times U(N_c)$ and corresponds
to giving a VEV to the respective fields in the adjoint
representation of the gauge group $U(N_a)$. The intersection
numbers on the torus remain unchanged by this operation. Of
course, the Green-Schwarz mechanism has to be performed again
after this gauge symmetry breaking.
\subsection{Brane recombination mechanisms}\label{cha:branerecombination}
The mechanism of adjoint higgsing certainly leaves the branes flat
and factorizable, so one still has a description by conformal
field theory. On the other hand, string theory also allows for
processes which are not of this type. A D-brane can be deformed
but still homologically wrap the same 3-cycles on the underlying
Calabi-Yau manifold. Unfortunately, the CFT description on the
orbifold point then does not apply anymore. To understand this
process, one has to distinguish in between sectors that preserve
$\mathcal{N}$=2 and those that just preserve $\mathcal{N}$=1
supersymmetry.
\subsubsection{Sectors preserving $\mathcal{N}$=2 supersymmetry}
The mechanism being described in this first section can also be
understood as a Higgs mechanism in low energy field theory,
although its string theoretical realization is different to the
one of the preceding section. Again, one gives VEV to some of the
open string massless fields and leaves the closed string
background unchanged.

This mechanism in our class of models is possible if two stacks of
D-branes preserve a common ${\cal N}=2$ supersymmetry, meaning
that they have to be parallel on one of the three 2-tori. In this
case, there exists a massless hypermultiplet, $H$, being localized
on the intersection of these two branes. $H$ then signals a possible
deformation of the two stacks of D-branes into recombined
D-branes, this can be seen for a specific example in figure
\ref{fig:z4recombined_branes}. These still wrap complex cycles and
have the same volume as the sum of volumes of the two D-branes
before the recombination process occurs. There exists a flat
direction $\langle h_1\rangle=\langle h_2\rangle$ in the  D-term
potential
\begin{equation}\label{eq:flat_dir_Dterm}
    V_D={1\over 2 g^2}\left( h_1 \bar{h}_1 - h_2 \bar{h}_2\right)^2\ ,
\end{equation}
along which the $U(N)\times U(N)$ gauge symmetry is broken to the
diagonal subgroup $U(N)$, but supersymmetry remains
unbroken.\footnote{This is not possible for the special case that
on one of the two stacks there only sits a single D6-brane: then
the F-term potential $\phi h_1 h_2$ forbids the existence of a
flat direction with $\langle h_1\rangle=\langle h_2\rangle$. In
string theory, the explanation for this fact is that there do not
exist large instantons in the $U(1)$ gauge group.} In this
equation, $h_1$ and $h_2$ denote the two complex bosons inside of
one hypermultiplet. So there exists an open string massless field
acting like a low energy Higgs field. In the T-dual picture, this
process is nothing but the deformation of a small instanton into
an instanton of finite size. In intersecting brane orientifold
models, such ${\cal N}=2$ Higgs sectors are coupled at brane
intersections to chiral ${\cal N}=1$ sectors. One still has to be
careful, because the brane recombination in the effective gauge
theory cannot simply be described by the renormalizable couplings.
In order to get the correct light spectrum, one also has to take
into account higher dimensional couplings from string theory, that
might alter the qualitative picture.

This process is also possible if the two stacks of branes $U(N_a)$
and $U(N_b)$ have a different number of branes $N_a\neq N_b$.
Then, a gauge breaking of the following kind might occur:
\begin{equation}\label{eq:gaugebreak_branerecomb}
U(N_a)\times U(N_b) \rightarrow U\left(\text{min}\{N_a, N_b\}\right)\times
U\left(\text{max}\{N_a, N_b\}-\text{min}\{N_a, N_b\}\right)\ ,
\end{equation}
where equally many branes from the one and the other stack
recombine and the rest stays unaltered.
\subsubsection{Sectors preserving $\mathcal{N}$=1 supersymmetry}\label{cha:gauge_sysm_break_tach}

The situation is different for the case that two D-branes only
preserve a common ${\cal N}=1$ supersymmetry and support a
massless chiral supermultiplet $\Phi$ on the intersection
\cite{Kachru:1999vj, Mihailescu:2000dn, Witten:2000mf}. In this
case, the analogous D-term potential to (\ref{eq:flat_dir_Dterm})
schematically is of the following form
\begin{equation}\label{eq:dtermpot2}
    V_D={1\over 2 g^2}\left( \phi  \bar{\phi} \right)^2\ .
\end{equation}
It is not possible to obtain a flat direction in this potential
by just giving a VEV to the massless boson $\phi$, if there are no
other chiral fields involved in the process.

Nevertheless, the massless fields still indicate for isolated brane
intersections that the complex structure moduli are chosen on a
line of marginal stability. If one moves away from this line in
one direction, the intersecting branes will break supersymmetry
without the appearance of a closed string tachyon, indicating that
the intersecting brane configuration is stable. But if one moves
away from the line in the other direction, the former massless
chiral fields will become tachyonic.

The question now arises if it is possible to find another new
supersymmetric ground state of the system on the line of marginal
stability, where some of the original gauge symmetries are broken,
in general involving non-flat D-branes that nevertheless wrap
special Lagrangian 3-cycles.

On the compact 6-torus, bifundamental chiral multiplets do at
least locally indicate the existence of a recombined brane, having
the same volume but broken gauge groups. Gauge breaking now is
possible if only certain bifundamental fields between different
stacks of branes would become tachyonic under an enforced
continuous complex structure deformation. If at the same time, one
gives a VEV to these fields, this actually does not happen, but
the system stays on the line of marginal stability. Nevertheless,
the gauge groups of the potentially tachyonic fields are broken
for the new system with the recombined branes.

We will make this discussion now more precise: from general
arguments for open string models with ${\cal N}=1$ supersymmetry,
it is known that the complex structure moduli only appear in the
D-term potential, whereas the K\"ahler moduli only appear in the
F-term potential \cite{Brunner:1999jq, Kachru:2000ih,
Kachru:2000an, Aganagic:2000gs}, the sum of which yielding the total
scalar potential for the field $\phi$.

Therefore, one has to shown that the D-terms allow for a flat
direction in the potential. A small variation of the complex
structure on the field theory side corresponds to including a
Fayet-Iliopoulos term $r$ into the D-term potential
\begin{equation}\label{eq:dtermwithFI}
    V_D={1\over 2 g^2}\left( \phi \bar{\phi} +r\right)^2\ .
\end{equation}
To see the precise form of $r$, we take another look at the
Chern-Simons couplings of the D-brane effective action
(\ref{eq:chern_simons_explicit}) which has the form of an
$F$-term. The corresponding D-term involves a coupling of the
auxiliary field $D_a$
\begin{equation}\label{eq:fayet_iliopoulos}
S_\text{FI}=  \sum_{i=1}^{b_3} \sum_{a=1}^k \int d^4 x\,  M_{ai}\,
{\partial{\cal K}\over \partial \phi_i} \,
               {1\over N_a}\,{\rm tr}(D_a)\ .
\end{equation}
where the $\phi_i$ are the superpartners of the Hodge duals of the
RR 2-forms. ${\cal K}$ denotes the K\"ahler potential.

These couplings give rise to Fayet-Iliopoulos terms depending on
the complex structure moduli which we simply parameterize by
$A_i={\partial{\cal K}/\partial \phi_i}$. We do not need the
precise form of the K\"ahler potential as long as the map from the
complex structure moduli $\phi_i$ to the new parameters $A_i$ is
one to one. This indeed is the case due to the positive
definiteness of the metric on the complex structure moduli space.
The D-term potential including only the chiral matter and the
FI-terms  in general reads
\begin{align}\label{eq:Dterm_pot}
     V_\text{D}&=\sum_{a=1}^k
 \sum_{r,s=1}^{N_a} {1\over 2 g_a^2}(D^{rs}_a)^2\ ,
\end{align}
where the indices $(r,s)$ numerate the $N_a^2$ gauge fields in the
adjoint representation of the gauge factor $U(N_a)$.
For a specific configuration, after inserting the $U(1)$ charges
$q_{ai}$ and the Green-Schwarz couplings $M_{ai}$, one now has to
find specific complex structure deformations $A_i$ which just
alter some of the 3-cycles. The branes (and consequently the
bifundamental fields) wrapping these specific cycles are getting
deformed, the bifundamental fields would then become tachyonic and
to prevent this, one gives a VEV to these fields. Consequently,
they drop out of the massless spectrum and on the side of string
theory, the involved branes have recombined into a new in general
non-factorizable one, which still wraps a special Lagrangian
3-cycle of the underlying Calabi-Yau manifold. This new brane in
homology is equivalent to the sum of the branes before the
recombination process.

\chapter{The Standard Model on the $\mathbb{Z}_3$-orientifold}\label{cha:Z3}
In this chapter, the main concern will be the construction of a
non-supersymmetric orientifold model, which is stable with
respect to the complex structure moduli. The
$\mathbb{Z}_3$-orbifold turns out to be particularly useful for
this purpose, therefore it will be discussed in great detail. The
chapter ends with a discussion of two phenomenologically
interesting three generation models, a standard-like model with
gauge groups $SU(3)\times SU(2)\times U_{B-L}(1)\times U_{Y}(1)$
and another flipped $SU(5)\times U(1)$ model.
\section{The $\mathbb{Z}_3$-orbifold}

The left-right symmetric $\mathbb{Z}_3$ orbifold should act by
\begin{equation}\label{eq:z3action}
    \Theta:  Z_I\to e^{2\pi i/ 3}\, Z_I
\end{equation}
on all three complex coordinates. Comparing this action with table
\ref{tab:Orbifold_classsification}, one has to take into account that the
action on the third torus $v_3=-2/3$ geometrically is equivalent
to $v_3=1/3$. This action in the closed string sector preserves
$\mathcal{N}$=2 in 4 dimensions without regarding the orientifold projection.
Together with the orientifold projection, this yields a model
\begin{equation}\label{eq:z3model}
    {{\text{Type\ IIA}}\ {\text{on}}\ T^{6} \over
                  \{\mathbb{Z}_3+\Omega R \mathbb{Z}_3\} } .
\end{equation}
All three complex structures are frozen, on a certain
torus they are fixed to be
\begin{equation}\label{eq:c_Struct_z3}
    U^I_\text{A}={1\over 2} +i {\sqrt{3}\over 2}\qquad \text{or}\qquad
    U^I_\text{B}={1\over 2} +i {1\over 2\sqrt{3}} ,
\end{equation}
for the A- or B-torus, respectively. It should be stressed
again that this is just the case for left-right symmetric $\Omega
R$ orientifolds, as these models are mapped by T-duality to
asymmetric type I orbifolds where the K{\"a}hler moduli are frozen.

For the A-torus, the angle between the two axes is fixed to be
$\pi/2$ and for the B-torus, the angle is fixed as well for given
radii $R^{(I)}_x$ and $R^{(I)}_y$. Together with the frozen
complex structures (\ref{eq:c_Struct_z3}), this means that the two
radii have to be equal, $R^{(I)}_x=R^{(I)}_y\equiv R_I$. The
K{\"a}hler modulus then is given by
\begin{equation}\label{eq:Kaeheler_z3}
    T^I_\text{A}=i{\sqrt{3} \over 2} {R_I}^2\qquad \text{or}\qquad
    T^I_\text{B}=i{1 \over 2\sqrt{3}} {R_I}^2 .
\end{equation}

\begin{figure}
\centering
\includegraphics[width=8cm,height=8cm]{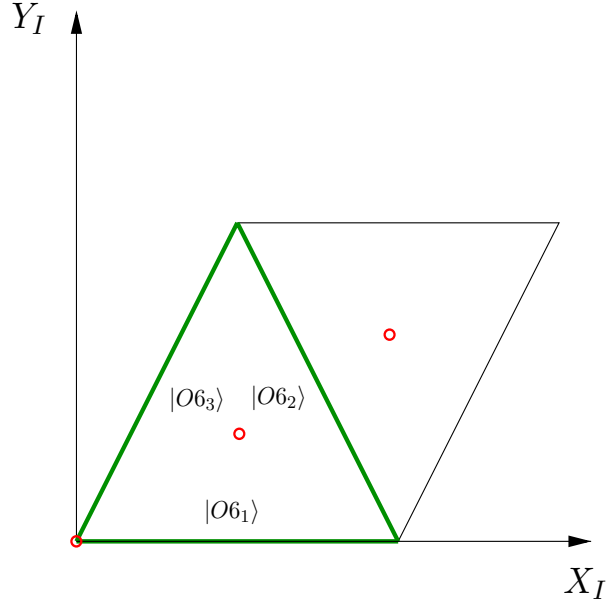}
\caption{The tree distinct O6-planes in the $\mathbb{Z}_3$ orientifold,
together with the three $\mathbb{Z}_3$ fixed points on the A-torus.}\label{fig:o6planes_z3}
\end{figure}

\begin{figure}
\centering
\includegraphics[width=8cm,height=8cm]{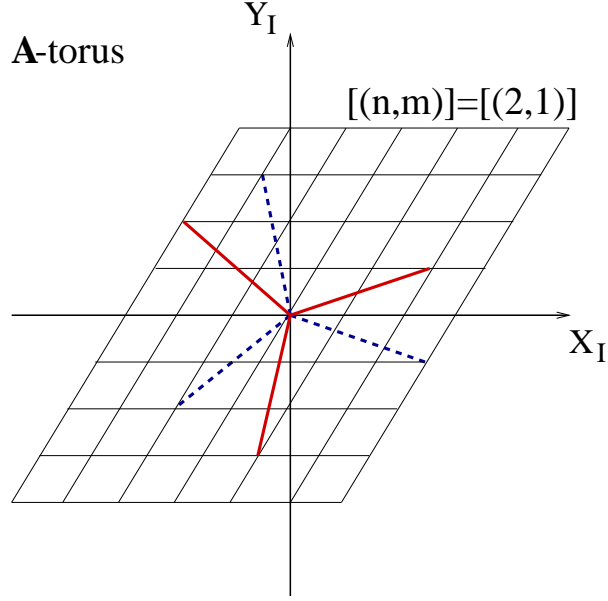}
\caption{The orbit of the exemplary brane $(2,1)$ on the A-torus.}\label{fig:equivalence_class}
\end{figure}

Turning onto the topological data, the $\mathbb{Z}_3$ Orbifold has
Hodge numbers $h_{2,1}=0$ and $h_{11}=36$, where 9 K{\"a}hler
deformations come from the untwisted sector and the remaining 27
are the blown-up modes of the fixed points. The model contains
three distinct O6-planes that are identified under the geometric
action (\ref{eq:z3action}), they are shown in figure
\ref{fig:o6planes_z3}.
The two different fundamental cycles, around which the branes are
wrapping $n_I$ and $m_I$ times, shall be defined as
follows\footnote{The difference as compared to the toroidal
$\Omega R$-orientifold is an unavoidable consequence of the
crystallographic $\mathbb{Z}_3$-symmetry.}:
\begin{equation}\label{eq:fund_cycles_z3}
    e_1^{\bf A} = e_1^{\bf B} = R , \quad
e_2^{\bf A} = {R \over 2} + i {\sqrt{3} R \over 2}, \quad
e_2^{\bf B} = {R \over 2} + i {R \over 2\sqrt{3} },
\end{equation}
for each $T^2$. The action of $\Omega R$ for both A- and B-torus
on the wrapping numbers is given by the map
\begin{align}\label{eq:ORmap_nm_Z3}
            &n_I'=n_I+m_I\ ,&\\
            &m_I'=-m_I\ .& \nonumber
\end{align}
Accordingly, the $\mathbb{Z}_3$-action is expressed by the map
(differing by $b_I=0$ or $b_I=1/2$ for the A- and B-torus,
respectively)
\begin{align}\label{eq:Z3map_Z3}
    &n_I'=-\left(1+2 b_I\right)n_I-m_I\ ,&\\
    &m_I'=\left(1+4 b_I\right)n_I+2 b_I m_I\ .& \nonumber
\end{align}
The six branes contained in the equivalence class
$[(n_I,m_I)]$ for either A- or B-torus are given by:
\begin{equation}\label{eq:z3_orbit}
\end{equation}
\begin{tabular}{p{55pt}p{1pt}p{125pt}p{1pt}p{120pt}}
$\bigg(\begin{array}{c} n_I\\m_I \end{array}\bigg)$
      &${\buildrel \mathbb{Z}_3 \over \Rightarrow}$
      &$\bigg(\begin{array}{c} -(1+2b_I)n_I-m_I\\ (1+4b_I)n_I+2b_I m_I\end{array}\bigg)$
      &${\buildrel \mathbb{Z}_3 \over \Rightarrow}$
      &$\bigg(\begin{array}{c} 2b_I n_I+m_I\\-(1+4b_I)n_I-(1+2b_I)m_I \end{array}\bigg)$\\
      \\
      $\ \Omega R\Downarrow$ &  &$\qquad \qquad \ \Omega R \Downarrow$ &  &$\qquad \qquad \ \ \Omega R \Downarrow$ \\
      \\
      $\bigg(\begin{array}{c}n_I+m_I\\-m_I \end{array}\bigg)$
      &${\buildrel \mathbb{Z}_3 \over \Leftarrow}$
      &$\bigg(\begin{array}{c} 2b_I n_I+(2b_I-1)m_I\\-(1+4b_I)n_I-2b_I m_I \end{array}\bigg)$
      &${\buildrel \mathbb{Z}_3 \over \Leftarrow}$
      &$\bigg(\begin{array}{c} -(1+2b_I)n_I-2b_I m_I\\ (1+4b_I)n_I+(1+2b_I)m_I \end{array}\bigg)\ .$
\end{tabular}
\newline
\newline
\noindent The equivalence class
$[(2,1)]$ is shown for for the A-torus as an example in figure
\ref{fig:equivalence_class}.
There is a very helpful simplification in the
$\mathbb{Z}_3$-model, being that only untwisted sector fields
couple to the orientifold planes. This is due to the relation
\begin{equation}\label{eq:relationZ3}
    \Theta\, (\Omega R) = (\Omega R)\,  \Theta^{-1} ,
\end{equation}
stating the non-commutativity of the two operators $\Omega R$ and
$\Theta$ which appear in the trace. Another simple argument
is that the orientifold planes are of codimension one on each
$T^2$ and therefore can avoid a blown-up $\mathbb{P}^1$ from an
orbifold fixed point. Similarly, the D6-branes can not wrap around
the blown-up cycles\footnote{Such branes are called fractional
branes and will become very important for the
$\mathbb{Z}_4$-orientifold.} and thus they are not charged under the
twisted sector RR-fields. As a direct consequence, there will be
only untwisted tadpoles in the $\mathbb{Z}_3$-orientifold
(corresponding to untwisted sectors in the tree channel). Note,
that this is a special feature of this model and will not be the
case for the $\mathbb{Z}_4$-orientifold.

The R-R and NS-NS tadpoles can be calculated following the
procedure from the two sections \ref{cha:RtadpolesOR} and
\ref{cha:NStadpoles_OR} with the alterations that are being described in
section \ref{cha:tadpolesR_NS_ZN}.

As the betti number divided by two is equal to one,
$\text{b}_3/2=1+h_{2,1}=1$, there will be only one R-R and one
NS-NS tadpole for the $\mathbb{Z}_3$-orientifold.
\section{R-R tadpole}\label{cha:RRz3}
\subsubsection{Klein bottle}
Let us start with the Klein-bottle amplitude. The amplitude in the
loop channel is given by
\begin{equation}\label{eq:kb_amplitudeRZ3}
\mathcal{K}^\text{(NS-NS,$-$)}=4c \int\limits_{0}^{\infty}\frac{dt}{t^3}
\text{Tr}_\text{U+T}^\text{(NS-NS,$-$)}\left(\frac{\Omega R}{2}\
\frac{1+\Theta+\Theta^{2}}{3}\frac{1+(-1)^F}{2}\ e^{-2\pi t
\mathcal{H}_{\text{closed}}}\right) ,
\end{equation}
In the untwisted sector, all insertions $1$, $\Theta$ and
$\Theta^2$ yield the same contribution, because the lattice
Hamiltonian $\mathcal{H}_{\text{lattice}}$ is
$\mathbb{Z}_3$-invariant. Using $\mathcal{H}_{\text{lattice}}$
(\ref{eq:ABHamiltonianZ3_lat}), the untwisted contribution takes
the form:
\begin{equation}\label{eq:Kleinbottle_amplitude_OR_loop}
\mathcal{K}^{\text{(NS-NS,$-$)}}_\text{U}=c
\int\limits_{0}^{\infty}\frac{dt}{t^3}\frac{-\tfkto{0}{1/2}^4}{\eta^{12}}
\prod_{I=1}^{3}\left[\sum_{r_I,s_I} e^{-\pi t \left(4\frac{{s_I}^2\alpha'}
{{R_I}^2}+\left(3-\frac{16}{3}\,b_I\right)\frac{{r_I}^2{R_I}^2}{\alpha'}\right)}\right] ,
\end{equation}
where $b_I$ can be again chosen separately for every torus to be $0$ or
$1/2$. The modular
transformation to the tree channel leads to
\begin{multline}\label{eq:Kleinbottle_amplitude_z3 tree erster}
\widetilde{\mathcal{K}}^{\text{(R-R,$+$)}}_\text{U, 1st part}=-24 \sqrt{3} c
\int\limits_{0}^{\infty}{dl}\frac{\tfkto{1/2}{0}^4}{\eta^{12}}\\
\cdot\left[\prod_{I=1}^{3}\frac{1}{\sqrt{9-16\, b_I}}
\sum_{r_I,s_I} e^{-4\pi l\left(\frac{{R_I}^2}
{4 \alpha'}+\frac{3\alpha'}{\left(9-16\,b_I\right){R_I}^2}\right)}\right]\ ,
\end{multline}
where we have to keep in mind that this is just the first part of
the untwisted tree channel R-R sector, as the twisted loop channel
sectors contribute as well,
\begin{equation}\label{eq:KB_amp_z3_R_bothterms}
\widetilde{\mathcal{K}}^{\text{(R-R,$+$)}}_\text{U}=\widetilde{\mathcal{K}}^{\text{(R-R,$+$)}}_\text{U, 1st part}
+\widetilde{\mathcal{K}}^{\text{(R-R,$+$)}}_\text{U, 2nd part}\ .
\end{equation}
The contribution of the twisted loop channel sectors transformed
to the tree channel is given by
\begin{multline}\label{eq:Kleinbottle_amplitude_z3 tree zweiter}
\widetilde{\mathcal{K}}^{\text{(R-R,$+$)}}_\text{U, 2nd part}=-24 \sqrt{3} c
\int\limits_{0}^{\infty}{dl}\\
\left(\gamma_1 e^{3\pi i}\,\frac{\tfkto{1/2}{0} \tfkto{1/2}{-1/3} \tfkto{1/2}{-1/3} \tfkto{1/2}{-1/3}}
{\eta^{3} \tfkto{1/2}{2/3-1/2} \tfkto{1/2}{2/3-1/2} \tfkto{1/2}{2/3-1/2}}\right.\\
\left.+\gamma_2 e^{3\pi i}\,\frac{\tfkto{1/2}{0} \tfkto{1/2}{1/3} \tfkto{1/2}{1/3} \tfkto{1/2}{1/3}}
{\eta^{3} \tfkto{1/2}{4/3-1/2} \tfkto{1/2}{4/3-1/2} \tfkto{1/2}{4/3-1/2}}\right)\ ,
\end{multline}
where
\begin{equation}\label{eq:}
    \gamma_1=2^3 \sin(2\pi/3)=3\sqrt{3}\qquad\text{and}\qquad \gamma_2=2^3\sin(4\pi/3)=-3\sqrt{3}
\end{equation}
has to be required in order to cancel
the bosonic zero modes. The Klein bottle tadpole can be obtained
by the zeroth order term in the series expansion of
(\ref{eq:KB_amp_z3_R_bothterms}) using $q=e^{-4\pi l}$:
\begin{equation} \label{eq:tadpole_z3_KB_R}
T^{\text{R-R}}_{\widetilde{\mathcal{K}}}=
-288{\frac {\sqrt {3}c}{\sqrt {9-16\,b_1}\sqrt {9-16\,b_2}
\sqrt {9-16\,b_3}}}\ .
\end{equation}
The normalization of a single orientifold plane
(\ref{eq:crosscapstate_OR_normdef}) can be obtained from only
$\widetilde{\mathcal{K}}^{\text{(R-R,$+$)}}_\text{U, 1st part}$.
This term has to be divided by three, as it contains the sum
of the three distinct O6-planes. It is given by
\begin{equation}\label{eq:Norm_O6_z3}
\mathcal{N}_\text{O6}=2\,{\frac {\sqrt [4]{3}\sqrt {c}}{\sqrt [4]{9-16\,b_1}\sqrt [4]{9
-16\,b_2}\sqrt [4]{9-16\,b_3}}}\ .
\end{equation}
\subsubsection{Cylinder}
The cylinder amplitude for a single stack of $N$ $D6$-branes of
the equivalence class $[i]$ contains 36 different contributions,
six for each open string end which can end on any of the branes
within the equivalence class of the model,
\begin{equation}\label{eq:cylinder_amp_z3_allcontrib}
    \mathcal{A}=\sum_{j\in [i]} \sum_{k\in [i]}\mathcal{A}_{jk}\ .
\end{equation}
In order to find the correct normalization factor for a single brane
within an equivalence class, we have to calculate the cylinder
amplitude only for the sector $\mathcal{A}_{ii}$, where an open
string goes from this single brane onto itself. At this time, the
brane location should not be specified, so the wrapping numbers
$n_I$ and $m_I$ on each torus are kept arbitrarily and the
amplitude of interest for a stack of $N$ branes is given by
\begin{equation}\label{eq:Annulus_aii_general_z3}
    \mathcal{A}_{ii}^\text{(NS,$-$)}=c\int\limits_{0}^{\infty}\frac{dt}{t^3}
\text{Tr}_{\text{D6i-D6i}}^\text{(NS,$-$)}\left(\frac{1}{2\cdot3}\frac{1+(-1)^F}{2}e^{-2\pi t
\mathcal{H}_{\text{open}}}\right)\ .
\end{equation}
Using the open string lattice Hamiltonian
(\ref{eq:ABHamiltonian_lat_annulus}), this easily can be
calculated to be
\begin{equation}\label{eq:Annulus_amplitude_z3_loop}
\mathcal{A}_{ii}^\text{(NS,$-$)}=\frac{c}{12}N^2
\int\limits_{0}^{\infty}\frac{dt}{t^3}\frac{-\tfkto{0}{1/2}^4}{\eta^{12}}
\prod_{I=1}^{3}\left[\sum_{r_I,s_I} e^{-\pi t
\left(\frac{\frac{2{r_I}^2}{{R_I}^2}+\left(\frac{3}{2}-\frac{8}{3}\,b_I\right){s_I}^2{R_I}^2}{L_I^i}\right)
}\right]\ ,
\end{equation}
where $L_I^i$ is the length of the D6-brane on the $I$-th torus
\begin{align}\label{eq:ABlength_brane z3}
L_I^i=\sqrt{{n_I^i}^2+n_I^i\,m_I^i+\left(1-\frac{4}{3}b_I\right){m_I^i}^2}\ .
\end{align}
Transforming (\ref{eq:Annulus_amplitude_z3_loop}) into the tree
channel R-sector leads to the following normalization of a single
brane
\begin{equation}\label{eq:Norm_D6_z3}
\mathcal{N}_\text{D6}=\frac{\sqrt [4]{3}}{2}\,\frac{\,N\sqrt
{c}L_1^i\,L_2^i\,L_3^i}{\sqrt [4]{9-16\,b_1}\sqrt
[4]{9-16\,b_2}\sqrt [4]{9-16\,b_3}}\ .
\end{equation}
The sectors of a string between another D-brane of the equivalence
class and itself can be obtained by mapping the wrapping numbers
$n_I$ and $m_I$ on every torus according to (\ref{eq:z3_orbit}),
and the result is that all contributions are equal. The cylinder
contributions between two D-branes at a non-vanishing do not have
to be calculated explicitly in order to determine the tadpole.
\subsubsection{M{\"o}bius strip}\label{cha:moebius_z3}
Next, we are going to determine the M{\"o}bius strip amplitude in
the tree channel. For one stack of branes, the relevant sectors
are in between one of the three O6-planes and one of the 6 branes
contained within its equivalence class. Due to the
$\mathbb{Z}_3$-symmetry, it is sufficient to calculate the sector
between a certain O6-brane and the sum over the orbit of one
arbitrary stack of branes,
\begin{multline}\label{eq:Moebius_amplitude_OR_tree}
\widetilde{\mathcal{M}}_{[i]}^\text{(R,$+$)}=\pm N
\int\limits_{0}^{\infty}{dl}\ 2\cdot2\cdot3\ \mathcal{N}_\text{D6}\ \mathcal{N}_\text{O6}\ \\
 \cdot \sum_{j\in [i]}\gamma_j\
\frac{\tfkto{1/2}{0}\tfkto{1/2}
{-\kappa_1^j}\tfkto{1/2}{-\kappa_2^j}\tfkto{1/2}{-\kappa_3^j}}{\tfkto{1/2}{1/2-\kappa_1^j}
\tfkto{1/2}{1/2-\kappa_2^j}\tfkto{1/2}{1/2-\kappa_3^j}\eta^{3}}\ .
\end{multline}
In this amplitude, the one factor of 2 cones from the two spin
structures, one factor of 2 from the exchangeability of the bra-
and cat-vectors and the factor 3 from the three O6-planes.
The contribution of every summand in the sum over the orbit at lowest order is given by
\begin{equation}
16\prod_{I=1}^{3}\cos (\pi \kappa_I^j)=16\prod_{I=1}^{3}
{\frac{n_I^j\,n_I^\text{O}+ \left( 1-\frac{4}{3}\,b_I \right) m_I^j\,m_I^\text{O}
+\frac{1}{2}\,n_I^j\,m_I^\text{O}+\frac{1}{2}\,m_I^j\,n_I^\text{O}}{L_I^j\,L_I^\text{O}}}\ ,
\end{equation}
Taking the sum over the orbit in
(\ref{eq:Moebius_amplitude_OR_tree}) and using the property of the
perfect square together with (\ref{eq:tadpole_z3_KB_R}),
(\ref{eq:Norm_O6_z3}) and (\ref{eq:Norm_D6_z3}), finally leads to the
R-R tadpole cancellation equation
\begin{multline}\label{eq:Rtadpol_z3}
  \sum_{a=1}^k\,N_a \Big[ { n_1^a}{ n_2^a}{ n_3^a}\\+\frac{1}{2}{ n_1^a}{ n_2^a}{ m_3^a}
+\frac{1}{2}{ n_1^a}{ n_3^a}{ m_2^a}+\frac{1}{2}{ n_2^a}{ n_3^a}{
m_1^a}-\frac{1}{6}{ n_1^a}{ m_2^a}{ m_3^a} \left(3-6{ b_2}-6{
b_3}+8{ b_2}{ b_3} \right)\\-\frac{1}{6}{ n_2^a}{ m_1^a}{ m_3^a}
 \left(3-6{ b_1}-6{ b_3}+8{ b_1}{ b_3} \right)-\frac{1}{6}
{ n_3^a}{ m_1^a}{ m_2^a} \left(3-6{ b_1}-6{ b_2}+8{ b_1}{ b_2}
\right) \\-\frac{1}{3}{ m_1^a}{ m_2^a}{ m_3^a}
 \left(3-3{ b_1}-3{ b_2}-3{ b_3}+2{ b_1}{ b_3}+2{ b_1}{ b_2}+2{ b_2}{ b_3}\right)  \Big]=2\ .
\end{multline}
Here, the tadpole cancellation condition already has been
generalized for the case of $k$ different stacks, each containing
$N_a$ branes. For the four tori AAA, AAB, ABB and BBB, this result
matches exactly the explicit expressions given in
\cite{Blumenhagen:2001te}. Later it will turn out to be useful to
define the following two quantities for any equivalence class
$[(n^I_a,m^I_a)]$ of D6$_a$-branes:
\begin{align}\label{eq:def_zy}
&Z_{[a]}={2 \over 3} \sum_{({n}_I^b,{m}_I^b)\in [a]}
                           \prod_{I=1}^3  \left( {n}_I^b +{1\over 2}\,
                         {m}_I^b \right)\ , \\
&Y_{[a]}= -{1\over 2} \sum_{({n}_I^b,{m}_I^b)\in [a]}
                          (-1)^M\, \prod_{I=1}^3
                         {m}_I^b\ ,\nonumber
\end{align}
where $M$ is defined to be odd for a mirror brane and otherwise even. The sums
are taken over all the individual D6$_b$-branes that are elements of the
orbit $[a]$. Then the R-R tadpole equation can be written as
\begin{equation}\label{eq:Rtadpol_z3_kurz}
    \sum_{a=1}^K N_a\, Z_{[a]} =2\ ,
\end{equation}
because $Z_{[a]}$ is nothing but the expression within the bracket
$[\ldots]$ in (\ref{eq:Rtadpol_z3}). A simple interpretaion for
this quantity is the projection of the entire orbit of
D6$_a$-branes onto the $X_I$ axes, i.e. the sum of their RR
charges with respect to the dual D9-brane charge. The RR-charges
of all D6-branes have to cancel the RR-charges of the orientifold
O6-planes.

Comparing this result with the supersymmetric solutions of
\cite{Blumenhagen:1999md, Blumenhagen:1999ev}, an important
difference can be seen. Whereas in the supersymmetric case all
$Z_{[a]}$ are positive (implying a very small rank of the  gauge
group), here the $Z_{[a]}$ may also be negative such that gauge
groups of a higher rank (as for instance three) might be realized.
\section{NS-NS tadpole}\label{cha:nsnsz3}
The NS-NS tadpole can be calculated in a similar manner, if one
makes the changes compared to the R-R tadpole that are being
described in much detail within section \ref{cha:NStadpoles_OR}.
The result is simply given by
\begin{equation}\label{eq:dilatontadpole_z3}
     \langle \phi \rangle_D \sim {\partial V \over \partial \phi}\
\end{equation}
with a scalar potential
\begin{equation}\label{eq:z3_scalar_pot}
    V(\phi)=e^{-\phi} \left( \sum_{a} N_a \prod_{I=1}^3  \sum_{j\in [a]}L_I^j -2
                         \right)\ ,
\end{equation}
and the length of a certain D6-brane given in
(\ref{eq:ABlength_brane z3}). That there is just one NS-NS tadpole
can directly be understood by the fact that all scalars related to
the complex structure moduli are projected out under
$\mathbb{Z}_3$. Consequently, only the dilaton itself can have a
disc tadpole. Similar to the toroidal $\Omega R$-orientifold of
chapter \ref{cha:NStadpoles_OR}, whenever the D-branes do not lie
on top of the orientifold planes the dilaton tadpole does not
vanish. This means that the local cancellation of the R-R charge
is in one to one correspondence with supersymmetric vacua and the
cancellation of NS-NS tadpoles. The only exception to this rule
appears to be a parallel displacement of orientifold planes and
D-branes, or in other words a Higgs mechanism breaking $SO(2N_a)$
to $U(N_a)$.
\section{Massless spectrum}
\begin{figure}
\centering
\includegraphics[scale=0.5]{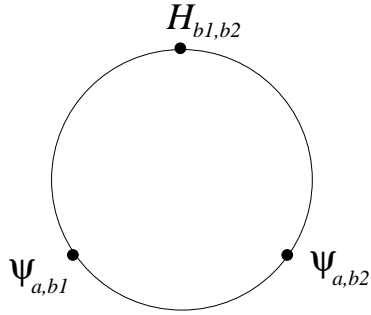}
\caption{The appearance of a dirac mass term in the $\mathbb{Z}_3$
orientifold.}\label{fig:diracmassterm}
\end{figure}

The massless spectrum consists of a chiral part, localized at the
brane intersections, and a non-chiral part. The chiral spectrum
can be determined by methods of topology as described in section
\ref{cha:chiral_spectrum_general}. Applying table
\ref{tab:massless_chiral_spectrum} to the
$\mathbb{Z}_3$-Orientifold leads to the spectrum shown in table
\ref{tab:chiral_spectrum_z3}.
\begin{table}
\centering
\sloppy
\renewcommand{\arraystretch}{1.5}
\begin{tabular}{|c||c|}
  \hline
  Representation & Multiplicity\\
  \hline
  \hline
  $[{\bf A_a}]_{L}$ & $Y_{[a]}$ \\
  \hline
  $[{\bf A_a+S_a}]_{L}$ & $Y_{[a]}\left( Z_{[a]}-{1\over 2}\right)$ \\
  \hline
  $[{\bf (\overline{N}_a,N_b)}]_{L}$ & $Z_{[a]}\, Y_{[b]}- Y_{[a]}\, Z_{[b]}$ \\
  \hline
  $[{\bf (N_a, N_b)}]_{L}$ & $Z_{[a]}\, Y_{[b]}+ Y_{[a]}\, Z_{[b]}$ \\
  \hline
\end{tabular}
\caption{The $d=4$ massless chiral spectrum for the $\mathbb{Z}_3$-Orientifold.}
\label{tab:chiral_spectrum_z3}
\end{table}
In addition to these chiral matter fields, the open strings stretching in between the a
certain brane $D6_a$ and its two $\mathbb{Z}_3$ images yield massless
fermions in the adjoint representation
\begin{equation}\label{eq:z3_adjoint}
     ({\rm
Adj})_L:\quad  3^{n_{\bf B}} \prod_{I=1}^3
          \left(L_I^{[a]}\right)^2\ ,
\end{equation}
where $n_{\bf B}$ stands for the number of {\bf B}-tori in $T^6$.
This non-chiral sector is ${\cal N}=1$ supersymmetric, as the
$\mathbb{Z}_3$ rotation on its own preserves supersymmetry.
Nevertheless, it will be very important for our phenomenological
discussions: for instance, let D6$_{b1}$ and D6$_{b2}$ be two
different branes in the orbit $[b]$ and D6$_a$ another one in the
orbit $[a]$. If there is a chiral fermion $\psi_{a,b1}$ in the
$(\o{N}_a,N_b)$ representation within the D$6_a$-D$6_{b1}$ sector
and a fermion $\psi_{a,b2}$ in the conjugate $({N}_a,\o{N}_b)$
representation within the D$6_a$-D$6_{b2}$ sector, then these
might pair up to yield a Dirac mass term with a mass of the order
of the string scale. The three-point coupling on the disc diagram
as shown in figure \ref{fig:diracmassterm} indeed does exist,
because in the sector D$6_{b1}$-D$6_{b2}$ there is this massless
scalar $H_{b1,b2}$ in the adjoint representation
(\ref{eq:z3_adjoint}) of $U(N_b)$.
If one by hand gives a vacuum expectation value to the $SU(N_b)$
singlet in the adjoint representation of $U(N_b)$, then the gauge
symmetry is left unbroken the fermions receive a mass. Such a
deformation in string theory corresponds to the situation, where
two intersecting branes of the orbit $[b]$ are deformed into a
single brane wrapping a supersymmetric cycle
\cite{Blumenhagen:2000eb}.
\subsection{Anomaly cancellation}
The $SU(N_a)$ gauge anomalies cancel in the same manner as shown
in the general case by the R-R tadpole equation, being described
in chapter \ref{cha:nonab_anomalies}. For the
$U(1)_a-g_{\mu\nu}^2$, $U(1)_a-U(1)_b^2$ and $U(1)_a-SU(N)_b^2$
anomalies, the Green-Schwarz mechanism has to be employed. Its
success in cancelling all these anomalies has been proven
generally in chapter \ref{cha:greenschwarz}. Here, these cancelled
$U(1)_a-g_{\mu\nu}^2$ anomalies are proportional to
\begin{equation}\label{eq:grav_anomaliesz3}
    3\, N_a\, Y_{[a]}\ ,
\end{equation}
and the mixed $U(1)_a-U(1)_b^2$ anomalies are given by
\begin{equation}\label{eq:mixed_anomaliesz3}
    2\, N_a\, N_b\, Y_{[a]} Z_{[b]}\ .
\end{equation}
There is only one anomalous $U(1)$,
\begin{equation}\label{eq:anomalousu1_z3}
    F_{\rm mass}=\sum_a  (N_a\, Y_{[a]})\, F_a\ ,
\end{equation}
becoming massive due to the second Chern-Simons coupling in
(\ref{eq:chern_simons_explicit}).

\subsection{Stability}\label{cha:z3stability}
The complex structure moduli are getting projected out of the
scalar potential (\ref{eq:z3_scalar_pot}), which consequently is
flat for the remaining moduli, only the dilaton is not being
stabilized, as for the toroidal $\Omega R$-orientifold. Still in
higher loop diagrams, we expect that the K\"ahler moduli to
contribute and correct the geometry.

Another stability issue concerns the existence of open string
tachyons, which also may spoil stability at the open string
loop-level. In general, the bosons of lowest energy in a
non-supersymmetric open string sector can have negative mass
squared. Here one has to distinguish two different cases. Either
the two D-branes in question intersect under a non-trivial angle
on all three two-dimensional  tori or the D-branes are parallel on
at least one of the 2-tori. In the latter case one can get
rid of the tachyons at least classically by making the distance
between the two D-branes on the $I$-th 2-torus large enough. In the
former case, it depends on the three angles, $\varphi_{ab}^I$,
between the  branes D$6_a$ and D$6_b$  whether there appear
tachyons or not.

Defining as usual $\kappa^{ab}_I=\varphi_{ab}^I/\pi$ and let $P_{ab}$ be
the number of  $\kappa^{ab}_I$ satisfying $\kappa^{ab}_I>1/2$,
to compute the ground state energy in this twisted open string
sector, one has to distinguish the following three cases:
\begin{equation}\label{eq:zeropoint_z3}
    E^0_{ab}=\begin{cases}
               {1\over 2} \sum_I |\kappa^{ab}_I|-{\rm max}\left\{
                      |\kappa^{ab}_I| \right\}
                  &\text{for}\ P_{ab}=0,1\ , \vspace{0.3cm}\\
            1+{1\over 2}\left( |\kappa^{ab}_I|-|\kappa^{ab}_J|-
           |\kappa^{ab}_K|\right)\\-
           {\rm max}\left\{ |\kappa^{ab}_I|,1-|\kappa^{ab}_J|,
                  1-|\kappa^{ab}_K| \right\}
                  &\text{for}\ P_{ab}=2\ \text{and}\ |\kappa^{ab}_I|\le{1\over 2}\ ,
                 \vspace{0.3cm} \\
           1-{1\over 2} \sum_I |\kappa^{ab}_I|
              &\text{for}\ P_{ab}=3\ .
              \end{cases}
\end{equation}
In order for a brane model  to be free of tachyons, for all open
string sectors $E^0_{ab}\ge 0$ has to be satisfied. Since in the
orbifold model each brane comes with a whole equivalence class of
branes, and the angles between two branes do not depend on any
moduli, freedom of tachyons is a very strong condition. We will
come back to this point in the next section.
\section{Phenomenological model building}\label{cha:std_model}
Every potentially interesting model at first has to satisfy the
R-R tadpole condition (\ref{eq:Rtadpol_z3_kurz}). Amazingly, even
in this fairly constrained orbifold set-up it is not too difficult
to get three generation models with $SU(3)\times SU(2)_L\times
U(1)_Y\times U(1)_{B-L}$ or $SU(5)\times U(1)$ gauge groups and
standard model matter fields enhanced by  a right-handed neutrino.
The open string tachyon, being present in many explicit examples,
has been suggested to have the interpretation of a Higgs particle,
although this seems problematic because this is surely an
off-the-mass-shell process and better should be described by open
string field theory. In the context of intersecting D-branes, not
much is known about these processes.

There are several papers \cite{Aldazabal:2000dg,
Blumenhagen:2000ea, Ibanez:2001nd} where three generation
intersecting brane worlds have been realized on four stacks of
D6-branes with gauge group  $U(3)\times U(2)\times U(1)\times
U(1)$ and chiral matter only in the bifundamental representations
of the gauge factors.

This realization is not possible for the $\mathbb{Z}_3$
orientifold: requiring that there does not exist any matter in the
antisymmetric representation of $U(3)$ forces $Y_1=0$. This
implies on the other hand using table \ref{tab:chiral_spectrum_z3} that
there are the same number of chiral fermions in the $({\bf 3,2})$
and in the $({\bf \bar 3, 2})$ representation of $U(3)\times
U(2)$, leading to an even number of left-handed quarks. So we will
have to find a different way.
\subsection{Semi-realistic extended standard model}
In order to find a standard-like model, we have to implement two
important requirements:
\begin{enumerate}
    \item The right-handed $(u,c,t)$-quarks have to be in the
antisymmetric representation of $U(3)$. For the specific
$\mathbb{Z}_3$-spectrum, it is similar to the anti-fundamental
representation $\bf\bar 3$.
    \item There should not appear any chiral matter in the
symmetric representation of $U(3)$ and $U(2)$. This requirement
demands $Z_3=Z_2=1/2$.
\end{enumerate}
From these two conditions, it seems to be possible to approach the
standard model with only three initial stacks of D6-branes with
gauge group $U(3)\times U(2)\times U(1)$ and the specific choices
for the parameters $Y_a$ and $Z_a$ as given in table
\ref{tab:stacks_std_z3}, which indeed fulfills the R-R tadpole
cancellation condition.
\begin{table}
\centering
\sloppy
\renewcommand{\arraystretch}{1.5}
\begin{tabular}{|c||c|c|}
  \hline
  Initial stack $D6_a$ & $Y_{a}$ & $Z_{a}$\\
  \hline
  \hline
  $U(3)$ & 3&$\frac{1}{2}$  \\
  \hline
  $U(2)$ & 3&$\frac{1}{2}$  \\
  \hline
  $U(1)$ & 3&$-\frac{1}{2}$ \\
  \hline
\end{tabular}
\caption{The initial stacks for the extended standard model.}
\label{tab:stacks_std_z3}
\end{table}
The chiral massless 3 generation spectrum can be found in table \ref{tab:std_modz3_chiralspec}.
\begin{table}
\centering
\sloppy
\renewcommand{\arraystretch}{1.1}
\begin{tabular}{|c||c|c|c|c|}
  \hline
 Matter & Representation & $SU(3)\times SU(2)\times U(1)^3$ & $U(1)_Y$ & $U(1)_{B-L}$\\
  \hline\hline
  $\left( Q_L\right)_i$ & bifundamental &$({\bf 3},{\bf 2})_{(1,1,0)}$ &  ${1\over 3}$ &
${1\over 3}$ \\

  $\left(u^c_L\right)_i$& $\bar A[U(3)]$ & $(\bar{\bf 3},{\bf 1})_{(2,0,0)}$ &
        $-{4\over 3}$ &  $-{1\over 3}$  \\

 $\left(d^c_L\right)_i$& anti-fundamental & $ (\bar{\bf 3},{\bf 1})_{(-1,0, 1)}$ &
       ${2\over 3}$ & $-{1\over 3}$   \\
\hline
 $\left(l_L\right)_i$ & bifundamental &$({\bf 1},{\bf 2})_{(0,-1,1)}$ & ${-1}$ &
           ${-1}$   \\

 $\left(e^+_L\right)_i$ & $A[U(2)]$ & $({\bf 1},{\bf 1})_{(0,2,0)}$ & ${2}$ &
            ${1}$   \\

 $\left(\nu^c_L\right)_i$ & $\bar S[U(1)]$ &$({\bf 1},{\bf 1})_{(0,0,-2)}$ & ${0}$ &
                  ${1}$    \\
\hline
\end{tabular}
\caption{Left-handed fermions of the 3 stack Standard-like model.}
\label{tab:std_modz3_chiralspec}
\end{table}
The non-chiral spectrum will not be treated in this section, as it
is assumed that the non-chiral fermions have paired up and have
decoupled. By the Green-Schwarz mechanism, the expected anomalous
U(1) is given by
\begin{equation}\label{eq:anomu1_z3_stdmod}
    U(1)_{\rm mass}=3\, U(1)_1 +2\, U(1)_2 + U(1)_3\ ,
\end{equation}
and between the anomaly-free ones, one can choose the specific linear
combination
\begin{align}
    U(1)_Y&=-{2\over 3}\, U(1)_1 + U(1)_2 , \\
    U(1)_{B-L}&=-{1\over 6}\left( U(1)_1 -3\, U(1)_2+3\, U(1)_3\right)\ .\nonumber
\end{align}
to be also anomaly-free,  which coincidentally are the hypercharge
and another extended $B-L$ symmetry. The $B-L$ decouples, but
still survives as a global symmetry. The possible Yukawa couplings
by this fact are more constrained than in the standard model.

Since the one-loop consistency of the string model requires  the
formal cancellation of the $U(2)$ and $U(1)$ (non-abelian) gauge
anomalies, the possible models are fairly constrained and require
the introduction of right-handed neutrinos, being a strong
prediction of these type of models. Because  the lepton number is
not a global symmetry of the model, there exists the possibility
to obtain Majorana mass terms and invoke the see-saw mechanism for
the neutrino mass hierarchy.

Of course, so far, we have not given any concrete realization of
these models in terms of their winding numbers $n_I$ and $m_I$. In
order to do so, a computer program has been set up, searching for
winding numbers in between -10 and 10 that realize the given
$Y_{a}$ and $Z_{a}$. The tachyon conditions, being described in
chapter \ref{cha:z3stability}, have been checked for every model,
and it has turned out that one could find 36 solutions for every
single one of the 3 stacks, independently of the choice of torus.
Another rather mysterious observation could be made: for all
models all winding numbers just range in between -3 and 3, and
only for the {\bf BBB} torus from $-5$ and $5$. The actual number
of inequivalent string models with the above mentioned
standard-like model features then is $4\cdot 36^3$.

\subsubsection{Gauge breaking}
The discussed version of the standard model extended by a gauged
$B-L$ symmetry together with right-handed neutrinos requires a two
step gauge symmetry breaking in order to serve as a realistic
model, see for instance \cite{Chikashige:1980ht}. In order to
avoid conflicts with various experimental facts, one has to
require a hierarchy of Higgs vacuum expectation values.

First of all, the $U(1)_{B-L}$ has to be broken at a scale at
least $\sim 10^{4-6}$ above the electroweak scale. This requires a
Higgs field charged under this group but a singlet otherwise. This
could be met by a tachyon from a sector of strings stretching
between two branes in the orbit that supports the $U(1)_3$. We
will embark this strategy in the following section, although it
has to be said that this process should better be described by
string field theory, as it is an off-shell process and the
endpoint of this tachyon condensation process so far cannot be
determined on general grounds.

Secondly, the familiar electroweak symmetry breaking which needs a
bifundamental Higgs doublet has to find an explanation within the
discussed model.

\subsubsection{Tachyons}
It has to be distinguished between tachyons in different sectors
of our model: when the two branes are in different equivalence
classes $[a]$ and $[b]$, the tachyonic Higgs field is in the
bifundamental representation of the $U(N_{a}) \times U(N_{b})$
gauge group and the condensation resembles the Higgs mechanism of
electroweak symmetry breaking. On the contrary, when the two
branes are elements of the same orbit $[a]$, the Higgs field will
be in the antisymmetric, symmetric or adjoint representation of
the  $U(N_{a})$ and thus affect only this factor.

A study among all the $4\cdot 36^3$ models looking for a suitable
tachyon spectrum has been performed. In any sector of open strings
stretching between two D6-branes $a$ and $b$ the lightest physical
state has a mass given by (\ref{eq:zeropoint_z3}). By expressing
the angle variables in terms of winding numbers, a computer
program has been set up to search for models with a Higgs scalar
in the $({\bf 2,1})$ and/or another one in the `symmetric'
representation of $U(1)_{B-L}$. All other open string sectors need
to be free of tachyons. The results of this search are listed in
table \ref{tab:tachyons_z3}.
\begin{table}
\centering
\sloppy
\renewcommand{\arraystretch}{1.1}
\begin{tabular}{|c||c|c|c|c|c|c|c|}
  \hline
Model & $U(3)$ & $[U(3),U(2)]$ &$[U(3),U(1)]$
& $U(2)$ & $[U(2),U(1)]$ & $U(1)$ &$\#$\\
  \hline\hline
{\bf AAA}&$\times$&$\times$&$\times$&$\times$&$-$&$-$&384\\
{\bf AAA}&$\times$&$\times$&$\times$&$\times$&$\times$&$-$&384\\
{\bf AAA}&$\times$&$\times$&$\times$&$\times$&$\times$&$\times$&0\\
\hline
{\bf AAB}&$\times$&$\times$&$\times$&$\times$&$-$&$-$&16320\\
{\bf AAB}&$\times$&$\times$&$\times$&$\times$&$\times$&$-$&10944\\
{\bf AAB}&$\times$&$\times$&$\times$&$\times$&$\times$&$\times$&10944\\
\hline
{\bf ABB}&$\times$&$\times$&$\times$&$\times$&$-$&$-$&17472\\
{\bf ABB}&$\times$&$\times$&$\times$&$\times$&$\times$&$-$&9024\\
{\bf ABB}&$\times$&$\times$&$\times$&$\times$&$\times$&$\times$&9024\\
\hline
{\bf BBB}&$\times$&$\times$&$\times$&$\times$&$-$&$-$&768\\
{\bf BBB}&$\times$&$\times$&$\times$&$\times$&$\times$&$-$&768\\
{\bf BBB}&$\times$&$\times$&$\times$&$\times$&$\times$&$\times$&384\\
\hline
\end{tabular}
\caption[Freedom from tachyons in the extended standard-like
model.]{Freedom from tachyons for all sectors between either
branes of the same or different equivalence classes and the number $\#$ of complying models. $\times$
denotes freedom from tachyons in the specific sector, $-$ means
that the tachyon condition is unchecked.} \label{tab:tachyons_z3}
\end{table}
Some important observations can be made from this table:
\begin{enumerate}
    \item For the {\bf AAA} and the {\bf BBB} type tori one can get
D6-brane configurations that display only tachyons charged under
$U(1)_{B-L}$, but none of these models does have a suitable tachyon
in the $({\bf 2,1})$. Vice versa, the {\bf AAB} and {\bf ABB}
models do have Higgs fields in the $({\bf 2,1})$ but no singlets
charged under $U(1)_{B-L}$.
    \item For all tori except the ${\bf AAA}$ type, one can even
set up D6-brane configurations without tachyons at all.
    \item There are several hundred models having either a
tachyon in $({\bf 2,1})$ or a tachyon in the `symmetric'
representation of $U(1)_{B-L}$. On the other hand, none of the
models contains both tachyons, what seems to be discouraging at
first sight. Regarding the necessity to have a hierarchy of a high
scale breaking of $U(1)_{B-L}$ and a low scale electroweak Higgs
mechanism, we are forced to choose a model with a singlet tachyon
condensing at the string scale but without any Higgs field in the
$({\bf 2,1})$. This favors an alternative mechanism for
electroweak symmetry breaking. An explicit realization is for
example given by
\begin{align}\label{eq:examplez3}
    [(n_1^I, m_1^I)] &= [(-3,2),(0,1),(0,-1)], \\
 [(n_2^I, m_2^I)] &= [(-3,2),(0,1),(0,-1)] ,\nonumber \\
  [(n_3^I,m_3^I)] &= [(-3,2),(1,-1),(-1,0)].\nonumber
\end{align}
 This model has precisely 3 Higgs singlets
\begin{equation}
h_i : \qquad ({\bf 1,1})_{(0,0,-2)}\ ,
\end{equation}
carrying only $B-L$ charge but no hypercharge. They are former
`superpartners' of the right-handed neutrinos.

\item Another astonishing result turns out while examining the possible
solutions to the tadpole equations: those models displaying the
Higgs singlet charged under $U(1)_{B-L}$ and no tachyons
otherwise, result from a model with a gauge group $SU(5)\times
U(1)$, deformed by giving a vacuum expectation value to a scalar
in the adjoint of $SU(5)$. Geometrically, this is evident in the
fact that the stacks of branes that support the $SU(3)$ and
$SU(2)_L$ are always parallel, thus their displacement is a
marginal deformation at tree level. Of course, we have to expect
that quantum corrections will generate a potential for the
respective adjoint scalar.
\end{enumerate}
\subsubsection{Yukawa couplings}
There is one important deviation of the discussed model from the
standard model: due to the additional global symmetries, an
appropriate Yukawa coupling giving a mass to the $(u,c,t)$-quarks
is absent. This can be seen from the quantum numbers of the Higgs
field $\tilde H$. It can be determined regarding the relevant
Yukawa coupling
\begin{equation}\label{eq:yukawa_z3}
    \tilde H\,\bar{Q}_L\, u_R\ .
\end{equation}
Gauge invariance forces $\tilde H$ to have the quantum
numbers
\begin{equation}\label{eq:htilde_qnum}
    \tilde H:\qquad ({\bf 1,2})_{(3,1,0)}\ .
\end{equation}
Unfortunately, no microscopic open string state can transform in
the singlet representation of $U(3)$ and at the same time have a
$U(1)$ charge $q=3$. On the other hand, the relevant Yukawa
coupling for the $(d,s,b)$ quarks and the leptons is given by
\begin{equation}\label{eq:yukawa_alle_z3}
    H\, \bar{Q}_L\,d_R, \qquad H\, \bar{l}_L\, e_R,
    \qquad H^*\, \bar{l}_L\, \nu_R,
\end{equation}
leading to the quantum numbers $({\bf 1,2})_{(0,1,1)}$ for the
Higgs fields $H$. This is not in contradiction to the open string
origin of the model.

From this observation, one has to draw the conclusion that in open
string models where the $(u,c,t)$-quarks arise from open strings
in the antisymmetric representation of $U(3)$, the usual Higgs
mechanism does not apply. The only solution to this problem
appears to be replacing the fundamental Higgs scalar by an
alternative composite operator, possessing the quantum numbers
stated in (\ref{eq:htilde_qnum}).

In order to end this section, the generation of neutrino masses
shall be discussed. A scalar $h$ in the `symmetric' representation
of the $U(1)_3$ can break the $U(1)_{B-L}$ symmetry via the Higgs
mechanism. This does not directly lead to a Yukawa coupling of
Majorana type for the right moving neutrinos. However, the
dimension five coupling
\begin{equation}\label{eq:dimfive}
    {1\over M_s}\, \left(h^*\right)^2\, \left(\bar\nu^c\right)_L\, \nu_R
\end{equation}
is invariant under all global symmetries and leads to a Majorana
mass for the right moving neutrinos. Together with the predicted
composite Higgs mechanism for the standard Higgs field this might
allow the realization of the see-saw mechanism in order to
generate small neutrino masses.
\subsubsection{Weinberg angle}
The $U(N_a)$ gauge couplings can be obtained by simple dimensional
reduction. This leads to
\begin{equation}\label{eq:gaugecouplings_stdmod}
    {4\pi^2\over g_a^2}={M_s\over g_s} \prod_{I=1}^3 L_a^I\ ,
\end{equation}
where $g_s$ is the string coupling. The gauge coupling for the hypercharge
\begin{equation}\label{eq:gaugecoupling_hypercharge}
    Q_Y=\sum_a c_a\, Q_a
\end{equation}
generally is given by
\begin{equation}\label{eq:gaugecoupling_hyperchargeexplicit}
    {1\over g_Y^2}=\sum_a {1\over 4} {c_a\over g_a^2}\ ,
\end{equation}
if the the normalization tr$(Q_a^2)=1/2$ is used for the abelian
subgroups $U(1)_a\subset U(N_a)$. For our specific model, this
yields a Weinberg angle
\begin{equation}\label{eq:thetaweinberg_stdmod}
\sin^2 \vartheta_W={3\over 6+2{g_2\over g_1}}\ ,
\end{equation}
where the hypercharge in terms of the correctly normalized
$U(1)$s, denoted by a tilde, is given by
\begin{equation}\label{eq:hypercharge_nochmal}
Q_Y=-{2\over 3} U(1)_1 + U(1)_2 =
         -{2\over 3}\sqrt{6}\, \widetilde{U(1)}_1 +
2\,\widetilde{U(1)}_2\ .
\end{equation}
Since in all interesting cases  the $U(3)$ branes have the same
internal volumes like the $U(2)$ branes, (\ref{eq:thetaweinberg_stdmod})  reduces to
 the prediction $\sin^2 \vartheta_W=3/8$, which is precisely the
$SU(5)$ GUT result. This indeed gives a strong prediction for any
$\mathbb{Z}_3$ orbifold model, different as compared to the
examples on toroidal orbifolds \cite{Ibanez:2001nd}, where the
complex structure moduli also play an important role.
\subsubsection{Proton decay}
One of the most important obstacles any phenomenological
standard-like model has to overcome is a possible proton decay.
Indeed, in the discussed model only the combination $B-L$ appears
as a symmetry, so that there are potential problems with the
stability of the proton. Following \cite{Aldazabal:2000cn}, as
long as the quark fields appear in bifundamental representations
of the stringy gauge group, the proton is stable because effective
couplings with three quarks are forbidden. This argument does not
apply in the present case.
\begin{figure}
\centering
\includegraphics[scale=0.5]{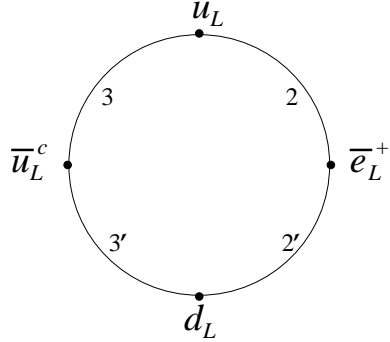}
\caption{Generation of a dimension six coupling.}\label{fig:dimsixcoupling}
\end{figure}
In figure \ref{fig:dimsixcoupling}, it is shown that a disc
diagram can generate a dimension six coupling
\begin{equation}\label{eq:dimsixcoupling}
    {\cal L}\sim {1\over M_s^2}
             \left(\bar{u}^c_L \, u_L\right)\,
             \left(\bar{e}^+_L \,  d_L\right),
\end{equation}
preserving $B-L$ but violating baryon and lepton numbers
separately. The numbers at the boundary indicate the D6$_a$-brane
to which the boundary of the disc is attached.

The conclusion which can be drawn from this observation is that as
long as the large extra dimension scenario is assumed yielding a
string scale $M_s\ll 10^{16}$GeV, there is a serious problem with
proton decay. This result certainly advocates a string scale at
the GUT scale opposite to one at the TeV scale.

\subsection{Flipped $SU(5)\times U(1)$ GUT model}
This section gives a reinterpretation of the extended
standard-like model of the last section as a GUT scenario. Similar
models can be found in \cite{Axenides:2003hs}. The unified model
in our case can be obtained from the standard-like model by moving
the two stacks for the $U(2)$ and $U(3)$ sector on top of each
other, or in other words, by giving a vanishing VEV to the adjoint
Higgs {\bf 24}.

In this realization, the common GUT gauge group $SU(5)$ is
extended by a single gauged $U(1)$ symmetry. On two stacks of
branes with $N_{5}=5$ and $N_{1}=1$ the model is realized by
picking again
\begin{equation}\label{eq:realisation_GUT}
(Y_5,Z_5)=\left(3,{1\over 2}\right),\qquad
(Y_1,Z_1)=\left(3,-{1\over 2}\right) .
\end{equation}
The task of expressing these effective winding numbers in terms of
$[(n_a^I,m_a^I)]$ quantum numbers is identical to that for the
previously discussed extended standard model. The number of
solutions is again 36 for each stack, yielding a total set of
$4\cdot 36^2$ inequivalent models. The chiral fermion spectrum is
given in table \ref{tab:GUTchiralspec}.
\begin{table}
\centering
\sloppy
\renewcommand{\arraystretch}{1.3}
\begin{tabular}{|c||c|c|}
  \hline
 Number & $SU(5)\times U(1)^2$ & $U(1)_\text{free}$\\
  \hline\hline
$3$ & $(\bar{\bf 5},{\bf 1})_{(-1,1)}$ &  $-{6\over 5}$\\
$3$ & $({\bf 10},{\bf 1})_{(2,0)}$ & ${2\over 5}$ \\
$3$ & $ ({\bf 1},{\bf 1})_{(0,-2)}$ & ${-2}$\\
\hline
\end{tabular}
\caption{Left-handed fermions of the flipped $SU(5)\times U(1)$
model.} \label{tab:GUTchiralspec}
\end{table}
The Green-Schwarz mechanism yields an anomalous $U(1)$ which is given by
\begin{equation}\label{eq:anomu1_GUT}
    U(1)_\text{mass}=5\, U(1)_5 + U(1)_1\ ,
\end{equation}
in accordance with (\ref{eq:anomalousu1_z3}), and the anomaly-free one is
\begin{equation}\label{eq:nonanomu1_GUT}
    U(1)_\text{free} = {1\over 5}\, U(1)_5 - U(1)_1\ ,
\end{equation}
This is the desired field content of a grand unified standard
model with additional right-handed neutrinos. Consequently, the
model fits into $SO(10)$ representations. The usual minimal Higgs
sector consists of the adjoint {\bf 24} to break $SU(5)$ to
$SU(3)\times SU(2)_L \times U(1)_Y$ and a $({\bf 5,1})$ which
produces the electroweak breaking. In addition we now also need to
have a singlet to break the extra $U(1)_{\rm free}$ gauge factor.
The adjoint scalar is present as being a part of the
vectormultiplet of the formerly ${\cal N}=4$ supersymmetric sector
of strings starting and ending on identical branes within the
stack $[5]$. Turning on a VEV in the supersymmetric theory means
moving on the Coulomb branch of the moduli space, which
geometrically translates into separating the five $D6_5$-branes
into parallel stacks of two and three branes. The form of the
potential generated for this modulus after supersymmetry breaking
is not known, and the existence of a negative mass term, being
required for the spontaneous condensation, remains speculative.

After having identified the $SU(5)$ GUT as a standard model where
two stacks of branes are pushed upon each other, it can be
referred to the former analysis of the scalar spectrum for the
other two needed Higgs fields. The results of the search for Higgs
singlets and bifundamentals done for the $SU(3) \times SU(2)_L
\times U(1)_Y \times U(1)_{B-L}$ model within the previous chapter
apply without modification, as the two stacks for $SU(3)\times
SU(2)_L$ are parallel in all cases. In \cite{Ellis:2002ci} it was
realized that the discussed model actually is a flipped $SU(5)$
model.

Similarly to the standard-like model, here the $U(1)_\text{mass}$
does not allow Yukawa couplings of the type ${\bf 10\cdot 10\cdot
5}$. Again the standard mass generation mechanism does not work.
Gauge symmetry breaking and mass generation have to be achieved
proposing a composite Higgs fields with the right quantum numbers.

\chapter{The MSSM on the $\mathbb{Z}_4$-orientifold}\label{cha:Z4}
In this chapter, we will concentrate on the construction of
supersymmetric orientifold models. In recent times, it often has
been stressed (see for instance \cite{Witten:2002wb}) that a
concrete realization of a model from string theory, having exactly
the matter content of the MSSM without any additional exotic
matter and at the same time $\mathcal{N}$=1 supersymmetry, still
is missing. On the other hand, such a model might be much more
predictive than just effective supergravity. Therefore, maybe it
could already be tested within the next generation of particle
accelerators, or concretely, the LHC.

The $\mathbb{Z}_4$-orbifold, containing also fractional D-branes,
is particularly useful for this purpose. In the end, a
3-generation Pati-Salam model will be discussed in detail. It is
shown that it can be broken down to a MSSM-like model, involving
several brane recombination mechanisms.
\section{The $\mathbb{Z}_4$-orbifold}

The following model will be discussed in this chapter:
\begin{equation}\label{eq:z4model}
    {{\text{Type\ IIA}}\ {\text{on}}\ T^{6} \over
                  \{\mathbb{Z}_4+\Omega R \mathbb{Z}_4\} } .
\end{equation}
The left-right symmetric $\mathbb{Z}_4$ orbifold shall act as
\begin{align}\label{eq:z4action}
    &\Theta_1:\qquad  Z_1\to e^{\pi i/ 2}\, Z_1\\
    &\Theta_2:\qquad  Z_2\to e^{\pi i/ 2}\, Z_2\nonumber\\
    &\Theta_3:\qquad  Z_3\to e^{-\pi i}\, Z_3\nonumber
\end{align}
on the three complex coordinates, assuming again a factorization
of the 6-torus by $T^6=T^2 \times T^2 \times T^2$. Indeed,
(\ref{eq:z4action}) corresponds to one case given in the cyclic
orbifold classification in table \ref{tab:Orbifold_classsification},
so this action in the closed string sector preserves
$\mathcal{N}$=2 supersymmetry in four dimensions without and
$\mathcal{N}$=1 together with the orientifold projection $\Omega R$.
Consequently, the orbifold can be seen as a singular limit of a
Calabi-Yau threefold.

An important observation from (\ref{eq:z4action}) is that the
orbifold action on the first two tori is a $\mathbb{Z}_4$-
rotation, whereas that on the last torus is a
$\mathbb{Z}_2$-rotation. As a direct consequence, the first two
tori have to be rectangular and therefore, the complex structure
on these tori is fixed to $U_2=1$.

\begin{figure}
\centering
\includegraphics{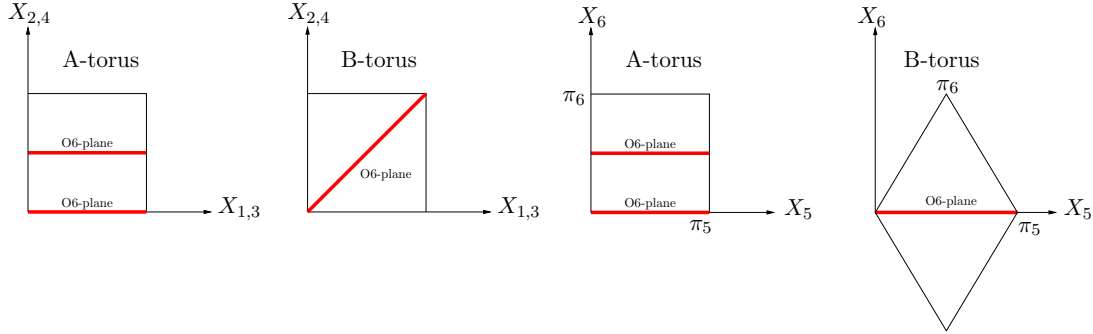}
\caption[The anti-holomorphic involutions for the  $\mathbb{Z}_4$
orientifold.]{The anti-holomorphic involutions and conventions on
the three 2-tori for the $\mathbb{Z}_4$
orientifold.}\label{fig:tori_z4}
\end{figure}

There are again two inequivalent choices for the each 2-torus, in
the T-dual picture corresponding to a vanishing or non-vanishing
constant NS-NS 2-form flux. Again, they shall be called A- and
B-torus. In the preceding chapters, the orientifold projection always
has been chosen for both tori in such a way that it was reflecting
along the X-axis and besides, the B-torus was tilted. Instead, one
can choose the B-torus to be rectangular, but at the same time take a
different choice for the anti-holomorphic involution $R$:
\begin{align}\label{eq:z4antiholomorphic_involutions}
& {\bf A}:\ Z_i{\buildrel R \over \longrightarrow}  \overline{Z}_i\\
& {\bf B}:\ Z_i{\buildrel R \over \longrightarrow} e^{\frac{\pi i}{2}} \overline{Z}_i\ .\nonumber
\end{align}
For reasons of simplicity, we will choose this convention on the
first two tori and on the third torus the usual one (meaning a tilted
B-torus). This choice is shown together with the $\Omega R$ fixed
point sets in figure \ref{fig:tori_z4}.
As a consequence, the complex structure on the 3rd torus is given
by $U=i U_2$ for the A-torus with $U_2$ unconstrained and by
$U=1/2+i U_2$ for the B-torus.

By combining all possible choices of complex conjugations, one
obtains eight possible orientifold models. However, taking into
account that the orientifold model on the $\mathbb{Z}_4$ orbifold
does not only contain the orientifold planes related to $\Omega R$
but likewise the orientifold planes related to $\Omega R \Theta$,
$\Omega R \Theta^2$ and $\Omega R \Theta^3$, only the four models
AAA, ABA, AAB and ABB are different.

The hodge numbers of this threefold are given by $h_{2,1}=7$ and
$h_{1,1}=31$, where the number $h_{2,1}=7$ corresponds to the number
of complex structure and $h_{1,1}=31$ to the number of K\"ahler
deformations. These moduli are coming from the following sectors:
\begin{itemize}
    \item {\bf untwisted sector:} As two complex structures are
fixed, another unconstrained one from the third torus remains in
the untwisted sector. Furthermore, there are 5 K\"ahler moduli in
this sector, where in each case two are coming from the first and
second trorus, one from the last. \item {\bf $\Theta$ and
$\Theta^3$ twisted sectors:} These two sectors together give rise
to 16 K\"ahler moduli, because they do contain 16
$\mathbb{Z}_4$-fixed points. \item {\bf $\Theta^2$ twisted
sector:} In this sector, there are 16 $\mathbb{Z}_2$-fixed points
from which 4 are also $\mathbb{Z}_4$-fixed points. Just the
$\mathbb{Z}_4$-fixed points give rise to one K\"ahler modulus
each. The remaining twelve $\mathbb{Z}_2$-fixed points are
organized in pairs under the $\mathbb{Z}_4$ action and so give
rise to 6 complex structure and 6 K\"ahler moduli.
\end{itemize}
The $\mathbb{Z}_2$-twisted sector contributes $h^{tw}_{2,1}=6$
elements to the number of complex structure deformations and
therefore contains twisted 3-cycles, which is the important new
feature of this $\mathbb{Z}_4$ orbifold compared to the
$\mathbb{Z}_3$ of the preceding chapter.

\section{An integral basis for $H_3(M,\mathbb{Z})$}

It would be very tedious to determine the tadpoles and the
massless chiral spectrum of this model by a pure CFT calculation,
as in the present case twisted sector tadpoles contribute as well.
So we will take a different route and use the description in terms
of homology that has been introduced in the two chapters
\ref{cha:RtadpolesOR} and \ref{cha:chiral_spectrum_general}. The
R-R tadpole reads in terms of homology
\begin{equation}\label{eq:RRtadpoleZ4}
\sum_a  N_a\, (\pi_a + \pi'_a)-4\, \pi_{O6}=0.
\end{equation}
As a first step, we have to determine the independent 3-cycles on the
$\mathbb{Z}_4$ orbifold space. The third betti number is given by
$b_3=2+2 h_{21}=16$, so one expects to get precisely this number
of independent 3-cycles, if one does not consider the $\Omega R$
projection at this point.

\begin{figure}
\centering
\includegraphics{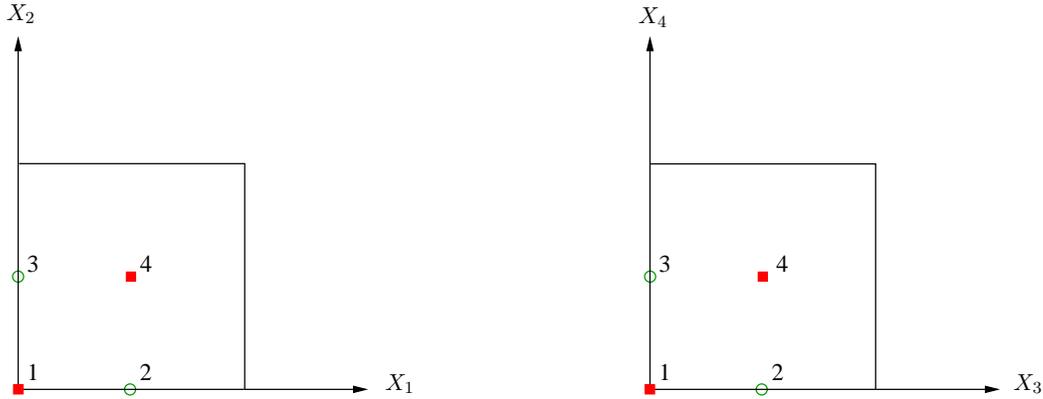}
\caption{The $\mathbb{Z}_4$ orbifold fixed points on the first two
$T^2$.}\label{fig:z4_fixedpoints}
\end{figure}

The first set of these cycles can be obtained easily, as it
descends from the ambient toroidal space $T^6$. The two
fundamental cycles on the torus $T^2_I$ ($I=1,2,3$) shall be
denoted as $\pi_{2I-1}$ and $\pi_{2I}$. Then the
toroidal 3-cycles can be defined by
\begin{equation}\label{eq:toroidal3cycles}
    \pi_{ijk} \equiv \pi_i\otimes\pi_j\otimes\pi_k\ .
\end{equation}
If we take the orbits under the $\mathbb{Z}_4$ action, we can
immediately deduce four $\mathbb{Z}_4$-invariant 3-cycles,
\begin{align}\label{eq:invariantz4cycles}
\rho_1 &\equiv 2 (\pi_{135}-\pi_{245} ),   \qquad
\bar{\rho}_1   \equiv 2 (\pi_{136}-\pi_{246} ), \\
\rho_2 &\equiv 2 (\pi_{145}+\pi_{235} ), \qquad
\bar{\rho}_2  \equiv 2 (\pi_{146}+\pi_{236} ) \ .\nonumber
\end{align}
The factor of two comes from the fact that $1$ and $\Theta^2$ as
well as $\Theta$ and $\Theta^3$ act equivalently on the toroidal
cycles. The intersection matrix on the orbifold space for these
3-cycles can be obtained from equation
(\ref{eq:intersection_orbifold_space}). It is given by
\begin{equation}\label{eq:intersectionmatrixz4_nonfractional}
     I_{\rho}=\bigoplus_{i=1}^2  \left(\begin{matrix} 0 & -2 \\
     2 & 0 \\ \end{matrix}\right)\ .
\end{equation}
We are still missing twelve 3-cycles, these do arise in the
$\mathbb{Z}_2$ twisted sector of the orbifold. Since $\Theta^2$
acts non-trivially only onto the first two 2-tori, the sixteen
$\mathbb{Z}_2$-fixed points do appear in the $\mathbb{Z}_2$
twisted sector on these tori, this is shown in figure
\ref{fig:z4_fixedpoints}.
The boxes in the figure indicate the $\mathbb{Z}_2$-fixed points
which are also fixed under the $\mathbb{Z}_4$ symmetry, the
circles indicate the ones that are just fixed under the $\mathbb{Z}_2$ and get
exchanged under the $\mathbb{Z}_4$ action.

After blowing up the orbifold singularities, each of these fixed
points gives rise to an exceptional 2-cycle $e_{ij}$ with the
topology of $S^2$. These exceptional 2-cycles can be combined with
the two fundamental 1-cycles on the third torus (indicated by
$\pi_5$ and $\pi_6$ in figure \ref{fig:tori_z4}) to form what
might be called exceptional 3-cycles with the topology of
$S^2\times S^1$. But the $\mathbb{Z}_4$ action leaves only four
fixed points invariant and arranges the remaining twelve in six
pairs. Since it acts by a reflection on the third torus, its
action on the exceptional cycles $e_{ij}\otimes \pi_{5,6}$ is
given by
\begin{equation}\label{eq:z4action_except_cycles}
    \Theta\left(e_{ij}\otimes \pi_{5,6}\right)=-e_{\theta(i)\theta(j)} \otimes \pi_{5,6}\ ,
\end{equation}
where
\begin{equation}\label{eq:z4action_except_cycles_ergaenzung}
    \theta(1)=1,\quad  \theta(2)=3,\quad \theta(3)=2,\quad
                      \theta(4)=4\ .
\end{equation}
The $\mathbb{Z}_4$-invariant fixed points drop out because of the
minus sign in (\ref{eq:z4action_except_cycles}) and exactly twelve
3-cycles remain, being
\begin{align}
    \varepsilon_{1}       & \equiv (e_{12}-e_{13})\otimes\pi_5\ ,  \qquad
\bar{\varepsilon}_{1}  \equiv (e_{12}-e_{13})\otimes\pi_6\ , \\
\varepsilon_{2}       & \equiv (e_{42}-e_{43})\otimes\pi_5\ , \qquad
\bar{\varepsilon}_{2}  \equiv (e_{42}-e_{43})\otimes\pi_6\ ,\nonumber \\
\varepsilon_{3}       & \equiv (e_{21}-e_{31})\otimes\pi_5\ , \qquad
\bar{\varepsilon}_{3}  \equiv (e_{21}-e_{31})\otimes\pi_6\ ,\nonumber \\
\varepsilon_{4}       & \equiv (e_{24}-e_{34})\otimes\pi_5\ , \qquad
\bar{\varepsilon}_{4}  \equiv (e_{24}-e_{34})\otimes\pi_6\ ,\nonumber \\
\varepsilon_{5}       & \equiv (e_{22}-e_{33})\otimes\pi_5\ , \qquad
\bar{\varepsilon}_{5}  \equiv (e_{22}-e_{33})\otimes\pi_6\ ,\nonumber \\
\varepsilon_{6}       & \equiv (e_{23}-e_{32})\otimes\pi_5\ , \qquad
\bar{\varepsilon}_{6}  \equiv (e_{23}-e_{32})\otimes\pi_6\ .\nonumber
\end{align}
Using (\ref{eq:intersection_orbifold_space}), the resulting
intersection matrix for the exceptional cycles takes the simple form
\begin{equation}\label{eq:intersectionmatrixz4_fractional}
    I_{\varepsilon}=\bigoplus_{i=1}^6  \left(\begin{matrix}0 & -2 \\
    2 & 0 \\ \end{matrix}\right) \ .
\end{equation}
These cycles do not form an integral basis of the free module,
although being part of the homology group $H_3(M,\mathbb{Z})$. The
reason is that their intersection matrix is not unimodular.

On the other hand, it is possible to obtain an integral basis:
Concerning the physical interpretation, what we are missing are
the so-called fractional D-branes that first have been described
in \cite{Douglas:1996sw, Douglas:1997xg}. Regarding our model,
these are D-branes which only wrap one-half times around any
toroidal cycle $\rho_1$, $\bar\rho_1$, $\rho_2$ or $\bar\rho_2$
and at the same time wrap around some of the exceptional cycles in
the blown-up manifold. In the orbifold limit they
are stuck at the fixed points and at least two such branes have to
combine to form a normal D-brane that can move into the bulk.

At this point, we have to clarify which combinations of toroidal
and exceptional cycles are possible for a fractional D-brane. This
can be figured out as follows: if a D-brane wraps around
a toroidal cycle, it passes through one or two of the
$\mathbb{Z}_2$-fixed points. It seems reasonable that only these
fixed points are able to contribute to the fractional brane. For
instance, if a brane wraps the toroidal cycle $1/2\,\rho_1$, then
only the exceptional cycles $\epsilon_1$, $\epsilon_3$ or
$\epsilon_5$ should be allowed. The total homological cycle then
could be for instance
\begin{equation}\label{eq:example_frac_brane}
\pi_a={1\over 2}\rho_1+{1\over 2}(
                \varepsilon_1+\varepsilon_3+\varepsilon_5)\ .
\end{equation}
From these simple considerations, we cannot fix the relative signs
of the different terms in (\ref{eq:example_frac_brane}). An
interpretation for these signs is given by the correspondence
to Wilson lines along an internal direction of the D-brane, which
we are free to turn on. Indeed, this construction is completely
analogous to the construction of boundary states for fractional
D-branes carrying also a charge under some $\mathbb{Z}_2$ twisted
sector states \cite{Sen:1998ii, Diaconescu:1998br,
Diaconescu:1999dt, Gaberdiel:2000jr}.

Therefore, only unbarred respectively barred cycles can be
combined into fractional cycles, as they wrap the same fundamental
1-cycle on the third 2-torus. Then the only non-vanishing
intersection numbers are in between barred and unbarred cycles.

Any unbarred fractional D-brane can be expanded as
\begin{equation}\label{eq:unbarred_frac_brane}
    \pi_a=v_{a,1} \rho_1 + v_{a,2}
\rho_2 + \sum_{i=1}^6 v_{a,i+2}\,
         \varepsilon_i\ ,
\end{equation}
where the coefficients $v_{a,i}$ are half-integer valued. We can
associate to it  a barred brane by simply exchanging the two
fundamental cycles on the third 2-torus
\begin{equation}\label{eq:barred_frac_brane}
\bar\pi_a=v_{a,1} \bar\rho_1 + v_{a,2} \bar\rho_2 + \sum_{i=1}^6 v_{a,i+2}\,
       \bar\varepsilon_i\ ,
\end{equation}
having the same coefficients $v_{a,i+8}=v_{a,i}$ for
$i\in\{1,\ldots,8\}$.

Now, all linear combinations with an intersection number $\pi\circ
\bar\pi=-2$ can be constructed if we respect the rules given above.
This provides a basis for the homology lattice. The
cycles respecting $\pi\circ \bar\pi=-2$ can be divided into three sets:

\begin{enumerate}
\item $\{(v_1,v_2;v_3,v_4;v_5,v_6;v_7,v_8) \\
\text{with}\ v_1+v_2=\pm \frac{1}{2},\,v_3+v_4=\pm \frac{1}{2},\,
v_5+v_6=\pm \frac{1}{2},\,v_7+v_8=\pm \frac{1}{2}, \\
\text{and}\ v_1+v_3+v_5+v_7=0\ {\rm mod}\ 1\}$

These combinations correspond to flat branes being parallel to the
fundamental cycles which intersect through certain fixed points.
They define $8\cdot 16=128$ different fractional 3-cycles.
\item $\{(v_1,v_2 ;v_3,v_4;0,0;0,0),\,(v_1,v_2 ;0,0;v_5,v_6;0,0),\,
(v_1,v_2 ;0,0;0,0;v_7,v_8), \\
(0,0;v_3,v_4;v_5,v_6 ;0,0),\, (0,0;v_3,v_4;0,0;v_7,v_8),\, (0,0;0,0;v_5,v_6;v_7,v_8)\\
\text{with}\ v_i={\pm 1/2} \}$

The first three types of elements of this set stretch along one of
the toroidal fundamental cycles on one 2-torus and diagonally on
the other one. The remaining three types arise from integer linear
combinations of the cycles introduced so far. In total this yields
$6\cdot 16 = 96 $ different 3-cycles.
\item $\{(v_1,v_2
;v_3,v_4;v_5,v_6;v_7,v_8)\ \text{with exactly one}\ v_i=\pm
  1\ \text{and the rest zero} \}$

Only the vectors with $v_1=\pm 1$ or $v_2=\pm 1$ can be derived from
untwisted branes. They are purely untwisted. The purely twisted
ones again arise from linear combinations. This third set contains
$2\cdot 8= 16$ 3-cycles.
\end{enumerate}
Adding up all possibilities, there are 240 distinct 3-cycles obeying
$\pi\circ \bar\pi=-2$. Intriguingly, this just corresponds to the
number of roots of the $E_8$ Lie algebra. Therefore, with the help
of a computer program, it is possible to search for a basis among
these 240 cycles such that the intersection matrix takes the
following form
\begin{equation}\label{eq:intersection_e8}
    I=\left(\begin{matrix}0 & C_{E_8} \\
                              -C_{E_8} & 0 \\
                              \end{matrix}\right)\ ,
\end{equation}
where $C_{E_8}$ denotes the Cartan matrix of $E_8$ which is given by
\begin{equation}\label{eq:cartan_matrix_e8}
     C_{E_8}=\left[\begin{matrix}
    -2  &   1 &   0 &  0 &  0 &  0 &  0 &   0 \\
    1  &  -2 &   1 &  0 &  0 &  0 &  0 &   0 \\
    0  &   1 &  -2 &  1 &  0 &  0 &  0 &   0 \\
    0  &   0 &   1 & -2 &  1 &  0 &  0 &   0 \\
    0  &   0 &   0 &  1 & -2 &  1 &  0 &   1 \\
    0  &   0 &   0 &  0 &  1 & -2 &  1 &   0 \\
    0  &   0 &   0 &  0 &  0 &  1 & -2 &   0 \\
    0  &   0 &   0 &  0 &  1 &  0 &  0 &  -2 \\
    \end{matrix}
    \right]\ .
\end{equation}
A typical solution that has been found is given by
\begin{align}\label{eq:simple_roots}
    \vec{v}_1 &= {1\over 2 }(-1,\phantom{-}0, -1,\phantom{-}0,-1,\phantom{-}0,-1,
\phantom{-}0)\nonumber \\
 \vec{v}_2 &= {1\over 2 } (\phantom{-}1,\phantom{-}0,\phantom{-}1,\phantom{-}0,
\phantom{-}1,\phantom{-}0,-1,\phantom{-}0)\nonumber \\
 \vec{v}_3 &= {1\over 2 } (\phantom{-}1,\phantom{-}0, -1,\phantom{-}0,-1,
\phantom{-}0,\phantom{-}1,\phantom{-}0)\nonumber \\
 \vec{v}_4 &= {1\over 2 } (-1,\phantom{-}0,\phantom{-}1,\phantom{-}0,
\phantom{-}0,\phantom{-}1,\phantom{-}0,\phantom{-}1) \\
\vec{v}_5 &= {1\over 2 } (\phantom{-}0,\phantom{-}1,-1,
\phantom{-}0,\phantom{-}1,\phantom{-}0,\phantom{-}0,-1)\nonumber \\
 \vec{v}_6 &= {1\over 2 } (\phantom{-}0,-1,\phantom{-}1,\phantom{-}0,-1,
  \phantom{-}0,\phantom{-}0,-1)\nonumber \\
\vec{v}_7 &= {1\over 2 } (\phantom{-}0,\phantom{-}1,\phantom{-}0,\phantom{-}1,
\phantom{-}0,-1,\phantom{-}0,\phantom{-}1)\nonumber \\
\vec{v}_8 &= {1\over 2 } (\phantom{-}0,-1, \phantom{-}0,-1,\phantom{-}0, -1,
 \phantom{-}0,\phantom{-}1).\nonumber
\end{align}
Since the Cartan matrix is unimodular, we indeed have constructed
by (\ref{eq:simple_roots}) an integral basis for the homology
lattice $H_3(M,\mathbb{Z})$.

In the following, it nevertheless turns out to be more convenient
to work with the non-integral orbifold basis which also allows for
half-integer coefficients. But one has to be careful in doing so:
not all such cycles are part of $H_3(M,\mathbb{Z})$, one has to
ensure for every apparent fractional 3-cycle that it is indeed
part of the unimodular lattice $H_3(M,\mathbb{Z})$. This can be
achieved by checking if it can be expressed as an integer linear
combination of the basis (\ref{eq:simple_roots}).

\section{The $\mathbb{Z}_4$-orientifold}

In the last section, we have determined an integer basis for the
3-cycles in the $\mathbb{Z}_4$ orbifold on the $T^6$, but have not
treated the orientifold projection $\Omega R$ so far. A first step
in doing so is to find the homological 3-cycles which the
O6-planes are wrapping, or in other words, determine the fixed
point sets of the four relevant orientifold projections $\Omega
R$, $\Omega R \Theta$, $\Omega R\Theta^2$ and $\Omega R\Theta^3$.
Here, the exemplary case of the ABB-model will be discussed in
detail, as it will be particularly useful for model building.
The results for the other cases are listed in the
table \ref{tab:O6planes_z4} of appendix \ref{cha:app_oplanes_z4}.
\subsection{The O6-planes for the ABB-orientifold}
The fixed points sets can be expressed in terms of the toroidal
3-cycles, as shown in table \ref{tab:O6planes_ABB}.
\begin{table}
\centering
\sloppy
\renewcommand{\arraystretch}{1.5}
\begin{tabular}{|c||c|}
  \hline
  Projection & Fixed point set\\
  \hline
  \hline
  $\Omega R$ & $2\, \pi_{135} + 2\, \pi_{145}  $ \\
  $\Omega R \Theta$ & $2\, \pi_{145} + 2\, \pi_{245}- 4\, \pi_{146}- 4\, \pi_{246}$ \\
  $\Omega R \Theta^2$ & $2\, \pi_{235} - 2\, \pi_{245} $ \\
  $\Omega R \Theta^3$ & $-2\, \pi_{135} + 2\, \pi_{235}+ 4\,\pi_{136}- 4\, \pi_{236}$ \\
  \hline
\end{tabular}
\caption{The O6-planes for the ABB-torus.}
\label{tab:O6planes_ABB}
\end{table}
\noindent
By adding up all contributions, one obtains
\begin{align}\label{eq:aab_z4_oplane}
    \pi_{O6}&=4\, \pi_{145} +4\, \pi_{235} +4\, \pi_{136}-4\,\pi_{246} -4\, \pi_{146} -4\, \pi_{236} \\
                      &=2\, \rho_2+2\, \bar\rho_1 -2\, \bar\rho_2\ .\nonumber
\end{align}
Strikingly, the total O-plane can be expressed in terms of bulk
cycles only. This reflects the fact that in the conformal field
theory the orientifold planes carry only charge under untwisted
R-R fields, as mentioned in chapter \ref{cha:Zn_orientifolds} already.

Next, we are trying to determine the action of $\Omega R$ on the
homological cycles. In terms of the orbifold basis, this is just
given by
\begin{align}
    &\rho_1\to \rho_2,& &\bar\rho_1\to \rho_2- \bar\rho_2,& \\
    &\rho_2\to \rho_1,& &\bar\rho_2\to \rho_1- \bar\rho_1.&  \nonumber
\end{align}
for the toroidal cycles and by
\begin{align}
    &\varepsilon_1\to -\varepsilon_1, & &
    \bar\varepsilon_1\to -\varepsilon_1+\bar\varepsilon_1,& \nonumber\\
    &\varepsilon_2\to -\varepsilon_2, & &
    \bar\varepsilon_2\to -\varepsilon_2+\bar\varepsilon_2,& \nonumber \\
    &\varepsilon_3\to \varepsilon_3, \phantom{-} & &
    \bar\varepsilon_3\to \varepsilon_3-\bar\varepsilon_3,& \\
    &\varepsilon_4\to \varepsilon_4, \phantom{-}& &
    \bar\varepsilon_4\to  \varepsilon_4-\bar\varepsilon_4,& \nonumber \\
    &\varepsilon_5\to \varepsilon_6, \phantom{-} & &
    \bar\varepsilon_5\to \varepsilon_6-\bar\varepsilon_6,& \nonumber \\
    &\varepsilon_6\to \varepsilon_5,\phantom{-} & &
    \bar\varepsilon_6\to \varepsilon_5-\bar\varepsilon_5.& \nonumber
\end{align}
for the exceptional cycles.
\subsection{Supersymmetric cycles}\label{cha:susy_cyclesABB}
In this chapter, we are interested in supersymmetric models.
Therefore, we do not only need to have control over the
topological data of the D6-branes, but also over the nature of the
special Lagrangian cycles.

The metric at the orbifold point is flat up to some isolated
orbifold singularities. Therefore, flat D6-branes in a given
homology class are definitely special Lagrangian. At this time, we
restrict the D6-branes to be flat and factorizable in the sense
that they can be described by the usual six wrapping numbers along
the fundamental toroidal cycles, $n_I$ and $m_I$ with $I=1,2,3$,
which have to be relatively coprime on every 2-torus. Given such a
bulk brane, one can easily compute the homology class that it wraps
in the orbifold basis
\begin{align}\label{eq:bulk_z4_brane}
\pi^\text{bulk}_a=&\left[ (n_1^a\,n_2^a -m_1^a\,m_2^a) n_3^a\right]\, \rho_1 +
                    \left[(n_1^a\,m_2^a + m_1^a\,n_2^a)
                   n_3^a\right]\, \rho_2 \\
                    +&\left[(n_1^a\,n_2^a - m_1^a\,m_2^a)
                    m_3^a\right]\, \bar\rho_1 +
                    \left[(n_1^a\,m_2^a + m_1^a\,n_2^a)
                m_3^a\right]\, \bar\rho_2\ . \nonumber
\end{align}

The supersymmetry that the orientifold plane preserves is
preserved by a specific D-brane as well if it fulfills the angle
criterion: the angles on all tori have to add up to the total
angle that the O-plane spans on the three 2-tori. This concretely
reads for the ABB-orientifold
\begin{equation}\label{eq:angle_crit_z4_abb}
    \varphi_1^a+ \varphi_2^a+ \varphi_3^a={\pi\over 4}\ {\rm mod}\ 2\pi\ ,
\end{equation}
and the three angles are given by
\begin{equation}\label{eq:winkelz4_ABB}
    \tan\varphi_1^a={m_1^a\over n_1^a}\ , \qquad
    \tan\varphi_2^a={m_2^a\over n_2^a}\ , \qquad
    \tan\varphi_3^a={U_2\, m_3^a\over n_3^a+{1\over 2} m_3^a }\ ,
\end{equation}
where we have to keep in mind that only the complex structures on
the first two 2-tori are fixed. It is possible to reformulate the
criterion (\ref{eq:angle_crit_z4_abb}) in terms of wrapping
numbers, if one simply takes the tangent on both sides of the
equation. But we have to be careful in doing so, as this only
yields a necessary condition: the tangent has a periodicity ${\rm
mod}\ \pi$. Solving for $U_2$ leads to
\begin{equation}\label{eq:u2_susy_z4}
    U_2={ \left(n_3^a+{1\over 2} m_3^a\right) \over m_3^a}
                  {\left(n_1^a\,n_2^a- m_1^a\,m_2^a- n_1^a\,m_2^a-
               m_1^a\,n_2^a\right)
           \over \left(n_1^a\,n_2^a- m_1^a\,m_2^a+ n_1^a\,m_2^a+
            m_1^a\,n_2^a \right)}\ .
\end{equation}
We can interpret this equation in such a way that already one
supersymmetric stack of D-brane fixes the complex structure on the
third torus. The introduction of additional D6-branes gives rise
to non-trivial conditions on the wrapping numbers of these
D-branes. For the other three orientifold models, these
supersymmetry conditions are listed in appendix
\ref{cha:susy_cycles_z4}.

Later we want to construct a supersymmetric model containing also
fractional branes in order to enlarge the model building
possibilities.  The explicit expression in terms of wrapping
numbers takes the following form
\begin{equation}\label{eq:fractional_z4_brane}
    \pi_a^\text{frac}={1\over 2} \pi_a^\text{bulk} + {n_3^a \over 2}
                        \left[ \sum_{j=1}^6  w_{a,j} \varepsilon_j\right]
                  +  {m_3^a \over 2}
                        \left[ \sum_{j=1}^6  w_{a,j}
\bar\varepsilon_j\right]\qquad\text{with}\ w_{a,j} \in \{-1, 0, 1\}\ .
\end{equation}
In terms of the formerly introduced
coefficients $v_{a,j}$, this reads
\begin{equation}\label{eq:va_na_wa}
    v_{a,j}={n_3^a \over 2} \, w_{a,j},\qquad\quad
    v_{a,j+8}={m_3^a \over 2} \, w_{a,j}
\qquad\quad\text{for}\ j\in\{1,\ldots,8\}\ .
\end{equation}
In \ref{eq:fractional_z4_brane} we have taken into account that
all the $\mathbb{Z}_2$-fixed points are located on the first two
2-tori: on the third torus fractional D-branes do have winding
numbers along the two fundamental 1-cycles. Moreover, the
coefficients of $\varepsilon_j$ and $\bar\varepsilon_j$ in
(\ref{eq:fractional_z4_brane}) must be equal, because they only
differ by the cycle on the third torus.

Using the same determination principles for the normalization
factors as for the toroidal $\Omega R$- and $\mathbb{Z}_3$
orientifolds\footnote{These are determined by the Cardy condition,
stating that the result for the annulus partition function must
coincide for  the loop and the tree channel computation.}, these
fractional D6-branes correspond to the following boundary states
in the conformal field theory of the $T^6 /\mathbb{Z}_4$ orbifold
model
\begin{multline}\label{eq:boundary_state_z4}
|D^f;(n_I,m_I),\alpha_{ij}\rangle={1\over 2\sqrt 2} \sqrt{n_3^2+n_3 m_3 +{\textstyle{m_3^2\over 2}}}\\
\shoveleft{\cdot\left[\bigg(\frac{1}{2}\prod_{j=1}^2 \sqrt{n_j^2+m_j^2}\bigg)
\bigg( \bigl|D;(n_I,m_I)\bigr\rangle_U +\bigl|D;\Theta(n_I,m_I)\bigr\rangle_U\bigg)\right.}\\
\left.+\bigg(
     \sum_{i,j=1}^{4} \alpha_{ij} \bigl|D;(n_I,m_I),e_{ij}\bigr\rangle_T +
    \sum_{i,j=1}^{4} \alpha_{ij} \bigl|D;\Theta(n_I,m_I),\Theta(e_{ij})\bigr\rangle_T \bigg)\right]\ .
\end{multline}
In this boundary state, there are contributions from both the
untwisted and the $\mathbb{Z}_2$-twisted sector and we have taken the
orbit under the $\mathbb{Z}_4$ action, being for the winding numbers
\begin{align}\label{eq:thetaaction_z4_winding}
    &\Theta\, n_{1,2}=-m_{1,2}\ ,& &\Theta\, m_{1,2}=n_{1,2}\ ,&\\
    &\Theta\, n_{3}=-n_3\ ,& &\Theta\, m_{3}=-m_{3}\ .&\nonumber
\end{align}
implying that $\Theta^2$ acts like the identity on the boundary
states. This explains why only two and not four untwisted boundary
states do appear in (\ref{eq:boundary_state_z4}). In the sum over
the $\mathbb{Z}_2$-fixed points, for each D6-brane precisely  four
coefficients take values $\alpha_{ij}=\pm 1$ and the remaining
ones are vanishing, as these coefficients are directly related to
the $w_i$ appearing in the description of the corresponding
fractional 3-cycles. It has been mentioned already that changing
the signs of an exceptional cycle does correspond to turning on a
discrete $\mathbb{Z}_2$ Wilson line along one internal direction
of the brane \cite{Gaberdiel:2000jr}. The action of $\Theta$ on
the twisted sector ground states $e_{ij}$ has been given
in (\ref{eq:z4action_except_cycles}). Furthermore, the elementary
boundary states $|D;(n_I,m_I)\rangle_U$ are the usual ones for a
flat $D6$-brane with wrapping numbers $(n_I,m_I)$ on a 6-torus
factorizing in three 2-tori. They are given in appendix
\ref{cha:boundary_states_z4}.
\begin{table}
\centering
\sloppy
\renewcommand{\arraystretch}{1.2}
\begin{tabular}{|c||c|c|c|c|}
  \hline
   & $n_1$ odd, $m_1$ odd & $n_1$ odd, $m_1$ even &$n_1$ even, $m_1$ odd\\
  \hline\hline
  $n_2$ odd & & $\varepsilon_3$, $\varepsilon_4$& $\varepsilon_3$, $\varepsilon_4$\\
  $m_2$ odd & & $\varepsilon_5$, $\varepsilon_6$& $\varepsilon_5$, $\varepsilon_6$\\
  \hline
  $n_2$ odd& $\varepsilon_1$, $\varepsilon_2$& $\varepsilon_1$, $\varepsilon_3$, $\varepsilon_5$&
  $\varepsilon_1$, $\varepsilon_3$, $\varepsilon_6$\\
   $m_2$ even& $\varepsilon_5$, $\varepsilon_6$& $\varepsilon_1$, $\varepsilon_4$, $\varepsilon_6$&
  $\varepsilon_1$, $\varepsilon_4$, $\varepsilon_5$\\
  & & $\varepsilon_2$, $\varepsilon_3$, $\varepsilon_6$&
  $\varepsilon_2$, $\varepsilon_3$, $\varepsilon_5$\\
  & & $\varepsilon_2$, $\varepsilon_4$, $\varepsilon_5$&
  $\varepsilon_2$, $\varepsilon_4$, $\varepsilon_6$\\
  \hline
   $n_2$ even&$\varepsilon_1$, $\varepsilon_2$& $\varepsilon_1$, $\varepsilon_3$, $\varepsilon_6$&
  $\varepsilon_1$, $\varepsilon_3$, $\varepsilon_5$\\
   $m_2$ odd& $\varepsilon_5$, $\varepsilon_6$& $\varepsilon_1$, $\varepsilon_4$, $\varepsilon_5$&
  $\varepsilon_1$, $\varepsilon_4$, $\varepsilon_6$\\
  & & $\varepsilon_2$, $\varepsilon_3$, $\varepsilon_5$&
  $\varepsilon_2$, $\varepsilon_3$, $\varepsilon_6$\\
   & & $\varepsilon_2$, $\varepsilon_4$, $\varepsilon_6$&
  $\varepsilon_2$, $\varepsilon_4$, $\varepsilon_5$\\
\hline
\end{tabular}
\caption{The allowed exceptional cycles for fractional branes on
the torus.} \label{tab:allowed_excep_cycles_z4}
\end{table}
At this point, we want to implement the constraint that only those
fractional cycles are able to contribute to the fractional brane
(\ref{eq:fractional_z4_brane}) where the bulk brane is passing through.
This can be easily implemented by hand, the result is given in table
\ref{tab:allowed_excep_cycles_z4}. At first glance, there seems to be a
mismatch between the number of parameters describing a 3-cycle and
the corresponding  boundary state: there are three non-vanishing
parameters $w_i$ but four $\alpha_{ij}$ for every D-brane. But
this is misleading, as a flat fractional brane and its
$\mathbb{Z}_4$-image always intersect in precisely one
$\mathbb{Z}_4$-fixed point times a circle on the third 2-torus.
This twisted sector effectively drops out of the boundary state,
because $\Theta$ acts on this fixed point with a minus sign.

\begin{figure}[t!]
\centering
\includegraphics[width=9cm,height=9cm]{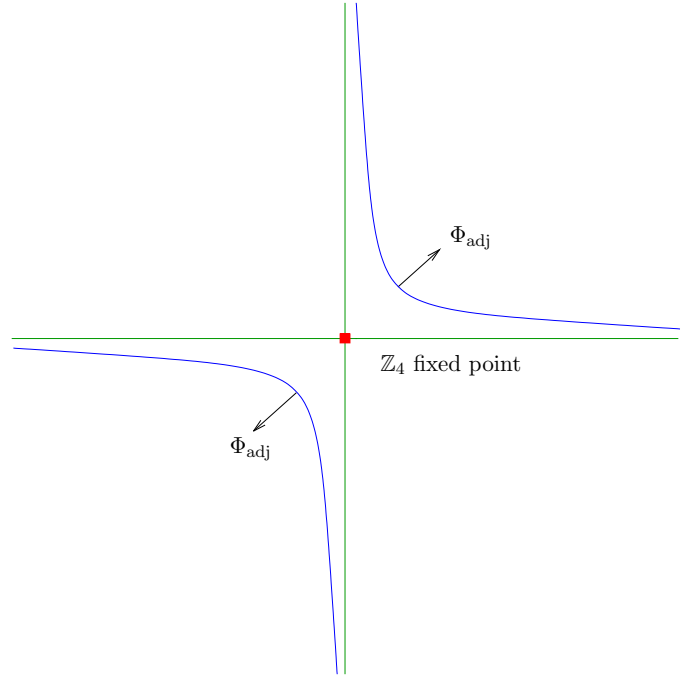}
\caption{The recombination process of two $\mathbb{Z}_4$ mirror
branes into a sigle non-flat D-brane.}\label{fig:z4recombined_branes}
\end{figure}

This is not merely a coincidence but has a far-reaching physical
interpretation: between a brane and its $\Theta$
mirror, there generally lives a hypermultiplet $\Phi_\text{adj}$ in the
adjoint representation, see equation (\ref{eq:z3_adjoint}). But
this sector is non-chiral in four dimensions and $\mathcal{N}$=2
supersymmetric, such that there exists a flat direction in the D-term
potential. This implies that the two branes recombine into a
single one which no longer passes through the $\mathbb{Z}_4$-fixed
point, an illustration of this process is given in figure
\ref{fig:z4recombined_branes}.
A non-trivial test for these considerations is given by the check if a
fractional brane (\ref{eq:fractional_z4_brane}), transformed to the
$E_8$-basis, yields integer coefficients. To see this, we write
the $8\times8$ matrix (\ref{eq:cartan_matrix_e8}) and a second
identical copy as the two diagonal blocks of a $16\times16$
matrix, and then act with the inverse of the transposed matrix
onto a general vector (\ref{eq:fractional_z4_brane}). Then we have
to investigate the different cases according to table \ref{tab:allowed_excep_cycles_z4}
separately. For instance for the case $n_1$ odd, $n_2$ odd, $m_1$
even, $m_2$ odd and fractional cycles $\varepsilon_3$,
$\varepsilon_4$ with signs $w_{3}$, $w_{4}$ respectively, we
substitute $m_1=2k_1$ and  obtain the following vector in the
$E_8$-basis:
\begin{equation}\label{eq:proof_integer}
\bigg[ \Big({1\over 2 }(n_{1}
m_{2}-w_{3})+k_{1} n_{2}\Big)n_{3},\
 \Big({1 \over 2 }(n_{1} n_{2}-w_{3})-k_{1} m_{2}+n_{1} m_{2}+2
k_{1} n_{2}\Big)n_{3},\  ...\bigg].
\end{equation}
Already for the first two components we can see what generally
happens for all cases and all components: since $n_1$, $n_2$, $m_2$
and $w_{3}$ are non-vanishing and because products of odd numbers
are odd as well, just sums and differences of two odd numbers occur.
These are always even or zero and therefore can be divided by
two and still lead to integer coefficients. This ends our
considerations about supersymmetric fractional D6-branes, and we
are ready to search for phenomenologically interesting
supersymmetric models.
\section{Phenomenological model building}\label{cha:z4modbuild}
In this section, we are interested in supersymmetric extensions of
the standard model. The most obvious attempt in model
building would be to directly search for a supersymmetric model
that has the stacks and intersection numbers as given in chapter
\ref{cha:std_model} or \cite{Ibanez:2001nd, Ibanez:2001dj}. This
in general could be achieved for any one of the four models AAA, ABA,
AAB and ABB  and for a fixed complex structure on the third torus
$U_2$. The choice of complex structure then by
(\ref{eq:u2_susy_z4}) restricts the choice for supersymmetric
branes. At least fixing $U_2$ to a specific value is a
legitimate action, because the scalar potential
(\ref{eq:scalar_pot_OR_NS}) for this modulus is vanishing. It
corresponds to a stable minimum as we will see from the NS-NS
tadpole condition.

In order to do so, a computer program has been set up to search
for such a model with wrapping numbers up to ten. This model was
required to satisfy both the R-R tadpole equation
(\ref{eq:RRtadpoleZ4}) and the supersymmetry condition
(\ref{eq:u2_susy_z4}) for a flexible number of stacks.
Unfortunately, a model with the correct intersection numbers in
all sectors in this way could not be found, although its existence
cannot be excluded on general grounds. For this reason, in the
following we will concentrate on a three generation Pati-Salam
model. This can even be broken down by certain mechanisms to a
supersymmetric standard-like model. In more detail, this process
involves two important steps of first breaking a 7 stack
Pati-Salam model down to a 3 stack model using a certain complex
structure deformation and then breaking this further down to the
supersymmetric standard-like model. This second breaking can be
achieved in two different ways, either by an adjoint or by a
bifundamental Pati-Salam breaking, both possibilities will be
discussed.

Non-supersymmetric Pati-Salam models are discussed in
\cite{Kokorelis:2002ip, Kokorelis:2002ns}. Another supersymmetric
four generation Pati-Salam model has been constructed in
\cite{Blumenhagen:2002gw}, but due to its minor phenomenological
relevance, it will not be treated in this work.
\subsection{Seven stack Pati-Salam model}
\begin{figure}
\centering
\includegraphics[width=7cm,height=7cm]{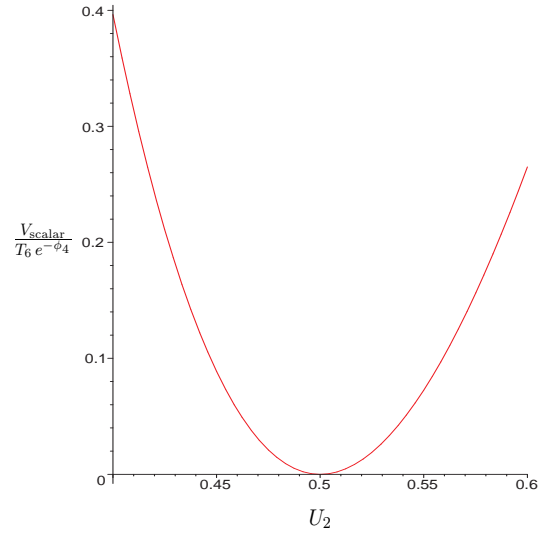}
\caption{The total scalar potential for $U_2$ in the discussed
model of table \ref{tab:pati_salam3gen_z4}.}\label{fig:scalarPot_patisalam}
\end{figure}

For the ${\bf ABB}$ model with a complex structure of the third
torus fixed to be $U_2=1/2$, the only mutual intersection numbers that
have been found by the computer program are $(\pi_a\circ
\pi_b,\pi'_a\circ \pi_b)=(0,0),(\pm 1,0),(0,\pm 1)$. This still allows
for a construction of a four generation supersymmetric Pati-Salam
model with initial gauge groups $U(4)\times U(2)^3\times U(2)^3$. One
typical example is realized by the wrapping number as shown in
table \ref{tab:pati_salam3gen_z4}.
\begin{table}
\centering
\sloppy
\renewcommand{\arraystretch}{1.2}
\begin{tabular}{|c||c|l|}
  \hline
   Stack & $(n_I,m_I)$  & Homology cycle\\
  \hline\hline
    U(4)  & $(1,0;1,0;0,1)$ & $\pi_1={1\over 2}\left(
         \bar\rho_1-\bar\varepsilon_1-\bar\varepsilon_3-\bar\varepsilon_5\right)$\\

    & &$\pi'_1={1\over 2}\left(\rho_2-\bar\rho_2+\varepsilon_1-\varepsilon_3-\varepsilon_6-
             \bar\varepsilon_1+\bar\varepsilon_3+\bar\varepsilon_6     \right)$ \\
\hline\hline
U(2)  &$(1,0;1,0;0,1)$ &$\pi_2={1\over 2}\left(
         \bar\rho_1-\bar\varepsilon_1+\bar\varepsilon_3+\bar\varepsilon_5\right)$\\
 & &$\pi'_2={1\over 2}\left(
          \rho_2-\bar\rho_2+\varepsilon_1+\varepsilon_3+\varepsilon_6-
             \bar\varepsilon_1-\bar\varepsilon_3-\bar\varepsilon_6     \right)$ \\
\hline
U(2)  & $(1,0;1,0;0,1)$ & $\pi_3={1\over 2}\left(
         \bar\rho_1-\bar\varepsilon_2+\bar\varepsilon_3+\bar\varepsilon_6\right)$\\
 & & $\pi'_3={1\over 2}\left(
          \rho_2-\bar\rho_2+\varepsilon_2+\varepsilon_3+\varepsilon_5-
             \bar\varepsilon_2-\bar\varepsilon_3-\bar\varepsilon_5     \right)$\\
\hline
U(2)  & $(1,0;1,0;0,1)$ & $\pi_4={1\over 2}\left(
         \bar\rho_1+\bar\varepsilon_2+\bar\varepsilon_3+\bar\varepsilon_6\right)$\\
 & & $\pi'_4={1\over 2}\left(
          \rho_2-\bar\rho_2-\varepsilon_2+\varepsilon_3+\varepsilon_5+
             \bar\varepsilon_2-\bar\varepsilon_3-\bar\varepsilon_5     \right)$\\
\hline\hline
U(2)  & $(1,0;1,0;0,1)$ & $\pi_5={1\over 2}\left(
         \bar\rho_1+\bar\varepsilon_1-\bar\varepsilon_3+\bar\varepsilon_5\right)$\\
 & & $\pi'_5={1\over 2}\left(
          \rho_2-\bar\rho_2-\varepsilon_1-\varepsilon_3+\varepsilon_6+
             \bar\varepsilon_1+\bar\varepsilon_3-\bar\varepsilon_6     \right)$\\
\hline
U(2)  & $(1,0;1,0;0,1)$ & $\pi_6={1\over 2}\left(
         \bar\rho_1+\bar\varepsilon_1+\bar\varepsilon_4-\bar\varepsilon_6\right)$\\
 & & $\pi'_6={1\over 2}\left(
          \rho_2-\bar\rho_2-\varepsilon_1+\varepsilon_4-\varepsilon_5+
             \bar\varepsilon_1-\bar\varepsilon_3+\bar\varepsilon_5     \right)$\\
\hline
U(2)  & $(1,0;1,0;0,1)$ & $\pi_7={1\over 2}\left(
         \bar\rho_1+\bar\varepsilon_1-\bar\varepsilon_4-\bar\varepsilon_6\right)$\\
 & & $\pi'_7={1\over 2}\left(
          \rho_2-\bar\rho_2-\varepsilon_1-\varepsilon_4-\varepsilon_5+
             \bar\varepsilon_1+\bar\varepsilon_3+\bar\varepsilon_5    \right)$\\
\hline
\end{tabular}
\caption{The wrapping numbers and homology cycles of the D6-branes
in the 7 stack Pati-Salam-model.}
\label{tab:pati_salam3gen_z4}
\end{table}
The next step is to determine the scalar potential for the modulus
$U_2$. This cannot be obtained from topological considerations,
only from the NS-NS tadpole calculation. For the specific
$\mathbb{Z}_4$-model in table \ref{tab:pati_salam3gen_z4}, we get
a contribution from the O6-plane tension which is given by
\begin{equation}\label{eq:z4oplane_contrib}
{V}_{O6}=-T_6\, e^{-\phi_4} 16 \sqrt{2}\left(
                    {1\over \sqrt{U_2}} +2 \sqrt{U_2} \right)\ ,
\end{equation}
and a contribution from the seven stacks of D6-branes:
\begin{align}\label{eq:z4d6brane_contrib}
    {V}_{1}&=T_6\, e^{-\phi_4} 16 \sqrt{ {1\over 4\, U_2}+ U_2}\ ,\\
          {V}_{2,\ldots,7}&=T_6\, e^{-\phi_4} 8 \sqrt{{1\over 4\, U_2}+ U_2}\ . \nonumber
\end{align}
The sum over all terms,
\begin{equation}\label{eq:tot_scalar_pot}
{V}_\text{scalar}={V}_{O6}+\sum_{i=1}^{7}{V}_{i}\ ,
\end{equation}
is plotted in figure \ref{fig:scalarPot_patisalam}.
It can be seen in this plot that the scalar potential indeed has a
minimum at the value $U_2=1/2$. This can be understood as a
consistency check for our procedure: supersymmetry really fixes
the complex structure on the third torus. This freezing of moduli
for supersymmetric backgrounds is very similar to what happens for
instance in compactifications with non-vanishing R-R fluxes
\cite{Dasgupta:1999ss, Giddings:2001yu, Kachru:2002he}.

Next, we have to determine the massless open string spectrum of
the model. In terms of ${\cal N}=2$ supermultiplets, the model
contains vector multiplets in the gauge group $U(4)\times
U(2)^3\times U(2)^3$. Beside this, there are two
hypermultiplets in the adjoint representation of each unitary
gauge factor. The complex scalar in the vector multiplet
corresponds to the unconstrained position of each stack of
D6-branes on the third 2-torus.

The chiral spectrum can be determined using table
\ref{tab:massless_chiral_spectrum}, where the mutual intersection
numbers of the cycles (\ref{tab:pati_salam3gen_z4}) have to be
computed. The result is listed in table
\ref{tab:pati_salamchiralspec}, and one can see that the
non-abelian anomalies cancel for this spectrum (including formally
also the $U(2)$ anomalies). Neither there are symmetric nor
antisymmetric representations, and the bifundamental representations
of the type ${\bf (\overline{N}_a,N_b)}$ also do
not give any net contributions as the corresponding intersection
numbers $\pi_a\circ\pi_b$ vanish.
\begin{table}
\centering
\sloppy
\renewcommand{\arraystretch}{1.1}
\begin{tabular}{|c||c|c|}
  \hline
 Field & Number & $U(4)\times U(2)^3\times U(2)^3$ \\
  \hline\hline
 $\Phi_{1'2}$ & 1 & $(4;2,1,1;1,1,1)$ \\
 $\Phi_{1'3}$ & 1 & $(4;1,{2},1;1,1,1)$ \\
 $\Phi_{1'4}$ & 1 & $(4;1,1,{2};1,1,1)$ \\
\hline
 $\Phi_{1'5}$ & 1 & $(\bar{4};1,1,1;\bar{2},1,1)$ \\
 $\Phi_{1'6}$ & 1 & $(\bar{4};1,1,1;1,\bar{2},1)$ \\
 $\Phi_{1'7}$ & 1 & $(\bar{4};1,1,1;1,1,\bar{2})$ \\
\hline
 $\Phi_{2'3}$ & 1 & $(1;\bar{2},\bar{2},1;1,1,1)$ \\
 $\Phi_{2'4}$ & 1 & $(1;\bar{2},1,\bar{2};1,1,1)$ \\
 $\Phi_{3'4}$ & 1 & $(1;1,\bar{2},\bar{2};1,1,1)$ \\
\hline
 $\Phi_{5'6}$ & 1 & $(1;1,1,1;2,{2},1)$ \\
 $\Phi_{5'7}$ & 1 & $(1;1,1,1;2,1,{2})$ \\
 $\Phi_{6'7}$ & 1 & $(1;1,1,1;1,{2},{2})$ \\
\hline
\end{tabular}
\caption{Chiral spectrum of the 7 stack Pati-Salam-model.}
\label{tab:pati_salamchiralspec}
\end{table}

This model of course at first glance is a one generation model,
but a three generation model can be obtained if the two triplets
$U(2)^3$ are broken down to their diagonal subgroups. Potential
gauge symmetry breaking candidates in this way are the chiral
fields $\{\Phi_{2'3},\Phi_{2'4},\Phi_{3'4}\}$ and
$\{\Phi_{5'6},\Phi_{5'7},\Phi_{6'7}\}$ from table
\ref{tab:pati_salamchiralspec}. However, one has to remember that
these are chiral ${\cal N}=1$ supermultiplets living on the
intersection of two D-branes in every case. In order to see which
gauge breaking mechanisms are possible, it is necessary to
determine the non-chiral spectrum. As  explained in chapter
\ref{cha:nonchiral_openstring_specrum}, this spectrum cannot be
calculated by homology, only by the conformal field theory
calculation. This is unproblematic, as we just have to calculate
the overlap of two boundary states of the type
(\ref{eq:boundary_state_z4}) and then transform the result to the
loop channel to get the cylinder amplitude
(\ref{eq:cylinder_amp_z3_allcontrib}), where we can read off the
massless states immediately. This also provides a check for the
chiral spectrum.

\begin{figure}
\centering
\includegraphics[width=6cm,height=6cm]{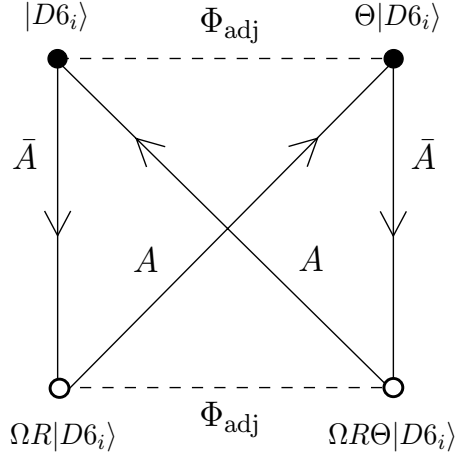}
\caption{Adjoint higgsing for the 7 stack Pati-Salam model.}\label{fig:quivernr1}
\end{figure}

The non-chiral spectrum first for one stack of $U(2)$ branes
contains the usual hypermultiplet
$\Phi_\text{adj}=(\phi_{adj},\tilde\phi_{adj})$ in the adjoint
representation of $U(2)$, being localized at the intersection
between a brane an its $\Theta$ image (see figure
\ref{fig:z4recombined_branes}). Additionally, there are two chiral
multiplets $\Psi_A$ and $\Psi_{\bar A}$, in the ${\bf A}$
respectively ${\bf \bar{A}}$ representation arising from the
sector between a brane and its $\Omega R$-mirror. These two fields,
carrying the conjugate representations of the gauge group, cannot
be seen from the topological intersection number which actually
vanishes, $\pi_i\circ\pi'_i=0$. The quiver diagram for these three
types of fields is given in figure \ref{fig:quivernr1}.
Every closed polygon that is contained in the diagram corresponds
to a possible term in the superpotential, that is then made up
of the product of the associated fields. In the present case, the following two terms
can appear
\begin{equation}\label{eq:terms_holo_superpot}
    W=\phi_\text{adj} \Psi_A \Psi_{\bar A} + \tilde\phi_\text{adj}\Psi_A\Psi_{\bar A}\ ,
\end{equation}
which will generate a mass for the anti-symmetric fields when the
adjoint multiplet gets a VEV. But as discussed in chapter
\ref{cha:susy_cyclesABB}, after giving a VEV to the adjoint field,
the brane recombines with its $\Theta$ mirror and consequently,
this recombined brane does not pass any longer through the fixed
point. So the only remaining fields are given by the
bifundamentals which are listed in table
\ref{tab:pati_salamnonchiralspec}.
\begin{table}
\centering
\sloppy
\renewcommand{\arraystretch}{1.1}
\begin{tabular}{|c||c|c|}
  \hline
 Field & Number & $U(4)\times U(2)^3\times U(2)^3$ \\
  \hline\hline
 $H_{12}$ & 1 & $(4;\bar{2},1,1;1,1,1)+c.c.$ \\
 $H_{13}$ & 1 & $(4;1,\bar{2},1;1,1,1)+c.c.$ \\
 $H_{14}$ & 1 & $(4;1,1,\bar{2};1,1,1)+c.c.$ \\
\hline
 $H_{15}$ & 1 & $(\bar{4};1,1,1;{2},1,1)+c.c.$ \\
 $H_{16}$ & 1 & $(\bar{4};1,1,1;1,{2},1)+c.c.$ \\
 $H_{17}$ & 1 & $(\bar{4};1,1,1;1,1,{2})+c.c.$ \\
\hline
 $H_{25}$ & 1 & $(1;{2},1,1;\bar{2},1,1)+c.c.$ \\
 $H_{26}$ & 1 & $(1;{2},1,1;1,\bar{2},1)+c.c.$ \\
 $H_{27}$ & 1 & $(1;{2},1,1;1,1,\bar{2})+c.c.$ \\
 $H_{35}$ & 1 & $(1;1,{2},1;\bar{2},1,1)+c.c.$ \\
 $H_{36}$ & 1 & $(1;1,{2},1;1,\bar{2},1)+c.c.$ \\
 $H_{37}$ & 1 & $(1;1,{2},1;1,1,\bar{2})+c.c.$ \\
 $H_{45}$ & 1 & $(1;1,1,{2};\bar{2},1,1)+c.c.$ \\
 $H_{46}$ & 1 & $(1;1,1,{2};1,\bar{2},1)+c.c.$ \\
 $H_{47}$ & 1 & $(1;1,1,{2};1,1,\bar{2})+c.c.$ \\
\hline
\end{tabular}
\caption{Non-chiral spectrum of the 7 stack Pati-Salam-model.}
\label{tab:pati_salamnonchiralspec}
\end{table}
\subsection{Three stack Pati-Salam model}
It is obvious from table \ref{tab:pati_salamnonchiralspec} that
although there are bifundamentals which generally could break down
the gauge groups $SU(4)\times SU(2)\times SU(2)$ down to
$SU(3)\times SU(2)\times U(1)$ by a Higgs mechanism, there are
none to break the two pairs of $U(2)^3$ down to the diagonal
subgroup. Therefore, another mechanism has to be involved.

However, there are the massless chiral bifundamental fields
$\{\Phi_{2'3},\ldots,\Phi_{6'7}\}$ living on  intersections and
preserving ${\cal N}=1$ supersymmetry. If it is possible to find a
continuous complex structure deformation on the line of marginal
stability in the D-term potential where just the four fields
$\{\Phi_{2'3}, \Phi_{2'4}, \Phi_{5'6}, \Phi_{5'7}\}$ have to get a
VEV in order not to become tachyonic. Then the system proceeds to a new
supersymmetric ground state where the gauge symmetries $U(2)^3$
are broken down to the diagonal subgroups. This mechanism has been
described in chapter \ref{cha:gauge_sysm_break_tach} in some more
detail.

We have to evaluate the general expression (\ref{eq:Dterm_pot})
for our specific case. As there are only chiral fields in
bifundamental representations, we obtain
\begin{equation}\label{eq:DtermpatiSalam}
     V_\text{D}=\sum_{a=1}^k \sum_{r,s=1}^{N_a}
       {1\over 2 g_a^2}\left(\sum_{j=1}^k\sum_{p=1}^{N_j} q_{aj}
       \, \Phi^{rp}_{aj} \,
          \bar\Phi^{sp}_{aj}+
            g_a^2\sum_{i=1}^{b_3}  {M_{ai}\over N_a}\, A_i\ .
                 \delta^{rs}\right)^2\ ,
\end{equation}
$M_{ai}$ is given by the matrix
\begin{equation}\label{eq:mass_matrix}
     M_{ai}= N_a (v_{a,i} - v'_{a,i} )\ ,
\end{equation}
and the indices $(r,s)$ numerate the $N_a^2$ gauge fields in the
adjoint representation of the gauge factor $U(N_a)$. The sum $j$
is taken over all chiral fields which are charged under $U(N_a)$. The gauge
coupling constants depend on the complex structure moduli as well,
but since we are only interested in the leading order effects, we
can set them to the constant values on the line of marginal
stability.  Since all branes have the same volume, in the
following we will simply set these volumes to one. The charges $q_{aj}$ are
just the corresponding $U(1)$ charges and can be directly read off
from table \ref{tab:pati_salamchiralspec}.

It is indeed possible to find a non-vanishing $\Omega R$-invariant
complex structure deformation which is only related to the four 3-cycles
$\{\varepsilon_1,\varepsilon_2,\varepsilon_3-2\bar\varepsilon_3,
\varepsilon_4-2\bar\varepsilon_4\}$. Explicitly, it is given by
\begin{align}
    &A_{3}=-{\kappa}\ ,& &A_{5}-2A_{13}=-{\kappa}\ , \\
    &A_{4}={\kappa}-2\lambda\ ,& &A_{6}-2A_{14}=2\mu-{\kappa}\ .\nonumber
\end{align}
A new supersymmetric ground state exists for the non-vanishing VEVs
\begin{align}
    &|\Phi^{rr}_{2'3}|^2={\lambda}\ ,&
    &|\Phi^{rr}_{2'4}|^2={\kappa-\lambda}\ ,\\
    &|\Phi^{rr}_{5'6}|^2={\mu}\ ,&
    &|\Phi^{rr}_{5'7}|^2={\kappa-\mu}\ ,\nonumber
\end{align}
with $r=1,2$, $\kappa>\lambda>0$ and $\kappa>\mu>0$, but otherwise
arbitrary. For these VEVs, both stacks of three $U(2)$ branes
$\{\pi_2,\pi'_3,\pi'_4\}$ and $\{\pi_5,\pi'_6,\pi'_7\}$ recombine
into a single stack of branes each within the same homology class, and
the gauge group in both cases is broken to the diagonal $U(2)$.
This configuration apparently is non-factorizable and maybe
non-flat, because its intersection numbers with a similar spectrum
to table \ref{tab:pati_salam3stackchiralspec} could not be found
from the beginning.

After this recombination process we are left with only three
stacks of D6-branes wrapping the homology cycles as they are shown in
table \ref{tab:pati_salam3stack_z4}.
\begin{table}
\centering
\sloppy
\renewcommand{\arraystretch}{1.2}
\begin{tabular}{|c||l|}
  \hline
   Stack & Homology cycle\\
  \hline\hline
    U(4)  &$\pi_a=\pi_1$\\
\hline
    U(2)  &$\pi_b=\pi_2+\pi'_3+\pi'_4$\\
\hline
    U(2)  &$\pi_c=\pi_5+\pi'_6+\pi'_7$\\
\hline
\end{tabular}
\caption{The homology cycles of the non-factorizable D6-branes in
the 3 stack Pati-Salam-model.} \label{tab:pati_salam3stack_z4}
\end{table}
The chiral spectrum of this model is listed in table
\ref{tab:pati_salam3stackchiralspec}, where the three $U(1)$
charges of the $U(4)$ and the two $U(2)$ groups are denoted with
subscripts.
\begin{table}
\centering
\sloppy
\renewcommand{\arraystretch}{1.2}
\begin{tabular}{|c||c|l|}
  \hline
 Field & Number & $SU(4)\times SU(2)\times SU(2)\times U(1)^3$ \\
  \hline\hline
$\Phi_{ab}$  &2 & $(4,2,1)_{(1,-1,0)}$\\
$\Phi_{a'b}$  &  1 & $(4,2,1)_{(1,1,0)}$\\
\hline
$\Phi_{ac}$  & 2 & $(\bar{4},1,2)_{(-1,0,1)}$\\
$\Phi_{a'c}$  &  1 & $(\bar{4},1,2)_{(-1,0,-1)}$\\
\hline
$\Phi_{b'b}$  & 1 & $(1,S+A,1)_{(0,2,0)}$\\
$\Phi_{c'c}$  & 1 & $(1,1,\bar{S}+\bar{A})_{(0,0,-2)}$\\
\hline
\end{tabular}
\caption{Chiral spectrum of the 3 stack Pati-Salam-model.}
\label{tab:pati_salam3stackchiralspec}
\end{table}
Interestingly, for the non-factorizable branes, the intersection
numbers $\pi'_{b,c}\circ \pi_{b,c}$ do not vanish any longer. So
they give rise to chiral multiplets in the symmetric and
anti-symmetric representations of the $U(2)$ gauge factors.
Certainly, these chiral fields are needed in order to cancel the
formal non-abelian $U(2)$ anomalies. These anti-symmetric fields
can be understood as the remnants of the chiral fields,
$\Phi_{3'4}$ and $\Phi_{6'7}$, which did not gain any VEV during
the brane recombination process. The mixed anomalies can be
computed from the chiral spectrum for dividing into the anomalous
and non-anomalous $U(1)$ gauge factors. As a result, the
only anomaly free combination is given by
\begin{equation}\label{eq:anomalyfreeu1patisalam}
U(1)=U(1)_a-3\, U(1)_b -3\, U(1)_c\ .
\end{equation}
The quadratic axionic couplings reveal
that the matrix $M_{ai}$ has a trivial kernel. Therefore,
all three $U(1)$ gauge groups become massive and survive
as global symmetries.

Finally, we have obtained a supersymmetric 3 generation
Pati-Salam model with gauge group $SU(4)\times SU(2)_L\times SU(2)_R$
which accommodates the standard model matter in addition
to some exotic matter in the (anti-)symmetric representation
of the $SU(2)$ gauge groups.

\begin{figure}
\centering
\includegraphics{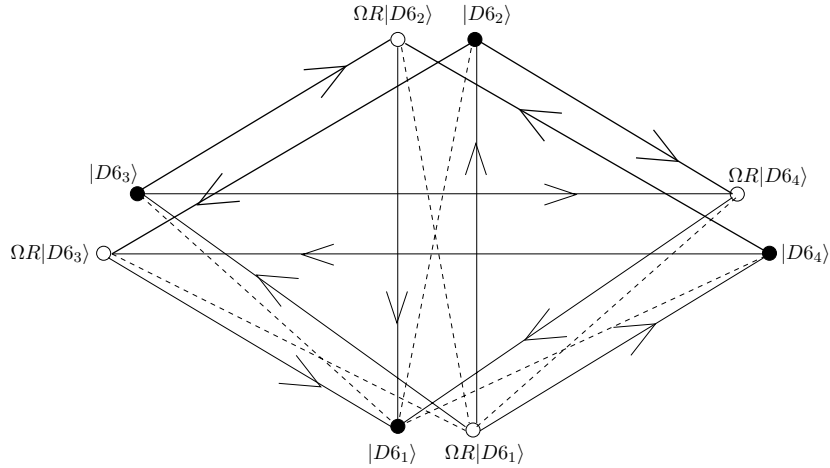}
\caption{Quiver diagram for the branes $\pi_1,\ldots,\pi_4$ in the 3
stack Pati-Salam model.}\label{fig:quivernr2}
\end{figure}

\begin{figure}
\centering
\includegraphics{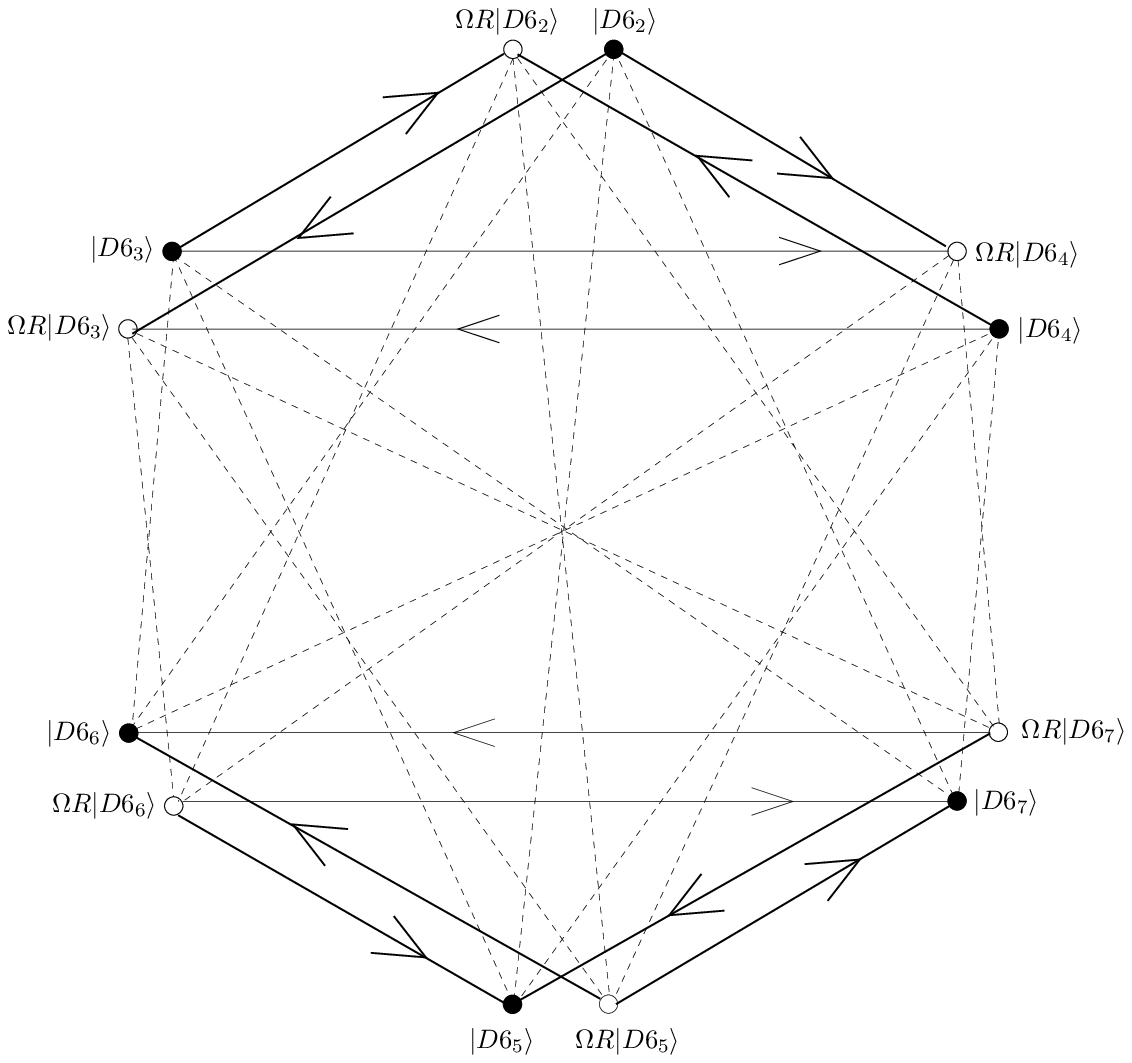}
\caption{Quiver diagram for the branes $\pi_2,\ldots,\pi_7$ in the 3
stack Pati-Salam model.}\label{fig:quivernr3}
\end{figure}

The next step again is to compute the massless non-chiral matter
after the brane recombination. As there is no CFT description for
the non-factorizable branes any longer, we have to determine which
Higgs fields receive a mass from couplings with the chiral
bifundamental fields having gained a non-vanishing VEV. One has to
admit that the applicability of the low energy effective field
theory is limited, but indeed, it is the only information. We
first consider the sector of the branes $\{\pi_1, \ldots, \pi_4\}$
in the quiver diagram of figure \ref{fig:quivernr2}.
The chiral fields are indicated by an arrow and non-chiral fields
by a dashed line. The fields which receive a VEV after the
discussed small complex structure deformations are depicted by a
fat line. The Higgs fields inside one
hypermultiplet will be decomposed into its two chiral components
$H_{1j}=(h^{(1)}_{1j}, h^{(2)}_{1j})$ for $j=2,3,4$. We observe a
couple of closed triangles in the quiver diagram in figure
\ref{fig:quivernr2} that give rise to the following Yukawa
couplings in the superpotential
\begin{align}\label{eq:yukawacouplingspatsal}
    &\Phi_{2'3}\, \Phi_{1'2}, h^{(2)}_{13}\ ,&
    &\Phi_{2'3}\, \Phi_{1'3}, h^{(2)}_{12}\ ,& \\
    &\Phi_{2'4}\, \Phi_{1'2}, h^{(2)}_{14}\ ,&
    &\Phi_{2'4}\, \Phi_{1'4}, h^{(2)}_{12}\ .&\nonumber
\end{align}
The VEV of the chiral fields $\Phi_{2'3}$ and $\Phi_{2'4}$ leads
to a mass matrix for the six fields $\{\Phi_{1'2}, \Phi_{1'3},
\Phi_{1'4},h^{(2)}_{12},h^{(2)}_{13}, h^{(2)}_{14}\}$ of rank
four. Therefore, one combination of the three fields $\Phi$, one
combination of the three fields $h^{(2)}$ and the
three fields $h^{(1)}$ remain massless. These modes just fit into
the three chiral fields of table
\ref{tab:pati_salam3stackchiralspec} in addition to another
hypermultiplet in the $(4,{2},1)$ representation of the Pati-Salam
gauge group $U(4)\times U(2)\times U(2)$. The condensation for the
second triplet of $U(2)$s is completely analogous and leads to a
massless hypermultiplet in the $(4,1,{2})$ representation. The
quiver diagram involving the six $U(2)$ gauge groups is shown in
figure \ref{fig:quivernr3}.
In this diagram there are closed polygons like
\begin{equation}\label{eq:closed_polygons}
    |D6_2\rangle\rightarrow\Omega R|D6_4\rangle
    \rightarrow\Omega R|D6_7\rangle\rightarrow|D6_6\rangle\ .
\end{equation}
The corresponding terms in the superpotential after giving the
VEVs generate a mass term for one chiral component from every of
the the nine hypermultiplets
$\{H_{25},H_{26},\ldots,H_{46},H_{47}\}$. Remember that a
hypermultiplet consists of two chiral multiplets of opposite
charge, $H=(h^{(1)},h^{(2)})$. The mass matrix for these nine
chiral fields is of rank six, so that three combinations of the
four chiral fields, $h^{(1)}$, in
$\{H_{36},H_{37},H_{46},H_{47}\}$ remain massless. Since the
intersection numbers in table \ref{tab:pati_salam3stackchiralspec}
tell us that there are no chiral fields in the $(1,2,2)$
representation of the $U(4)\times U(2)\times U(2)$ gauge group,
the other chiral components of the hypermultiplets, $h^{(2)}$,
must also gain a mass during the brane recombination. This
indicates that this process also might involve the condensation of
massive string modes, allowing correct mass terms in the quiver
diagram \cite{Cremades:2002cs}. Nevertheless, the quiver diagram
induces the non-chiral spectrum as listed in table
\ref{tab:pati_salam3nonchiralspec}.
\begin{table}
\centering
\sloppy
\renewcommand{\arraystretch}{1.1}
\begin{tabular}{|c||c|l|}
  \hline
 Field & Number & $U(4)\times U(2)\times U(2)$ \\
  \hline\hline
$H_{aa}$  & 1 & $({\rm Adj},1,1)+c.c.$\\
$H_{bb}$  & 1 & $(1,{\rm Adj},1)+c.c.$\\
$H_{cc}$  & 1 & $(1,1,{\rm Adj})+c.c.$\\
\hline
$H_{a'b}$  & 1 & $(4,{2},1)+c.c.$\\
\hline
$H_{a'c}$  & 1 & $(4,1,{2})+c.c.$\\
\hline
$H_{bc}$  & 3 & $(1,2,\bar{2})+c.c.$\\
\hline
\end{tabular}
\caption{Non-chiral spectrum of the 3 stack Pati-Salam-model.}
\label{tab:pati_salam3nonchiralspec}
\end{table}
\subsection{Supersymmetric standard-like model}
In the following, there will not be any detailed discussion of the
further phenomenological details of the 3-stack Pati-Salam model.
Instead, two different ways will be discussed how to obtain a
supersymmetric standard-like model by a further breaking of gauge
symmetries. Of course, first one has to find a model with the
right gauge groups. This will be possible in our approach. But
such a model cannot be called MSSM unless the spectrum is exactly
that of the standard model in its supersymmetric extension.
Unfortunately, this is not the case for the two models which will
be discussed in the following.

\subsubsection{Adjoint Pati-Salam breaking}
In the open string spectrum, there are also the moduli scalar
fields in the adjoint representation characterizing the
unconstrained positions of the branes on the third 2-torus. A
simple gauge breaking mechanism lies in giving a VEV to the
distance between two parallel stacks of branes. This mechanism is
described in \ref{cha:adjointhiggsing}. In the present setting, we
can simply move one of the four D6-branes away from the $U(4)$
stack, this breaks the gauge group down to $U(3)\times U(2)\times
U(2)\times U(1)$. The resulting spectrum looks like that of a
three generation left-right symmetric extension of the standard
model. It is shown in table \ref{tab:MSSM_lrsym_chiralspec}.
\begin{table}
\centering
\sloppy
\renewcommand{\arraystretch}{1.1}
\begin{tabular}{|c||l|c|}
  \hline
 Number &$SU(3)_c\times SU(2)_L\times SU(2)_R\times U(1)^4$  &
             $U(1)_{B-L}$ \\
  \hline\hline
 1 & $(3,2,1)_{(1,1,0,0)}$ & ${1\over 3}$ \\
 2 & $(3,2,1)_{(1,-1,0,0)}$ & ${1\over 3}$\\
\hline
 1 & $(\bar{3},1,2)_{(-1,0,-1,0)}$ & $-{1\over 3}$ \\
 2 & $(\bar{3},1,2)_{(-1,0,1,0)}$ & $-{1\over 3}$ \\
\hline
 1 & $(1,2,1)_{(0,1,0,1)}$ & ${-1}$ \\
 2 & $(1,2,1)_{(0,-1,0,1)}$ & ${-1}$ \\
\hline
 1 & $(1,1,2)_{(0,0,-1,-1)}$ & ${1}$ \\
 2 & $(1,1,2)_{(0,0,1,-1)}$ & ${1}$ \\
\hline
 1 & $(1,S+A,1)_{(0,2,0,0)}$ & $0$ \\
 1 & $(1,1,\bar{S}+\bar{A})_{(0,0,-2,0)}$ & $0$\\
\hline
\end{tabular}
\caption{Chiral spectrum of the 4 stack left-right symmetric supersymmetric standard model.}
\label{tab:MSSM_lrsym_chiralspec}
\end{table}
Again calculating the mixed anomalies, one finds two anomaly free
$U(1)$s, of which the combination ${1/3}(U(1)_1-3U(1)_4)$ remains
massless even after the Green-Schwarz mechanism. This linear
combination in fact can be interpreted as the $U(1)_{B-L}$
symmetry, which is expected to be anomaly-free in a model with
right-handed neutrinos.

We can apply this gauge breaking mechanism a second time and give
a VEV to the distance between the two branes in the
$U(2)_R$-stack, thus breaking this gauge groups down to
$U(1)_R\times U(1)_R$. Therefore, we obtain the total gauge symmetry
$SU(3)\times SU(2)_L\times U(1)_R\times U(1)_R\times U(1)$. In this
case the following two $U(1)$ gauge factors remain massless after
checking the Green-Schwarz couplings
\begin{align}
    &U(1)_{B-L}={1\over 3}U(1)_1-U(1)_5\ , \\
    &U(1)_Y={1\over 3}U(1)_1 +U(1)_3-U(1)_4-U(1)_5\ .\nonumber
\end{align}
This means that in this model, we really obtain a massless
hypercharge. The final supersymmetric chiral spectrum is listed in
table \ref{tab:MSSMchiralspec} with respect to the unbroken gauge
symmetries.
\begin{table}
\centering
\sloppy
\renewcommand{\arraystretch}{1.1}
\begin{tabular}{|c||c|l|l|}
  \hline
Number & Field & $SU(3)\times SU(2)\times U(1)^3$ & $U(1)_Y \times U(1)_{B-L}$  \\
  \hline\hline
 1 &  $q_L$ & $(3,2)_{(1,1,0,0,0)}$ &  $\left({1\over 3},{1\over
    3}\right)$ \\
 2 &  $q_L$ & $(3,2)_{(1,-1,0,0,0)}$ &  $\left({1\over 3},{1\over
    3}\right)$ \\
\hline
 1 &  $u_R$ & $(\bar{3},1)_{(-1,0,-1,0,0)}$ &  $\left(-{4\over 3},-{1\over
    3}\right)$ \\
 2 &  $u_R$ & $(\bar{3},1)_{(-1,0,0,1,0)}$ &  $\left(-{4\over 3},-{1\over
    3}\right)$ \\
 2 &  $d_R$ & $(\bar{3},1)_{(-1,0,1,0,0)}$ &  $\left({2\over 3},-{1\over
    3}\right)$ \\
 1 &  $d_R$ & $(\bar{3},1)_{(-1,0,0,-1,0)}$ &  $\left({2\over 3},-{1\over
    3}\right)$ \\
\hline
 1 &  $l_L$ & $(1,2)_{(0,1,0,0,1)}$ &  $\left(-{1},-{1}\right)$ \\
 2 &  $l_L$ & $(1,2)_{(0,-1,0,0,1)}$ &  $\left(-{1},-{1}\right)$ \\
\hline
 2 &  $e_R$ & $(1,1)_{(0,0,1,0,-1)}$ &  $\left({2},{1}\right)$ \\
 1 &  $e_R$ & $(1,1)_{(0,0,0,-1,-1)}$ &  $\left({2},{1}\right)$ \\
 1 &  $\nu_R$ & $(1,1)_{(0,0,-1,0,-1)}$ &  $\left({0},{1}\right)$ \\
 2 &  $\nu_R$ & $(1,1)_{(0,0,0,1,-1)}$ &  $\left({0},{1}\right)$ \\
\hline
 1 &  $$ & $(1,S+A)_{(0,2,0,0,0)}$ & $\left({0},{0}\right)$  \\
 1 &  $$ & $(1,1)_{(0,0,-2,0,0)}$ & $\left(-{2},{0}\right)$  \\
 1 &  $$ & $(1,1)_{(0,0,0,-2,0)}$ & $\left({2},{0}\right)$  \\
 2 &  $$ & $(1,1)_{(0,0,-1,-1,0)}$ & $\left({0},{0}\right)$  \\
\hline
\end{tabular}
\caption{Chiral spectrum of the 5 stack supersymmetric standard model with some additional exotic matter.}
\label{tab:MSSMchiralspec}
\end{table}
The one anomalous $U(1)_1$ can be identified with the baryon number
operator and survives the Green-Schwarz mechanism as a global
symmetry. Therefore, in this model the baryon number
is conserved and the proton is stable.
Similarly, $U(1)_5$ can be identified with the lepton number
and also survives as a global symmetry.
To break the gauge symmetry $U(1)_{B-L}$, one can furthermore recombine
the third and the fifth stack of D6 branes, which is expected to
 correspond to giving a VEV to the Higgs field $H_{3'5}$, thus
giving a mass to the right handed neutrinos.

Besides these $U(1)$ gauge factors, there is a deviation from the
standard model matter: there are some additional symmetric and
antisymmetric representations of the $SU(2)$ gauge factor.
Nevertheless, this model is very interesting and deserves a
further phenomenological analysis. Some additional issues, as the
electroweak symmetry breaking and gauge coupling ratios, are
discussed in \cite{Blumenhagen:2002gw}. In the following, we will
switch to the other possible gauge breaking mechanism for the 3
stack Pati-Salam model.
\subsubsection{Bifundamental  Pati-Salam breaking}
Instead of the fields in the adjoint representation, it is
possible to break the gauge symmetry using the non-chiral Higgs
fields of the type $H_{a'c}$ from table
\ref{tab:pati_salam3nonchiralspec}. This mechanism corresponds to
another brane recombination of the four branes wrapping $\pi_a$
with one of the branes wrapping $\pi'_c$, so afterwards one
obtains the homology cycles as shown in table
\ref{tab:ssdtmod4cycles}.
\begin{table}
\centering
\sloppy
\renewcommand{\arraystretch}{1.2}
\begin{tabular}{|c||l|}
  \hline
   Stack & Homology cycle\\
  \hline\hline
    U(3)  &$\pi_A=\pi_a$\\
\hline
    U(2)  &$\pi_B=\pi_b$\\
\hline
    U(1)  &$\pi_C=\pi_a+\pi'_c$\\
\hline
    U(1)  &$\pi_D=\pi_c$\\
\hline
\end{tabular}
\caption{The homology cycles of the non-factorizable D6-branes in
the 4 stack supersymmetric standard model.} \label{tab:ssdtmod4cycles}
\end{table}
For these cycles, the tadpole cancellation conditions are still satisfied.
The chiral spectrum can be obtained by computing the homological
intersection numbers, it is shown in table \ref{tab:MSSM2chiralspec}.
\begin{table}
\centering
\sloppy
\renewcommand{\arraystretch}{1.1}
\begin{tabular}{|c||c|l|l|c|}
  \hline
Number & Field & Sector &$SU(3)_c\times SU(2)_L\times U(1)^4$ & $U(1)_Y$  \\
  \hline\hline
 2 &   $q_L$ &  $(AB)$ & $(3,2)_{(1,-1,0,0)}$  & ${1/3}$ \\
 1 & $q_L$ &  $(A'B)$ & $(3,2)_{(1,1,0,0)}$  & ${1/3}$ \\
\hline
 1 & $u_R$ &  $(AC)$ & $(\bar{3},1)_{(-1,0,1,0)}$  & $-{4/3}$ \\
 2 & $d_R$ &  $(A'C)$ & $(\bar{3},1)_{(-1,0,-1,0)}$   & ${2/3}$  \\
\hline
 2 & $u_R$ &  $(AD)$ & $(\bar{3},1)_{(-1,0,0,1)}$  & $-{4/3}$ \\
 1 & $d_R$ &  $(A'D)$ & $(\bar{3},1)_{(-1,0,0,-1)}$ & ${2/3}$  \\
\hline
 2 & $l_L$ &  $(BC)$ & $(1,2)_{(0,-1,1,0)}$  & $-{1}$  \\
 1 & $l_L$ &  $(B'C)$ & $(1,2)_{(0,1,1,0)}$  & $-{1}$  \\
\hline
 1 & $e_R$ &  $(C'D)$ & $(1,1)_{(0,0,-1,-1)}$   & ${2}$ \\
 1 & $e_R$ & $(C'C)$   & $(1,1)_{(0,0,-2,0)}$ & $2$ \\
 1 & $e_R$ & $(D'D)$   & $(1,1)_{(0,0,0,-2)}$ & $2$ \\
\hline
 1 & $S$  & $(B'B)$   & $(1,S+A)_{(0,2,0,0)}$ & $0$ \\
\hline
\end{tabular}
\caption{Chiral spectrum of the 4 stack supersymmetric standard model with some additional exotic matter.}
\label{tab:MSSM2chiralspec}
\end{table}
Again, we have to compute the mixed anomalies. As a result, there
are two anomalous $U(1)$ gauge factors and two anomaly free ones,
explicitly given by
\begin{align}\label{eq:secondPSmodanomalousu1}
    U(1)_Y&=\frac{1}{3}U(1)_A-U(1)_C - U(1)_D\ ,\\
    U(1)_{K}&=U(1)_A-9\, U(1)_B +9\, U(1)_C-9\,U(1)_D\ .\nonumber
\end{align}
This model again yields a massless hypercharge. Finally, the gauge
group is then given by $SU(3)_C\times SU(2)_L\times U(1)_Y$. In
this model, only the baryon number generator can be identified with
$U(1)_1$, whereas the lepton number is broken. Therefore, the
proton is stable and lepton number violating couplings as Majorana
mass terms are possible. There are no massless right-handed
neutrinos in this model.

Actually, this model is related to the model discussed in the
previous section by a further brane recombination process,
affecting the mass of the right handed neutrinos. This additional
brane recombination can be considered as a string-theoretical
mechanism to generate GUT scale masses for the right handed
neutrinos \cite{Cremades:2002cs}.

There are some deviations from the usual standard model: Only one
of the right handed leptons is realized as a bifundamental field,
the remaining two are given by symmetric representations of
$U(1)$. This behavior surely has consequences for the allowed
couplings, in particular for the Yukawa couplings and the
electroweak Higgs mechanism. Furthermore, there are again
additional symmetric and antisymmetric representations of the
$U(2)_L$ gauge factor.

This ends our discussion of this second model, but further
phenomenological aspects surely should be investigated for this
model, too. Some of the most burning questions surely are the
gauge coupling ratios and gauge coupling unification, some
progress has been made in \cite{Blumenhagen:2003jy}. Furthermore,
it might be instructive to calculate the gauge threshold
corrections \cite{Lust:2003ky}.

\chapter{Inflation in Intersecting Brane Models}\label{cha:Inflation}
In both the $\mathbb{Z}_3$-Orientifold intersecting brane models
of chapter \ref{cha:Z3} and the $\Omega R$-orientifold of chapter
\ref{cha:ORorientifold}, we have encountered several perturbative
instabilities coming from non-vanishing NS-NS tadpoles. We have
investigated the leading order perturbative potential for the
closed string moduli fields of the simple $\Omega R$-orientifold
in some detail and have found out that the consequence of the
uncancelled tadpoles is a run-away behavior of some scalar fields,
which are finally getting pushed to a degenerate limit. At least a
partial freezing of some of the complex structure moduli has been
achieved for instance by the orbifolding procedure in chapter
\ref{cha:Zn_orientifolds}. Another way of stabilizing the
geometric moduli within type $0'$-string theory has been proposed
in \cite{Blumenhagen:2002mf}, where the reader should be reminded
that type $0'$ string theory is a non-supersymmetric theory right
from the outset.

Nevertheless, one unstable field always remains in these
constructions: the dilaton. This surely is a crucial problem for
string perturbation theory, mainly because the expectation value
of the dilaton determines the perturbative string coupling constant. If
this constant is very large, the whole perturbation expansion
looses its significance. From the perspective of string theory,
there seems to be a great problem.

Indeed, such instabilities generally occur in non-supersymmetric
models, where the vanishing of the R-R tadpole does not
automatically imply the vanishing of the NS-NS tadpoles, as it
does in supersymmetric models like the $\mathbb{Z}_4$-models of
chapter \ref{cha:Z4}. On the other hand, from the point of view of
the effective phenomenological theory, models with non-vanishing
NS-NS, but cancelled R-R tadpoles still seem to be acceptable, as
they do not suffer from gauge and gravitational anomalies and,
like it is described in chapter \ref{cha:Z3}, one has come even
close to the standard model of elementary particle physics that of
course is non-supersymmetric.

As already mentioned in the introductory chapter, the ultimate
goal of string theory would be to provide a unified description of
particle physics and cosmology. Standard cosmology has its
shortcomings in many respects, too, and a very successful
resolution of many of its problems has been given by inflation,
that initially has been introduced by Guth \cite{Guth:1981zm,
Guth:1982ec}, Linde \cite{Linde:1982mu} and Albrecht et al.
\cite{Albrecht:1982wi} in the beginning of the 80th. One cannot
actually speak of a model, because too many formulations do exist
differing in detail, one should better speak of 'the inflationary
scenario'. The feature common to all these scenarios is
the existence of a short inflationary phase within the evolution
of the universe, which usually is described by an additional
scalar field that couples to gravity and behaves like an effective
cosmological constant, forcing spacetime to be of de Sitter type
during the inflationary phase.

Although no inflaton to date has been directly detected in an
accelerator experiment, it must be contained in a unified theory.
Thus, it is natural to ask the question if the available unstable
moduli of string theory could play the role of the inflaton
$\psi$. The key criterion to decide this question is the slow
rolling condition which a candidate inflaton scalar field has to
fulfill. This requirement commonly is rephrased in terms of the
potential which then must obey the two relations
\begin{equation}\label{eq:slowroll}
    \epsilon = {M_\text{pl}^2 \over 2} \left( {V'(\psi) \over V(\psi)}\right)^2 \ll 1\ ,
    \qquad\quad \eta = M_\text{pl}^2 {V''(\psi) \over V(\psi)}\ll 1\ ,
\end{equation}
where $V(\psi)$ is the effective potential of the field $\psi$ and
a prime denotes the derivative with respect to $\psi$. We will
investigate this possibility for the $\Omega R$-orientifold model
containing $D6$-branes at general angles in this chapter. Thereby,
we follow a different philosophy than in the preceding chapters.
The apparent stability problems of the non-supersymmetric models
will be transformed into an advantage, if possible.

There also have been various other attempts to realize inflation
within string theory \cite{Bailin:1998tu, Dvali:1998pa,
Alexander:2001ks, Dvali:2001fw, Burgess:2001fx, Shiu:2002xp,
Shiu:2001sy, Kachru:2003aw, Dasgupta:2002ew, Herdeiro:2001zb,
Kyae:2001mk, Garcia-Bellido:2001ky, Burgess:2001vr} mostly in a
similar manner, useful reviews are given by \cite{Quevedo:2002xw,
Kraniotis:2000ut}.
\section{Inflation and the shortcomings of standard cosmology}\label{cha:intro_inflation}
In this section, a short summary about inflation will be given,
where also the conditions will be stated, which a successful
inflationary model has to fulfill.

All the discussed scenarios start with the assumption that a big
bang has occurred in the beginning of the universe. Standard
cosmology gives a description of the universe by the combination
of general relativity with a simplistic hydrodynamical ansatz for
the matter energy momentum tensor. According to this model, the
universe is dominated by highly relativistic radiation until the
redshift of approximately $1100\leq1+z\leq1200$. Radiation has an
equation of state, where the pressure equals one third of the
energy density, $p=\rho/3$. Afterwards, the universe is dominated
by pressureless, non-relativistic matter. In standard cosmology,
this transition actually happens at a time $\sim 10^{13} \text{s}$
after the big bang \cite{Weinberg:1972, Kolb:1990, Peacock:1999}.

Two major shortcomings of standard cosmology are the so-called
horizon and flatness problems: the particle horizon is
time-dependent, and indeed has grown in time. Today, we measure a
highly homogenous cosmic microwave background radiation (CMBR).
But if one now carefully interpolates back the evolution equations
for the horizon in time, according to standard cosmology, then one
realizes that the regions, where the CMBR has been sent out,
cannot have been in causal contact at the time of its radiation.
So, the thermal equilibrium, which is a requirement for
homogeneity, seems unexplainable.

The second problem is the flatness problem, as follows. The total
energy density parameter $\Omega$ is defined by
\begin{equation}\label{eq:def_omega}
\Omega(t)\equiv\frac{\rho(t)}{\rho_\text{c}}\ ,
\end{equation}
where the critical energy density corresponding to a flat universe
is given by
\begin{equation}\label{eq:def_critedens}
\rho_\text{c}=\frac{3H^2{M_\text{pl}^2}}{8\pi}\ .
\end{equation}
In this equation, $H$ denotes the Hubble constant. Using the
Friedmann-equation,
\begin{equation}\label{eq:friedmann}
    H^2+\frac{k}{R^2}=\frac{8\pi}{3M_\text{pl}^2}\rho\ ,
\end{equation}
$\Omega(t)$ can be expressed by
\begin{equation}\label{eq:total_energy_density}
    \Omega(t)=\frac{1}{1-\frac{3 k M_\text{pl}^2}{8\pi R^2(t) \rho(t)}}\ .
\end{equation}
In this equation, $k$ is a dimensionless parameter which is
connected to the 3-dimensional curvature scalar. In standard
cosmology, a universe with a critical energy density $\Omega=1$ thus is
unstable, because the energy density for radiation depends on the
cosmic scale factor like $\rho \propto R^{-4}$ $R(t)$.

On the other hand, it is a fact that the total energy density
today lies within the region $0.3\leq \Omega \leq 1.1$
\cite{Kolb:1990}. Consequently, at the Planck time ($\sim 10^{-43}
\text{s}$), the deviations must have been very much smaller,
\begin{equation}\label{eq:flatness_problem}
    |\Omega(10^{-43} \text{s})-1|\leq \mathcal{O}(10^{-60})\ .
\end{equation}
In order to satisfy this inequality, an enormous amount of
finetuning is required, which seems very unnatural.

\subsubsection{Solution to the flatness and horizon problems}
Inflation mainly has been introduced to solve these two crucial
problems. The basic idea is that within the very early universe,
meaning at a time of about $10^{-34} \text{s}$, there was a short
period $[t_I, t_R]$, in which the expansion of the universe was
'faster than light', or more precisely in terms of the the cosmic
scale factor, $R(t) \propto t^\alpha$ with $\alpha>1$. In order
not to destroy all the merits of standard cosmology, inflation
must end at the so-called reheating time $t_R$, without leaving
any trace. Inflation implies an equation of state of the type
\begin{equation}\label{eq:equation_of_state_Inflation}
    \rho+3p<0\ .
\end{equation}
The only known kind of energy that fulfils this equation is vacuum
energy, as coming from a fundamental cosmological constant
$\Lambda$ or at least an effective cosmological constant. This energy has
an equation of state
\begin{equation}\label{eq:eos_vacuum}
    \rho_{\Lambda}+p_{\Lambda}=0\ .
\end{equation}
The cosmological constant energy rapidly becomes dominant, and
both matter and radiation redshift away in comparison to this.
This implies that the scale factor grows exponentially
\begin{equation}\label{eq:exp_scale_factor}
    R(t)\sim e^{H(t) t}\qquad \text{with}\ t\in [t_I, t_R]\ ,
\end{equation}
where the Hubble parameter $H(t)$ quickly approaches the constant
value $H=\sqrt{\Lambda/3}$. This exponential expansion is called
'de Sitter space'. Inflation solves the horizon problem in a
rather elegant way: for an exponential expansion of the universe,
the integral of the particle horizon diverges:
\begin{equation}\label{eq:part_horizon}
    d_H(t)=\int_{0}^{r_H(t)}\sqrt{g_{rr} dr}=R(t)\int_{0}^t \frac{dt'}{R(t')}\ ,
\end{equation}
so the lightcone in the forward direction increases exponentially,
whereas the one in the backward direction remains unchanged. This
consequently means that if inflation has last for a long enough
time, then a region that already has been in causal contact at a
time $t_I$, today is our observable universe. Concretely, this
requires that
\begin{equation}\label{eq:efoldings}
    e^{\Delta t H}\geq \frac{T_R}{T_0}\ ,
\end{equation}
or if we insert a typical reheating temperature $T_R\sim 10^{14}$
GeV and todays CMBR temperature $T_0\sim 10^{-13}$ GeV,
\begin{equation}\label{eq:efoldings2}
    \Delta t \gg 60 H^{-1}.
\end{equation}
Inflation therefore must have last at least for 60 e-foldings.
This is a strong requirement for inflation.

The flatness problem is solved by inflation, too. In order to see
this, we shall again take a look at the total energy density
(\ref{eq:total_energy_density}). $\rho(t)$ stays approximately
constant in an inflationary universe, because of the dominating
vacuum energy. Consequently, even if $k$ was non-vanishing before
inflation and the universe highly curved, during inflation
$\Omega(t)$ approaches 1, corresponding to a flat universe. This
again requires $\Delta t \gg 60 H^{-1}$, and one might already
suspect that indeed both problems are always solved at the same
time.
\subsubsection{Realization of inflation with a scalar field $\psi$}
In general, there are different possibilities how to achieve
inflation. The simplest formulation is given in positing an
additional scalar field $\psi$ with the standard field equations
\cite{Weinberg:1972}
\begin{align}
&\ddot \psi+3 H\dot \psi-\nabla^2 \psi=-V',
\end{align}
which weakly couples to matter. At the time $t_I$, this field
shall not sit exactly at the minimum of its shallow potential, but
nearby roll down slowly. More precisely, this means that $|\ddot
\psi|$ shall be negligible in comparison with $|3 H \dot \psi|$
and $|\frac{\text{d}V}{\text{d}\psi}|$. Written as a restriction
for the potential, this exactly corresponds to the slow roll
conditions (\ref{eq:slowroll}). This is the strongest condition
for a candidate inflaton field.
\subsubsection{Reheating}
We are not living in an exponentially accelerating universe, so
inflation must end at a certain cosmological time. Some mechanism
to explain this fact is needed. One possibility to achieve this
for a scalar field is given by an explicit coupling of this field
to matter, effectively acting as a damping term in the field equations.
At some time $t_R$, the slow-roll conditions are not fulfilled
anymore and $\psi$ undergoes oscillations of declining amplitude,
where the energy density of the scalar field is pumped int the
matter fields. This is called reheating and corresponds to a
cosmological phase transition.
\subsubsection{Density perturbations and spectral index}
Inflation leads to a scale invariant perturbation spectrum, the
reason can easily be seen in the fact that de Sitter space is
invariant under time translations. From COBE observations of the
temperature fluctuations in the CMBR it is known that the
magnitude of the perturbations (which stays constant for vacuum
energy) should be of the order \cite{Smoot:1998jt}
\begin{equation}\label{eq:mag_perturb}
{\delta_H}^2=\frac{1}{75\pi^2
M_{\text{Pl}}}\frac{V^3}{{V^{\prime}}^2}\approx \left(1.91\times 10^{-5}\right)^2\ .
\end{equation}
This again demands a restriction on the potential $V$.

The spectral index is defined by
\begin{equation}\label{eq:spectral_index}
    n(k)-1\equiv \frac{\text{d} P_k}{\text{d}(\ln k)}=2\eta_h-6\epsilon_h \ ,
\end{equation}
where the index $h$ refers to the horizon exit. The boomerang and
maxima experiments give bounds \cite{Jaffe:2000tx}
\begin{equation}\label{eq:boomerang_maxima}
    0.8 \leq n \leq 1.2\ ,
\end{equation}
inducing restrictions on $\epsilon$ and $\eta$, too.
\subsubsection{Hybrid inflation}
\begin{figure}
\centering
\includegraphics[width=10cm,height=8cm]{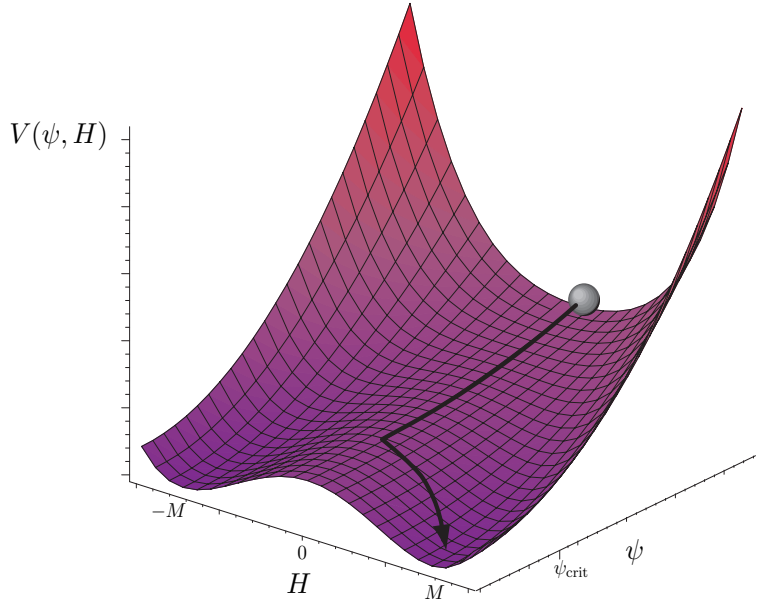}
\caption{The typical potential in hybrid inflation, $\psi$ rolls
down on the line $H=0$ until it eventually reaches the critical value
$\psi_{\text{crit}}$, then falls off to the true minimum at $H=\pm
M$.}\label{fig:hybrid_inflation}
\end{figure}

One well elaborated realization of inflation is the so-called
hybrid inflation, which was first introduced by Linde
\cite{Linde:1994cn} in 1993. In this model, there are two scalar
fields, the inflaton field $\psi$ and the field $H$ which
both play an important role. In the scalar potential of the gauge
theory sector, there then is a term
\begin{equation}\label{eq:pot_hybrid}
    V^{\text{YM}}(\psi,H)\sim\left(M(\psi)^2-\frac{1}{4}H^2 \right)^2\ .
\end{equation}
This means that the slow rolling of the inflaton field $\psi$
affects the mass of $H$ such that a phase transition occurs as
soon as $M^2(\psi)$ becomes negative and inflation ends. This is
schematically shown in figure \ref{fig:hybrid_inflation}.
This is a smart way to achieve reheating and it is especially
appropriate in the context of intersecting brane models. There,
open string tachyons usually do appear after some evolution of the
closed string moduli, as discussed in detail in chapter
\ref{cha:NStadpoles_OR}, and then could serve as the $H$ field.
This also fits in the sense that the tachyons signal a phase
transition in string theory \cite{Sen:1998ii, Sen:1998sm}, namely,
the condensation of higher dimensional D-branes into lower
dimensional ones or the condensation of two intersecting D-branes
into a single one wrapping a non-trivial supersymmetric 3-cycle
\cite{Blumenhagen:2000eb}. These tachyonic scalars are even well
suited to serve as Higgs fields and drive the gauge theory
spontaneous symmetry breaking mechanisms \cite{Bachas:1995ik,
Blumenhagen:2000wh, Aldazabal:2000cn, Ibanez:2001nd}. As a
speculation, this could link the exit from inflation to a phase
transition in the gauge sector of the theory, maybe even to the
electroweak phase transition itself.

The possibility of a slowly rolling tachyon by itself also has
been discussed in some recent work by Sen \cite{Sen:2002nu} that
has attracted much attention in the context of cosmology
\cite{Mukohyama:2002cn, Feinstein:2002aj, Gibbons:2002md,
Shiu:2002xp, Sami:2003qx}.
\section{Tree level scalar potential for the moduli}
Generally speaking, there are two different possibilities of which
geometric moduli fields could represent the inflaton, this being
the closed and open string moduli.

The closed string moduli are related to the background geometry.
If one assumes a factorization of the total space into the
4-dimensional Minkowski space and a compact 6-dimensional
subspace, as in the intesecting brane model that we have discussed
in all this work, the closed string moduli just come from this
subspace. Concretely, these are the K{\"a}hler moduli $T^I$
(\ref{eq:Torus_T}), being related to the size of the tori, and the
complex structure moduli $U^I$ (\ref{eq:Torus_U}) that is related
to its shape. Furthermore, in the orbifold models there are the
twisted moduli, being localized at the fixed point singularities
of these orbifolds.

In contrast to this, the open string moduli are related to the
concrete locations of the D-branes in the internal space, or in
particular the distance of certain parallel D-branes and the
Wilson lines of gauge fields along the branes. It has been
discussed in several papers \cite{Burgess:2001fx, Burgess:2001vr,
Garcia-Bellido:2001ky} if the open string moduli could satisfy the
slow-rolling conditions and the result has been that this is
possible if one makes the severe simplification to assume that the
closed string moduli are frozen. Geometrically speaking, this
means that if the background space is fixed and no backreaction on
the presence of the branes takes place, then their motion along
this space can be very slow for a certain time. After this time,
they start approaching each other faster and at a critical
distance a tachyon appears to signal their condensation. On the
other hand, the dynamics of the entire setting is determined by
the fastest rolling field. Therefore, this assumption implies that
the closed string moduli have to roll slower than the open string
moduli. Otherwise, the space could for instance shrink very
quickly and bring the two branes within their critical distance
much faster than originally estimated from the simplified analysis
with frozen volume. Only in the simplified brane-anti-brane model
the closed string moduli have not been ignored completely
\cite{Burgess:2001vr}. Of course, at first sight, an argument in
favor of ignoring the closed string moduli can be found: in a
brane-anti-brane setting, the tree level tadpoles are proportional
to the inverse volume of the transversal space to the branes.
These are driving the dynamics of the transverse geometry, so in a
large extra dimension scenario \cite{Arkani-Hamed:1998rs,
Antoniadis:1998ig}, the volume is large and consequently, the
tadpoles are suppressed. On the other hand, the evolution of the
transverse volume under consideration still can be fast on
cosmological scales, so the argument is not solid.

Because of this reason, the main concern in the following will be
given to the closed string moduli, their stability or slow-roll
behavior being a requirement such that slow rolling in the open
sting moduli might be possible as well. It has to be distinguished
carefully for these considerations between the so-called string
frame, being the usual frame in which perturbative string theory
is being performed, and the Einstein-frame that usually is taken
for general relativity and cosmology. In this second coordinate
system, the dilaton has been transformed away via a spacetime Weyl
transformation, which seems reasonable as no phenomenological
observations of the dilaton have been made so far.
\subsection{The potential in the string frame}
The potential for the closed string moduli in the string frame has
been computed already for the $\Omega R$ orientifold, containing
D6-branes intersecting at general angles in equation
(\ref{eq:scalar_pot_OR_NS}), in which only the imaginary part of
the three complex structure moduli and the dilaton play a role.
Therefore, it already occurs at open string tree level, but still
is exact in all orders of $\alpha'$. Regarding the orbifold models
of chapters \ref{cha:Z3} and \ref{cha:Z4}, the potential for the
untwisted moduli is similar, although one has to carefully
consider if some of these moduli are fixed by the orbifold
projection. If instead, one is interested in intersecting D-branes
of type IIA or type 0$'$ string theory, the contribution from the
orientifold planes is absent. This implies in type IIA that any
net R-R charge due to supersymmetry is absent, but not so in type
0$'$. In this non-supersymmetric string theory the orientifold
planes are rather exotic objects that carry charge but no tension.

The dilaton $\phi$ in the potential (\ref{eq:scalar_pot_OR_NS})
actually is the 4-dimensional dilaton, that here will be denoted
as $\phi_4$, because the difference to the 10-dimensional dilaton
$\phi_{10}$ will be crucial,
\begin{equation}\label{eq:10dim_dilaton}
    e^{-\phi_{4}}=M_\text{s}^3\, e^{-\phi_{10}} \prod_{I=1}^{3} \sqrt{R_x^{(I)}\, R_y^{(I)}} ,
\end{equation}
where $M_\text{s}$ is the string mass that should be written down
explicitly in this context. Only the $A$-torus with $b_I=0$ will
be discussed here, as the equations then simplify and the
qualitative behavior will be similar to the $B$-torus. Then the
potential simplifies:
\begin{equation}\label{eq:pot_string_frame}
V_\text{s}(\phi_4,U^I_2)=   M_\text{s}^4\, e^{-\phi_4}\,
\left( \sum_{a=1}^k  N_a\, \prod_{I=1}^3
   \sqrt{\left( n_a^I\right)^2 {1\over U^I_2}+
                          \left( m_a^I\, \right)^2
             U^I_2}
          -16\, \prod_{I=1}^3 \sqrt{\frac{1}{U^I_2}} \right) .
\end{equation}
and the NS-NS-tadpole cancellation conditions can then be expressed by
the wrapping numbers
\begin{equation}\label{eq:NStadpoleAtorus}
    \sum_{a=1}^k{ N_a\, \prod_{I=1}^3{n_I^a} } -16 =
\sum_{a=1}^k{ N_a\, n^a_I m^a_J m^a_K } =0 \qquad \text{with}\qquad I\neq J \neq K \ .
\end{equation}
At this point we recall that the potential for the imaginary part
 of the K\"ahler structures $T^I_2 = M_\text{s}^2 R_x^{(I)}
R_y^{(I)}$ is flat at tree-level and thus can be neglected at this
order. In the following, we will refer to the two quantities
$\phi_4$ and $U^I_2$ by the name 'Planck coordinates'. From these
considerations this appears to be the natural choice of variables
for expressing the string frame leading order potential.

From 4-dimensional ${\cal N}=1$ supersymmetric effective field
theories, it is known that only the particular combinations of scalars
\begin{align}\label{eq:def_ui_s}
    s&=M_\text{s}^3\, e^{-\phi_{10}}\, \prod_I R_x^{(I)}=
              e^{-\phi_{4}} \prod_I \frac{1}{\sqrt{U^I_2}} , \\
                     u^I&= M_\text{s}^3\, e^{-\phi_{10}}\, R_x^{(I)}\, R_y^{(J)}\, R_y^{(K)}=
                   e^{-\phi_{4}}  \sqrt{U^J_2\, U^K_2\over U^I_2}\nonumber
\end{align}
appear in chiral superfields such that the effective gauge
couplings can be expressed as a linear function of these variables
\cite{Cremades:2002te}. In terms of these 'gauge coordinates' the
string frame scalar potential can be simply expressed as
\begin{multline}
    V_\text{s}(s,u^I)= M_\text{s}^4\,\sum_{a=1}^k  N_a\,
         \Biggl[ \left(n^a_1 n^a_2 n^a_3\right)^2\, s^2+
                \sum_{I=1}^3 \left(n^a_I m^a_J m^a_K\right)^2\,
            \left( u^I \right)^2 \\
     + \left(m^a_1 m^a_2 m^a_3\right)^2 \left({u^1 u^2 u^3 \over s}\right) +
         \sum_{I=1}^3 \left(m^a_I n^a_J n^a_K\right)^2\,
               \left({s u^J u^K \over u^I}\right) \Biggr]^\frac{1}{2} -
        16\, M_\text{s}^4\,s\ ,
\end{multline}
where the last term  is the contribution from the O6-planes.
\subsection{The potential in the Einstein frame}
We now make the transition to the Einstein frame by performing the
Weyl rescaling
\begin{equation}\label{eq:weyl_Rescaling}
g^{(4),\text{s}}_{\mu \nu}\rightarrow g^{(4),\text{E}}_{\mu \nu}
=e^{2 \phi_{4}}g^{(4),\text{s}}_{\mu \nu}\ ,
\end{equation}
which results in the following potential in the Einstein frame:
\begin{equation}\label{eq:pot_scalar_Einstein_string}
    V_\text{E}(\phi_4,U^I)={M^4_\text{Pl}\over M^4_\text{s}}\, e^{4\phi_{4}}
             \, V_\text{s}(\phi_4,U^I) ,
\end{equation}
or explicitly,
\begin{multline}\label{eq:scalar_pot_Einstein}
    V_\text{E}(s,u^I) = M_\text{Pl}^4 \, \sum_{a=1}^k  N_a\,
              \biggl[ \left(n^a_1 n^a_2 n^a_3\right)^2\, \left({1\over
                  u^1 u^2 u^3} \right)^2\\+
                \sum_{I=1}^3 \left(n^a_I m^a_J m^a_K\right)^2\,
            \left({1\over s\, u^J\, u^K}  \right)^2
          +\left(m^a_1 m^a_2 m^a_3\right)^2
        \left({1 \over (s)^3\, u^1\, u^2\, u^3 }\right)\\ +
         \sum_{I=1}^3 \left(m^a_I n^a_J n^a_K\right)^2\,
         \left({1 \over s\, (u^I)^3\, u^J\, u^K}\right) \biggr]^{1\over 2}
         -16\, M_\text{Pl}^4\, \left({1\over u^1 u^2 u^3}\right) .
\end{multline}
The fact, that there is just one fundamental scale in string
theory, implies the following relation between the string scale
$M_\text{s}$ and the Planck scale $M_\text{Pl}$
\begin{equation}\label{eq:relation_String_planck_scale}
    {M_\text{s}\over M_\text{Pl}}= e^{\phi_4} = (s\, u^1\, u^2\, u^3)^{-{1/4}} .
\end{equation}
This means that a running of any single one of the four fields
$s,u^I$ at fixed $M_\text{Pl}$ implies a dynamical evolution of the
fundamental string scale $M_\text{s}$.

After a dimensional reduction down to four dimensions, rewriting
in terms of the 4-dimensional dilaton and the appropriate Weyl
rescaling, the kinetic terms for the scalar fields have the form
\begin{equation}\label{eq:kinetic_phi_U}
     S_\text{kin}= M_\text{Pl}^2 \int d^4 x\, \left[
           - (\partial^\mu \phi_4) (\partial_\mu \phi_4) -
           {1\over 4}\sum_{I=1}^3
           {(\partial^\mu \log U^I ) (\partial_\mu \log U^I)}\right]\ ,
\end{equation}
being explicitly calculated in appendix \ref{cha:kinetic_terms}.
In terms of the 'gauge coordinates' this reads as
\begin{equation}\label{eq:kinetic_s_u}
    S_\text{kin}= M_\text{Pl}^2 \int d^4 x\, {1\over 4}\,\left[
            -(\partial^\mu \log s) (\partial_\mu \log s) -
           \sum_{I=1}^3
           {(\partial^\mu \log u^I ) (\partial_\mu \log u^I)}\right] .
\end{equation}
Besides the transformation into the Einstein frame, these kinetic
terms have to be canonically normalized to be comparable to
standard cosmological results. The fields
$s,u^I$ have a logarithmic derivative in (\ref{eq:kinetic_phi_U})
and (\ref{eq:kinetic_s_u}). Therefore, the correctly normalized
fields $\tilde{s},\tilde{u}^I$ are defined by
\begin{equation}\label{eq:kanonically_normalized}
    {s}=e^{\sqrt 2 {\tilde{s}/M_\text{Pl}}} \quad \text{and} \qquad
    {u}^I=e^{\sqrt 2 {\tilde{u}^I/M_\text{Pl}}} .
\end{equation}
\section{Inflation from dilaton and complex structure}
A crucial assumption that has been adopted throughout the whole
work is that it is possible to work within perturbative string
theory. This implies that the 10-dimensional string coupling has
to be small and, as it is nothing but the expectation value of the
10-dimensional dilaton, $e^{\phi_{10}}\ll 1$. Another assumption
that has to be made for convenience is that numbers of stacks $N_a$,
and the wrapping numbers $n_I^a$ and $m_I^a$ are not extremely
large. This seems unproblematic, as very large numbers would
result in an unacceptable particle spectrum anyway. But compared
to the former work of Burgess et al.\cite{Burgess:2001vr}, we do
not have to impose that the internal radii are small compared to
the string scale, as our potential is exact to all orders of
$\alpha'$.

As mentioned in the introduction, the most important constraint are
the slow-rolling conditions (\ref{eq:slowroll}), which in the
Einstein frame for the two fields $s$ and $u^I$ correspond to
\begin{equation}\label{eq:slow_roll_einstein}
\epsilon = {M_\text{Pl}^2 \over 2} \left(
{V_\text{E}'(s,u^I) \over V_\text{E}(s,u^I)} \right)^2 \ll 1, \qquad \eta =
M_\text{Pl}^2 {V_\text{E}''(s,u^I) \over V_\text{E}(s,u^I)} \ll 1\ .
\end{equation}
The derivatives of $V(s,u^I)$ have to be taken with respect to the
canonically normalized fields (\ref{eq:kanonically_normalized}).
Instead of this set of coordinates, one could as well try to
satisfy the slow-rolling conditions using the two fields $\phi_4$
and $U^I$. At first sight, it seems to be irrelevant which set of
coordinates is beeing used as the physical result should not
depend on this. But here, the situation is different, as we shall
see that inflation is not possible if none of the four moduli
are assumed to be frozen. But this freezing then distinguishes
among different physical situations: for the set $(s,u^I)$, due to
(\ref{eq:relation_String_planck_scale}), the string scale is
always forced to change during inflation if not all moduli are
being frozen at the same time. In contrast to this, for the set
$(\phi_4, U^I)$, the string scale can be made constant by just
freezing $\phi_4$.

In the following, we will discuss inflation in both coordinate
systems, as it is unclear which of these two physical
situations is the right one.
\subsection{Discussion for the coordinates $(s,u^I)$}
According to \cite{Burgess:2001vr}, the coordinates $(s,u^I)$ are
the natural coordinates if one assumes ${\cal N}=1$ supersymmetric
dynamics at some higher energy scale. But one has to be careful,
as this assumes that there occurs a spontaneous breaking of
supersymmetry which can be achieved via a continuous deformation
of the theory. Indeed, the vanishing of the supertrace Str$({\cal
M}^2)=0$ in the open string spectrum of toroidal type I
intersecting brane world models suggests a spontaneous
supersymmetry breaking \cite{Bachas:1995ik}. On the other hand, the potential
$V_\text{E}(s,u^I)$ is not generally of the type which can occur
as the scalar potential in a supersymmetric theory. But this means
that the only possibility to make the transition would be a
discontinuous phase transition, separating the supersymmetric
vacuum from the non-supersymmetric theory. This theory then will
be non-supersymmetric at all scales for a given set of winding
numbers.

It has to be mentioned that under special circumstances, also a
continuous phase transition is possible: in a supersymmetric
theory, the scalar potential should have the form of a D-term
potential. For certain choices of the complex structure moduli,
this indeed can be achieved, as being described in
\cite{Cremades:2002te}. The spontaneous supersymmetry breaking
then in a ${\cal N}=1$ supersymmetric theory occurs by adding a
Fayet-Iliopoulos term, that is supposed to be the low-energy
manifestation of a small change of the complex structure as
described in chapter \ref{cha:Z4}.

In our case, such a continuous deformation cannot be guaranteed in
general, such that it remains somehow doubtful why $(s,u^I)$
should be the right variables.

It is a first, somehow disappointing observation from the
potential (\ref{eq:scalar_pot_Einstein}) that slow rolling
definitely is impossible, if not 3 of the parameters $s$ and
$u^I$ are frozen at the same time. The reason is that the fast
rolling scalars destabilize the background before slow-rolling in
the other moduli can become relevant on a cosmological scale. This
result is almost trivial from a mathematical point of view.
Qualitatively, the potential looks like a four-dimensional
generalization of a similar potential as shown in figure
\ref{fig:potential_inflation}. At any point, there always does
exist a direction in which the potential is constant, just simply
the contour line $V=\text{const}$. On the other hand, it is a non-trivial
question if these lines of constant $V$ do correspond or are at
least close to certain specific variables like $s$ or $u_I$.

Therefore, we will now discuss in analogy to
\cite{Burgess:2001vr} what happens if three moduli are getting
frozen 'by hand'. In doing this, we will only discuss the generic
situation, neglecting very specific choices of the wrapping numbers
or very special regions in parameter space where new features
generally might appear.
\subsubsection{Inflation in $s$, all $u_I$ frozen}
For this case, the scalar potential for only the D-branes
has the schematic form:
\begin{equation}\label{eq:s_inflation_potential}
V_\text{E}^{\rm D6}=M_\text{Pl}^4 \sum_{a=1}^k N_a\, \sqrt{
                \alpha_a+{\beta_a\over s^2} + {\gamma_a\over s}
                  +{\delta_a\over s^3}} ,
\end{equation}
where the coefficients can be read off from
(\ref{eq:scalar_pot_Einstein}) and involve the fixed scalars $u^I$
and some numbers of order one. Due to the appearing squares,
$\alpha_a > 0$. For large values of $s$, we have
\begin{equation}\label{eq:large_s_pot}
    V_\text{E}^{\rm D6}=M_{Pl}^4 \left[ \left(\sum_a N_a\,
                \sqrt{\alpha_a} \right)+
                    {1\over 2} \left( \sum_a N_a\,
               {\gamma_a\over \sqrt{\alpha_a}}\right) {1\over s} +\ \cdots
\right] \qquad \text{for}\ s\gg\ 1\ .
\end{equation}
The orientifold planes contribute
\begin{equation}\label{eq:orientifold_contibution_s_inflation}
V_\text{E}^{\rm O6}=-M_\text{Pl}^4\, 16 \left[\prod_{I=1}^3 u^I\right]^{-1}.
\end{equation}
to the potential. We have to distinguish between two cases:
\begin{enumerate}
    \item {\bf all $\prod_I n_a^I$ positive:}\\
    If we choose all wrapping numbers $\prod_I
n_a^I$ to be positive, then the constant term in the D-brane
contribution cancels precisely against the O6-planes contribution
due to the R-R tadpole cancellation conditions. In this case, one
simply gets $V\sim 1/s$, which implies $V' \sim V$, with a
constant of proportionality of order one \footnote{The derivative
of course has to be taken with respect to the canonically
normalized field $\tilde{s}$.} and $s$ consequently does not show
slow-rolling behavior.
    \item{\bf some $\prod_I n_a^I$ negative:}\\
If instead some of the $\prod_I n_a^I$ are negative, then the
potential takes the form
\begin{equation}\label{eq:pot_Einstein_sinflation}
    V_\text{E}= V_\text{E}^{\rm D6} + V_\text{E}^{\rm O6} = M_\text{Pl}^4\left( A + B\,
           e^{-\sqrt 2{ \tilde{s}/M_\text{Pl}}} + \ \cdots \right),
\end{equation}
The slow rolling parameters then read
\begin{equation}\label{eq:slow_rolling_s_inflation}
    \epsilon={B^2\over A^2}{1\over s^2}, \quad\quad
            \eta={2B\over A}{1\over s}\ ,
\end{equation}
meaning that $\eta\ll 1$ directly implies  $\epsilon\ll 1$.
If one inserts the explicit expressions for $\alpha_a$ and $\gamma_a$, one gets
\begin{equation}\label{eq:eta_s_inflation}
     \eta=\sum_I \zeta_I {u^J\, u^K \over u^I\, s}
           =\sum_I {\zeta_I \over \left( U^I \right)^2}\ ,
\end{equation}
If all complex structure moduli satisfy $U^I\gg 1$ with their
relative ratios fixed, then $s$ is slow rolling. This is
self-consistent with the assumption $s\gg 1$, as it is evident
from the definition (\ref{eq:def_ui_s}). So for the given
assumptions, $s$ is a reasonable inflaton candidate. One
interesting observation for $B>0$ is that the length of the string
in this scenario also inflates.
\end{enumerate}
In type II or type 0$'$ string theory, the distinction between the
two cases does not apply as no negative orientifold contribution
appears in the potential, the potential generally is of the second
form (\ref{eq:pot_Einstein_sinflation}).\footnote{Actually, this
potential is identical to the one in \cite{Burgess:2001vr}. The
configuration of D$9$- and D$5$-branes in this paper is just a
very specific choice of D9-branes with magnetic flux, or more
precisely, infinite magnetic flux for a D5-brane. This is the
T-dual situation of what is studied here.}

\subsubsection{Inflation in one $u_i$, $s$ and other $u_I$ frozen}
If one freezes $s$ together with all but one $u_I$, one again gets
a potential of the form (\ref{eq:pot_Einstein_sinflation}) in a
$1/u^I$ expansion. Here, the constant term
$A$ never vanishes, not even in type I models, so that the
slow-rolling parameter $\eta$ always can be written as
\begin{equation}\label{eq:eta_u_inflation}
    \eta =
\zeta_1 {u^J\, u^k \over u^I\, s}
            +\zeta_2 {u^J\, s \over u^I\, u^K}
           +\zeta_3 {u^K\, s \over u^I\, u^J}
          ={\zeta_1\over  \left( U^I\right)^2} +\zeta_2
                         \left( U^J\right)^2 +\zeta_3 \left
                           ( U^K\right)^2,
\end{equation}
with all $\zeta_i$ of order one. Slow rolling is possible if one
requires $U^I\gg 1$ and $U^J,U^K\ll 1$, again being
self-consistent with the assumption $u^I\gg 1$. The constant $A$
is positive, but now the constant $B$ generally can become negative
in type I theory, as the orientifold planes contribute as well.
In case $B$ is negative, the evolution leads towards smaller values
of $u^I$ until the slow-rolling condition is no longer satisfied
or open string tachyons do appear.
\subsubsection{Phenomenological Discussion}
In this section, we are going to look at several phenomenological
questions that follow the first step of fulfilling the slow-rolling
conditions.
\begin{enumerate}
    \item {\bf Inflationary exit scenario}\\
In the case of $s$-inflation, we had to require $A>0$ and $B>0$
for slow-rolling, so it follows that $\eta>0$ in
(\ref{eq:slow_rolling_s_inflation}) and one faces a positive
cosmological constant. Hence $s$ rolls towards larger values,
meaning deeper into the slow-rolling region. This means that the
naive picture from reheating that at some point the slow-rolling
conditions are not fulfilled anymore, does not apply. But there
still is another possibility: depending on the angle in between
the D-branes, the open string sector can have tachyons, localized
at the intersection locus, see for instance equation
(\ref{eq:zeropoint_z3}). For fixed wrapping numbers $n_I$ and
$m_I$ of the two branes $a$ and $b$, the angle only depends on the
complex structure moduli:
\begin{equation}\label{eq:angle}
    \theta^I={1\over \pi}\, \arctan\left({\left({m_I^b\over n_I^b}+
          {m_I^a\over n_I^a}\right) U^I \over 1+\left({m_I^a\, m_I^b \over
                n_I^a\, n_I^b}\right)\, \left(U^I\right)^2 }\right) ,
\end{equation}
The mass of the lowest bosonic mode is given by
\begin{equation}\label{eq:mass_lowest_bos_mode}
    M^2_\text{scal} = {1 \over 2} \sum_{I=1}^3
\theta^I -  {\rm max}\{ \theta^I:I=1,2,3  \} ,
\end{equation}
where one has to assume that $0<\theta^I<1/2$, such that this
equation is correct. Tachyons do appear, if this squared mass
becomes negative, and this triggers the decay of the intersecting
brane configuration to different ones, finally to a stable one. So
it might be possible that $s(U^I,\phi_4)$ first rolls slowly and
then, depending on the behavior of the complex structure moduli
$U^I$, suddenly tachyon condensation is triggered off, leading to
a phase transition. If this phase transition is understood to be a
cosmic phase transition, then this realizes the hybrid inflation
scenario.

The decay process itself can be understood via boundary string
field theory, not by simple perturbative string theory
\cite{Sen:1998sm, Kutasov:2000qp, Kutasov:2000aq, Andreev:2001ak}.
Nevertheless, the suggestion has been given (and treated in more
detail within chapter \ref{cha:Z3}) that the tachyon might act as
a Higgs field in the effective gauge theory, based on the simple
observation that the tachyon being localized at the brane
intersection, carries a bifundamental representation of the
unitary gauge groups of the two stacks. This would link the cosmic
phase transition with a spontaneous breaking of gauge symmetry.

In order to better understand the appearing of tachyons, we should
take a look at figure \ref{fig:tachyons}.
\begin{figure}
\centering
\includegraphics[width=9cm,height=7cm]{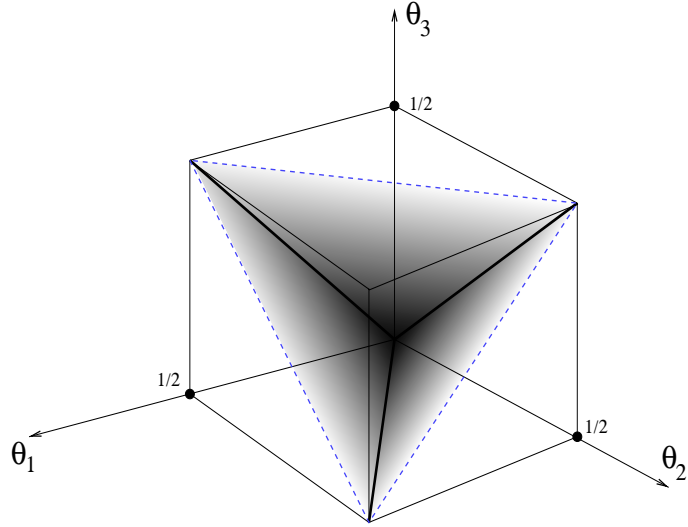}
\caption{The appearance of tachyons.}\label{fig:tachyons}.
\end{figure}
Within the interior of the shaded cone (with edges given by the
bold diagonal lines), no tachyons do appear. The system preserves
${\cal N}=1$ supersymmetry on the shaded faces of the cone and
${\cal N}=2$ supersymmetry on the edges. The origin corresponds to
parallel branes which preserve the maximal ${\cal N}=4$
supersymmetry. During $s$-inflation the background geometry is
driven towards larger values of all three complex structures $U^I$
but with their ratios fixed. Using (\ref{eq:angle}), we conclude that if
none of the two D6-branes is parallel to the $X^I$-axis, i.e.
$m_a^I\not= 0$ for $a=1,2$, then the intersection angle $\theta^I$
is driven to $0$. If in fact any one of the two is parallel to the
$X^I$-axis, then $\theta^I$ goes to $1/2$. To summarize, up to
permutations we have the following four possible endpoints of the
flow
\begin{align}
                    (\theta^1,\theta^2,\theta^3)&\to (0,0,0),& \quad\quad
                    &{\cal N}=4\ {\rm SUSY,\ no\ tachyons}& \\
                     (\theta^1,\theta^2,\theta^3)&\to (0,1/2,1/2),& \quad\quad
                    &{\cal N}=2\ {\rm SUSY,\ no\ tachyons}& \nonumber \\
                    (\theta^1,\theta^2,\theta^3)&\to (0,0,1/2),& \quad\quad
                    &{\cal N}=0\ {\rm  SUSY,\ tachyons}& \nonumber \\
                    (\theta^1,\theta^2,\theta^3)&\to (1/2,1/2,1/2),& \quad\quad
                    &{\cal N}=0\ {\rm SUSY,\ no\ tachyons}.& \nonumber
\end{align}
This classification actually leaves out the fact that the points
that the parameters are driven to cannot be reached within a given
set of winding numbers for any finite value of $U^I$. For instance
in the first case, two branes may approach vanishing intersection
angles very closely, but only if their $(n^a_I,m^a_I)$ were
proportional, they could become parallel. Thus, there may occur a
situation where a set of branes evolves towards an ${\cal N}=4$
supersymmetric setting dynamically, approaching it arbitrarily
well, but never reaching it without tachyon condensation. In fact,
tachyons can then no longer be excluded for such a brane setting
of the first type, as with three very small relative angles, the
mass of the NS ground state may still become negative. But one
thing clearly can be deduced: Whenever the model contains two
intersecting D-branes, where one of the D-branes is parallel to
exactly one of the $X^I$-axes, the system evolves to a region
where tachyons do appear. Unfortunately,  it is difficult to
determine in general the precise end-point of inflation, i.e. the
point where the model crosses one of the faces in figure \ref{fig:tachyons}.

    \item {\bf Number of e-foldings, density perturbations and spectral index}\\
    As discussed in chapter \ref{cha:intro_inflation} in detail,
    inflation must last for a long enough time, or more precisely,
    for 60 e-foldings, see (\ref{eq:efoldings2}). This number can
    be easily calculated if one makes the simplifying assumption
    that the potential $V_\text{E}$ stays approximately constant
    during inflation at a value $V_\text{inf}$ and one obtains the
    following criterion:
    \begin{equation}\label{eq:efoldings}
    N=-\int_{\tilde{s}_\text{I}}^{\tilde{s}_\text{R}} d\tilde{s} {1\over M_\text{Pl}^2}\,
             {V_\text{E} \over V_\text{E}'}\simeq {A\over 2B}\,
             (s_\text{R}-s_\text{I})\simeq
          60-\log\left({10^{16}\ {\rm GeV} \over V_\text{inf}^{1/4} }\right) \ .
    \end{equation}
    In this equation, the index I refers to the start and
    the index R to the end of inflation. The magnitude of the
    primordial density fluctuations (\ref{eq:mag_perturb}) also can be calculated:
    \begin{equation}\label{eq:density_perturbations}
        \delta_H\sim {1\over 5 \sqrt{3}\pi} \left({V_\text{E}^{3/2} \over
                      M^3_\text{Pl}\, V_\text{E}'}\right) \simeq {1\over 5 \sqrt{6}\pi}
                       \left({A^{3/2} s_\text{I} \over B} \right),
    \end{equation}
    As mentioned earlier, from COBE observations one knows that it
should be of the size $\delta_H=1.91\times 10^{-5}$. The spectral
index (\ref{eq:spectral_index}) is calculable as well, it is given
by
\begin{equation}\label{eq:spectral_index_model}
n-1=-6\epsilon_h+2\eta_h\simeq 2\eta_h\simeq {4B\over A s_\text{I}}\ ,
\end{equation}
which has to match the Maxima and Boomerang bounds
(\ref{eq:boomerang_maxima}).

For $s$-inflation, one has $A,B>0$
such that $s_e\gg s_h$, which then implies $N=\eta_e^{-1}$. But is
impossible to make any more detailed prediction, if $s_e$ and
$s_h$ are unknown.

For $u^I$-inflation, one instead assumes $u^I_I \gg u^I_R$ and
using $u_\text{I}^I=-2\,B\,N/A$, it is possible to express the
density fluctuations and the spectral index in terms of $N$ and $A$ only:
\begin{equation}\label{eq:u_Iinflation_spectralindex}
    \delta_H \simeq {2\over 5
\sqrt{6}\pi}{A^{1/2}N},\qquad
            n-1=-{2\over N}.
\end{equation}
As $n$ by Boomerang and Maxima is bounded in between $0.8\leq
n\leq 1.2$, one gets a direct the prediction for $A\leq 0.45\times
10^{-10}$ in this case.
\end{enumerate}
\subsection{Discussion for the coordinates $(\phi_4, U^I)$}
\begin{figure}
\centering
\includegraphics[width=10cm,height=10cm]{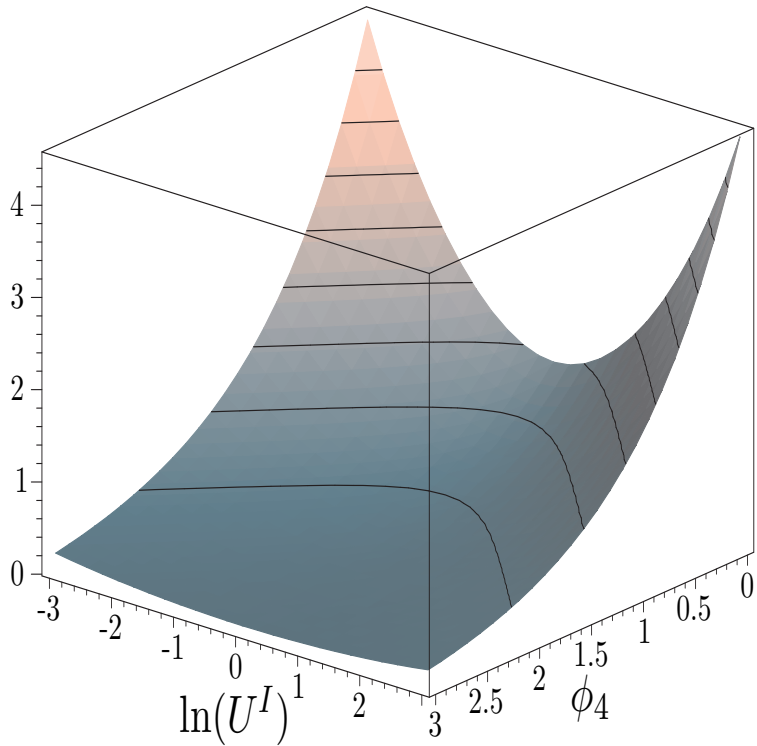}
\caption{The schematic scalar potential for the coordinates
$(\phi_4, U^I)$.}\label{fig:potential_inflation}
\end{figure}

The potential in the string frame for the coordinates $(\phi_4,
U^I)$ is given by (\ref{eq:pot_string_frame}). It can be
immediately noticed that for $\phi_4$ slow-rolling is not
possible. On the other hand, for $\phi_4$ and two complex
structure moduli $U^I$ frozen, depending on the the specific winding
numbers (and assuming non-trivial intersection angles), the potential
is of the type
\begin{equation}\label{eq:pot_planck_coord}
    V_\text{E}(U^I)=M_\text{Pl}^4\, {A\, \sqrt{U^I}}\qquad
\text{or}\qquad V_\text{E}(U^I)=M_\text{Pl}^4\, {A \over \sqrt{U^I}}\ ,
\end{equation}
for $U^I\gg 1$, which in both cases does not fulfil the
slow-rolling conditions. It is schematically shown in figure
\ref{fig:potential_inflation}.
The result for the case $U^I\ll 1$ is similar, so the only
possibility would be near local minima of the potential, that have
not been found so far: for all studied examples, $\epsilon \ll 1$
near an extremum, but then $\eta=\mathcal{O}(1)$, although no
general proof does exist. Concluding this section, it can be
stated that slow rolling seems to be impossible for the choice of
coordinates $\phi_4$ and $U^I$.
\section{Inflation from the K{\"a}hler structure}
In the previous section, we have discussed the possibility to
achieve slow-rolling for any one of the fields $\phi _{4}$ or
$U^{I}$. We have seen that it does not seem to be possible; even
more, the only chance to avoid fast rolling is to freeze the
complex structures $U^{I}$, while the general stabilization of the
dilaton remains an open problem. The complex structure can be
frozen by orbifolding, as shown in chapter \ref{cha:Z3}. The are
also other possibilities:\ in type 0' backgrounds, the complex
structures are getting dynamically frozen at values of order one
\cite{Blumenhagen:2002mf}, the same is true in type I models with
some negative wrapping numbers. A recent example for type IIA
orientifold models on Calabi-Yau spaces is given in
\cite{Blumenhagen:2003vr}. There, a three-form G-flux has been
turned on, and the freezing of complex structures takes place by means
of a F-term scalar potential\cite{Taylor:1999ii}. Actually, a
stabilization of the complex dilaton takes place as well, but one
has to keep in mind that this statement is only valid in
leading order in string perturbation theory, as $\alpha ^{\prime
}$ corrections to the K\"{a}hler potential may alter the F-term
potential. Although it is assumed that this alteration still
leaves the dilaton constant, the fact that the D-branes with
G-flux and the orientifold planes do not chancel their R-R charges
locally, or in other words, D-branes and orientifold planes do not
lie on top of each other, invalidates the assumption.

In this section, the assumption will be made that all complex structure
moduli $U^{I}$ will be frozen by any such mechanism and the leading order
potential for the K\"{a}hler moduli will be studied. As the dilaton is
not frozen, the tree level potential in the Einstein frame is then given by
\begin{equation}
V_\text{E}^{\text{tree}}=M_\text{Pl}^{4}\,\ C\,\ e^{3\phi _{4}},
\label{eq:tree_pot_Einstein}
\end{equation}
where $C$ is a constant of order one. A non-trivial dependence of
the potential on the K\"{a}hler moduli at the earliest arises at
one-loop level. If the closed string sector preserves
supersymmetry, the torus and Klein-bottle amplitude vanish,
because the R-R and NS-NS tadpoles are similar. The remaining
cylinder and M{\"o}bius strip amplitudes for a non-vanishing angle
on a certain 2-torus do depend on its K\"{a}hler modulus through
the lattice contributions in non-supersymmetric open string
sectors. These Kaluza-Klein and winding contributions can also
depend on open string moduli, being the distance $x$ between
the branes and their relative Wilson line $y$. The Hamiltonian for
these modes on one 2-torus is given by
\begin{equation}\label{eq:Hamiltonian_lat_annulus_generalization}
\mathcal{H}_{\text{lattice, op.}}=\sum_{r,s}
\frac{\left|(r+x_I)+(s+y_I)T^I\right|^2}{T^I_2}\frac{U^I_2}{|n_I+m_I U^I|^2}\ ,
\end{equation}
which is a generalization of (\ref{eq:Hamiltonian_lat_annulus}).
More precisely, $0\le y_I\le 1$ denotes the relative transversal
distance between the two D-branes and $0\le x_I\le 1$ the relative
Wilson line along the longitudinal direction of the two D-branes
on the specific 2-torus. The full potential can then be understood as a
function on $T^I$, $x_I$ and $y_I$.

\begin{figure}
\centering
\includegraphics[width=14cm,height=6cm]{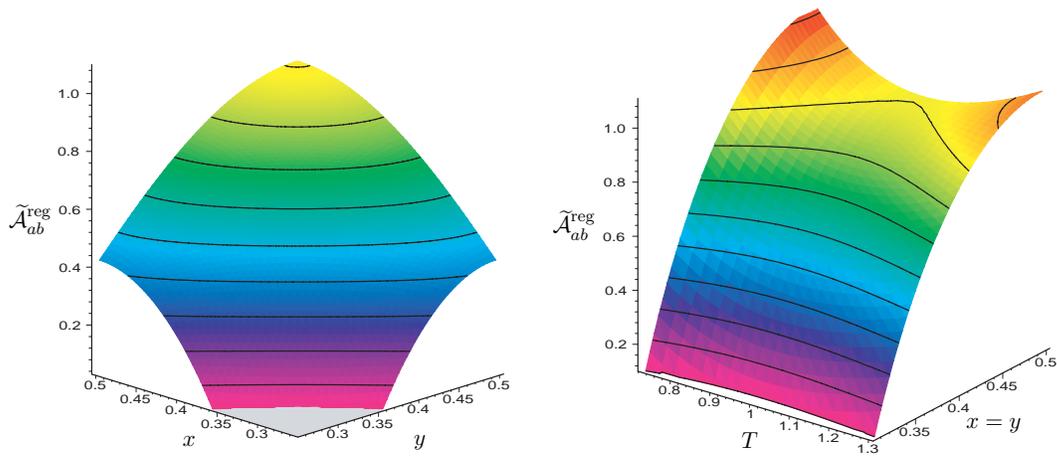}
\caption[The integrated regularized cylinder amplitude
$\widetilde{{\cal A}}_{ab}^{\rm reg}$ for a typical example.]{The
numerically integrated regularized cylinder amplitude for the
typical example $\kappa_1=1/3$, $\kappa_2=-1/3$,
$\Delta=\sqrt{3}/14$, where the off-set of the amplitude has been
chosen arbitrarily. The left plot shows the dynamical stability of
$x=y$ for $T=1$, where the grey area at the bottom indicates the
appearance of tachyons when the numerical integration is
diverging. The right plot shows the minimum at $T=1$ for $x=y$,
slow-rolling is not possible as $V''$ is not small.}\label{fig:zweiplots_inflation}
\end{figure}

For simplicity, we will now restrict to the case of two stacks of
branes that intersect on two tori but are parallel on the
remaining one. Furthermore, we assume that the branes are not
parallel to any one of the O6-planes in order to break
supersymmetry. Then, the only relevant amplitude at the one-loop
level is given by the cylinder diagram for open strings stretching
between the two stacks. The potential up to one-loop order is
given by
\begin{equation}\label{eq:one_loop_pot_inflation}
    V_\text{E}(T,x,y)=M_\text{Pl}\, M_\text{s}^3\, C_0  -
                        M_\text{s}^4\, C_1\, -
                        {\cal A}_{ab}(T,x,y)+\ldots \ ,
\end{equation}
where the first two terms stand for all constant contributions
that are independent of $T$,$x$ and $y$. For our simple case, we
have just a dependence of ${\cal A}_{ab}$ on one $T^I\equiv T$.
The three variables are related to the canonical normalized open string
moduli by
\begin{equation}\label{eq:canonical_normalized_open_str}
    Y^2={1\over M_\text{s}^2}\, \Delta\, T\, y^2,\qquad X^2={1\over M_\text{s}^2}\, {\Delta\over  T}\, x^2 \ ,
\end{equation}
where $\Delta$ is a constant of order one, depending on the
specific wrapping numbers of the D-branes and the choice of model $b=0$ or $b=1/2$.
From (\ref{eq:Annulus_amplitude_OR_loopij}) and
(\ref{eq:Annulus_amplitude_OR_loopNS_ij}), we directly obtain the
complete cylinder loop channel amplitude
\begin{multline}\label{eq:loop_amplitude_inflation}
    {\cal A}_{ab}(T,x,y)
={M_\text{Pl}^4\over (8\pi^2)^2}\, e^{4\phi_4}\,
      N_a\, N_b\, I_{ab}
                 \int_0^\infty {dt\over t^3}\,
              \left(\sum_{r,s\in \mathbb{Z}} e^{-2\pi t\, \Delta\, \left[
                 {(r+x)^2/T} + T (s+y)^2 \right]}\right)\\
\shoveleft{  \cdot
 \left( \frac{\tfkto{0}{0}^2 \tfkto{\kappa_1}{0} \tfkto{\kappa_2}{0}-
         e^{-\pi i (\kappa_1+\kappa_2)}\,
                    \tfkto{0}{{1\over 2}}^2 \tfkto{\kappa_1}{{1\over 2}}
                  \tfkto{\kappa_2}{{1\over 2}}}{
                   \eta^6\, \tfkto{\kappa_1+{1\over 2}}{{1\over 2}}
                    \tfkto{\kappa_2+{1\over 2}}{{1\over 2}}
          e^{-\pi i (\kappa_1+\kappa_2+1)} }\right.}\\
         \left. -\frac{\tfkto{{1\over 2}}{0}^2 \tfkto{\kappa_1+{1\over 2}}{0}
             \tfkto{\kappa_2+{1\over 2}}{0}}{
                   \eta^6\, \tfkto{\kappa_1+{1\over 2}}{{1\over 2}}
                    \tfkto{\kappa_2+{1\over 2}}{{1\over 2}}
          e^{-\pi i (\kappa_1+\kappa_2+1)} } \right).
\end{multline}
The argument of the $\vartheta$-functions are given by $q=\exp(-2\pi
t)$ and for the NS-sector ground state energy, one obtains
\begin{equation}\label{eq:NSgroundstate_energy}
    M^2_\text{scal}=\left[
     \left(\Delta {x^2\over T} + \Delta\, T\, y^2\right) +
                {1\over 2}\left(\kappa_1+\kappa_2\right)-{\rm max}
             \{\kappa_I:I=1,2\}\right]\ \text{for}\ 0<\kappa_I<1/2\ .
\end{equation}
The modular transformation to the tree channel via $l=1/(2t)$ leads to
\begin{multline}\label{eq:tree_amplitude_inflation}
    \widetilde{{\cal A}}_{ab}(T,x,y)
={M_\text{Pl}^4\over (8\pi^2)^2}\, e^{4\phi_4}\,
      N_a\, N_b\, I_{ab}
                 \int_0^\infty {dl}\,
     \left({1\over \Delta} \sum_{r,s\in\mathbb{Z}} e^{-\frac{\pi l}{\Delta}\, \left[
                 {T\, r^2} + \frac{s^2}{T} \right]}\,
              e^{-2\pi i (r\,x+s\, y)}\right)\\
  \shoveleft{\cdot   \left( \frac{\tfkto{0}{0}^2 \tfkto{0}{\kappa_1} \tfkto{0}{\kappa_2}-
                   \tfkto{0}{{1\over 2}}^2 \tfkto{0}{\kappa_1+{1\over 2}}
             \tfkto{0}{\kappa_2+{1\over 2}}}{
                   \eta^6\, \tfkto{{1\over 2}}{\kappa_1+{1\over 2}}
                    \tfkto{{1\over 2}}{\kappa_2+{1\over 2}}  }\right. }\\
                    \left.-\frac{
                    \tfkto{{1\over 2}}{0}^2 \tfkto{{1\over 2}}{\kappa_1}
                  \tfkto{{1\over 2}}{\kappa_2}}{
                   \eta^6\, \tfkto{{1\over 2}}{\kappa_1+{1\over 2}}
                    \tfkto{{1\over 2}}{\kappa_2+{1\over 2}}  } \right) ,
\end{multline}
where the argument of the $\vartheta$-functions is $q=\exp(-4\pi l)$. This amplitude contains divergencies, coming
from the uncancelled NS-tadpoles for the non-supersymmetric vacua,
we are interested in. In order to make cosmological calculations,
one has to regularize (\ref{eq:tree_amplitude_inflation}) by
subtracting the divergencies
\begin{equation}\label{eq:regularization}
    \widetilde{{\cal A}}_{ab}^{\rm reg}(T,x,y)
=\widetilde{{\cal A}}_{ab}(T,x,y)- \widetilde{K}_{ab},
\end{equation}
which are just given by
\begin{equation}\label{eq:divergencies_NS}
    \widetilde{K}_{ab}={M_\text{s}^4\over (8\pi^2)^2}\,
      N_a\, N_b\, I_{ab}
                 \int_0^\infty {dl}\, {4\over \Delta}\,
             {\sin^2\left({\pi(\kappa_1+\kappa_2) \over 2}\right)\,
              \sin^2\left({\pi(\kappa_1-\kappa_2) \over 2}\right) \over
                \sin(\pi \kappa_1)\,  \sin(\pi \kappa_2) }\ .
\end{equation}
The first observation we can make is that the K{\"a}hler modulus
gets stabilized dynamically, the argument being as follows: if
$x=y$, then (\ref{eq:Hamiltonian_lat_annulus_generalization}) is
invariant under T-duality, exchanging $x$ and $y$. Therefore,
there must be at least a local extremum around the self-dual point
$T_\text{sd}=1$, fixing the internal radii at values of order of
the string scale. In figure \ref{fig:zweiplots_inflation}, a
numerically calculated example of the potential
(\ref{eq:one_loop_pot_inflation}) is shown.

Assuming $T$ is frozen, then still the open string moduli $x=y$
could show a slow-rolling behavior if $x=y$ is dynamically stable
(what indeed is the case, see the left plot in figure
\ref{fig:zweiplots_inflation}). In a similar situation, such a
result indeed has been found \cite{Burgess:2001fx} for $T\gg 1$ in
the neighborhood of the instable antipodal point. In this example,
the second derivative $V''$ of the potential vanishes at the
antipodal point, therefore both $\epsilon$ and $\eta$ become
small, what is sufficient for slow-rolling.

It is an important question, if this statement stays true when T
approaches its true minimum at $T=1$, where also massive
contributions contribute to the force between the two D-branes. In
order to clarify this point, we are going to expand the annulus
amplitude (\ref{eq:tree_amplitude_inflation}) in $q$. Taking the
zeroth order term $q^0$ in the $\vartheta$- and $\eta$-functions,
but summing over all Kaluza-Klein and winding modes, means that
one has to evaluate the integral
\begin{equation}\label{eq:integral_KKW}
    \int_0^\infty dl \left[\left( 1+2\sum_{r\ge 1}
                  e^{-\frac{\pi l}{\Delta}\,
                 {T\, r^2} } \cos (2\pi  r x) \right)
                 \left( 1+2\sum_{s\ge 1}
                  e^{-\frac{\pi l}{\Delta}\,
                 { {s^2\over T}} } \cos (2\pi s y) \right)-1\right] .
\end{equation}
After expanding the result around the 'symmetric antipodal' point,
by using $x=1/2-\o x$, $y=1/2-\o y$, we find that the linear and
quadratic terms in the fluctuations $\o x$ and $\o y$ precisely
vanish, confirming the large distance result of
\cite{Burgess:2001fx}. On the other hand, we have found that out
minimum in $T$ is at distances of the order of the string scale,
$T=1$. If one now also takes higher order terms like $q^1$ into
account, one finds that although the linear terms in
 $\o x,\ \o y$ still vanish, the quadratic terms do not. This
destroys the slow-rolling property, because $\eta$ is not small
any longer. The same result has been found in the numerical
integration of the complete amplitude (\ref{eq:regularization}),
where also numerically the first and second field derivative have
been calculated. The potential is shown for a typical example in
figure \ref{fig:zweiplots_inflation}.

Concluding this section, it can be stated that no slow-rolling
properties could be found for the open string moduli $x$ and $y$
near the minimum of the dynamically stabilized K{\"a}hler modulus.
Consequently, inflation seems to be impossible using any of
these moduli fields.

\chapter{Conclusions and Outlook}\label{cha:Conclusions}
The main concern of this work has been to elaborate on possible
phenomenological model building approaches within orientifolded
type II string theory containing intersecting D6-branes. This has
been done with respect to both particle accelerator physics and
cosmology. A particularly important guiding principle has been the
issue of stability.

Chapter \ref{cha:IntersectingBraneWorlds} has provided a detailed
survey of the construction principles for these particular models.
The string theoretical conformal field theory construction has
been discussed in great detail, using the toroidal $\Omega R$
orientifold as an example. The R-R tadpole cancellation condition,
being the most important consistency requirement, has been derived
for this model. The scalar moduli potential for the toroidal model
has been obtained from the NS-NS tadpole for a first time, being
of most significance for stability. Indeed, it has been shown that
toroidal orientifold models generally suffer from complex
structure and dilaton instabilities already at one-loop level.

Furthermore, it has been discussed how to obtain the open and
closed string massless spectra, which are essential for low energy
model building. The absence of quantum anomalies in spacetime has
been addressed, mainly being a consequence from R-R tadpole
cancellation for the non-abelian anomalies, and involving a
generalized Green-Schwarz mechanism as a string theoretical tool
in the case of mixed anomalies. Finally, different gauge breaking
mechanisms have been discussed.

In chapter \ref{cha:Z3}, it has been searched for a stable
non-supersymmetric standard model. The $\mathbb{Z}_3$-orientifold
has emerged to be particularly useful in this respect as all
complex structures are fixed, therefore ensuring that the
background geometry is not driven to the degenerate limit of
collapsed 2-tori. Two particular examples containing 3 generations
of chiral matter are discussed in detail, a standard-like model
with gauge groups $SU(3)\times SU(2)_L\times U(1)_Y \times
U(1)_{B-L}$ and a flipped $SU(5)\times U(1)$ model. The
standard-like model shows some deviation from the usual standard
model: first, there are right handed neutrinos, this is definitely
a strong prediction. Secondly, there exists an additional
$B-L$-symmetry which survives as a global symmetry after a
discussed gauge breaking mechanism. The additional global symmetry
prohibits the standard model Yukawa couplings of the $(u,c,t)$
quarks with electroweak Higgs doublets. As a consequence, the
standard mass generation mechanism with fundamental Higgs scalars
does not work and a composite Higgs particle has to be
incorporated to cure this problem. On the other hand, this model
yields a massless hypercharge, being a non-trivial result.

This model is related to the second flipped $SU(5)$ model by an
adjoint breaking if the initial $U(3)$ and $U(2)$ stacks of branes
are parallel. The flipped $SU(5)$-model yields the GUT result for
Weinberg angle $\sin^2\theta_W=3/8$. Unfortunately, there is a
problem with proton decay and gauge coupling unification if the
string scale is of the same order as the GUT scale. Therefore, the
natural scale for the discussed model appears to be the GUT scale
instead of the TeV scale.

Chapter \ref{cha:Z4} approaches the hierarchy problem by directly
constructing a $\mathcal{N}$=1 supersymmetric model. The
$\mathbb{Z}_4$-orbifold has proved to be a well suited example to
do so, intrinsically containing also exceptional cycles from the
fixed points. Fractional D-branes, i.e. branes which also wrap
around these twisted 3-cycles, are being constructed for the first
time explicitly for this type of models. Finally, a three
generation Pati-Salam model with gauge groups $SU(4)\times
SU(2)\times SU(2)$ is being obtained, involving a brane
recombination mechanism. As a consequence, some of the branes are
non-flat and non-factorizable and one can only use a description
by homology and some tools from effective field theory like quiver
diagrams to discuss phenomenological aspects.

In Chapter \ref{cha:Inflation} the question has been discussed if
the unstable moduli, which are often unavoidable in
non-supersymmetric intersecting brane scenarios, might play the
role of the inflaton in cosmology. A candidate field would have to
fulfill in particular the slow-rolling condition. Two different
possibilities have been discussed, the unstable moduli coming from
either the open or closed string sector.

For the closed string moduli, the result was negative if there is
no assumed stabilization mechanism for some of the moduli. It has
to be distinguished between two different physical situations.
First of all, there has been a discussion for the introduced
'gauge coordinates' which yielded the result that the potential
fulfills slow-rolling conditions for both $s$- or $u^I$-Inflation
if the three remaining moduli are frozen. Then, the scenario of
the 'Planck coordinates' has been discussed in detail. Here,
slow-rolling was generally not possible.

For the open string leading order moduli scalar potential, it has
been found out that the K\"ahler moduli are being stabilized
dynamically in the discussed type of models, such that they cannot
lead to inflation. Also at the antipodal points of the open string
moduli, the slow-rolling properties are not fulfilled for small
values of the internal radii at the minimum of the potential.

In order to conclude, one can say that intersecting D-branes in
type IIA theory are very successful for phenomenological model
building and a good alternative to the heterotic string.
Nevertheless, some of the most important problems have not been
generally solved so far. There is still no string-theoretical
realization of the exact matter content of the MSSM. In order to
obtain such a model, it might be interesting to extend the ideas
of chapter \ref{cha:Z4} also to $\mathbb{Z}_6$ or $\mathbb{Z}_N
\times \mathbb{Z}_M$ orbifolds. This already has been done for the
case of $\mathbb{Z}_4\times \mathbb{Z}_2$ \cite{Honecker:2003vq}.
Furthermore, it would be very interesting to also address the
question of gauge coupling unification in these supersymmetric
models, a first attempt is given by \cite{Blumenhagen:2003jy}.

Considering more fundamental problems, one has to understand the
dynamical mechanism of supersymmetry breaking for these models.
Some progress already has been made in \cite{Cvetic:2003yd}.
Another great problem is the instability of the dilaton, where one
has to think about possible stabilization mechanisms. Tachyon
condensation also deserves a better understanding within
intersecting D-branes.

As one can see already from these short outlook, there remains a
lot to be done in future work.

\appendix
\chapter{Superstring Coordinates and Hamiltonians}
In this chapter, we will give definitions for the superstring
coordinates and Hamilton operators that are used in the main text.
These are given for light cone quantization, which is generally
used in the field of intersecting branes.

In the closed string, the bosonic coordinates in terms of
oscillators can be separated into a left and right moving part:
\begin{equation}\label{eq:boscoord}
    X^\mu(\tau, \sigma) =X_R^\mu(\tau-\sigma)+X_L^\mu(\tau+\sigma) \ ,
\end{equation}
which are given by:
\begin{align}\label{eq:boscoordleftright}
&X^\mu_{L}(\tau+\sigma) = \frac{x^\mu}{2} + \frac{p_{L}^\mu}{2}
(\tau + \sigma) + \frac{i}{\sqrt{2}} \sum_{n \not=0}{\frac{\alpha^\mu_n}{n}
  e^{-in(\tau +\sigma)} }\ , \\
&X^\mu_{R}(\tau-\sigma) = \frac{x^\mu}{2} + \frac{p_{R}^\mu}{2}
(\tau - \sigma) + \frac{i}{\sqrt{2}} \sum_{n
  \not=0}{\frac{\tilde{\alpha}^\mu_n}{n} e^{-in(\tau -\sigma)} }\ . \nonumber
\end{align}
In this equation, $x^\mu$ and $p^\mu=p_{L}^\mu+p_{R}^\mu$ are the
center of mass position and momentum, respectively. They are given in units of $\alpha'$. The same
division into left and right movers can be done for the fermionic
coordinates:
\begin{equation}\label{eq:boscoord}
    \Psi^\mu(\tau, \sigma) =\Psi_R^\mu(\tau-\sigma)+\Psi_L^\mu(\tau+\sigma) \ ,
\end{equation}
that are then given by:
\begin{align}\label{eq:fermcoordleftright}
&\Psi^\mu_{L}(\tau+\sigma) = \sum_{r\in\mathbb{Z}+\nu}{\psi_r^\mu e^{-ir(\tau + \sigma)}}\ , \\
&\Psi^\mu_{R}(\tau-\sigma) =
\sum_{r\in\mathbb{Z}+\tilde{\nu}}{\tilde{\psi}_{r}^\mu e^{-ir(\tau
- \sigma)}}\ , \nonumber
\end{align}
where the left and right moving $\nu$ and $\tilde\nu$ take the
value $0$ in the Ramond and $1/2$ in the Neveu-Schwarz sector.
This leads to commutation relations of the kind:
\begin{align}
&\left[ \alpha^\mu_n, \alpha^\nu_m \right]=\left[ \tilde\alpha^\mu_n, \tilde \alpha^\nu_m \right]=n \delta_{n+m,0}\,\delta^{\mu\nu}\ ,\nonumber\\
&\{ \psi_r^\mu, \psi_s^\nu \}=\{ \tilde \psi_r^\mu, \tilde \psi_s^\nu \}=\delta_{r+s,0}\,\delta^{\mu\nu} \ ,\\
&\left[ \alpha^\mu_n, \tilde \alpha^\nu_m \right]=\{\psi_r^\mu,\tilde \psi_s^\nu \}=0 \ . \nonumber
\end{align}
In the NS-sector, the ground state is unique, whereas in the
R-sector, the groundstate is degenerate and carries a
representation of the Clifford algebra.

The Hamilton operator of the closed string does not depend on the
specific placements of the D-branes, it is a sum of the left and
right moving parts and given by:
\begin{multline}\label{eq:Hamilton_closed}
    {\mathcal{H}}_{\text{closed}} = ({p^\mu})^2 +\sum_{\mu=1}^{10}{ \left( \sum_{n=1}^\infty{\left
      ( \alpha_{-n}^\mu \alpha_{n}^\mu + \tilde{\alpha}_{-n}^\mu
      \tilde{\alpha}_{n}^\mu \right)} \right. } \\
    \left. + \sum_{r\in\mathbb{Z}+\nu,\ r>0}{\left( r \psi^\mu_{-r}
        \psi^\mu_{r}\right)} +
  \sum_{s\in\mathbb{Z}+\tilde{\nu},\ s>0}{\left(s
  \tilde{\psi}^\mu_{-s} \tilde{\psi}^\mu_{s} \right) } \right)  +
  E_0^{L}+E_0^{R}\ .
\end{multline}
By way of contrast, the Hamilton operator of the open string does
depend on the placement of the D-branes. At this point, it is useful
to define complex oscillators on the three compact 2-tori by
\begin{equation}
    \alpha^I=\frac{1}{\sqrt{2}}\left(\alpha^{X_I}+i\alpha^{Y_I}\right)\ ,\qquad\qquad
    \alpha^{\bar I}=\frac{1}{\sqrt{2}}\left(\alpha^{X_I}-i\alpha^{Y_I}\right)\ .
\end{equation}
As an example the Hamiltonian is given for strings that stretch between
two certain D6-branes that intersect at an angle of $\kappa={(\varphi_2-\varphi_1)}/{\pi}$:
\begin{multline}\label{eq:Hamilton_open}
    {\mathcal{H}}_{\text{open}}=\frac{({p^\mu})^2}{2} +\sum_{\mu=1}^{4}{ \left( \sum_{n=1}^\infty{\left
      ( \alpha_{-n}^\mu \alpha_{n}^\mu\right)}+ \sum_{r\in\mathbb{Z}+\nu,\ r>0}{\left( r \psi^\mu_{-r}
        \psi^\mu_{r}\right)}\right)}\\
        +\sum_{I=1}^{3}{ \left[ \sum_{m\in\mathbb{Z}+\kappa, m>0}{\left(\alpha_{-m}^I\alpha_{m}^{\bar{I}}
        +\alpha_{-m}^{\bar{I}}\alpha_{m}^I\right)}+
        \sum_{s\in\mathbb{Z}+\nu+\kappa,\ s>0}{s\left(\psi^I_{-s}\psi^{\bar{I}}_{s}+ \psi^{\bar{I}}_{-s}\psi^{I}_{s}\right)}\right]}+ E_0\ .
\end{multline}
The zero point energies in light
cone gauge can be determined by the following general formulae for
in each case one complex particle:
\begin{align}\label{eq:zeropoint_energies}
&E_0^{\text{boson}}=-\frac{1}{12}+\frac{1}{2}\kappa\left(1-\kappa\right)&
&\text{boson with moding }\mathbb{Z}+\kappa\ ,\\
&E_0^{\text{fermion, R}}=-\frac{1}{24}+\frac{1}{2}\left(\frac{1}{2}-\kappa\right)^2&
&\text{R-sector fermion with moding }\mathbb{Z}+\kappa\ ,\nonumber\\
&E_0^{\text{fermion, NS}}=-\frac{1}{24}+\frac{1}{2}\kappa^2&
&\text{NS-sector fermion with moding }\mathbb{Z}+\kappa\ .\nonumber
\end{align}

\chapter{Modular functions}
In order to apply the modular transformations, the following
equations are important, they are valid for an argument $q=\exp(-2
\pi t)$. Many more useful formulas can be found in
\cite{MUbook:1982}.
\begin{equation}\label{eq:modulartrans_theta}
\tfkt{a}{b}{t}=e^{2 \pi i a b}\ t^{-1/2}\ \tfkt{b}{-a}{1/t}\ ,
\end{equation}
\begin{equation}\label{eq:modulartrans_eta}
\eta(t)=t^{-1/2}\ \eta(1/t)\ .
\end{equation}
Then there is the Poisson resummation formula
\begin{equation}
\sum\limits_{n\in\mathbb{Z}}e^{-\frac{\pi(n-c)^2}{t}}=\sqrt{t}\sum\limits_{m\in\mathbb{Z}}e^{2\pi i c m}\ e^{-\pi m^2 t}\ ,
\end{equation}
from which directly follows:
\begin{equation}\label{eq:simple_poissonresummation}
\sum\limits_{m\in\mathbb{Z}}e^{-\pi m^2 t}=t^{-1/2}\sum\limits_{n\in\mathbb{Z}}e^{-{\pi n^2}/{t}}\ .
\end{equation}
For a general argument $q$, the theta functions with
characteristics are defined as
\begin{equation}\label{eq:defthetafktsum}
\tfkt{a}{b}{q}=\sum\limits_{n\in\mathbb{Z}}q^{(n+a)^2/2}e^{2\pi i (n+a)b}\ ,
\end{equation}
or in product form:
\begin{equation}\label{eq:defthetafktprod}
\tfkt{a}{b}{q}=e^{2\pi i a
b}q^{a^2/2}\prod\limits_{m=1}^{\infty}\left(1-q^m\right)\left(1+e^{2\pi
i b}q^{m-1/2+a}\right)\left(1+e^{-2\pi i b}q^{m-1/2-a}\right)\ ,
\end{equation}
where $a$ has to be chosen within the range $-1/2<a\leq1/2$.
The theta functions have the following important
identities:
\begin{align}
&\tfkt{a\pm1}{b}{q}=\tfkt{a}{b}{q}\ ,\\
&\tfkt{a}{b\pm1}{q}=e^{\pm2 \pi i a}\ \tfkt{a}{b}{q}\ .\nonumber
\end{align}
Furthermore, the following product representation is useful:
\begin{equation}\label{eq:theta_over_eta}
\frac{\tfkto{a}{b}}{\eta}=e^{2\pi i a
b}q^{\frac{a^2}{2}-\frac{1}{24}}\prod\limits_{m=1}^{\infty}\left(1+e^{2\pi
i b}q^{m-1/2+a}\right)\left(1+e^{-2\pi i b}q^{m-1/2-a}\right)\ .
\end{equation}
Also very important is Jacobi's abstruse identity:
\begin{equation}\label{eq:abstruse_identity}
\tfktvier{0}{0}{q}-\tfktvier{0}{1/2}{q}-\tfktvier{1/2}{0}{q}=0\ .
\end{equation}
The Dedekind $\eta$ function in product form is defined as
\begin{equation}\label{eq:defeta}
\eta(q)=q^{1/24}\prod\limits_{m=1}^{\infty}(1-q^m) \ .
\end{equation}
$\eta$ also can be written as a sum:
\begin{equation}\label{eq:eta_def_Sum}
\eta(q)=q^{1/24}\left(1+\sum_{n=1}^{\infty}(-1)^n\left[q^{n(3n-1)/2}+q^{n(3n+1)/2}\right]\right)\ .
\end{equation}
Then, there is the useful relation:
\begin{equation}\label{eq:limit_theta}
    \lim_{\phi\rightarrow 0}\frac{2 \sin(\pi
\phi)}{\tfkt{1/2}{1/2+\phi}{q}}=-\frac{1}{\eta^3(q)}\ .
\end{equation}
\chapter{The cylinder amplitude for $\mathbb{Z}+\kappa$ moding}
In this section, a simple but general prescription is given, of
how to find the correct cylinder amplitude for any sector with a
non-trivial moding, without having to quantize the string again
from first principles. This can be applied to both branes at
non-vanishing angles as to twisted sectors in orbifold theories.
\section{One complex boson}
For $\mathbb{Z}$-moding, the contribution of one complex boson to
the trace of the loop channel one loop amplitude is simply given by
\begin{equation}\label{eq:one_complex_boson_trivmoding}
    f_\text{boson, loop}^\text{$\mathbb{Z}$-moding}=\frac{1}{{\eta(q)}^2}\ .
\end{equation}
In a sector with $\mathbb{Z}+\kappa$ moding, two things have to be
altered, the first one being the moding within the product
representation of the $\eta$-function (\ref{eq:defeta})and the
second being the zero point energy.

The correct change of the moding within the $\eta$-function is given by
\begin{equation}\label{eq:change_eta1}
    \eta^2\rightarrow
q^\frac{1}{12}\prod_{n=0}^{\infty}\left(1-q^{n+\kappa}\right)
\prod_{m=1}^{\infty}\left(1-q^{m-\kappa}\right)\ .
\end{equation}
This can be rewritten as
\begin{align}\label{eq:change_eta2}
&q^\frac{1}{12}\prod_{n=0}^{\infty}\left(1-q^{n+\kappa}\right)
\prod_{m=1}^{\infty}\left(1-q^{m-\kappa}\right)& \\
&=q^\frac{1}{12}\left(1-q^\kappa\right)\prod_{n=1}^{\infty}\left(1+e^{2\pi i\frac{1}{2}}
q^{n+\left(\kappa-\frac{1}{2}\right)+\frac{1}{2}}\right)
\prod_{m=1}^{\infty}\left(1+e^{-2\pi i\frac{1}{2}}
q^{m-\left(\kappa-\frac{1}{2}\right)-\frac{1}{2}}\right)&\nonumber\\
&=q^\frac{1}{12}e^{-\left(\kappa-\frac{1}{2}\right)\pi i}
q^{\frac{1}{24}-\frac{\left(\kappa-\frac{1}{2}\right)^2}{2}}\frac{\tfkt{\kappa-1/2}{1/2}{q}}{\eta(q)}\nonumber\\
&=e^{-\left(\kappa-\frac{1}{2}\right)\pi i}
q^{\frac{\kappa-\kappa^2}{2}}\frac{\tfkt{\kappa-1/2}{1/2}{q}}{\eta(q)}\ .\nonumber
\end{align}
As the next step, the zero point energy has to be corrected
according to equation (\ref{eq:zeropoint_energies}). A check that
such a procedure is justified is given by the series expansion of
(\ref{eq:change_eta2}) that gives a zero point energy $E_0=1/12$
which is just the correct one in the case of $\mathbb{Z}$-moding.
One has to include by hand the shift
\begin{equation}\label{eq:shift_bosonic_zero_point_energy}
    \Delta E_0=\frac{1}{2}\kappa\left(1-\kappa\right)\ ,
\end{equation}
which finally leads to the following loop channel trace
contribution:
\begin{equation}\label{eq:one_complex_boson_nontrivmoding}
    f_\text{boson, loop}^\text{($\mathbb{Z}+\kappa$)-moding}=
    e^{\left(\kappa-\frac{1}{2}\right)\pi i}
    \frac{\eta(q)}{\tfkt{\kappa-1/2}{1/2}{q}}\ .
\end{equation}
\section{One complex fermion}
For $\mathbb{Z}$-moding, the contribution of one complex fermion to
the trace of the loop channel amplitude is simply given by
\begin{equation}\label{eq:one_complex_fermion_trivmoding}
    f_\text{fermion, loop}^\text{$\mathbb{Z}$-moding}=\frac{\tfkt{a}{b}{q}}{\eta(q)},
\end{equation}
where $a$ and $b$ take the usual values of the type I theory in the R or NS sectors.
Changing the moding to $\mathbb{Z}+\kappa$ in the product
representation (\ref{eq:theta_over_eta}) simply means
\begin{equation}\label{eq:ansatz_ferm_nontrivmod}
   \frac{\tfkt{a}{b}{q}}{\eta(q)}\rightarrow e^{2\pi i a
b}q^{\frac{a^2}{2}-\frac{1}{24}}\prod\limits_{m=1}^{\infty}\left(1+e^{2\pi
i b}q^{m+\kappa-1/2+a}\right)\left(1+e^{-2\pi i b}q^{m-\kappa-1/2-a}\right)\ .
\end{equation}
In order to rewrite this equation correctly, one has to
distinguish between R- and NS-loop channels, which correspond to
$a=1/2$ and $a=0$ respectively, the reason being that the product
representation is just defined correctly for a first
characteristic of the theta function within the range from $-1/2$
to $1/2$.

Starting with the NS-sector, one rewrites
(\ref{eq:ansatz_ferm_nontrivmod}) in the following way:
\begin{align}\label{eq:fermionpartann}
    &e^{2\pi i a b}q^{\frac{a^2}{2}-\frac{1}{24}}\prod\limits_{m=1}^{\infty}\left(1+e^{2\pi
i b}q^{m+(a+\kappa)-1/2}\right)\left(1+e^{-2\pi i b}q^{m-(a+\kappa)-1/2}\right)&\\
    &=e^{-2\pi i b \kappa}q^{-a'\kappa+\frac{\kappa^2}{2}}\frac{\tfkt{a'}{b}{q}}
    {\eta(q)}=e^{-2\pi i b \kappa}q^{-\left(a\kappa+\frac{\kappa^2}{2}\right)}
    \frac{\tfkt{a+\kappa}{b}{q}}{\eta(q)}\ ,\nonumber
\end{align}
where $a'=a+\kappa$. Next, the zero point energy has to be
altered according to (\ref{eq:zeropoint_energies}), this induces a
shift
\begin{equation}\label{eq:shift_bosonic_zero_point_energy}
    \Delta E_0=\frac{1}{2}\kappa^2\ .
\end{equation}
Finally, setting $a=0$ leads to the result
\begin{equation}\label{eq:one_complex_fermion_NS}
    f_\text{fermion, NS-loop}^\text{($\mathbb{Z}+\kappa$)-moding}=
    e^{-2\pi i b \kappa}\frac{\tfkt{\kappa}{b}{q}}{\eta(q)}\ .
\end{equation}

Now switching to the R-sector, (\ref{eq:ansatz_ferm_nontrivmod})
one formally can use (\ref{eq:fermionpartann}), but has to act
with a transformation of the type $a\rightarrow a-1$ on the
equation, which is equivalent to setting $a \rightarrow -a$ in this sector, this then leads to:
\begin{equation}
\frac{\tfkt{a}{b}{q}}{\eta(q)}\rightarrow e^{-2\pi i b \kappa}q^{a\kappa-\frac{\kappa^2}{2}}
    \frac{\tfkt{-a+\kappa}{b}{q}}{\eta(q)}\ ,
\end{equation}
Changing the zero point energy induces a shift
\begin{equation}\label{eq:shift_bosonic_zero_point_energy}
    \Delta E_0=\frac{1}{2}\kappa\left(\kappa-1\right)\ ,
\end{equation}
and the final result therefore is given by:
\begin{equation}\label{eq:one_complex_fermion_R}
    f_\text{fermion, R-loop}^\text{($\mathbb{Z}+\kappa$)-moding}=
    e^{-2\pi i b \kappa}\frac{\tfkt{-1/2+\kappa}{b}{q}}{\eta(q)}\ .
\end{equation}
\section{Application to the $\Omega R$-orientifold}\label{cha:appendix_OR_annij}
In this section, we follow the notation of chapter
\ref{cha:cylinder_OR} and allow for a general angle $\varphi_{ab}$ in
between the two stacks of branes $a$ and $b$ with $N_a$ and $N_b$
parallel branes, within the $\Omega R$-orientifold model.
\subsection{Tree channel R-sector}
Starting with the tree channel R-sector, which corresponds to the
loop channel sector (NS,$-$), we have to be reminded that there
are 8 scalar bosons, or equivalently 4 complex bosons, of which 3
are coming from the 6-torus, or more exactly, one from each
2-torus. So one simply has to substitute the corresponding modular
functions (\ref{eq:one_complex_boson_trivmoding}) for each of the
3 bosons from the 2-tori by the expressions
(\ref{eq:one_complex_boson_nontrivmoding}), where the
corresponding intersection angle that the two branes have on the
specific torus has to be inserted, leading to a moding
$\kappa_I=\varphi_{ab}^I/\pi$. The same procedure has to be applied to
the 4 complex fermions, where
(\ref{eq:one_complex_fermion_trivmoding}) on the 2-tori has to be
substituted by (\ref{eq:one_complex_fermion_R}) and the resulting
amplitude is given by:
\begin{equation}\label{eq:Annulus_amplitude_OR_loopij}
\mathcal{A}_{ab}^\text{(NS,$-$)}=-\frac{c}{4} N_a N_b I_{ab}
\int\limits_{0}^{\infty}\frac{dt}{t^3}e^{-\frac{3}{2} \pi
i}\frac{\tfkto{0}{1/2}\tfkto{\kappa_1}{1/2}\tfkto{\kappa_2}{1/2}\tfkto{\kappa_3}{1/2}}
{\tfkto{\kappa_1-1/2}{1/2}\tfkto{\kappa_2-1/2}{1/2}\tfkto{\kappa_3-1/2}{1/2} \eta^{3}}
\ ,
\end{equation}
where the argument of the $\vartheta$- and $\eta$-functions is
given by $q=e^{-2\pi t}$. The additional factor $I_{ab}$ that
corresponds to the intersection number on the torus will be
derived in the tree channel by a comparison with the boundary
states. In this equation, it has been assumed that the angle in
between the two stacks of branes is non-vanishing \footnote{Later,
we will see that this assumption can be dropped.}. The modular transformation to the tree channel
by using $t=1/(2l)$ directly gives the amplitude
\begin{equation}\label{eq:Annulus_amplitude_ij_OR_tree}
\widetilde{\mathcal{A}}_{ab}^\text{(R,$+$)}=-\frac{c}{2} N_a N_b I_{ab}
\int\limits_{0}^{\infty}{dl}\frac{\tfkto{1/2}{0}\tfkto{1/2}
{-\kappa_1}\tfkto{1/2}{-\kappa_2}\tfkto{1/2}{-\kappa_3}}{\tfkto{1/2}{1/2-\kappa_1}
\tfkto{1/2}{1/2-\kappa_2}\tfkto{1/2}{1/2-\kappa_3}\eta^{3}}
\end{equation}
with an argument $q=e^{-4\pi l}$ of the $\vartheta$ and
$\eta$-functions. Of course, if we now want to derive the
expansion of the equation in $q$, the angles will already appear
in the zero-order term which leads to the tadpole. It is
reasonable to apply the following two simplifications:
\begin{align}\label{eq:expansion_theta_mitwinkel}
&\tfkto{1/2}{1/2-\kappa_I}=i\left(e^{-\pi i\kappa_I }-e^{\pi i\kappa_I }\right){q}^{1/8}+O \left(
{q}^{{9/8}} \right)=2 \sin \left( \pi {\kappa_I} \right) {q}^{1/8}+O \left(
{q}^{9/8} \right)\ ,&\\
&\tfkto{1/2}{-\kappa_I}=\left(e^{- \pi i\kappa_I}+e^{\pi i\kappa_I }\right){q}^{1/8}+O \left(
{q}^{{9/8}} \right)=2 \cos \left( \pi {\kappa_I} \right) {q}^{1/8}+O \left(
{q}^{9/8} \right)\ .&\nonumber
\end{align}
Furthermore, the angles can be expressed by the wrapping numbers
using equation \ref{eq:intangle in nm_general} and finally, the tadpole
is given by
\begin{equation}\label{eq:tadpole_OR_Aij}
\text{tp}^{\text{R}}_{\tilde{\mathcal{A}}_{ab}}=8 c N_a N_b{ I_{ab}}
\prod_{I=1}^{3}\left[{\frac {c{N}^{2} \left( { n_I^a}{{ R_x^{(I)}}}^{2}
{n_I^b}+{ m_I^a}{{ R_y^{(I)}}}^{2}{ m_I^b}+{ b_I}{{ R_x^{(I)}}}^{2}
\left({ m_I^b}{ n_I^a}+{ m_I^a}{ n_I^b}\right)
 \right) }{{ R_x^{(I)}}\sqrt{4{{ R_y^{(I)}}}^{2}-2{ b_I}{{ R_x^{(I)}
}}^{2}} \left( { m_I^a}{ n_I^b}-{ m_I^b}{ n_I^a} \right) }}\right]\ .
\end{equation}
The only possibility to reproduce the correct normalization of the
boundary states (\ref{eq:Norm_D6_OR}) is by assuming that
\begin{equation}\label{eq:Intersection_number_2tesmal}
    I_{a b}=\prod_{I=1}^{3}I_{a b}^{(I)}=\prod_{I=1}^{3}
    \left(n_I^a m^b_I-m^a_I n^b_I\right) \ .
\end{equation}
Interestingly, the case where $\kappa=0$ on a certain torus then
formally can just be obtained by simply setting $n_I^a=n_I^b$ and
$m_I^a=m_I^b$, although in this case there is a Kaluza-Klein and
winding contribution which contributes to the tadpoles. Therefore,
this is already encoded in the normalization of a single D-brane
boundary state. It also should be added that in order to proof this
equivalence, one has to make the substitution
\begin{equation}\label{eq:substitution_OR_ann}
    \frac{1}{\sqrt{1-b_I}\sqrt{\left(1+2b_I\right){R_y^{(I)}}^2-b_I{R_x^{(I)}}^2}}\rightarrow
    \frac{2}{\sqrt{4{{ R_y^{(I)}}}^{2}-2{ b_I}{{ R_x^{(I)}}}^{2}}} \ ,
\end{equation}
which is legitimate for the two only cases $b_I=0$ or $b_I=1/2$.
\subsection{Tree channel NS-sector}\label{cha:appendix_OR_annij_NS}
The tree channel NS-sector correspond to a combination of the
(NS,$+$) and (R,$+$) loop channels. In order to write down the
correct loop channel ansatz, the (NS,$+$) fermionic contribution
is given by (\ref{eq:one_complex_fermion_NS})while the other is
given by (\ref{eq:one_complex_fermion_R}), so altogether
\begin{multline}\label{eq:Annulus_amplitude_OR_loopNS_ij}
\mathcal{A}_{ab}^\text{(NS,$+$)}+\mathcal{A}_{ab}^\text{(R,$+$)}=\frac{c}{4} N_a N_b I_{ab}
\int\limits_{0}^{\infty}\frac{dt}{t^3}{e^{\pi i \left(\kappa_1+\kappa_2+\kappa_3-\frac{3}{2}\right)}}\\
\shoveleft{\cdot\left(\frac{\tfkto{0}{0}\tfkto{\kappa_1}{0}\tfkto{\kappa_2}{0}\tfkto{\kappa_3}{0}}
{\tfkto{\kappa_1-1/2}{1/2}\tfkto{\kappa_2-1/2}{1/2}\tfkto{\kappa_3-1/2}{1/2} \eta^{3}}\right.}\\
\left.-\frac{\tfkto{1/2}{0}\tfkto{\kappa_1-1/2}{0}\tfkto{\kappa_2-1/2}{0}\tfkto{\kappa_3-1/2}{0}}
{\tfkto{\kappa_1-1/2}{1/2}\tfkto{\kappa_2-1/2}{1/2}\tfkto{\kappa_3-1/2}{1/2} \eta^{3}}\right)\ .
\end{multline}
In the tree channel, this leads to
\begin{multline}\label{eq:Annulus_amplitude_ij_OR_treeNS}
\widetilde{\mathcal{A}}_{ab}^\text{NS}=\frac{c}{2} N_a N_b I_{ab}
\int\limits_{0}^{\infty}{dl}
\left(\frac{\tfkto{0}{0}\tfkto{0}{-\kappa_1}\tfkto{0}{-\kappa_2}\tfkto{0}{-\kappa_3}}
{\tfkto{1/2}{1/2-\kappa_1}\tfkto{1/2}{1/2-\kappa_2}\tfkto{1/2}{1/2-\kappa_3}\eta^{3}}\right.\\
\left.-\frac{\tfkto{0}{-1/2}\tfkto{0}{1/2-\kappa_1}\tfkto{0}{1/2-\kappa_2}\tfkto{0}{1/2-\kappa_3}}
{\tfkto{1/2}{1/2-\kappa_1}\tfkto{1/2}{1/2-\kappa_2}\tfkto{1/2}{1/2-\kappa_3}\eta^{3}}\right) .
\end{multline}
In addition to (\ref{eq:expansion_theta_mitwinkel}), one needs the following two simplifications
\begin{align}\label{eq:expansion_theta_mitwinkel2}
&\tfkto{0}{-\kappa_I}=1+2 \cos \left(2 \pi {\kappa_I} \right) {q}^{1/2}+O \left(
{q}^{3/2} \right)\ ,&\\
&\tfkto{0}{1/2-\kappa_I}=1-2\cos\left(2 \pi {\kappa_I} \right) {q}^{1/2}+O \left(
{q}^{3/2} \right)\ .& \nonumber
\end{align}
The tadpole coming from (\ref{eq:Annulus_amplitude_OR_loopNS_ij}) then can be written as
\begin{equation}\label{eq:tadpole_OR_AijNS}
\text{tp}^{\text{NS}}_{\tilde{\mathcal{A}}_{ab}}=-\frac{c}{2} N_a N_b{ I_{ab}}
\frac{\big( \cos^2 \left( \pi {\kappa_1}
 \right)+\cos^2 \left( \pi {\kappa_2}
 \right)+\cos^2 \left( \pi {\kappa_3} \right)-1\big) }{\sin \left( \pi {\kappa_1} \right) \sin
 \left( \pi { \kappa_2} \right) \sin \left( \pi {\kappa_3}
 \right) }\ .
\end{equation}
We omit to write down the explicit form in terms of the winding
numbers for this expression, but mention the two necessary relations for obtaining it:
\begin{multline}\label{eq:cos_phi2_Phi1}
\cos(\kappa_I \pi)=\cos(\varphi_I^b-\varphi_I^a)=\\ \frac {\left({b_I}^2 m_I^b m_I^a+\left(n_I^b
m_I^a-\frac{1}{2} m_I^b m_I^a + m_I^b n_I^a\right)b_I+n_I^b n_I^a\right){ R_x^{(I)}}^{2}
+{ m_I^a}{ m_I^b}{{ R_y^{(I)}}}^{2}}
{\sqrt {{{
n_I^b}}^{2}{{ R_x^{(I)}}}^{2}+{{
 m_I^b}}^{2}{{ R_y^{(I)}}}^{2}+2b{ n_I^b}{ m_I^b}{{
R_x^{(I)}}}^{2}} \sqrt {{{ n_I^a}}^{2}{{ R_x^{(I)}}}^{2}+{{
m_I^a}}^{2}{{ R_y^{(I)}}}^{2}+2b{  n_I^a}{ m_I^a}{{
R_x^{(I)}}}^{2}}}
\end{multline}
and
\begin{multline}\label{sin_phi2_Phi1}
\sin(\kappa_I \pi)=\sin(\varphi_I^b-\varphi_I^a)=\sqrt {\frac{1}{2}{{ R_y^{(I)}}}^{2}-\frac{b_I}{4}{{
R_x^{(I)}}}^{2}}{  R_x^{(I)}} \\
\cdot{\frac {{ m_I^b}\sqrt {2{{ n_I^a}}^{2}+4{
m_I^a}{ n_I^a}b_I+{{  m_I^a}}^{2}b_I}-{ m_I^a}\sqrt {2{{
n_I^b}}^{2}+4b_I{ n_I^b} { m_I^b}+{{ m_I^b}}^{2}b_I}
}{\sqrt {{{ n_I^b}}^{2}{{ R_x^{(I)}}}^ {2}+{{ m_I^b}}^{2}{{
R_y^{(I)}}}^{2}+2b_I{ n_I^b}{ m_I^b}{{ R_x^{(I)}}}^{2 }}\sqrt {{{
n_I^a}}^{2}{{ R_x^{(I)}}}^{2}+{{ m_I^a}}^{2}{{ R_y^{(I)}}}^{2}+2
b_I{ n_I^a}{ m_I^a}{{ R_x^{(I)}}}^{2}}}}\ ,
\end{multline}
where $b_I$ again can take on the discrete values $b_I=0$ or $b_I=1/2$ for the A- and B-torus.
\chapter{Lattice contributions}
\section{Klein bottle}\label{cha:appendix_latcontrib_KB}
In general, it is possible to obtain the
Kaluza-Klein and winding mode contributions for one certain torus
to the loop channel in the following way: Firstly, the lattice
vectors have to be specified. Usually, one takes two linearly
independent vectors $\mathbf{e}_1$ and $\mathbf{e}_2$ that are
normalized to $(\mathbf{e}_i)^2=2$. The basis of the lattice then
is given by the two vectors $\sqrt{1/2} R_x \mathbf{e}_1$ and $\sqrt{1/2} R_y \mathbf{e}_2$.
Furthermore, it is necessary to define the dual torus by the two
vectors $\mathbf{e}^\ast_1$ and $\mathbf{e}^\ast_2$ that can be obtained from
\begin{equation}\label{eq:dualtorus}
\mathbf{e}_i\ \mathbf{e}^\ast_j=\delta_{ij} \ ,
\end{equation}
Then the dual lattice is spanned by the vectors
$\sqrt{2}/R_x\mathbf{e}^\ast_1$ and $\sqrt{2}/R_y\mathbf{e}^\ast_2$.
The Kaluza-Klein momenta and winding modes take the following
form:
\begin{align}\label{eq:kkwindingstates}
    &\mathbf{P}=\sqrt{\alpha'}\left(\frac{s_1}{R_x} \mathbf{e}^\ast_1 +\frac{s_2}{R_y} \mathbf{e}^\ast_2\right) \ ,\\
    &\mathbf{L}=\frac{1}{\sqrt{\alpha'}}\left(R_x r_1 \mathbf{e}_1
+R_y r_2 \mathbf{e}_2\right) \ .\nonumber
\end{align}
The left and right moving momenta of the torus appearing in the Hamiltonian then take the form
\begin{equation}\label{eq:leftrightmoving_momenta}
    \mathbf{p}_{L,R}=\mathbf{P}\pm\frac{1}{2}\mathbf{L}\ .
\end{equation}
In a specific model, not all momenta and winding modes
(\ref{eq:kkwindingstates}) are allowed by the symmetries of the
model, so one must pick out just the invariant ones.
In the following, some specific models of the main text are treated.
\subsection{The toroidal $\Omega R$-orientifold}
For the $A$-torus, the lattice vectors and dual lattice vectors are given by:
\begin{align}
    &\mathbf{e}_1^\text{A}=\binom{\sqrt{2}}{0}\ ,& &\mathbf{e}_2^\text{A}=\binom{0}{\sqrt{2}}\ ,\\
    &\mathbf{e}_1^{\ast\text{A}}=\binom{1/\sqrt{2}}{0}\ ,& &\mathbf{e}_2^{\ast\text{B}}=\binom{0}{1/\sqrt{2}}\ .\nonumber
\end{align}
For the $B$-torus, they are given by:
\begin{align}
    &\mathbf{e}_1^\text{B}=\binom{\sqrt{2}}{0}\ ,&
    & \mathbf{e}_2^\text{B}=\frac{1}{\sqrt{2}R_y}\binom{{R_x}}{\sqrt{4{{R_y}^2}-{{R_x}^2}}}\ ,\\
    &\mathbf{e}_1^{\ast\text{B}}=\frac{1}{\sqrt{2}}\binom{1}{-R_x/\sqrt{4{{R_y}^2}-{{R_x}^2}}}\ ,&
    & \mathbf{e}_2^{\ast\text{B}}=\sqrt{2}\binom{0}{R_y/\sqrt{4{{R_y}^2}-{{R_x}^2}}}\ .\nonumber
\end{align}
The worldsheet parity transformation $\Omega$ acts as
\begin{equation}
\Omega : \qquad \mathbf{P} \xrightarrow{\Omega } \mathbf{P}, \qquad
\mathbf{L} \xrightarrow{\Omega} -\mathbf{L} \ ,
\end{equation}
whereas the reflection $R$ acts as:
\begin{equation}
R : \qquad P^1 \xrightarrow{R } P^1,\qquad P^2 \xrightarrow{R } -P^2,\qquad
L^1 \xrightarrow{R} L^1,\qquad L^2 \xrightarrow{R } -L^2\ .
\end{equation}
Therefore, the combined action is given by:
\begin{equation}\label{eq:OmegaR_action}
\Omega R : \qquad P^1 \xrightarrow{\Omega R} P^1,\qquad P^2 \xrightarrow{\Omega R} -P^2,\qquad
L^1 \xrightarrow{\Omega R} -L^1,\qquad L^2 \xrightarrow{\Omega R} L^2\ .
\end{equation}
Keeping just the invariant terms under (\ref{eq:OmegaR_action})
together with (\ref{eq:leftrightmoving_momenta}), leads to the
lattice contributions of the Hamiltonian for the $A$-torus:
\begin{equation}\label{eq:AHamiltonian_lat}
\mathcal{H}^\text{A}_{\text{lattice, cl.}}=\frac{(\mathbf{p}^\text{A}_{L})^2+(\mathbf{p}^\text{A}_{R})^2}{2}=\sum_{r_2,
s_1}\frac{1}{2}\left(\frac{\alpha'{s_1}^2}{{R_x}^2}+\frac{{R_y}^2{r_2}^2}{\alpha'}\right) \ .
\end{equation}
For the $B$-torus, the procedure is slightly more involved: the
$\Omega R$-symmetry imposes linear relations between $s_1$ and
$s_2$ and between $r_1$ and $r_2$ that can be solved easily. The
result is the invariant Hamiltonian:
\begin{equation}\label{eq:BHamiltonian_lat}
\mathcal{H}^\text{B}_{\text{lattice, cl.}}=\frac{(\mathbf{p}^\text{B}_{L})^2+(\mathbf{p}^\text{B}_{R})^2}{2}=\sum_{r_1,
s_2}\left(2\alpha'\frac{{s_2}^2}{{R_x}^2}+\frac{1}{2\alpha'}\left({4{{R_y}^2}-{{R_x}^2}}\right){r_1}^2\right) \ .
\end{equation}
Finally, the indices of $r$ and $s$ can be skipped. It is possible
to parameterize both possibilities (\ref{eq:AHamiltonian_lat}) and
(\ref{eq:BHamiltonian_lat}) in one equation:
\begin{equation}\label{eq:ABHamiltonian_lat}
\mathcal{H}_{\text{lattice, cl.}}=\sum_{r,
s}\left(\frac{\alpha'{s}^2}{2{R_x}^2}\left(1+6b\right)+\frac{r^2}{2\alpha'}\left(\left(1+6b\right){R_y}^2-2b{R_x}^2\right)\right)
\ ,
\end{equation}
where $b=0$ for the $A$-torus and $b=1/2$ for the $B$-torus.
\subsection{The $\mathbb{Z}_3$-orientifold}

For the $A$-torus, the lattice vectors and dual lattice vectors are given by:
\begin{align}
    &\mathbf{e}_1^\text{A}=\binom{\sqrt{2}}{0}\ ,& &\mathbf{e}_2^\text{A}=\binom{1/\sqrt{2}}{\sqrt{3/2}}\ ,\\
    &\mathbf{e}_1^{\ast\text{A}}=\binom{1/\sqrt{2}}{-1/\sqrt{6}}\ ,& &\mathbf{e}_2^{\ast\text{B}}=\binom{0}{\sqrt{2/3}}\ .\nonumber
\end{align}
For the $B$-torus, they are given by:
\begin{align}
    &\mathbf{e}_1^\text{B}=\binom{\sqrt{2}}{0}\ ,&
    & \mathbf{e}_2^\text{B}=\binom{1/\sqrt{2}}{1/\sqrt{6}}\ ,\\
    &\mathbf{e}_1^{\ast\text{B}}=\binom{1/\sqrt{2}}{-\sqrt{3/2}}\ ,&
    & \mathbf{e}_2^{\ast\text{B}}=\binom{0}{\sqrt{6}}\ .\nonumber
\end{align}
Both $\Omega$ and $R$ act in the same way as for the toroidal
$\Omega R$-orientifold, so we can use equation
(\ref{eq:OmegaR_action}) in order to determine the lattice
Hamiltonian. Due to its $\mathbb{Z}_3$-symmetry, this is not just
valid for the closed string trace insertion $1$, but for $\Theta$
and $\Theta^2$ as well. The resulting Hamiltonian, again
parameterized for both tori, is given by:
\begin{equation}\label{eq:ABHamiltonianZ3_lat}
\mathcal{H}_{\text{lattice, cl.}}=\frac{(\mathbf{p}_{L})^2+(\mathbf{p}_{R})^2}{2}=\sum_{r,
s}\left(2\frac{\alpha'{s}^2}{{R}^2}+\left(\frac{3}{2}-\frac{8}{3}\,b\right)\frac{{R}^2{r}^2}{\alpha'}\right) \ .
\end{equation}

\section{Cylinder}
The general equation for the lattice contribution to the Cylinder
amplitude in the D6-branes at angles picture for one torus is
given by the equation \cite{Blumenhagen:2000fp}:
\begin{equation}\label{eq:Hamiltonian_lat_annulus}
    \mathcal{H}_{\text{lattice, op.}}=\sum_{r,s}\frac{|r+sT|^2}{T_2}\frac{U_2}{|n+mU|^2} \ ,
\end{equation}
where $n$ and $m$ mean the two wrapping numbers of the brane in
consideration on the torus and $U$ and $T$ the complex structure
and K{\"a}hler moduli.
\subsection{The toroidal $\Omega R$-orientifold}
For the $A$- and $B$-torus, after having inserted the complex
structure and K{\"a}hler moduli, this explicitly means
\begin{align}\label{eq:Hamiltonian_lat_annulus}
&\mathcal{H}^\text{A}_{\text{lattice, op.}}=\sum_{r,s}\frac{r^2+s^2{R_x}^2{R_y}^2}{n^2{R_x}^2+m^2{R_y}^2}\ ,& \\
&\mathcal{H}^\text{B}_{\text{lattice, op.}}=\sum_{r,s}\frac{r^2+s^2\left(4{R_x}^2{R_y}^2-{R_x}^4\right)}{\left(2n^2+2mn\right){R_x}^2+2m^2{R_y}^2}\ .&\nonumber
\end{align}
These two possibilities again can be parameterized in one equation:
\begin{equation}\label{eq:ABHamiltonian_lat_annulus}
    \mathcal{H}_{\text{lattice, op.}}=\sum_{r,s}\frac{r^2\left(1-b\right)+s^2\left[\left(1+2b\right){R_x}^2{R_y}^2-b{R_x}^4\right]}{\left(n^2+2bnm\right){R_x}^2+m^2{R_y}^2}\ .
\end{equation}

\subsection{The $\mathbb{Z}_3$-orientifold}

Inserting the complex structure (\ref{eq:c_Struct_z3}) and the K{\"a}hler moduli
(\ref{eq:Kaeheler_z3}) into equation (\ref{eq:Hamiltonian_lat_annulus}) directly leads to
\begin{align}\label{eq:ABHamiltonian_lat_annulus}
\mathcal{H}_{\text{lattice, op.}}=\sum_{r,s}
&\frac{\frac{r^2}{R^2}+\left(\frac{3}{4}-\frac{4}{3}\,b\right)s^2R^2}{L_a}\\
&\text{with}\ L_a=\sqrt{{n_a}^2+n_a\,m_a+\left(1-\frac{4}{3}b\right){m_a}^2}\ ,\nonumber
\end{align}
for a brane $a$ on either A- or B-torus, differing by $b=0$ or $b=1/2$.

\chapter{The $\mathbb{Z}_4$-orientifold}

\section{Orientifold planes}\label{cha:app_oplanes_z4}
The results for the O6-planes and the action of $\Omega R$ on the
homology lattice are listed in this appendix for the cases not
being treated within the main text. The O6-planes can be found in
table \ref{tab:O6planes_z4}.
\begin{table}[h!]
\centering
\sloppy
\renewcommand{\arraystretch}{1.5}
\begin{tabular}{|c||p{120pt}|}
  \hline
  Model & O6-plane\\
  \hline
  \hline
  ${\bf AAA}$ & $4\, \rho_1 -2\, \bar\rho_2 $ \\
  ${\bf AAB}$ & $2\, \rho_1 + \rho_2 -2\, \bar\rho_2 $ \\
  ${\bf ABA}$ & $2\, \rho_1 + 2\, \rho_2+  2\, \bar\rho_1  -2\,\bar\rho_2 $ \\
  ${\bf ABB}$ & $2\, \rho_2+2\,\bar\rho_1 -2\, \bar\rho_2 $ \\
  \hline
\end{tabular}
\caption{The O6-planes of the four distinct $\mathbb{Z}_4$-orientifold models.}
\label{tab:O6planes_z4}
\end{table}
\noindent
The action of $\Omega R$ on the orbifold basis is given by:
\begin{description}
\item[{\bf AAA}:]
For the toroidal and exceptional 3-cycles we get
\begin{align}
    &\rho_1\to \phantom{-}\rho_1,& &\bar\rho_1\to -\bar\rho_1,& \nonumber \\
    &\rho_2\to -\rho_2,& &\bar\rho_2\to \phantom{-}\bar\rho_2,& \\
    &\varepsilon_i\to \phantom{-}\varepsilon_i& &\bar\varepsilon_i\to -\bar\varepsilon_i,& \nonumber
\end{align}
for all $i\in\{1,\ldots,6\}$.

\item[{\bf AAB}:]
For the toroidal and exceptional 3-cycles we get
\begin{align}
    &\rho_1\to \phantom{-}\rho_1,&  &\bar\rho_1\to \phantom{-}\rho_1 -\bar\rho_1,& \nonumber\\
    &\rho_2\to -\rho_2,&  &\bar\rho_2\to -\rho_2+\bar\rho_2,& \\
    &\varepsilon_i\to \phantom{-}\varepsilon_i&  &\bar\varepsilon_i\to \phantom{-}\varepsilon_i-\bar\varepsilon_i,& \nonumber
\end{align}
for all $i\in\{1,\ldots,6\}$.

\item[{\bf ABA}:]
For the toroidal and exceptional 3-cycles we get
\begin{align}
    &\rho_1\to \phantom{-}\rho_2,&  &\bar\rho_1\to -\bar\rho_2,& \nonumber \\
    &\rho_2\to \phantom{-}\rho_1,&  &\bar\rho_2\to -\bar\rho_1,& \nonumber\\
    &\varepsilon_1\to -\varepsilon_1,& &\bar\varepsilon_1\to \phantom{-}\bar\varepsilon_1,& \nonumber\\
    &\varepsilon_2\to -\varepsilon_2,& &\bar\varepsilon_2\to \phantom{-}\bar\varepsilon_2,& \\
    &\varepsilon_3\to \phantom{-}\varepsilon_3,& &\bar\varepsilon_3\to -\bar\varepsilon_3,& \nonumber\\
    &\varepsilon_4\to \phantom{-}\varepsilon_4,& &\bar\varepsilon_4\to -\bar\varepsilon_4,& \nonumber\\
    &\varepsilon_5\to \phantom{-}\varepsilon_6,& &\bar\varepsilon_5\to -\bar\varepsilon_6,& \nonumber\\
    &\varepsilon_6\to \phantom{-}\varepsilon_5,& &\bar\varepsilon_6\to -\bar\varepsilon_5.\nonumber
\end{align}
\end{description}

\section{Supersymmetry conditions}\label{cha:susy_cycles_z4}

The supersymmetry conditions for the three orientifold models not
being discussed in the main text are listed in this appendix.
\begin{description}
\item[{\bf AAA}:]
The condition that such a D6-brane preserves the same supersymmetry
as the orientifold plane is given by
\begin{equation}\label{eq:angle_z4_aaa}
    \varphi_1^a+ \varphi_2^a+ \varphi_3^a=0
                 \ {\rm mod}\ 2\pi\ ,
\end{equation}
with
\begin{equation}\label{eq:tanz4_aaa}
    \tan\varphi_1^a={m_1^a\over n_1^a}\ , \qquad
              \tan\varphi_2^a={m_2^a\over n_2^a}\ , \qquad
               \tan\varphi_3^a={U_2\, m_3^a\over n_3^a}\ .
\end{equation}
This implies the following necessary condition in terms of the wrapping
numbers
\begin{equation}\label{eq:u2_z4aaa}
    U_2=-{ n_3^a  \over m_3^a}
                  { \left(n_1^a\,m_2^a +
               m_1^a\,n_2^a \right)
           \over \left(n_1^a\,n_2^a - m_1^a\,m_2^a  \right)}\ .
\end{equation}

\item[{\bf AAB}:]
The condition that such a D6-brane preserves the same supersymmetry
as the orientifold plane is given by
\begin{equation}\label{eq:angle_z4_aab}
    \varphi_1^a+ \varphi_2^a+ \varphi_3^a=0
                 \ {\rm mod}\ 2\pi\ ,
\end{equation}
with
\begin{equation}\label{eq:tanz4_aab}
    \tan\varphi_1^a={m_1^a\over n_1^a}\ , \qquad
              \tan\varphi_2^a={m_2^a\over n_2^a}\ , \qquad
               \tan\varphi_3^a={U_2\, m_3^a\over n_3^a+{1\over 2} m_3^a }\ .
\end{equation}
This implies the following necessary condition in terms of the wrapping
numbers
\begin{equation}\label{eq:u2_z4aab}
   U_2=-{ \left(n_3^a+{1\over 2} m_3^a\right)  \over m_3^a}
                  { \left(n_1^a\,m_2^a +
               m_1^a\,n_2^a \right)
           \over \left(n_1^a\,n_2^a - m_1^a\,m_2^a  \right)}\ .
\end{equation}

\item[{\bf ABA}:]
The condition that such a D6-brane preserves the same supersymmetry
as the orientifold plane is given by
\begin{equation}\label{eq:angle_z4_aab}
    \varphi_1^a+ \varphi_2^a+ \varphi_3^a={\pi\over 4}
                 \ {\rm mod}\ 2\pi\ ,
\end{equation}
with
\begin{equation}\label{eq:tanz4_aab}
    \tan\varphi_1^a={m_1^a\over n_1^a}\ , \qquad
              \tan\varphi_2^a={m_2^a\over n_2^a}\ , \qquad
               \tan\varphi_3^a={U_2\, m_3^a\over n_3^a }\ .
\end{equation}
This implies the following necessary condition in terms of the wrapping
numbers
\begin{equation}\label{eq:u2_z4aab}
   U_2={ n_3^a \over m_3^a}
                  { \left(n_1^a\,n_2^a - m_1^a\,m_2^a - n_1^a\,m_2^a -
               m_1^a\,n_2^a \right)
           \over \left(n_1^a\, n_2^a - m_1^a\,m_2^a + n_1^a\,m_2^a +
            m_1^a\,n_2^a \right)}\ .
\end{equation}
\end{description}

\section{Boundary states}\label{cha:boundary_states_z4}
The unnormalized boundary states in light cone gauge for D6-branes at angles in the
untwisted sector are given by
\begin{multline}\label{eq:unnorm_boundstate_z4}
    |D;(n_I,m_I)\rangle_{U}=
                |D;(n_I,m_I),\text{NS-NS},\eta=1\rangle_{U}+
                |D;(n_I,m_I),\text{NS-NS},\eta=-1\rangle_{U}\\
                +|D;(n_I,m_I),\text{R-R},\eta=1\rangle_{U}+
                |D;(n_I,m_I),\text{R-R},\eta=-1\rangle_{U}
\end{multline}
with the coherent state
\begin{multline}\label{eq:z4coherent_state}
    |D;(n_I,m_I),\eta\rangle_{U}
   = \int dk_2 dk_3 \sum_{\vec r,\vec s} {\rm exp}
 \biggl(-\sum_{\mu=2}^3 \sum_{n>0} {1\over n} \alpha^\mu_{-n}
               \tilde \alpha^\mu_{-n}\\
       -\sum_{I=1}^3    \sum_{n>0} {1\over 2n} \left(e^{2i\varphi_I}
 \zeta^I_{-n} \tilde{\zeta}^I_{-n}  +
  e^{-2i\varphi_I} \bar\zeta^I_{-n} \tilde{\bar\zeta}^I_{-n} \right)
       + i \eta \bigl[ \text{fermions} \bigr] \biggr)
             |\vec r,\vec s,\vec k, \eta \rangle\, .
\end{multline}
Here $\alpha^\mu$ denotes the two real non-compact directions and
$\zeta^I$  the three complex compact directions. The angles
$\varphi_I$ of the D6-brane relative to the horizontal axis on
each of the three internal tori $T^2$ can be expressed by the
wrapping numbers $n_I$ and $m_I$ as listed in appendix
\ref{cha:susy_cycles_z4}. The boundary state
(\ref{eq:z4coherent_state}) involves a sum over the internal
Kaluza-Klein and winding ground states parameterized by $(\vec
r,\vec s)$ and the mass of  these Kaluza-Klein and winding modes
on each $T^2$ has been given already in terms of the complex
structure and K\"ahler moduli for the loop channel in
(\ref{eq:Hamiltonian_lat_annulus}). This reads for the loop
channel as
\begin{equation}\label{eq:lattice_hamiltonian_tree}
    M^2_{I}={ |r_I+s_I\,{U_I}|^2\over U_{I,2} }\,{ |n_I+m_I\, {T_{I}}|^2\over {T_{I,2}}}\ ,
\end{equation}
where the $r_I,s_I\in\mathbb{Z}$ take the same numbers as in
(\ref{eq:z4coherent_state}). If the brane carries some discrete
Wilson lines, $\vartheta=1/2$, appropriate factors of the form
$e^{i s R \vartheta}$ have to be introduced into the winding sum
in (\ref{eq:z4coherent_state}).

In the $\Theta^2$ twisted sector, the boundary state involves the
analogous sum over the fermionic spin structures as in
(\ref{eq:unnorm_boundstate_z4}) with
\begin{multline}\label{eq:z2_twisted_sector_boundary_state}
|D;(n_I,m_I),e_{ij}, \eta
                  \bigr\rangle_T=\int dk_2 dk_3  \sum_{r_3,s_3} {\rm exp}
 \biggl(-\sum_{\mu=2}^3 \sum_{n>0} {1\over n} \alpha^\mu_{-n}
               \tilde \alpha^\mu_{-n}  \\
    -\sum_{I=1}^2    \sum_{r\in\mathbb{Z}^+_0 +{1\over 2}} {1\over 2r} \left(e^{2i\varphi_I}
 \zeta^I_{-r} \tilde{\zeta}^I_{-r}  +
  e^{-2i\varphi_I} \bar\zeta^I_{-r} \tilde{\bar\zeta}^I_{-r} \right)
    -\sum_{n>0 } {1\over 2n} \left (e^{2i\varphi_3}
 \zeta^3_{-n} \tilde{\zeta}^3_{-n}  +
  e^{-2i\varphi_3} \bar\zeta^3_{-n} \tilde{\bar\zeta}^3_{-n} \right)\\
       + i \eta \bigl[ \text{fermions} \bigr] \biggr)
             |r_3,s_3,\vec k,e_{ij},\eta \rangle\ ,
\end{multline}
where the $e_{ij}$ denote the 16 $\mathbb{Z}_2$-fixed points. In this
equation, we have already taken into account that the twisted
boundary state can only have Kaluza-Klein and winding modes on the
third 2-torus and that the bosonic modes on the two other 2-tori
carry half-integer modes.

\chapter{The kinetic terms of $\phi_4$ and $U^I$ in the effective 4D theory}\label{cha:kinetic_terms}
The 10-dimensional effective spacetime supergravity action in the
NS-sector can be written as \cite{Polchinski:1998rr}:
\begin{equation}\label{eq:supergravityNS}
    \emph{S}_{\mathrm{NS}}=\frac{1}{{\kappa_{10}}^2}\int d^{10}x \sqrt{-G} e^{-2\phi_{10}}\left(-\frac{\mathcal{R}_{10}}{2}
    +2\, \partial_\mu \phi_{10}\, \partial^\mu \phi_{10}-\frac{1}{4}|H_3|^2 \right)\ .
\end{equation}
This definition uses the curvature conventions of Weinberg
\cite{Weinberg:1972}. In this calculation, the string scale $M_\text{s}$ will
not be taken along explicitly. Our compact manifold is given by
$T^6=T^2 \times T^2 \times T^2$, and here, every 2-torus is taken
to be an A-torus, then the 10-dimensional metric takes the
following form:
\begin{equation}\label{eq:10dmetrik}
g_{\mu \nu}^{(10)}=\begin{bmatrix}
      g_{\mu \nu}^{(4)} & 0 & 0 & 0 & 0 & 0 & 0 \\
      0 & {T^1}/{U^1} & 0 & 0 & 0 & 0 & 0 \\
      0 & 0 & U^1 T^1 & 0 & 0 & 0 & 0 \\
      0 & 0 & 0 & {T^2}/{U^2} & 0 & 0 & 0 \\
      0 & 0 & 0 & 0 & U^2 T^2  & 0 & 0 \\
      0 & 0 & 0 & 0 & 0 & {T^3}/{U^3}& 0 \\
      0 & 0 & 0 & 0 & 0 & 0 & U^3 T^3  \\
    \end{bmatrix}\ .
\end{equation}
In this ansatz, $g_{\mu \nu}^{(4)}$ denotes a general
4-dimensional metric in the non-compact space, which we do not
have to specify any further. Furthermore, to simplify notation,
$U^I$ and $T^I$ actually stand for the imaginary part of the
complex structure and K{\"a}hler moduli (\ref{eq:Torus_U_T_AB}),
as both real parts are fixed to be $0$. The term $-{1}/{4}|H_3|^2$
does not play any role for these considerations, so it will be
disregarded. Using the metric (\ref{eq:10dmetrik}), the Ricci
scalar $\mathcal{R}_{10}$ can be explicitly calculated,
\begin{equation}\label{eq:ricci10mricci4}
\mathcal{R}_{10}=\sum_{I=1}^{3}\frac{1}{2}\frac{\partial_\mu
U^I\, \partial^\mu U^I}{{U^I}^2}+f\left(T^I,\partial_\mu T^I,
\partial_\mu \partial^\mu T^I,g_{\mu \nu},\partial_\kappa g_{\mu
\nu}\right)+\mathcal{R}_{4}\ .
\end{equation}
In this expression, $f$ is a quite complicated function not
depending on the complex structure moduli. As a next step, the
10-dimensional dilaton $\phi_{10}$ in equation (\ref{eq:supergravityNS}) has
to be replaced by the 4-dimensional one,
\begin{equation}\label{eq:transf_10d_dilaton}
    \phi_{10}=\phi_4+\frac{1}{2}\ln(T^1 T^2 T^3)\ .
\end{equation}
Furthermore, a 4-dimensional Weyl-rescaling has to be applied in
order to transform to the Einstein frame, denoted by a tilde. In arbitrary dimension $d$, one has to
to make the transition
\begin{equation}\label{eq:weyl_Rescaling_gen}
\tilde{g}^{(d)}_{\mu \nu}
=e^{4\frac{\phi_{d}}{d-2}}\, g^{(d)}_{\mu \nu}\ ,
\end{equation}
implying for the curvature scalar
\begin{equation}\label{eq:curv_transform}
\tilde{\mathcal{R}}_{d}=e^{-4\frac{\phi_{d}}{d-2}}\,\left(
\mathcal{R}_{d}+4\frac{d-1}{d-2}\partial_\mu \partial^\mu
\phi_{d}+4\frac{d-1}{d-2}\partial_\mu \phi_{d}\,
\partial^\mu \phi_{d}\right)\ .
\end{equation}
Moreover, all covariant and contravariant derivatives for the
variables have to be rewritten in the Einstein metric, as for instance
\begin{equation}\label{eq:bsp_Einstein_trafo}
e^{-4\frac{\phi_{d}}{d-2}}\,\partial_\mu \phi_{d}\,\partial^\mu \phi_{d}
=e^{-4\frac{\phi_{d}}{d-2}}\,g^{\mu \nu, (d)} \partial_\mu \phi_{d}\,\partial_\nu \phi_{d}
=\tilde{g}^{\mu \nu, (d)}\partial_\mu \phi_{d}\,\partial_\nu \phi_{d}
=\tilde{\partial}_\mu \phi_{d}\,\tilde{\partial}^\mu \phi_{d}\nonumber\ .
\end{equation}
The 4-dimensional action in the Einstein frame is given by:
\begin{multline}\label{eq:supergravity4dim}
    \emph{S}=\frac{1}{{\kappa_{4}}^2}\int d^{4}x \sqrt{-\tilde{G}}\\
    \cdot\left[-\frac{\mathcal{R}_{4}}{2}
    -\sum_{I=1}^3\frac{1}{4}\tilde{\partial}_\mu (\ln U^I)\, \tilde{\partial}^\mu (\ln U^I)
    -\tilde{\partial}_\mu \phi_{4}\, \tilde{\partial}^\mu \phi_{4}
    +h\left(T^I,\tilde{\partial}_\mu T^I,
\tilde{\partial}_\mu \tilde{\partial}^\mu T^I,\tilde{g}_{\mu \nu},\tilde{\partial}_\kappa \tilde{g}_{\mu
\nu}\right)\right]\ ,
\end{multline}
containing the kinetic terms for the 4-dimensional dilaton and the
complex structure moduli $U^I$.

\chapter{Program Source}
Most of the programs used in this work have been programmed in C++
and C, using numerical routines from \cite{Press:1992}. All other
computations have been performed using the algebra system Maple.
Due to the lack of space, the programs are not printed here, but
they can be obtained from the author.

 \backmatter
\addcontentsline{toc}{chapter}{Bibliography}

\chapter*{Acknowledgements}
\markboth{Acknowledgements}{ }

I would like to thank Dieter L\"ust, Ralph Blumenhagen, Lars
G\"orlich and Boris K\"ors for the collaboration on parts of this
work. Furthermore, I would like to thank Oleg Andreev, Gabriel
Lopes Cardoso, Andrea Gregori, Stephan Stieberger, Andre Miemiec,
Georgios Kraniotis, Matthias Br\"andle and Volker Braun for
inspiring discussions and hints.
This work is supported by the Graduiertenkolleg
{\it The Standard Model of Particle Physics - structure, precision
tests and extensions}, maintained by the DFG.
\end{document}